\documentclass[aps,ropp,amstext,amsmath,twocolumn,superscriptaddress,showpacs]{revtex4-1}

\usepackage{graphicx}
\usepackage{color}

\begin{document}

\title{Using electron irradiation to probe iron - based superconductors}

\author{Kyuil~Cho}
\email[Corresponding author: ]{kcho@ameslab.gov}
\affiliation{Ames Laboratory, Ames, IA 50011}
\affiliation{Department of Physics $\&$ Astronomy, Iowa State University, Ames, IA 50011}

\author{M. Ko\'nczykowski}
\affiliation{Laboratoire des Solides Irradi\'es, Ecole Polytechnique, CNRS-UMR 7642, CEA, Universite Paris-Saclay, F-91128 Palaiseau, France}

\author{S. Teknowijoyo}
\affiliation{Ames Laboratory, Ames, IA 50011}
\affiliation{Department of Physics $\&$ Astronomy, Iowa State University, Ames, IA 50011}

\author{M. A. Tanatar}
\affiliation{Ames Laboratory, Ames, IA 50011}
\affiliation{Department of Physics $\&$ Astronomy, Iowa State University, Ames, IA 50011}

\author{R.~Prozorov}
\affiliation{Ames Laboratory, Ames, IA 50011}
\affiliation{Department of Physics $\&$ Astronomy, Iowa State University, Ames, IA 50011}

\date{\today}

\begin{abstract}
High energy electron irradiation is an efficient way to create vacancy-interstitial Frenkel pairs in crystal lattice, thereby inducing controlled non-magnetic point - like scattering centers. In combination with London penetration depth and resistivity measurements, the irradiation was particularly useful as a phase - sensitive probe of the superconducting order parameter in iron - based superconductors lending strongest support to sign - changing $s_{\pm}$ pairing. Here we review the key results on the effect of electron irradiation in iron-based superconductors. 
\end{abstract}
\maketitle

\section{Introduction}
Due to the unconventional pairing mechanism and relatively high superconducting transition temperatures, $T_c$, iron - based superconductors (FeSC)  remain in the focus of research activity even a decade after their discovery \cite{Kamihara2006JACS_LaOFeP, KamiharaHosono2008JACS}. Large body of experimental and theoretical works revealed a vast diversity of related compounds unified by the multi-band superconductivity and proximity to, or direct coexistence, with long - range magnetism. It is impossible to provide a comprehensive list of references here and we only give some key review articles on basic properties and models  \cite{Johnston2010AP_review, Canfield2010ARCMP, Paglione2010NP, Hirschfeld2011ROPP, Stewart2011RevModPhys, Carrington2011RPP_review, Abrahams2011JPCM_review, CARRINGTON2011CRPhys_review, Prozorov2011RPP_review, Wen2011ARCMP_review, Zhang2011FP_review, Chubukov2012ARCMP, Kordyuk2012LowTempPhys_review, SONG201339CuOpSSMS_review, Long2013CPB_review, Jiang2013CPB_review, Carretta2013PhysScr_review, Ye2013ChPB_review, Eremin2014JPSJ_review, Charnukha2014JPCM_review, Shibauchi2014ARCMP, Dai2015RMP_review, JasekKarpinski2015PhilMag_review, BOHMER2016CRP_review, PallecchiPutti2016SST_review, GALLAIS2016ComprenPhys_review, LiWu2016SST_review_FeSC, MARTINELLI2016CompRenPhys_review, Guterding2017PhysStatSol_review, Fernandes2017RPP_review, Yi2017NPJ_QuanMat_review} as well as applications \cite{Putti2010SST_FeSC_Application, Ma2012SST_FeSC_Application, TanabeHosono2012JJAP_FeSC_Application, MA2015PhysicaC_FeSC_Application, Hosono2017MatToday_application}. Some of key contributions come from the studies of the effect of controlled disorder induced by MeV - energy range electron irradiation. These relativistic electrons have enough energy to create vacancy - interstitial Frenkel pairs, but not too much energy to induce (undesirable) extended cascades of secondary defects produced by heavier particles, such as protons and $\alpha$-particles, or columnar tracks produced by heavy ions of GeV energy~\cite{Mizukami2014NatureComm, Cho2016ScienceAdvances_BaK122_e-irr}. While the superconducting energy gap and the critical temperature of an isotropic single - band s-wave superconductors are insensitive to nonmagnetic disorder (Anderson theorem)~\cite{Anderson1959JPCS, AbrikosovGorkov1961JETP}, multi-gap, anisotropic gaps and different gap symmetries are quite sensitive to such disorder, each with fairly unique signature in the behavior of thermodynamic and transport properties \cite{Balian1963PR, Openov1998PRB_impurity, SengaKontani2008JPSJ, OnariKontani2009PRL, EfremovHirschfeld2011PRB, FernandesChubukov2012PRB_TcEnhancement, WangHirschfeldMishra2013PRB, ChenMishraHirschfield2016PRB_impurity_theory, Korshunov2017Phy-Usp_impurity_theory_review}. Contrary to the high - temperature cuprates in which a single-gap d-wave superconducting state is firmly established \cite{TsueiKirtley2000RMP_review_cuprates}, several candidates of pairing symmetry are discussed for FeSCs due to multiple sheets of the Fermi surface supporting nesting and itinerant magnetism\cite{Hirschfeld2011ROPP,Chubukov2012ARCMP}. Among them, there are two dominant scenarios for superconducting ``glue", - spin fluctuations (repulsive interaction) and orbital fluctuations (attractive interaction). The former predicts the state that requires the sign-change between different sheets of the Fermi surface ($s_\pm$ pairing) \cite{Hirschfeld2011ROPP}, the other predicts no sign change ($s_{++}$ pairing) \cite{SengaKontani2008JPSJ, OnariKontani2009PRL}. Unlike high - $T_c$ cuprates where direct order parameter phase - sensitive experiments have proven $d-$wave state, similar methods cannot be applied to FeSCs due to complex multi-band electronic band structure. Some more complicated and difficult phase sensitive techniques (e.g. quasiparticle interference in STM measurements) were developed, but they are limited by surface quality and other issues related to tunneling. The alternative, based on the effects of a controlled non-magnetic disorder to distinguish $s_\pm$ and $s_{++}$ pairing states, were suggested \cite{EfremovHirschfeld2011PRB, WangHirschfeldMishra2013PRB} and implemented \cite{Mizukami2014NatureComm,Cho2016ScienceAdvances_BaK122_e-irr}. Traditionally, effect of any irradiation on superconductors was assessed by measuring the change of $T_c$, critical current and, sometimes, upper critical field. This is insufficient since the low-temperature quasiparticles to examine pairing mechanism need to be studied upon irradiation. Specific heat, thermal conductivity and London penetration depth are the direct probes that should be used in addition to other measurements. Indeed, the electron irradiation combined with London penetration depth measurement was used as an effective phase-sensitive tool to reveal superconducting gap structure of several iron - based superconductors. For examples, the $T$ - linear dependence of London penetration depth of isovalently substituted BaFe$_2$(As$_{1-x}$P$_x$)$_2$ was changed to exponential - like dependence using the 2.5 MeV electron irradiation, suggesting that the nodes are of accidental type and lifted upon irradiation \cite{Mizukami2014NatureComm}. Another example is Ba$_{1-x}$K$_x$Fe$_2$As$_2$ that shows the evolution of the gap structure across the superconducting ``dome", particularly near $x$ = 0.8 \cite{Cho2016ScienceAdvances_BaK122_e-irr}. 
In this review article, we summarize the key findings in studies of the effect of electron irradiation in iron - based superconductors. We limit our attention to superconductors derived upon substitution from BaFe$_2$As$_2$ and SrFe$_2$As$_2$ 
(referred in the following as 122 compounds).

\section{Experimental}
\subsection{Using irradiation to induce structural disorder}

\begin{figure}[htb]
\includegraphics[width=8.5cm]{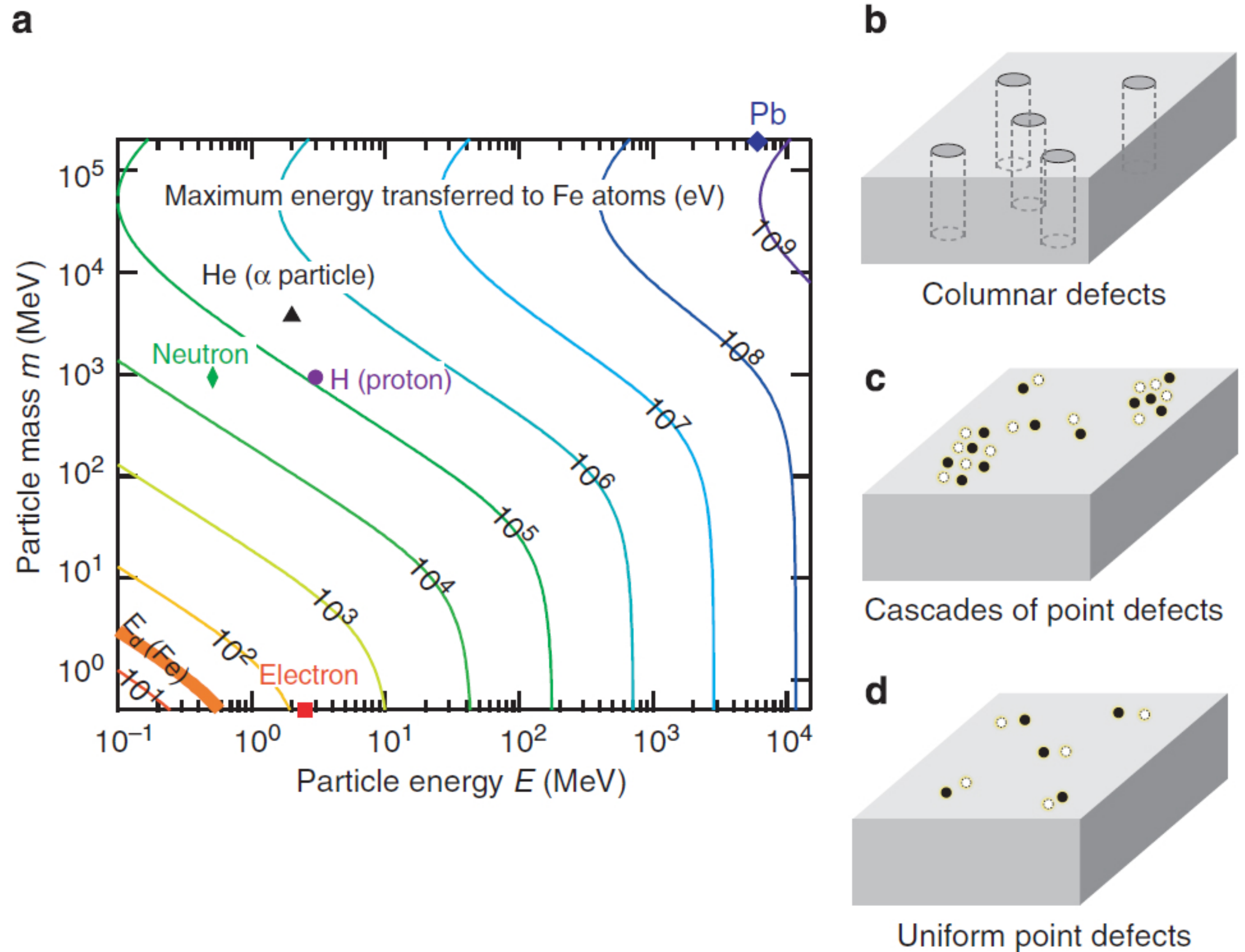}
\caption{(Color online) (a) Particle energy diagram of various energetic particles used for irradiation. (b, c, d) Different types of defects produced by diverse irradiation techniques. Reprinted with permission from Nature Communications, Ref.~\onlinecite{Mizukami2014NatureComm}, copyright Macmillan Publishers Ltd.}
\label{fig01_particle_energy}
\end{figure}

Apart from plastic deformation, perhaps the most common way to introduce controlled disorder in solid samples is chemical substitution. Indeed, many such studies were conducted in high - $T_c$ cuprates \cite{Alloul2009RMP_defect_cuprates} and FeSCs \cite{LiWu2016SST_review_FeSC}. However, the substitution changes not only the scattering but also electronic band structure, chemical potential, lattice parameters, and so on. \cite{Alloul2009RMP_defect_cuprates, LiWu2016SST_review_FeSC}. The alternative, technically more involved, way is the irradiation with energetic particles such as heavy-ions, protons, $\alpha$ - particles, neutrons, and electrons. The structure of the resultant disorder depends on the type of irradiation based on mass, charge and energy of the particles used \cite{Damask1963PointDefectsInMetals}. Different types of irradiation produce more desirable results depending on the goals. For example, some types of heavy-ion irradiation produce columnar defects \cite{Nakajima2009PRB_HeavyIon, CivaleHoltzberg1991PRL_YBCO_columnar_defects, KonczykowskiPRB1991_HeavyIon_YBCO}, which play very prominent role in vortex physics of layered materials, but are very difficult to analyze in terms of scattering centers. Yet, early experiments with heavy-ion irradiation in FeSCs have shown a strong violation of the Anderson theorem with saturating behavior of low-temperature London penetration depth, 
providing firm experimental support for multi-band $s_{\pm}$ pairing \cite{KimProzorov2010PRB_heavyIon, Prozorov2010PRB_HeavyIon,MurphyProzorov2013PRB_heavy_ion}. Proton \cite{Nakajima2010PRB_BaCo122_proton, Taen2013PRB_proton_irr, Smylie2016PRB_BaP122_proton_irr, Moroni2017PRB_proton-irr}, $\alpha$-particle \cite{Tarantini2010PRL_alpha_irr}, and neutron irradiation were also used in iron - based superconductors. While the results qualitatively indicate multi-band pairing, it is hard to achieve quantitative agreement due to difficulty of analyzing cascades or clusters of defects produced by these types of irradiation. A more detailed systematic investigation of the connection between the size of the defects and $T_c$ suppression rate was done theoretically in Ref.~\onlinecite{Yamakawa2013PRB_impurity_theory}.

Thanks to their small mass and large charge, electrons can be accelerated to relativistic speeds in a highly controlled way using relatively compact Van der Graaf type ``pelletron" accelerators. The effects of such irradiation on different systems, particularly metals and their compounds, was studied in great detail over more than half a century (See Refs.~\onlinecite{Mott1929ProRoySocLondon_electon_scattering, Damask1963PointDefectsInMetals}). Some MeV range electrons 
produce point-like defects with minimal impact on the material itself. The large penetration depth of electrons allows homogeneous damage of fairly thick samples (tens of $\mu$m). Following Mott's work in 1929~\cite{Mott1929ProRoySocLondon_electon_scattering}, Damask \textit{et al.} conducted analysis of the energy transfer from an accelerated particle smashing into the crystal lattice and found that only electrons with energies of 1$\sim$10 MeV produce point-like defects in form of interstitial ions and vacancies (Frenkel pairs) that form perfect scattering centers \cite{Damask1963PointDefectsInMetals}. The energy transferred to an ion due to head-on collision by the particle of rest mass $m$ and kinetic energy, $E$, is shown in Fig.~\ref{fig01_particle_energy}. The ion displacement energy needed to create a Frenkel pair is typically in the range of 10-50 eV, so it is clear from Fig.~\ref{fig01_particle_energy} that only electrons would produce such individual defects. Higher energy/mass particles lead to secondary impacts resulting in cascades. The interstitials are more mobile and migrate to various ``sinks", such as dislocations, grain boundaries and surfaces leaving metastable, but robust population of vacancies behind. The studies reviewed in this article were conducted using the 2.5 MeV electron irradiation which is known to generate point-like disorders in metals and compounds. 

\subsection{Low -temperature electron irradiation}
Electron irradiation reviewed in this article was conducted at SIRIUS facility operated by Laboratoire des Solides Irradi\'es at \'Ecole Polytechnique, Palaiseau, France. Its main elements are a pelletron type accelerator made by National Electrostatics Corporation (Wisconsin, USA) and a closed cycle cryo-cooler to maintain liquid hydrogen for cooling the sample. This cooling is required to efficiently channel the heat produced upon collisions between electrons and ions, and to prevent vacancy-interstitial on-site recombination. With a calculated head-on collision displacement energy for Fe ions of 22 eV and a cross section to create Frenkel pairs in BaFe$_2$As$_2$ at 2.5 MeV of 115 barn, a dose of 1 C/cm$^2$ result in about 0.07$\%$ of the defects per iron site. Similar numbers were obtained for other ion sites with cross - sections for Ba and As being 105 and 35 barn, respectively. The electron irradiation was conducted in liquid hydrogen at 22 K, and recombination of the vacancy-interstitial pairs upon warming up to room temperature varies depending on compounds, but in general 20 - 30 $\%$, as measured directly from the decrease of residual resistivity \cite{Prozorov2014PRX_e-irr}. After initial annealing, the defects remained stable for most of crystals, but some compounds show gradual slow annealing over the time (months). 

\subsection{Controlled disorder as a phase sensitive probe}
In most previous cases, only the suppression of $T_c$ with increased disorder was studied. When measurements of $T_c$ as a function of disorder are combined with measurement of London penetration depth, a phase-sensitive nature of impurity scattering enables distinguishing different scenarios for the superconducting pairing. This combination of measurements was used to identify accidental character of nodes in iso-electron substituted BaFe$_2$(As$_{1-x}$P$_x$)$_2$~\cite{Mizukami2014NatureComm} and SrFe$_2$(As$_{1-x}$P$_x$)$_2$~\cite{StrehlowProzorov2014PRB_SrP122}. The concomitant suppression of $T_c$ and closing of gap nodes in penetration depth study landed a strong support to s$_{\pm}$ pairing. More recently, the same idea was used to verify evolution of superconducting gap structure with composition in hole-doped Ba$_{1-x}$K$_x$Fe$_2$As$_2$~\cite{Cho2016ScienceAdvances_BaK122_e-irr}. 

For the unconventional superconductors, one of the important aspects
of the impurity effects is to mix gaps on different parts of the
Fermi surface and thereby smear out the momentum
dependence~\cite{Mishra2009PRB}. In the case of superconducting gap with
symmetry protected nodes such as d-wave, this averaging
mechanism leads to the suppression of the gap amplitude and creation of nodal quasi-particles. In penetration depth measurements this results in cross-over from $T$-linear temperature dependent penetration depth $\Delta \lambda (T)$ in clean limit to a $T^2$ dependence in dirty case. In addition to this, the sign change in the order
parameter gives rise to impurity-induced Andreev bound states, which lead to additional quasiparticle excitations~\cite{Hirschfeld1993PRB}. Such pair-breaking effects of nonmagnetic impurities have been
observed, for example, in Zn-doped YBa$
_2$Cu$_3$O$_7$ in the bulk
measurements of magnetic penetration depth, where the $T$-linear
temperature dependence in the clean-limit d-wave
superconductivity gradually changes to a $T^2$ dependence at low
temperatures with increasing Zn concentrations~\cite{BonnHardy1994PRB_YBCO_impurity}.

It is convenient to characterize the experimental data of low-temperature penetration depth  using a power law function $\Delta \lambda (T)=A+BT^n$. In the above example of superconductors with symmetry imposed line nodes, the exponent $n$ varies with increased disorder in the range between $n=1$ (clean limit) and $n=$2 (dirty limit).  For fully gapped s-wave superconductors, $\Delta \lambda (T)$ shows exponential $T$-dependence which can be described as a high power-law behavior $n > 3$, but $n = 2$ when it is in dirty limit.

In sharp contrast, when the nodal positions are not symmetry
protected, as in the nodal s-wave case, the averaging mechanism
of impurity scattering can displace the nodes, and at a certain
critical impurity concentration the nodes may be lifted if
intraband scattering dominates~\cite{Mishra2009PRB}, eliminating the low-energy
quasiparticle excitations. In power law analysis of the penetration depth data this crossover would lead to exponent $n$ acquiring values $n>$2. In the fully gapped state after lifting of node, we have two cases in the multiband superconductors. If
the signs of the order parameter on different bands are opposite,
residual interband scattering can give rise to midgap Andreev
bound states localized at nonmagnetic impurities that can
contribute to the low-energy excitations, provided that the
concentration of impurities is sufficient to create such states. If
there is no sign change, the gap and $T_c$ will be independent of disorder
at some high rate impurity/defect scattering: since no
Andreev states will be created, no significant change of the low-energy excitations is expected. Indeed, such a
difference between nodal sign-changing s$_{\pm}$ and sign-preserving
s$_{++}$ cases has been theoretically suggested by the recent
calculations for multiband superconductivity, considering the
band structure of FeSCs~\cite{WangHirschfeldMishra2013PRB}. Therefore, studying effects of impurity/defects on the gap nodes and low-energy excitations can be used as a powerful probe for the pairing symmetry of superconductors.

\subsection{London penetration depth}

\begin{figure}[htb]
\includegraphics[width=6.0cm]{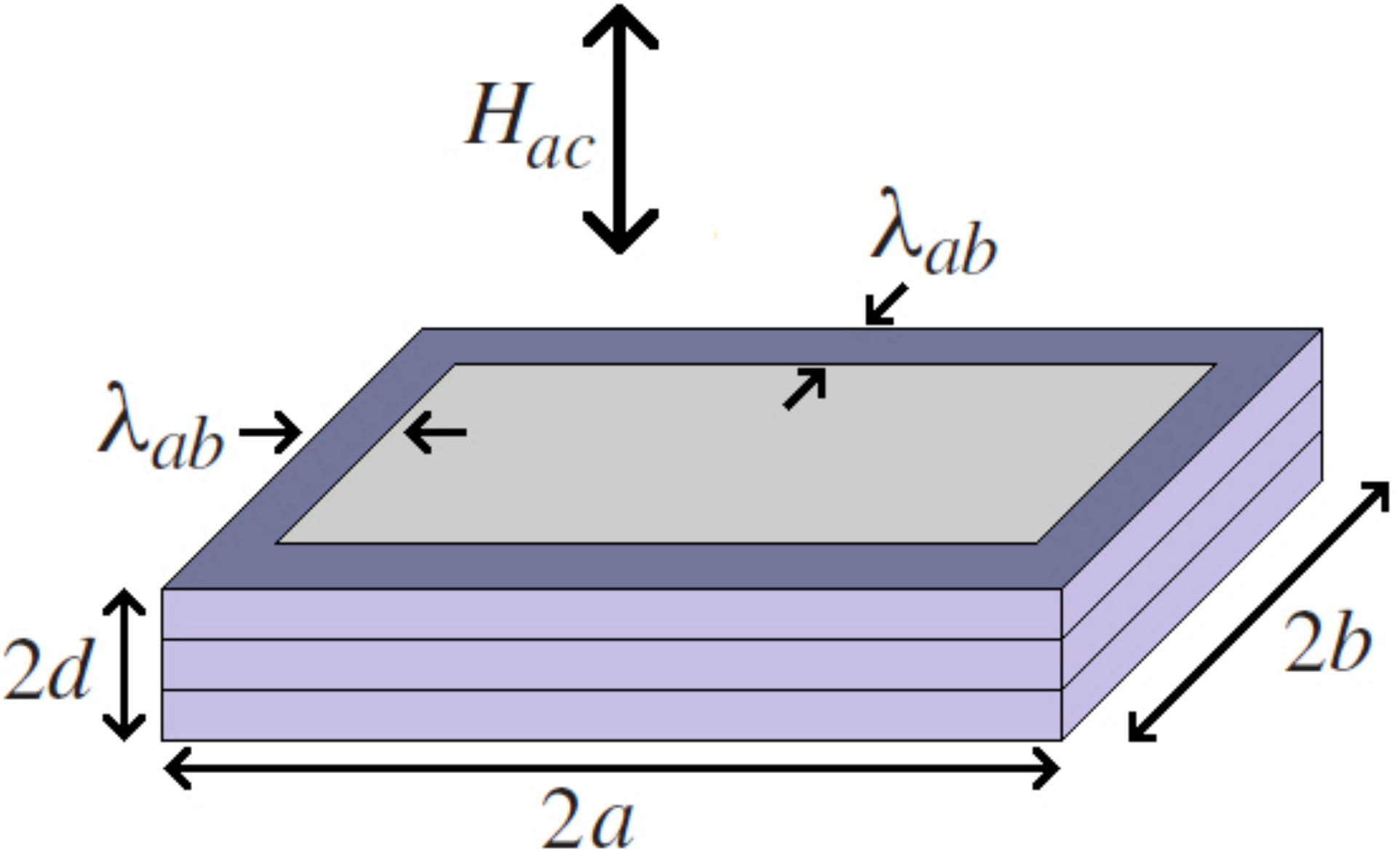}
\caption{(Color online) Typical sample dimension for measurement of in-plane penetration depth $\lambda_{ab}$. $2a \times 2b \times 2d \approx 500 {\mu}m \times 500 {\mu}m \times 50 {\mu}m$ The ac magnetic field of $H_{ac}$($\sim$ 20 mOe) is produced by TDR coil.}
\label{fig02_sample}
\end{figure}

The London penetration depth of FeSCs has been measured using a variety of techniques such as muon-spin rotation ($\mu$SR)~\cite{Luetkens2008PRL_uSR, Williams2010PRB_uSR, Sonier2011PRL_uSR}, frequency-dependent conductivity~\cite{ValdesAguilar2010PRB_terahertz, Wu2010PRB_infrared}, microwave cavity perturbation~\cite{Hashimoto2009PRL_Microwave, Hashimoto2009PRL_Microwave_PrFeAsO, Bobowski2010PRB_Microwave}, mutual inductance~\cite{YongLemberger2011PRB_Mutual_Induc}, magnetic force and superconducting quantum interference device (SQUID) microscopy~\cite{LuanMoler2010PRB_MFM, LuanMoler2011PRL_MFM_SQUID}, measurements of the first critical field using either global~\cite{Prozorov2009PhysicaC, SongKwon2011EPL_Hc1} or local probes~\cite{Okazaki2009PRB_Hc1_hall_sensor, Klein2010PRB_hall_probe}, Nitrogen-vacancy center in diamond magnetometry~\cite{JoshiProzorov2017_NV_Hc1}, and the self-oscillating tunnel-diode resonator (TDR)~\cite{Prozorov2009PhysicaC, Shibauchi2009PhysicaC, MaloneKarpinski2009PRB_dL, Gordon2009PRB}. There are pros and cons for each method. The most important advantage of the tunnel diode resonator (TDR) technique is that it provides the highest resolution of London penetration depth: sub-$\mathring{A}$ for sub-mm size sample. Since the technical details are available from the previous review articles~\cite{ProzorovKogan2011RPP, Prozorov2006SST,Prozorov2000PRB}, here we briefly describe some of key aspects of this technique. 

The tunnel diode resonator (TDR) is a self-oscillating tank circuit that resonates at its fundamental frequency ($f_0 = \frac{1}{2 \pi \sqrt{L_0 C}}$). In Ames Laboratory and other research labs, the researchers were able to make the TDR circuit ($f_0$ $\cong$ 14 MHz) with high stability of 1 part per $10^9$~\cite{Prozorov2006SST}. When a non-magnetic conducting sample is inserted into a TDR coil, it induces the change in frequency ($\Delta f$).  In case of a a finite size sample with magnetic susceptibility ($\chi$), the change of frequency can be described as 
\begin{equation}
\Delta f = - \frac{f_0}{2}\frac{V_s}{V_c} 4 \pi \chi \label{eq01}
\end{equation}
where $V_s$ and $V_c$ are the volumes of sample and TDR coil. For a a finite size sample of rectangular slab, the magnetic susceptibility ($\chi$) can be written as 
\begin{eqnarray}
- 4 \pi \chi &=& \frac{1}{1-N} \left[1 - \frac{\lambda_{ab}}{R} tanh \left( \frac{R}{\lambda_{ab}} \right) \right] \\
&\cong& \frac{1}{1-N} \left[1 - \frac{\lambda_{ab}}{R} \right] \text{, if } R \gg \lambda_{ab}.\label{eq02}
\end{eqnarray}
Here $R$ is the effective dimension and $N$ is a demagnetization factor. For a rectangular slab with dimensions of $2a \times 2b \times 2d$ (Fig.~\ref{fig02_sample}), $R$ can be approximated~\cite{Prozorov2000PRB} as
\begin{equation}
R \cong \frac{\omega}{2 \left[  1+ \left(1+(\frac{2d}{\omega} \right)^2) arctan \left( \frac{\omega}{2d} \right) - \frac{2d}{\omega} \right] } \label{eq03}
\end{equation}
with $\omega \approx \frac{2ab}{a+b}$. Combining equations~\ref{eq01} and \ref{eq02}, the relation between $\Delta f$ and $\lambda_{ab}$ is obtained as
\begin{eqnarray}
\Delta f &=& \frac{f_0}{2}\frac{V_s}{V_c} \frac{1}{1-N} \left[1 - \frac{\lambda_{ab}}{R} \right]\\
&=& G \left[1 - \frac{\lambda_{ab}}{R} \right] \label{eq04}
\end{eqnarray}
where G (= $\frac{f_0}{2}\frac{V_s}{V_c}\frac{1}{1-N}$) is a geometric calibration constant that can be directly measured by pulling the sample out of the coil. Thus, the variation of penetration depth ($\delta \lambda_{ab}$) from $T_{min}$ to $T$ is
\begin{eqnarray}
\delta \lambda_{ab} &=& \lambda_{ab}(T) - \lambda_{ab} (T_{min})\\
&=& \frac{R}{G} (\Delta f (T_{min}) - \Delta f(T)). \label{eq05}
\end{eqnarray}
Based on equation~\ref{eq05}, one can measure the change in London penetration depth ($\delta \lambda_{ab}$) from the change in the frequency. When a sub-mm scale sample is used,  the a part per billion resolution of a TDR frequency can be converted to sub-$\mathring{A}$ resolution in $\lambda_{ab}$. 

\section{Effect of electron irradiation on the 122 compounds}

\begin{figure}[htb]
\includegraphics[width=8.5cm]{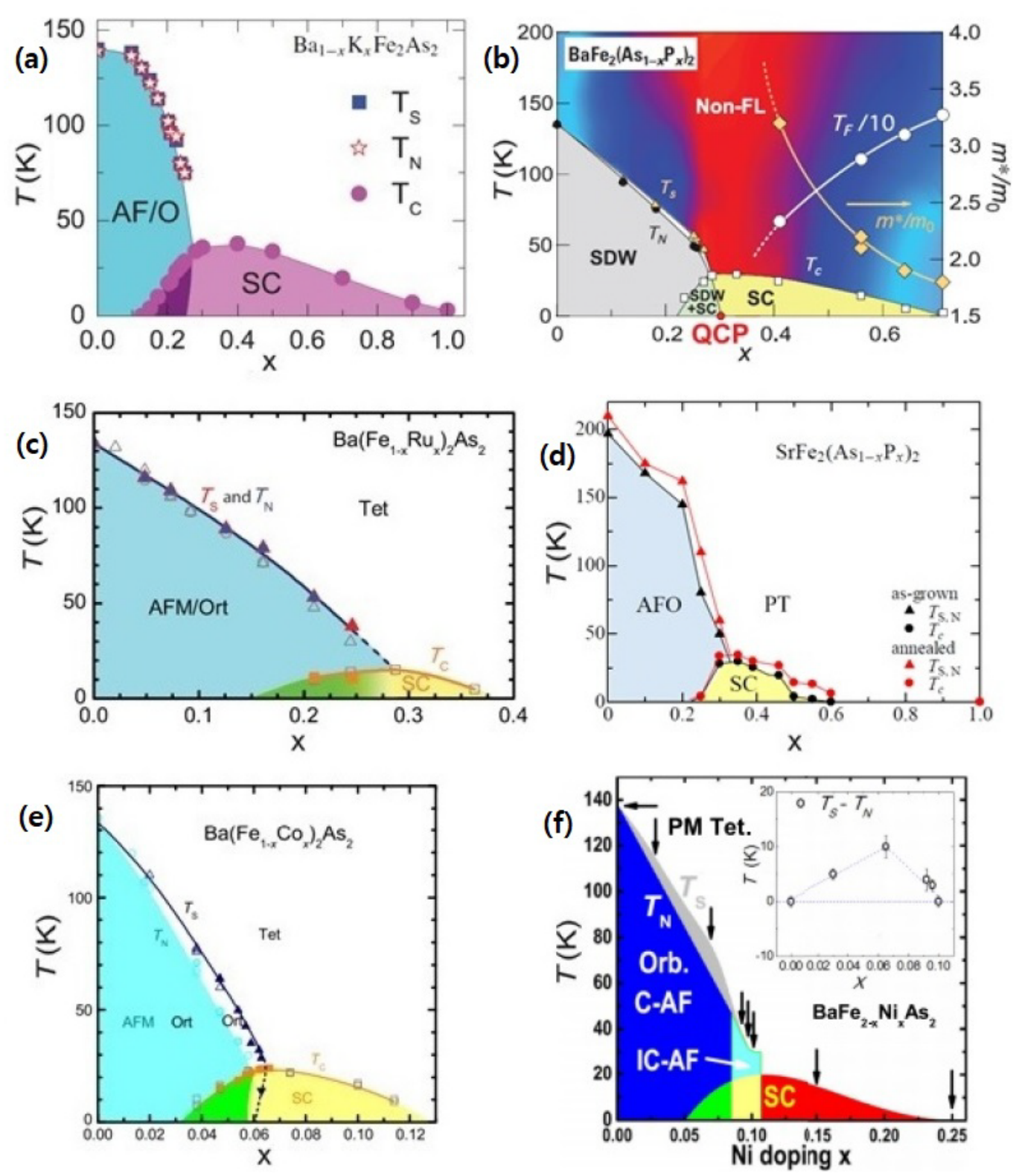}
\caption{(Color online) Phase diagrams of 122 family of FeSCs: (a) Hole-doped Ba$_{1-x}$K$_x$Fe$_2$As$_2$~\cite{Avci2012PRB}; (b) isovalently-doped BaFe$_2$(As$_{1-x}$P$_x$)$_2$)~\cite{HashimotoCho2012Science_BaP122_dL}, (c) Ba(Fe$_{1-x}$Ru$_x$)$_2$As$_2$~\cite{KimGoldman2011PRB_BaRu122} and (d) SrFe$_2$(As$_{1-x}$P$_x$)$_2$~\cite{Kobayashi2014JPSJ_SrP122}; (e) electron-doped Ba(Fe$_{1-x}$Co$_x$)$_2$As$_2$~\cite{NandiGoldman2010PRL} and (f) Ba(Fe$_{2-x}$Ni$_x$)As$_2$~\cite{Abdel-Hafiez2015PRB_BaNi122}. Panels~(a), (c), (e), and (f): reprinted with permission from Ref.~\onlinecite{Avci2012PRB, NandiGoldman2010PRL, Abdel-Hafiez2015PRB_BaNi122, KimGoldman2011PRB_BaRu122}, copyright 2010, 2011, 2012 APS. Panel~(b): reprinted with permission from Science, Ref.~\onlinecite{HashimotoCho2012Science_BaP122_dL}, copyright AAAS. Panel~(d): reprinted with permission from Ref.~\onlinecite{Kobayashi2014JPSJ_SrP122}, copyright JPSJ.}
\label{fig2-0-1_122_Family}
\end{figure}

\begin{figure}[htb]
\includegraphics[width=8.5cm]{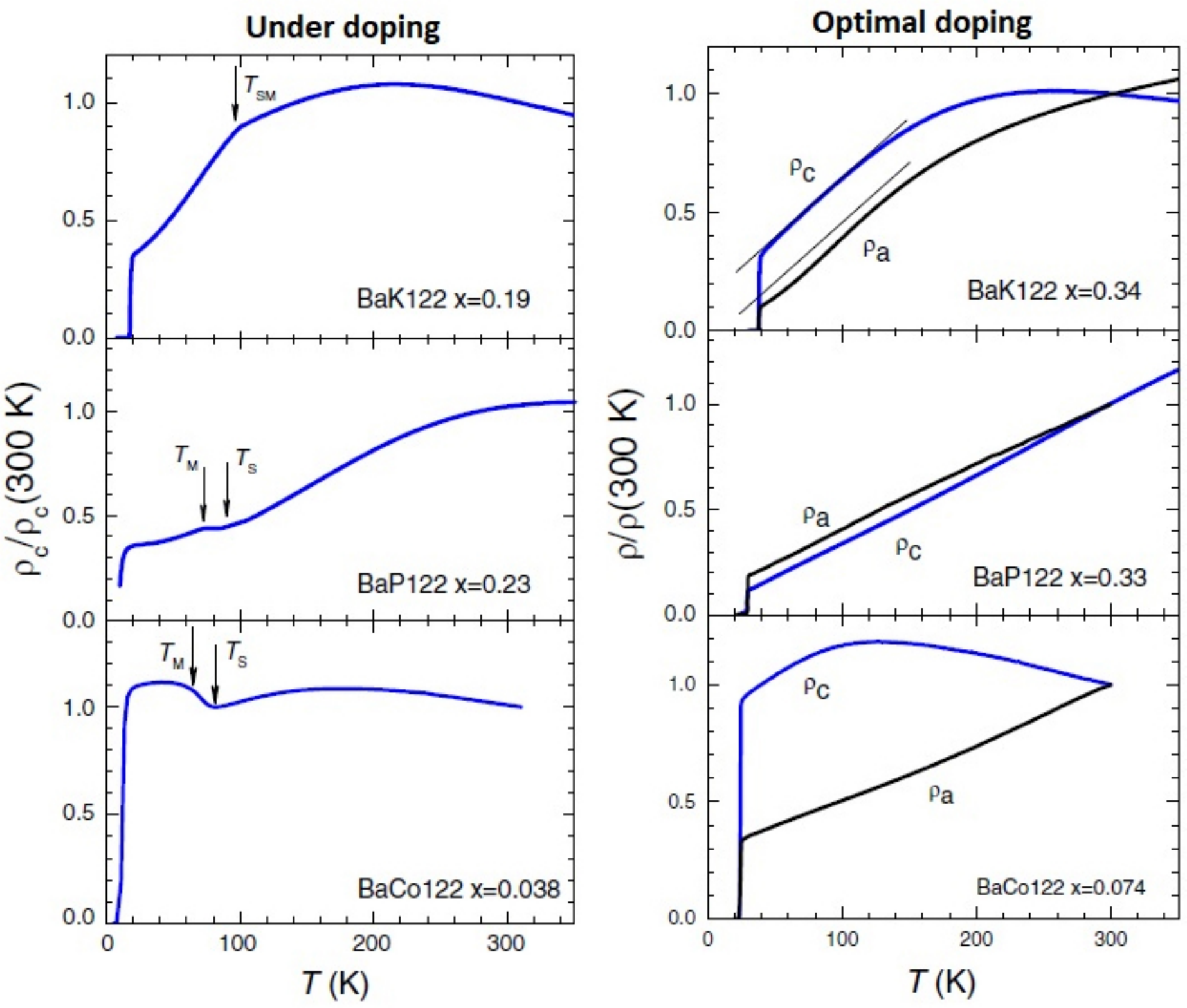}
\caption{(Color online) Temperature-dependent resistivity of under-doped and near optimally doped representative 122 FeSCs: (Ba$_{1-x}$K$_x$)Fe$_2$As$_2$, BaFe$_2$(As$_{1-x}$P$_x$)$_2$, and Ba(Fe$_{1-x}$Co$_x$)$_2$As$_2$. Reprinted with permission from Ref.~\onlinecite{TanatarProzorov2014PRB_BaK122}, copyright 2014 APS.}
\label{fig2-0-2_122_Family}
\end{figure}

\begin{figure}[htb]
\includegraphics[width=8cm]{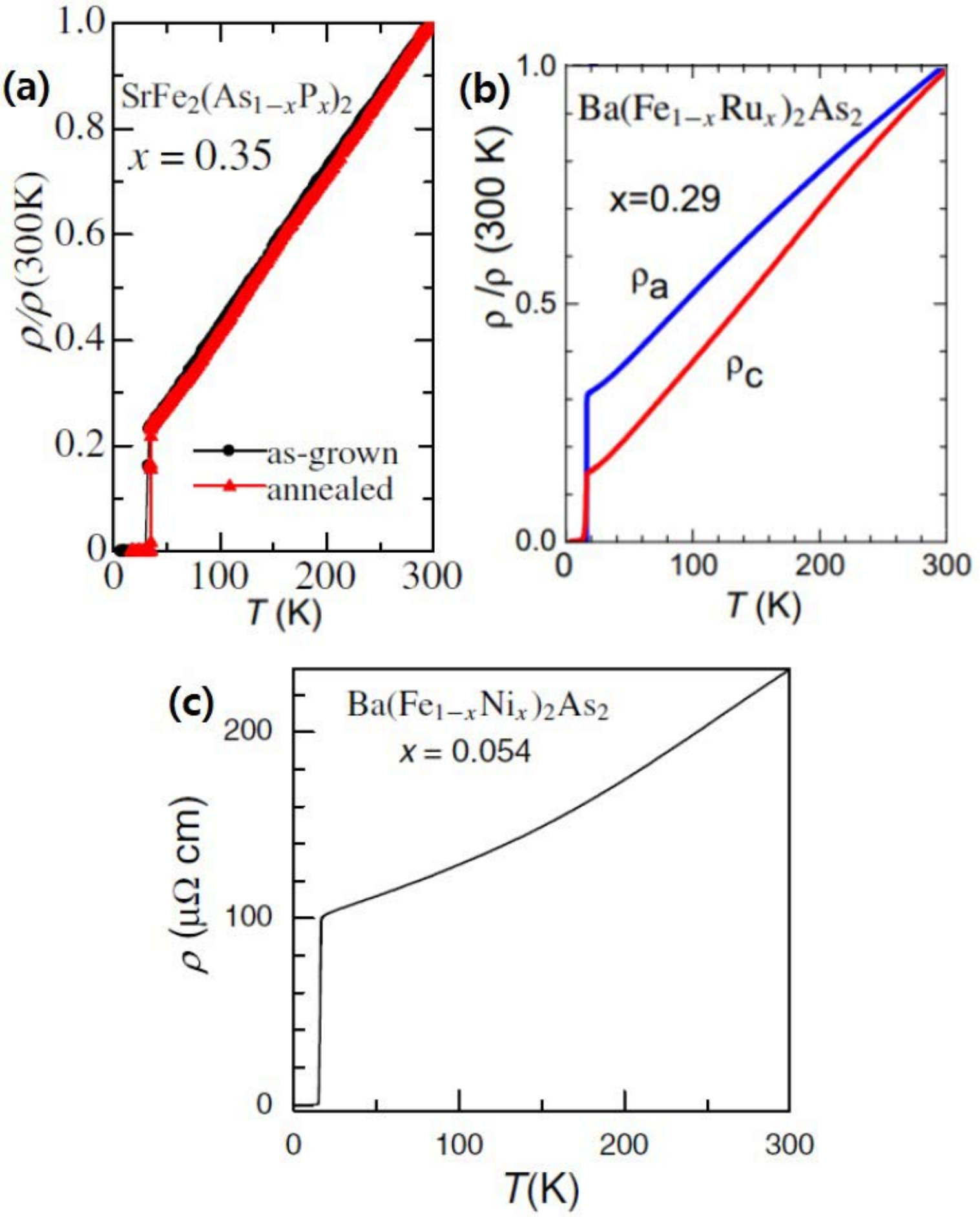}
\caption{(Color online) Temperature-dependent resistivity of near-optimally doped 122 FeSCs: SrFe$_2$(As$_{1-x}$P$_x$)$_2$~\cite{Kobayashi2014JPSJ_SrP122}, Ba(Fe$_{1-x}$Ru$_x$)$_2$As$_2$~\cite{TanatarProzorov2014PRB_BaRu122}, Ba(Fe$_{1-x}$Ni$_x$)$_2$As$_2$~\cite{MurphyProzorov2013PRB_over-doped_BaNi122}. Panel (a): reprinted with permission from Ref.~\onlinecite{Kobayashi2014JPSJ_SrP122}, copyright JPSJ. Panel (b) and (c): reprinted with permission from Ref.~\onlinecite{TanatarProzorov2014PRB_BaRu122, MurphyProzorov2013PRB_over-doped_BaNi122}, copyright 2013, 2014 APS.}
\label{fig2-0-3_122_Family}
\end{figure}

\subsection{Materials}
The FeSCs in the 122 family share several common characteristics. One of them is ubiquitous appearance of superconductivity with highest $T_c$ near the edge of domain of long range magnetic ordering in the phase diagram regardless of types of chemical substitution. As an example, Fig.~\ref{fig2-0-1_122_Family} shows phase diagrams of various 122 FeSCs: hole-doped Ba$_{1-x}$K$_x$Fe$_2$As$_2$~\cite{Avci2012PRB}; isovalently-doped BaFe$_2$(As$_{1-x}$P$_x$)$_2$~\cite{HashimotoCho2012Science_BaP122_dL}, Ba(Fe$_{1-x}$Ru$_x$)$_2$As$_2$~\cite{KimGoldman2011PRB_BaRu122}, and SrFe$_2$(As$_{1-x}$P$_x$)$_2$~\cite{Kobayashi2014JPSJ_SrP122}; electron-doped Ba(Fe$_{1-x}$Co$_x$)$_2$As$_2$~\cite{NandiGoldman2010PRL} and Ba(Fe$_{2-x}$Ni$_x$)As$_2$~\cite{Abdel-Hafiez2015PRB_BaNi122}. In all cases, the superconducting dome occurs with suppression of magnetic phase, and particularly, the maximum $T_c$ occurs where the anti-ferromagnetic order is expected to disappear. Gradual suppression of magnetic order with composition as a tuning parameter~\cite{Mathur1998Nature_HeavyFermion} suggests the existence of the quantum critical point suggesting the close relation between magnetic fluctuations and maximum $T_c$. The most clear case for quantum critical scenario is found in isovalently substituted  BaFe$_2$(As$_{1-x}$P$_x$)$_2$, for which both in-plane and inter-plane resistivities show $T$-linear dependence at optimal doping~\cite{TanatarProzorov2014PRB_BaK122}, see the middle panel in the right column of Fig.~\ref{fig2-0-2_122_Family}. Indeed, the quantum quantum critical point is observed beneath the superconducting dome in BaFe$_2$(As$_{1-x}$P$_x$)$_2$ by measuring zero-temperature penetration depth~\cite{HashimotoCho2012Science_BaP122_dL}. 

However, several important differences are observed between various types of doping in both normal and superconducting state. For example, the temperature dependent resistivity shows quite distinct behavior depending on the types of chemical substitution as shown in Fig.~\ref{fig2-0-2_122_Family} and \ref{fig2-0-3_122_Family}. Another significant difference comes from distinct superconducting gap order parameters. While similar Fermi surfaces are found among different types of substitutions, the superconducting gap structures vary from nodal gap in BaFe$_2$(As$_{1-x}$P$_x$)$_2$~\cite{HashimotoCho2012Science_BaP122_dL} to anisotropic full gaps in Ba(Fe$_{1-x}$Co$_x$)$_2$As$_2$~\cite{Tanatar2010PRL}. More interestingly, the order parameter of (Ba$_{1-x}$K$_x$)Fe$_2$As$_2$~\cite{Cho2016ScienceAdvances_BaK122_e-irr} is known to evolve with doping from full gap ($x<0.8$) to gap with accidental nodes for compositions $x>0.8$ where the Lifshitz transition of Fermi surfaces occurs \cite{Xu2013PRB_BaK122, RichardDing2015JPCM_APRES_IBS}. In this section, we will review the effect of electron irradiation on six 122 FeSCs: hole-doped Ba$_{1-x}$K$_x$Fe$_2$As$_2$; isovalently-doped BaFe$_2$(As$_{1-x}$P$_x$)$_2$, Ba(Fe$_{1-x}$Ru$_x$)$_2$As$_2$, and SrFe$_2$(As$_{1-x}$P$_x$)$_2$; electron-doped Ba(Fe$_{1-x}$Co$_x$)$_2$As$_2$ and Ba(Fe$_{2-x}$Ni$_x$)As$_2$.
 
\subsection{Hole-doped Ba$_{1-x}$K$_x$Fe$_2$As$_2$}

\begin{figure}[htb]
\includegraphics[width=7cm]{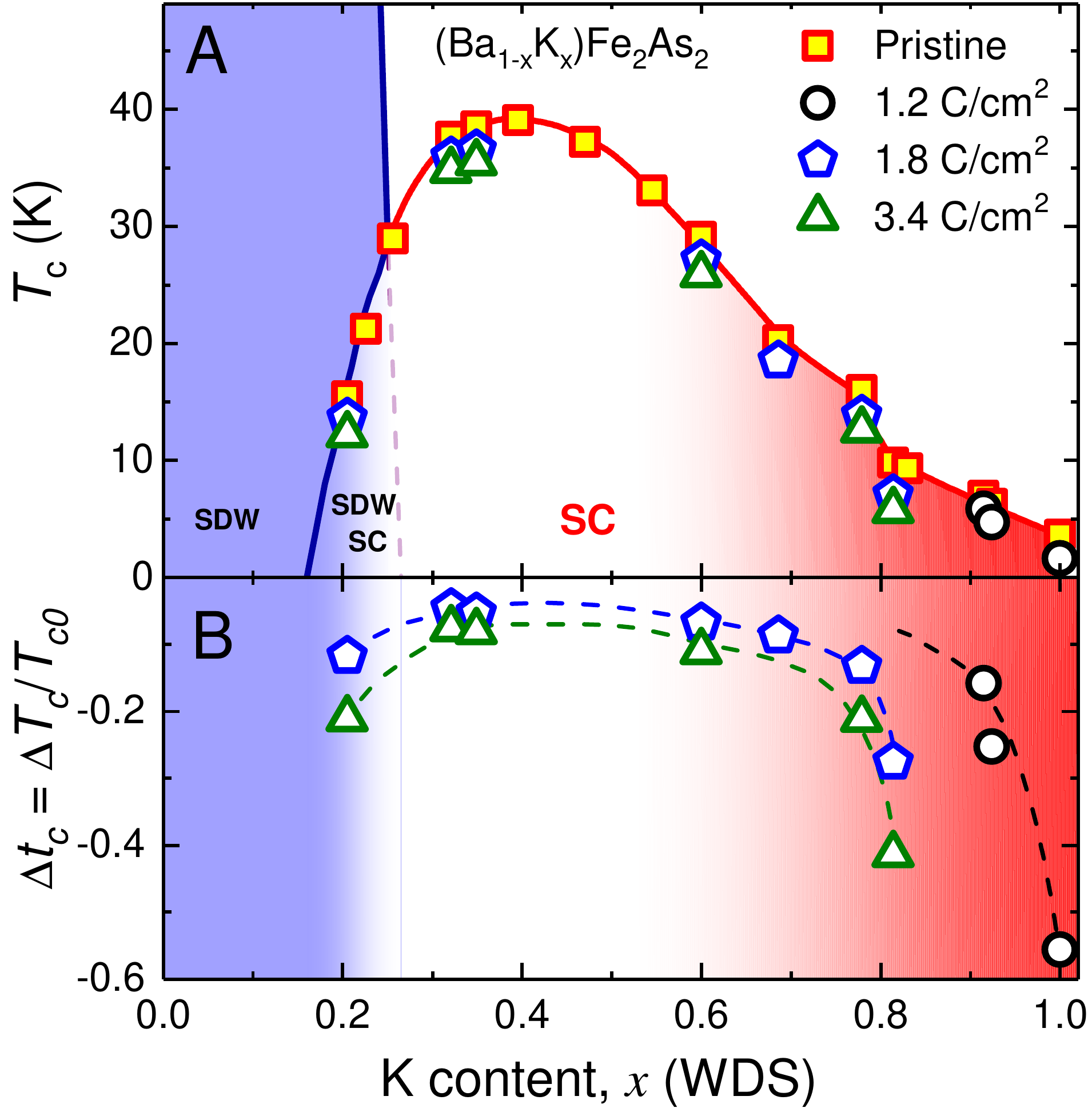}
\caption{(Color online) (A) $T_c$ - $x$ phase diagram of pristine (squares)
and electron-irradiated (other symbols, see legend) samples. SDW, spin-density wave; SC, superconducting phase. (B) Normalized supression of $T_c$ ($\Delta t_c = \Delta T_{c}/T_{c0}$) versus $x$. Reprinted with permission from Science Advances, Ref.~\onlinecite{Cho2016ScienceAdvances_BaK122_e-irr}, copyright AAAS.}
\label{fig2-1-1_BaK122}
\end{figure}

\begin{figure}[htb]
\includegraphics[width=7cm]{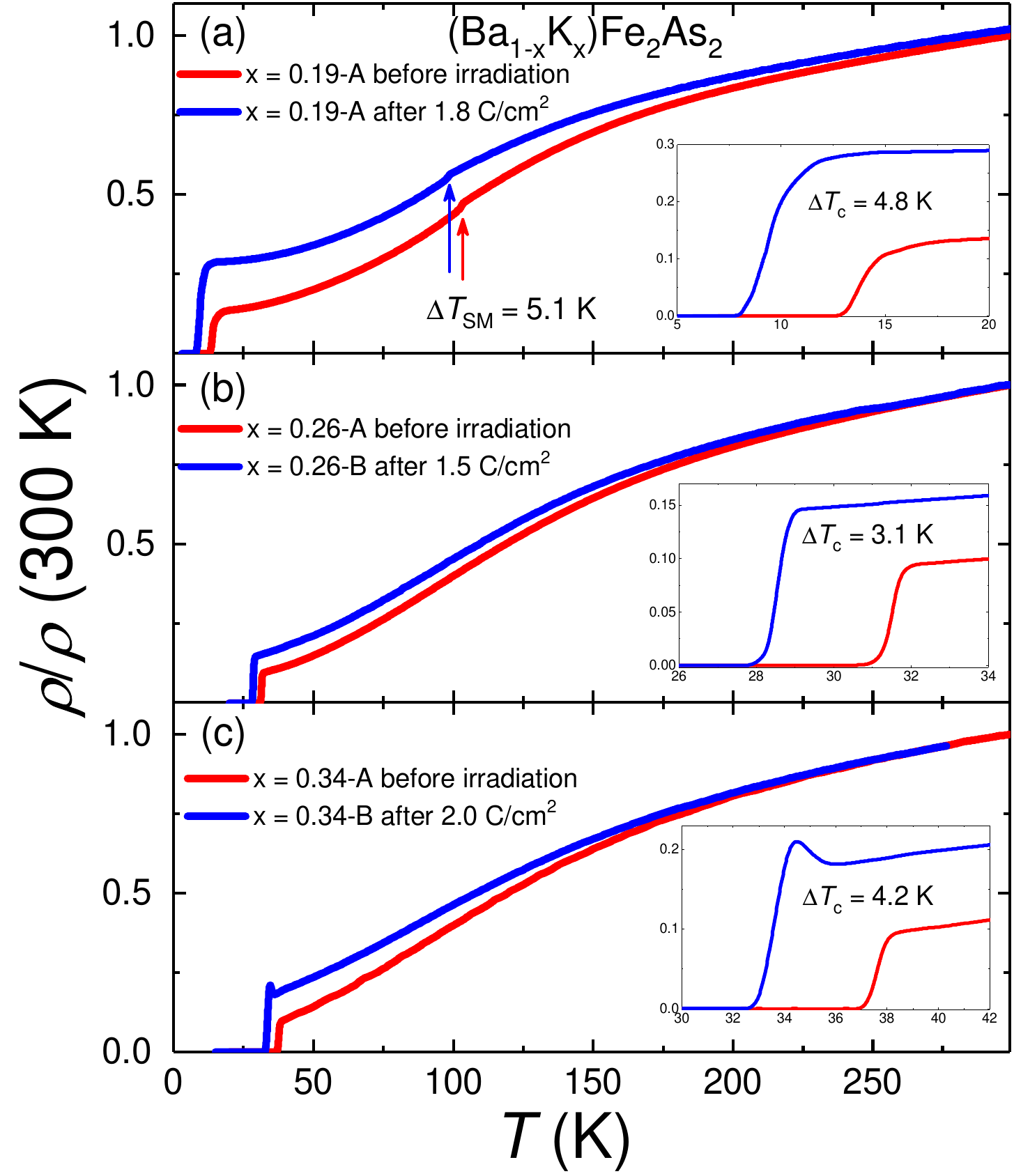}
\caption{(Color online) Evolution of the temperature-dependent resistivity (normalized by the value at 300 K) upon electron irradiation in Ba$_{1-x}$K$_x$Fe$_2$As$_2$: (a) $x$ = 0.19, (b) $x$ = 0.26, and (c) $x$ = 0.34. Reprinted from Ref.~\onlinecite{Cho2014PRB_e-irr}, copyright 2014 APS.}
\label{fig2-1-2_BaK122}
\end{figure}

\begin{figure}[htb]
\includegraphics[width=7cm]{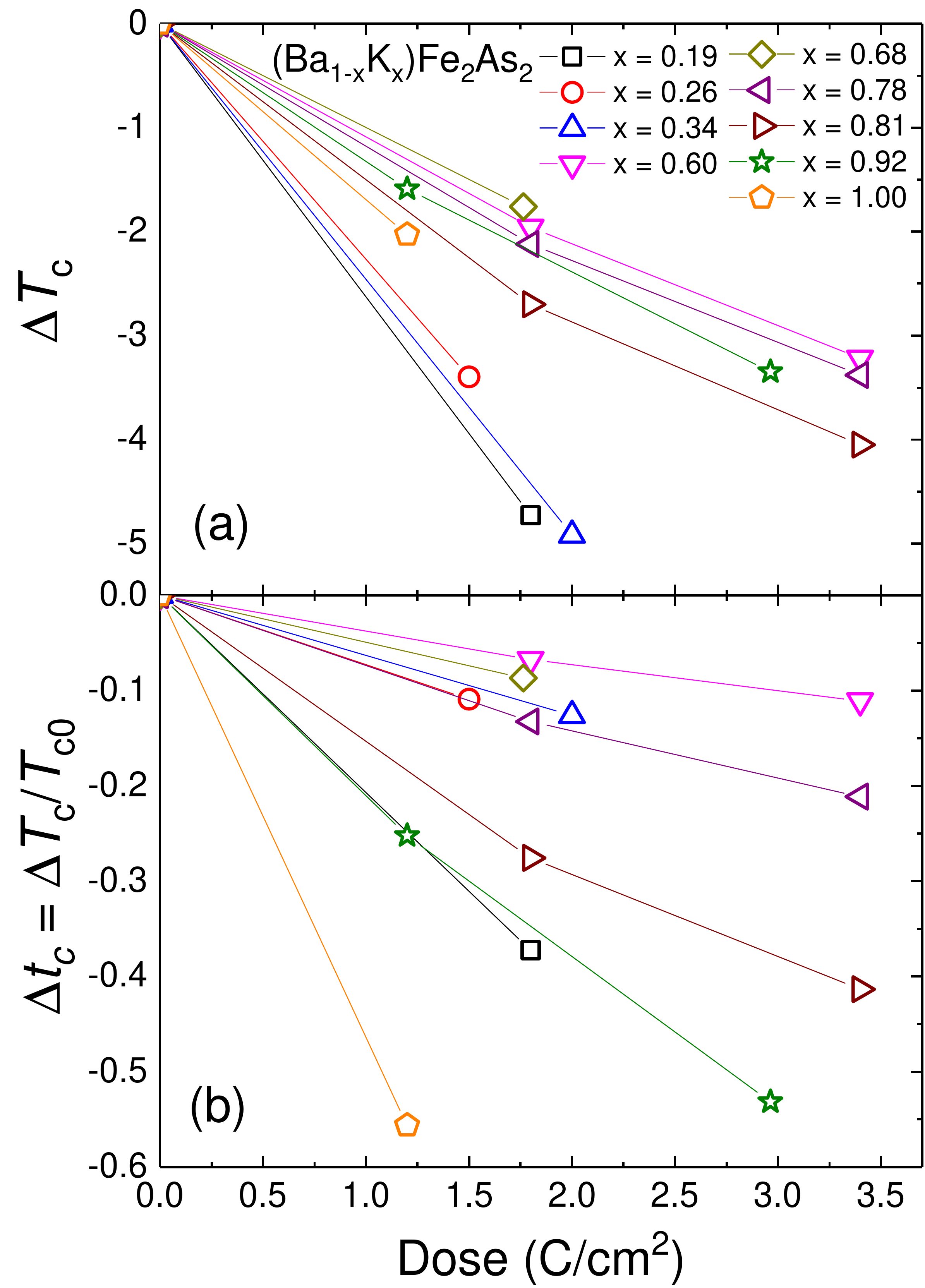}
\caption{(Color online) $T_c$ suppression in Ba$_{1-x}$K$_x$Fe$_2$As$_2$ upon electron irradiation: (a) $\Delta T_c$ and (b) $\Delta t_c = \Delta T_{c}/T_{c0}$ against dosage. Data from Ref.~\onlinecite{Cho2014PRB_e-irr, Cho2016ScienceAdvances_BaK122_e-irr}.}
\label{fig2-1-3_BaK122} 
\end{figure}

\begin{figure}[htb]
\includegraphics[width=8.5cm]{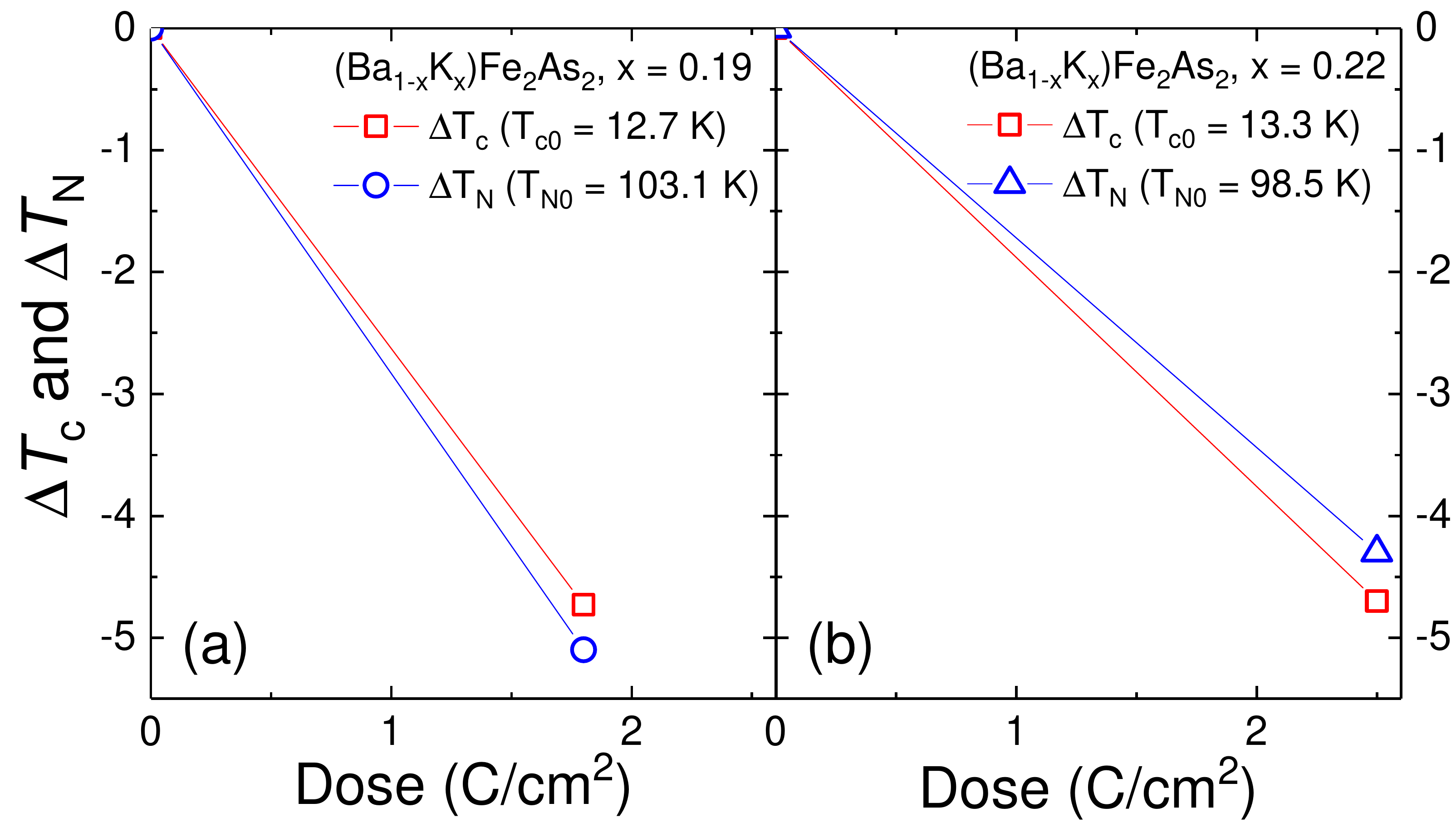}
\caption{(Color online) Comparison between  $\Delta T_c$ and $\Delta T_N$ upon electron irradiation for $x$ = 0.19~\cite{Cho2014PRB_e-irr} and 0.22~\cite{Konczykowski_unpublished} of Ba$_{1-x}$K$_x$Fe$_2$As$_2$.}
\label{fig2-1-4_BaK122} 
\end{figure}

\begin{figure}[htb]
\includegraphics[width=7cm]{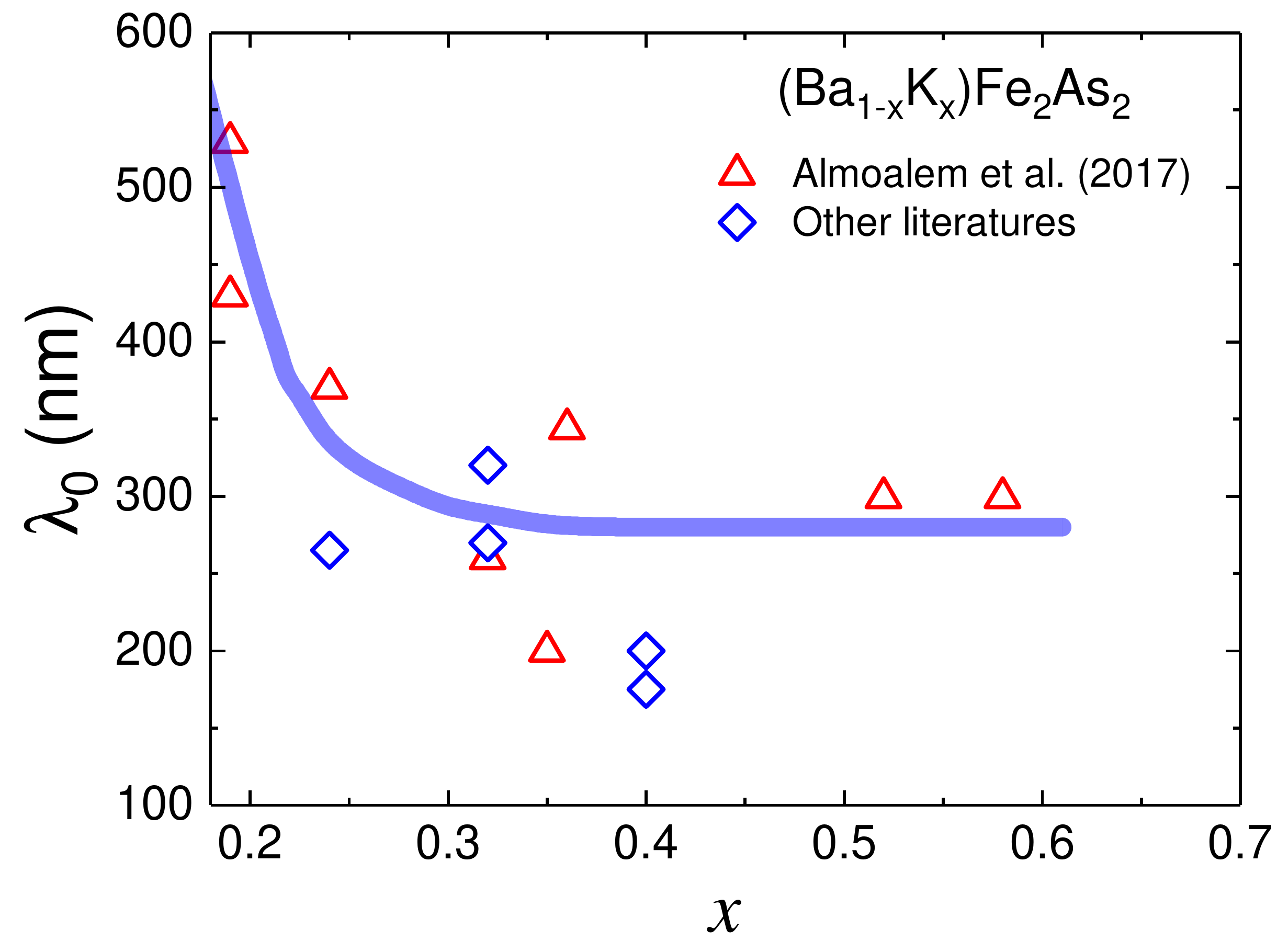}
\caption{(Color online) Zero-temperature penetration depth ($\lambda_0$) of Ba$_{1-x}$K$_x$Fe$_2$As$_2$ from literature~\cite{AlmoalemChoProzorovAuslaender2017arXiv_BaK122, EvtushinskyBoriesenko2009NJP_BaK122, Welp2009PRB_BaK122, LiWang2008PRL_BaK122}.}
\label{fig2-1-5_BaK122} 
\end{figure}

\begin{figure}[htb]
\includegraphics[width=7cm]{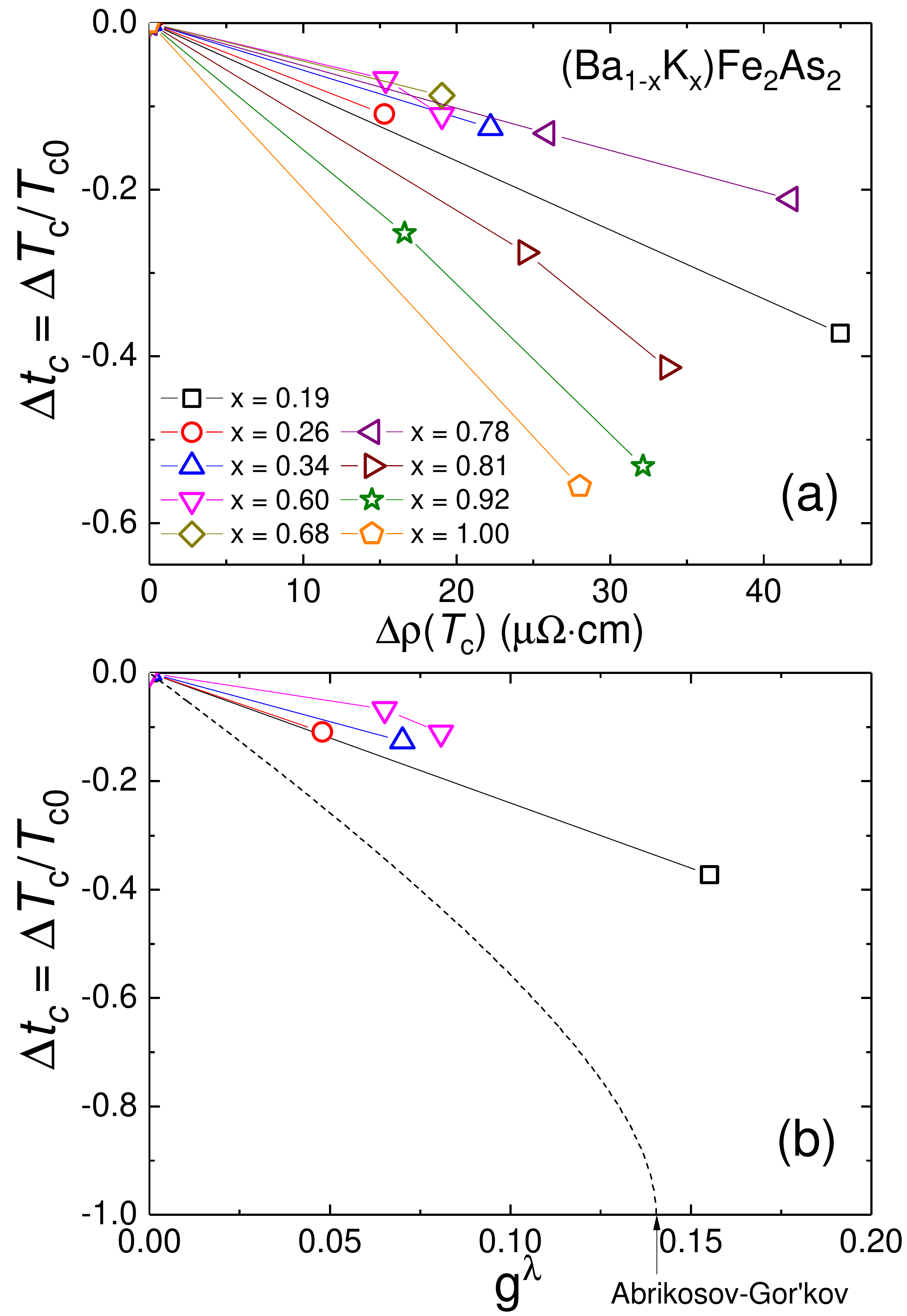}
\caption{(Color online) Normalized suppression, $t_c$ (= $\Delta T_{c}/T_{c0}$) as functions of (a) $\Delta \rho$ and (b) g$^\lambda$.  g$^\lambda$ is calculated only for $x < 0.6$ where experimental $\lambda_0$ is available from literatures.}
\label{fig2-1-6_BaK122} 
\end{figure}

\begin{figure*}[htb]
\includegraphics[width=15cm]{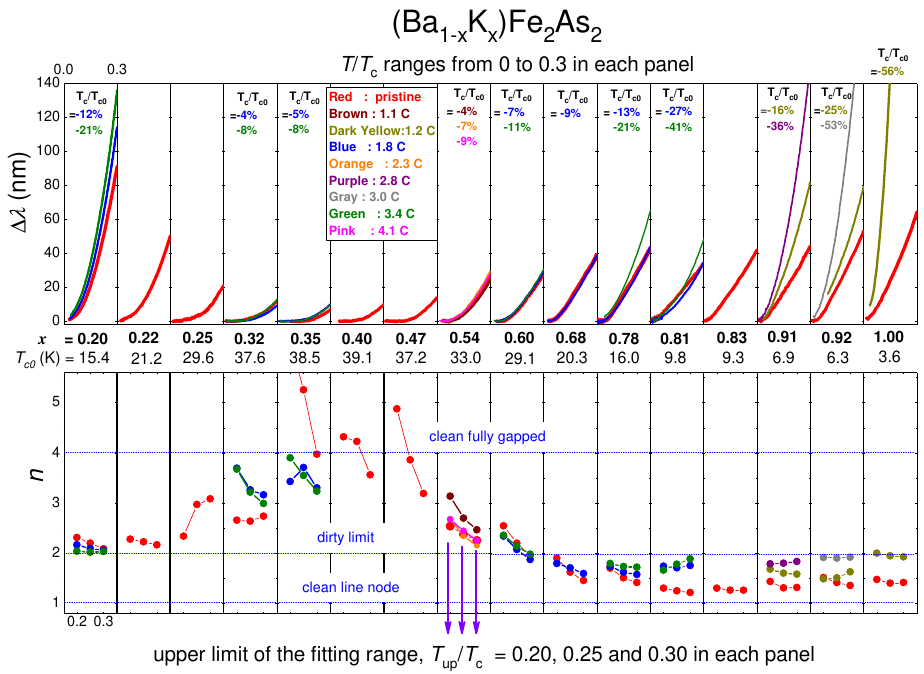}
\caption{(Color online) Evolution of temperature dependence of London penetration depth ($\Delta \lambda$). Upper panels: $\Delta \lambda$ versus $T/T_{c}$ for 16 different compositions before and after electron irradiation. Each individual panel shows a low-temperature region of $T/T_{c} < 0.3$ (full-range curves are shown in fig.S1 of Sci. Advances). Lower panels: Exponent $n$ obtained from the power-law fitting, $\Delta \lambda$ = $A(T/T_{c})^{n}$. For each curve, three different upper-limit temperatures were used, $T_{up}/T_{c}$ = 0.20, 0.25, and 0.30, whereas the lower limit was fixed by the lowest temperature. Reprinted with permission from Science Advances, Ref.~\onlinecite{Cho2016ScienceAdvances_BaK122_e-irr}, copyright AAAS.}
\label{fig2-1-7_BaK122} 
\end{figure*}

\begin{figure}[htb]
\includegraphics[width=7cm]{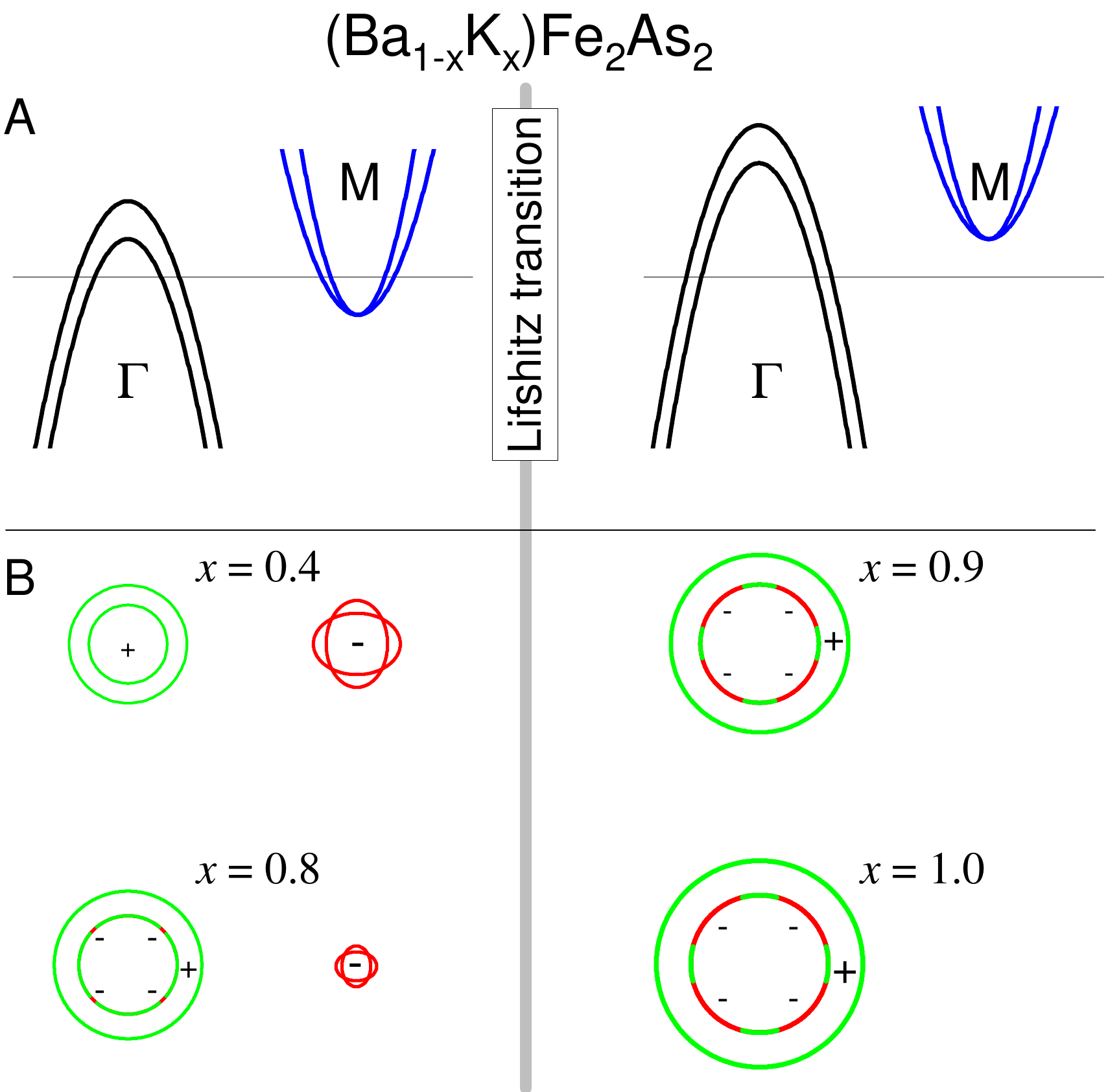}
\caption{(Color online) (a) Schematic change in the electronic band structure across the Lifshitz transition. (b) Hole ($\Gamma$) and electron (M) pockets relevant for calculations with the sign-changing order parameter. Signes are encoded by green (+) and red (-) colors. Reprinted with permission from Science Advances, Ref.~\onlinecite{Cho2016ScienceAdvances_BaK122_e-irr}, copyright AAAS.}
\label{fig2-1-8_BaK122} 
\end{figure}

\begin{figure}[htb]
\includegraphics[width=7cm]{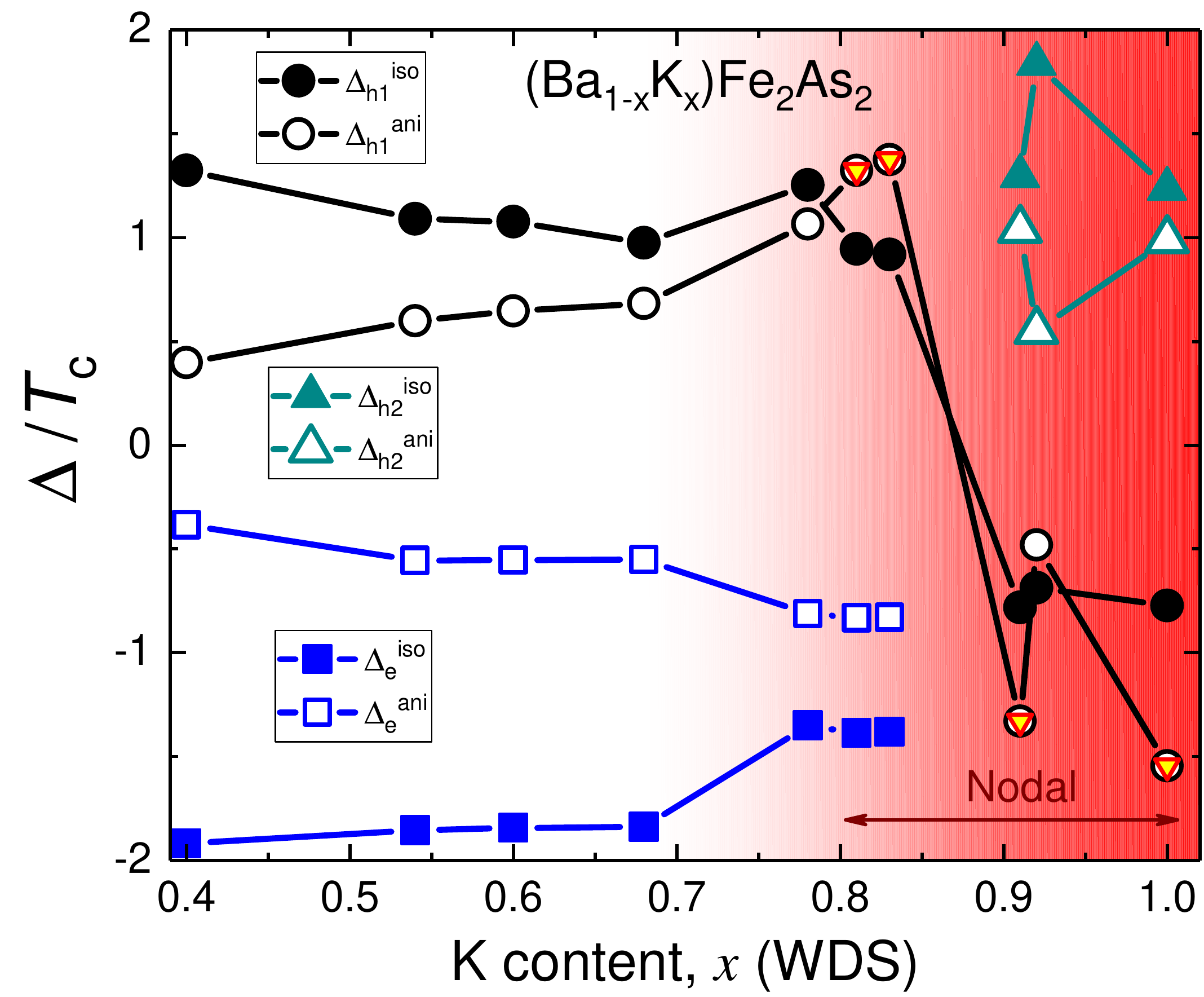}
\caption{(Color online) Evolution of the superconducting gaps obtained from self-consistent $t$-matrix fitting. The assumed electronic structure is described in Fig.~\ref{fig2-1-8_BaK122}. As long as the isotropic part is greater than the anisotropic one, the state is nodeless (that is, for $x < 0.8$). In the opposite limit, the nodes appear. This is shown by inscribed triangles in the figure for h1 contribution. Consequently, the s$_\pm$ pairing switches from hole-electron pockets below the Lifshitz transition to hole-hole above. For details, refer to Ref.~\cite{Cho2016ScienceAdvances_BaK122_e-irr}. Reprinted with permission from Science Advances, Ref.~\onlinecite{Cho2016ScienceAdvances_BaK122_e-irr}, copyright AAAS.}
\label{fig2-1-9_BaK122} 
\end{figure}

(Ba$_{1-x}$K$_x$)Fe$_2$As$_2$ is one of the most intensively studied compounds among FeSCs due to its evolution of the superconducting gap structure over composition ($x$). In the optimally doped region ($x$ = 0.35 - 0.4), two effective isotropic superconducting
gaps were identified in various experiments, such as thermal conductivity~\cite{LuoTaillefer2009PRB}, London penetration depth~\cite{Cho2014PRB_e-irr, EvtushinskyBoriesenko2009NJP_BaK122}, and angle-resolved photoemission spectroscopy (ARPES)~\cite{EvtushinskyBoriesenko2009NJP_BaK122, Ding2008EPL, Nakayama2011PRB, OtaOkazaki2014PRB_ARPES_BaK122}. However, a gap with line nodes was identified in the heavily overdoped region ($x \ge 0.8$) from thermal conductivity~\cite{ReidTaillefer2012PRL_KFe2As2_d-wave, ReidTanatarProzorovTaillefer2012SST_KFe2As2, Hong2015CPL_BaK122_HeatTransport, WatanabeMatsuda2014PRB_BaK122}, London penetration depth~\cite{Hashimoto2010PRB_KFe2As2}, and ARPES~\cite{Ding2008EPL, OtaOkazaki2014PRB_ARPES_BaK122}. This variation of the superconducting gap structure is likely to be connected to the Lifshitz transition near $x$ = 0.7 - 0.9 where the electron-like pockets at the M point changes to hole-like pockets~\cite{Xu2013PRB_BaK122, RichardDing2015JPCM_APRES_IBS}. The evolution of gap strucutre has been discussed in several models such as i) a crossover between two generalized s-wave states, where the usual configuration of isotropic gaps with opposite signs on the electron and hole pockets crosses over to a configuration with opposite signs on the hole pockets resulting in accidental nodes~\cite{WatanabeMatsuda2014PRB_BaK122}, ii) an intermediate time-reversal symmetry broken s + $i$s state~\cite{Maiti2015PRB_s+is}, iii) a transition from s$_\pm$ to d wave either directly~\cite{Thomale2011PRL} or with an intermediate s + $i$d state~\cite{Chubukov2012ARCMP, Platt2012PRB_IBS, FernandesMillis2013PRL}, and iv) the existence of too-small-to-measure but finite “Lilliputian” gaps~\cite{Hardy2014JPSJ, Hardy2016PRB_BaK122}. 

To resolve this unusual variation, the 2.5 MeV electron irradiation in combination with resistivity and London penetration depth measurements was used by Cho \textit{et al.}~\cite{Cho2014PRB_e-irr, Cho2016ScienceAdvances_BaK122_e-irr}.  First of all, the electron irradiation effectively suppresses $T_c$ over all compositions as shown in Fig.~\ref{fig2-1-1_BaK122}. The large suppression of $T_c$ occurs in under and over-doped compositions. For under-doped compositions shown in Fig.~\ref{fig2-1-2_BaK122} (a), the magnetic transition temperature $T_{N}$ (or $T_{SM}$) is also effectively suppressed in $x$ = 0.19. Interestingly, the amounts of decrease of $\Delta T_c = -4.8 K$ is comparable to $\Delta T_N = -5.1 K$ (Fig.~\ref{fig2-1-2_BaK122} (a)). This correlation between $\Delta T_c$ and $\Delta T_N$ also exists in another under-doped composition $x$ = 0.22 as shown in Fig.~\ref{fig2-1-4_BaK122} (b). One can test this correlation in isovalently doped BaFe$_2$(As$_{1-x}$P$_x$)$_2$ as will be shown later in Fig.~\ref{fig2-2-6_BaP122}, but there exists no particular correlation potentially due to the influence of the quantum critical point. For further analysis on $T_c$ suppression in Ba$_{1-x}$K$_x$Fe$_2$As$_2$, the changes in T$_c$ and normalized $T_c$ upon irradiation are summarized in Fig.~\ref{fig2-1-3_BaK122}. It clearly shows that heavily under and over-doped samples are most susceptible against irradiation. 

Since the effect of irradiation varies in different materials, the dosage is not a good parameter to indicate the amount of disorder. To avoid this problem we used an increase of the normal state residual resistivity upon irradiation as a measure of disorder, which is clearly seen for example in Fig.~\ref{fig2-1-2_BaK122}. Figure~\ref{fig2-1-6_BaK122} (a) summarizes $\Delta t_{c} = \Delta T_{c}/T_{c0}$ as a function of $\Delta \rho$. 

Experimentally determined values of resistivity increase ($\Delta \rho$) and the absolute value of London penetration depth ($\lambda_0$) enable us to define dimensionless scattering rate as~\cite{Kogan2009PRB-2, Prozorov2014PRX_e-irr}
\begin{equation}
g^{\lambda} = \frac{\hbar \Delta \rho}{2 \pi k_B \mu_0 T_{c0} \lambda_{0}^2}, \label{eq06}
\end{equation}
where $\lambda_0$ is the zero temperature London penetration depth, $T_{c0}$ is $T_c$ before irradiation, and $\Delta \rho$ is the variation of residual resistivity. The relative change of the superconducting transition temperature $\Delta t_c = \Delta T_{c}/T_{c0}$ as a function of resistivity change $\Delta \rho$ is summarized in Fig.~\ref{fig2-1-6_BaK122} (a). The values of $\lambda_0$ available from the literature \cite{AlmoalemChoProzorovAuslaender2017arXiv_BaK122, EvtushinskyBoriesenko2009NJP_BaK122, Welp2009PRB_BaK122, LiWang2008PRL_BaK122}  are plotted in Fig.~\ref{fig2-1-5_BaK122}. Since there are no reports on $\lambda_0$ in the over-doped region, we only consider the compositions with $x\leq$0.6. Based on these parameters, the variation of reduced transition temperature $t_c=T_c / T_{c0}$ is calculated as a function of dimensionless parameter ($g^\lambda$) as shown in Fig.~\ref{fig2-1-6_BaK122} (b). In general, $t_c$ shows substantial decrease with increasing $g^\lambda$, but much slower than Abrikosov-Gor'kov value. 

The London penetration depth was also measured for all compositions upon increasing dose of irradiation as shown in  Fig.~\ref{fig2-1-7_BaK122}. In optimally doped region, the exponent of the power law fit $n$ (bottom panel) is above 4, which is experimentally indistinguishable from exponential dependence. This is a clear signature of the full gap superconductivity. However, on moving away from the optimal doping, the exponent decreases toward $n$ = 2 for the under-doped region and below $n$ = 2 for over-doped region. The former finding is consistent with previous study in strongly underdoped compositions~\cite{KimProzorov2014PRB_underdoped_BaK122}, interpreted as anisotropy appearing due to coexistent magnetic order~\cite{ReidTanatarProzorov2016PRB_Underdoped_BaK122}. The $T$ - linear behavior of low-temperature penetration depth in the over-doped region is a signature of nodal gaps. To understand this doping dependent variation of the superconducting gap structure, a minimal two gap model is introduced to fit  the penetration depths of all pristine samples (See the Supplementary Materials of Ref.~\onlinecite{Cho2016ScienceAdvances_BaK122_e-irr} for details). 
\begin{eqnarray}
\Delta_{1} &=& \Delta_{01} (1.0 + r_1 \text{cos}4\phi)\\
\Delta_{2} &=& \Delta_{02} (1.0 + r_2 \text{cos}4\phi) \label{eq07}
\end{eqnarray}

Then, the interaction potentials were calculated and the impurity scattering upon electron irradiation was treated within self-consistent $t$-matrix approximation. Considering the Fermi surface change near the Lifshitz transition at $x\sim$0.8 (Figs.~\ref{fig2-1-8_BaK122}), all results of penetration depth were fitted with this model and  the superconducting gap evolution was found as shown in \ref{fig2-1-9_BaK122}. Interestingly, all experimental data are well explained assuming that the sign-change between hole and electron pockets (near optimal doped region) varies to sign-change within the same hole pockets in heavily over-doped region. This clearly supports that the nodes observed in $x > 0.8$ is not symmetry imposed but accidental nodes, which is consistent with various other experimental observations.

\subsection{Isovalent-substituted BaFe$_2$(As$_{1-x}$P$_{x}$)$_2$}

\begin{figure}[htb]
\includegraphics[width=8.5cm]{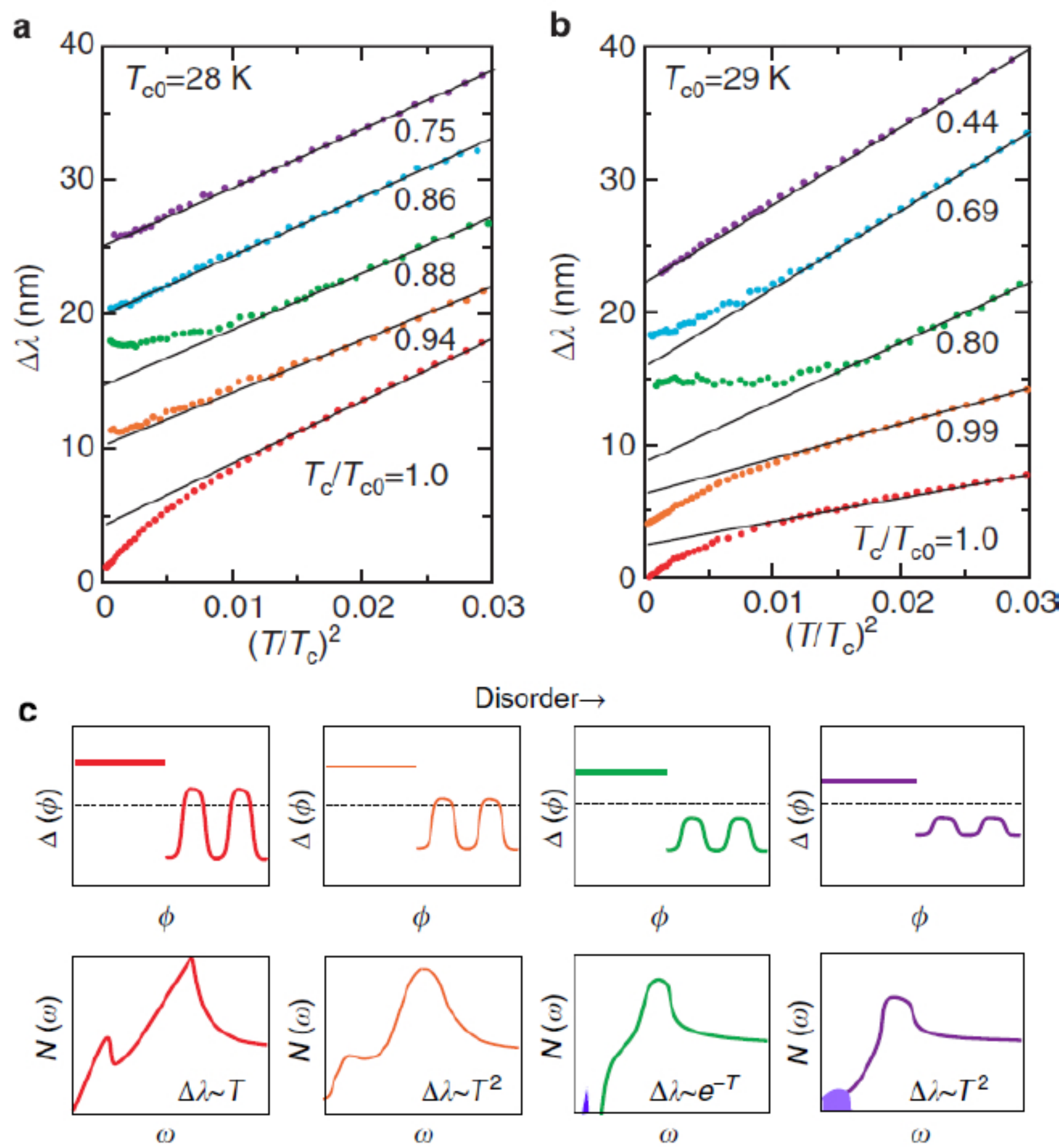}
\caption{(Color online) Effect of electron irradiation on the low-temperature penetration depth $\Delta \lambda$ of two samples of BaFe$_2$(As$_{1-x}$P$_{x}$)$_2$: (a) $T_{c0}$ = 28 K and (b) $T_{c0}$ = 29 K. Each curve is shifted vertically for clarity. Lines are the $T^2$ dependence fits at high temperatures. (c) Schematic of $s_{\pm}$ order parameter versus azimuthal angle $\phi$ (top row) and density of states $N$ versus energy $\omega$ (bottom row) with increasing irradiation dosage (from the left to right).
Reprinted with permission from Nature Communications, Ref.~\onlinecite{Mizukami2014NatureComm}, copyright Macmillan Publishers Ltd.}
\label{fig2-2-1_BaP122}
\end{figure}

\begin{figure}[htb]
\includegraphics[width=8.5cm]{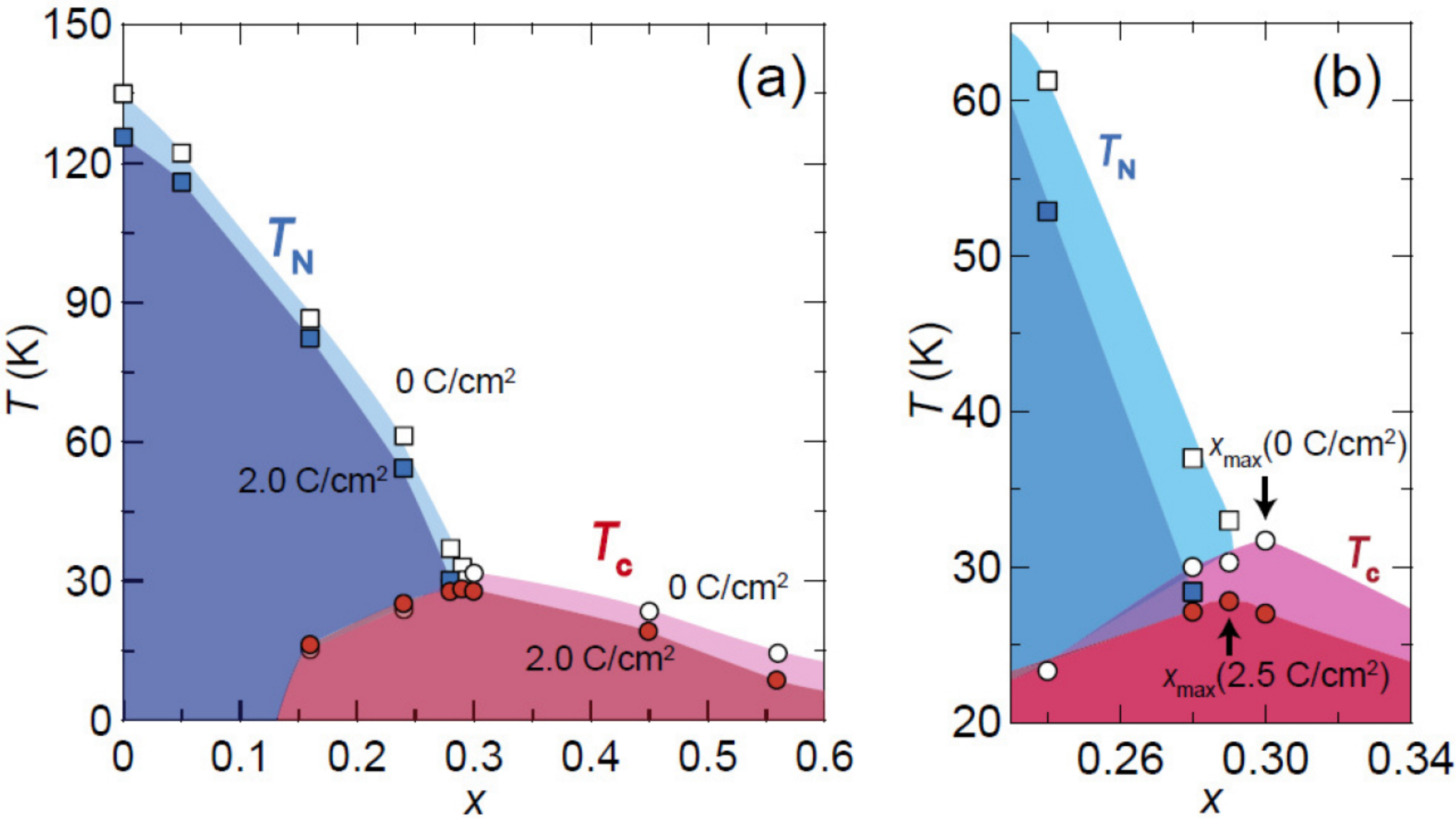}
\caption{(Color online) (a) The doping phase diagram of BaFe$_2$(As$_{1-x}$P$_{x}$)$_2$ for 0 and 2.0 C/cm$^2$ of electron irradiations. (b) The zoom of the region of  near optimally doped composition for 0 and 2.5 C/cm$^2$. The arrow (maximum $T_c$) moves toward lower composition indicating the shift of superconducting dome upon irradiation. Reprinted with permission from Ref.~\onlinecite{Mizukami2017JPSJ_BaP122_e-irr}, copyright JPSJ.}
\label{fig2-2-2_BaP122} 
\end{figure}

\begin{figure}[htb]
\includegraphics[width=7cm]{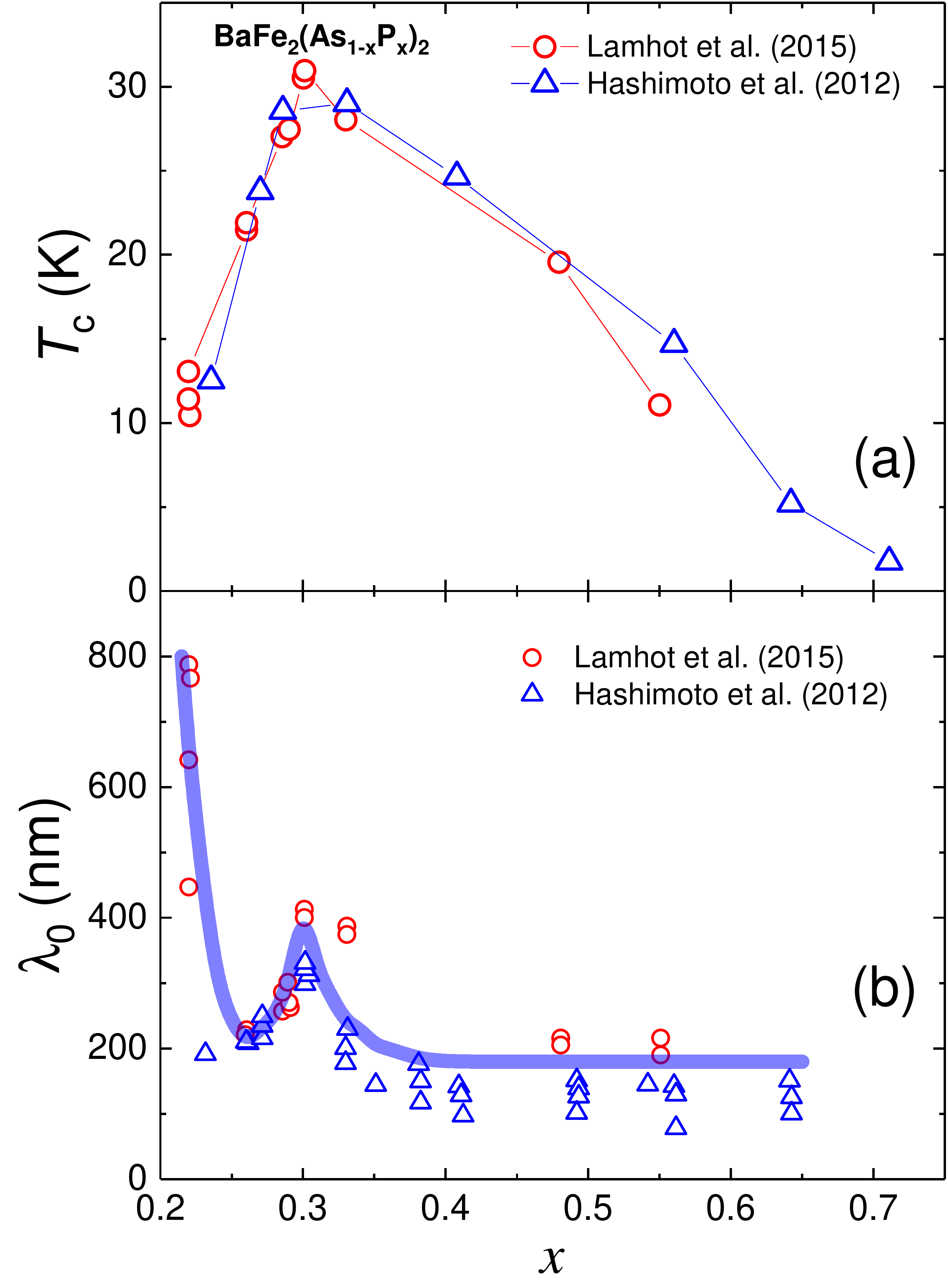}
\caption{(Color online) (a) $T_c$ and (b) zero-temperature penetration depth $\lambda_0$ versus $x$ in BaFe$_2$(As$_{1-x}$P$_x$)$_2$. Data from Ref.~\onlinecite{HashimotoCho2012Science_BaP122_dL, Lamhot2015PRB_BaP122_L(0)}}
\label{fig2-2-3_BaP122} 
\end{figure}

\begin{figure}[htb]
\includegraphics[width=7cm]{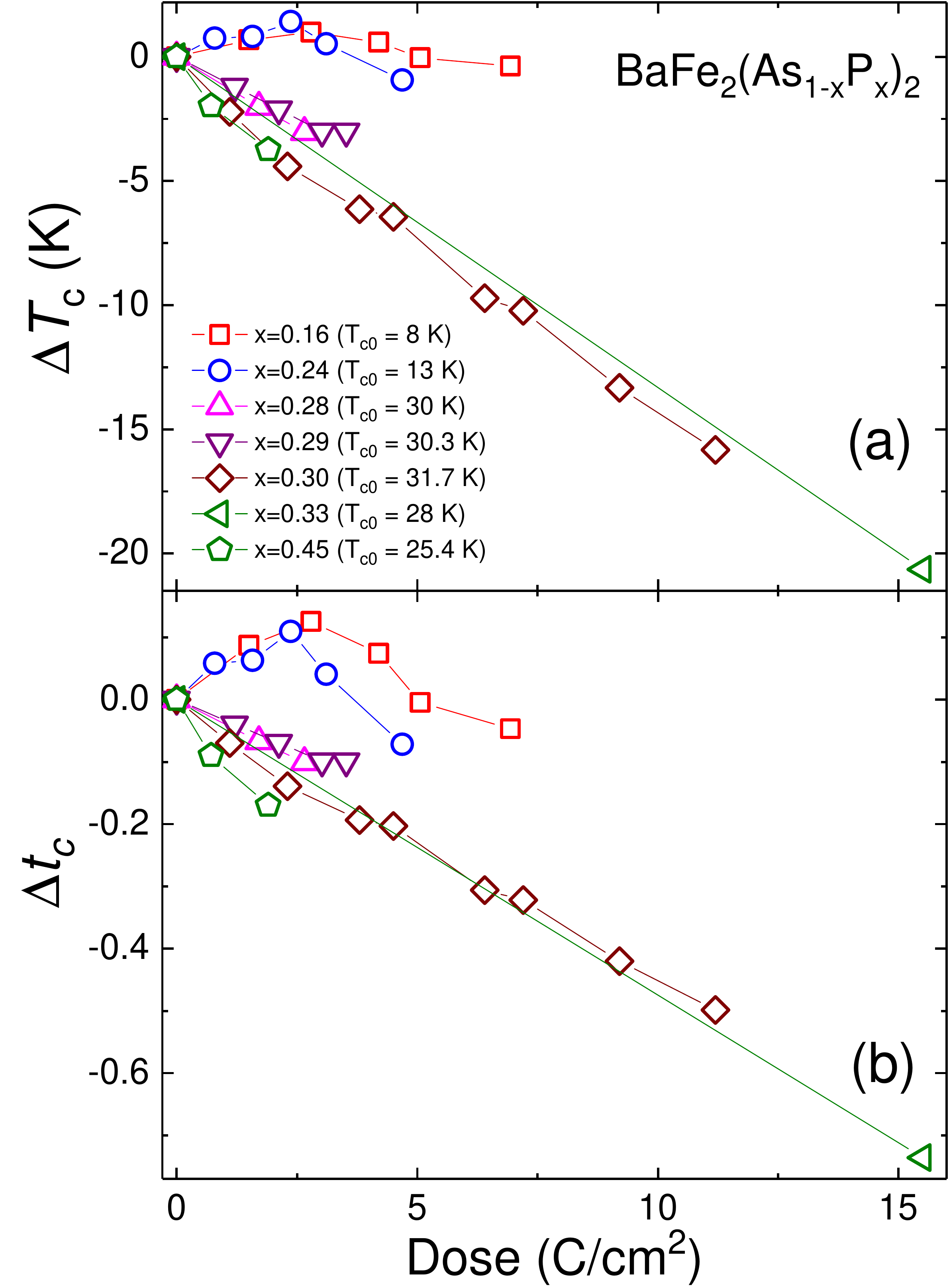}
\caption{(Color online) Suppression of $T_c$: (a) $\Delta T_c$ and (b) $\Delta t_c = \Delta T_c / T_{c0}$ of BaFe$_2$(As$_{1-x}$P$_x$)$_2$ upon electron irradiation. Data from Ref.~\onlinecite{Mizukami2017JPSJ_BaP122_e-irr}. Data of $x$ = 0.25 and 0.33, and extended data of $x$ = 0.30 (dose $>$ 6.4 C/cm$^2$) are directly obtained by authors of Ref.~\onlinecite{Mizukami2017JPSJ_BaP122_e-irr} and presented with permission.}
\label{fig2-2-4_BaP122} 
\end{figure}

\begin{figure}[htb]
\includegraphics[width=7cm]{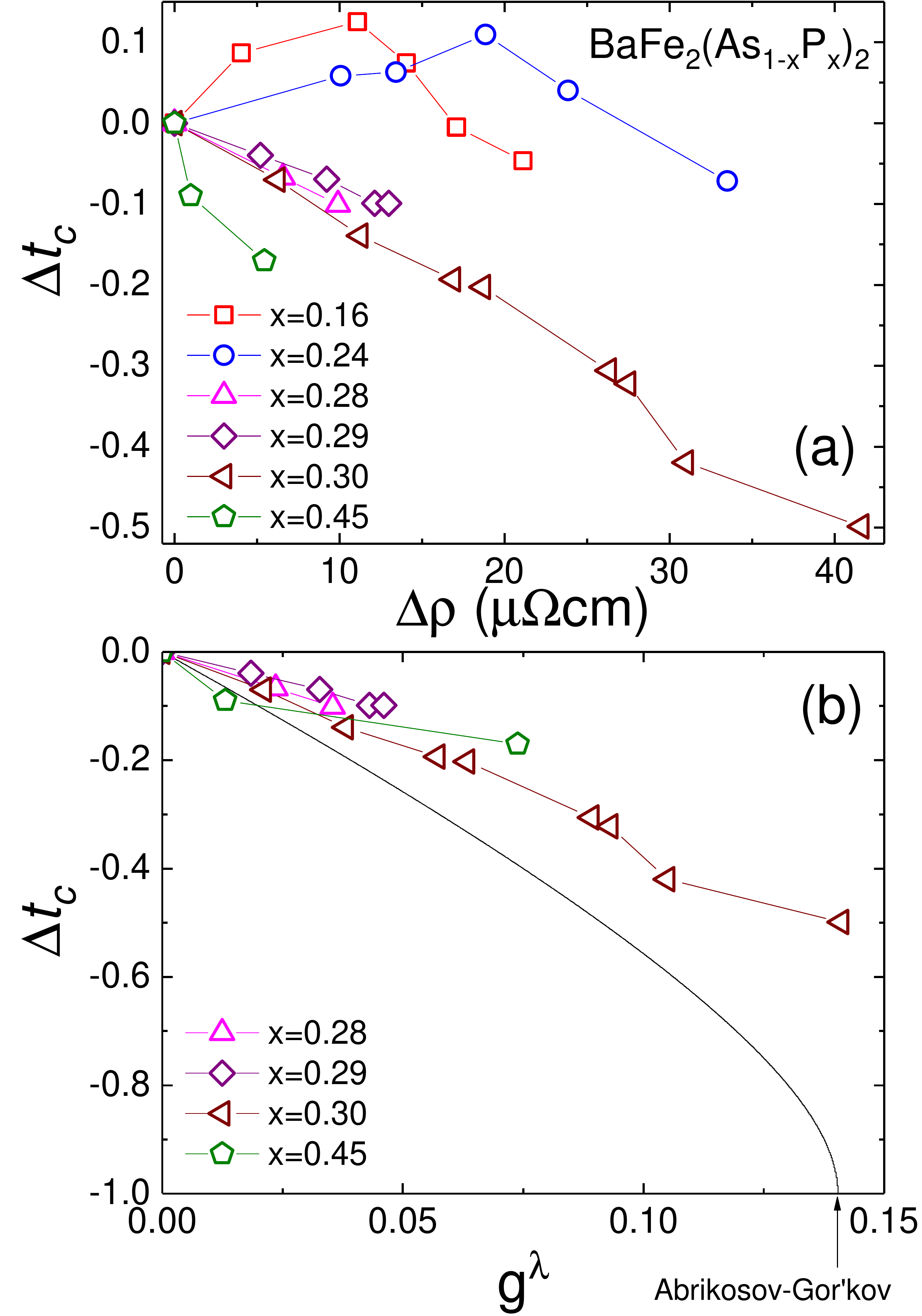}
\caption{(Color online) $\Delta t_c = \Delta T_c / T_{c0}$ of BaFe$_2$(As$_{1-x}$P$_x$)$_2$ versus (a) $\Delta \rho$~\cite{Mizukami2017JPSJ_BaP122_e-irr} and (b) the dimensionless scattering parameter ($g^\lambda$) calculated following equation~\ref{eq06}. The solid line is from Abrikosov-Gor'kov calculation.}
\label{fig2-2-5_BaP122} 
\end{figure}

\begin{figure}[htb]
\includegraphics[width=7cm]{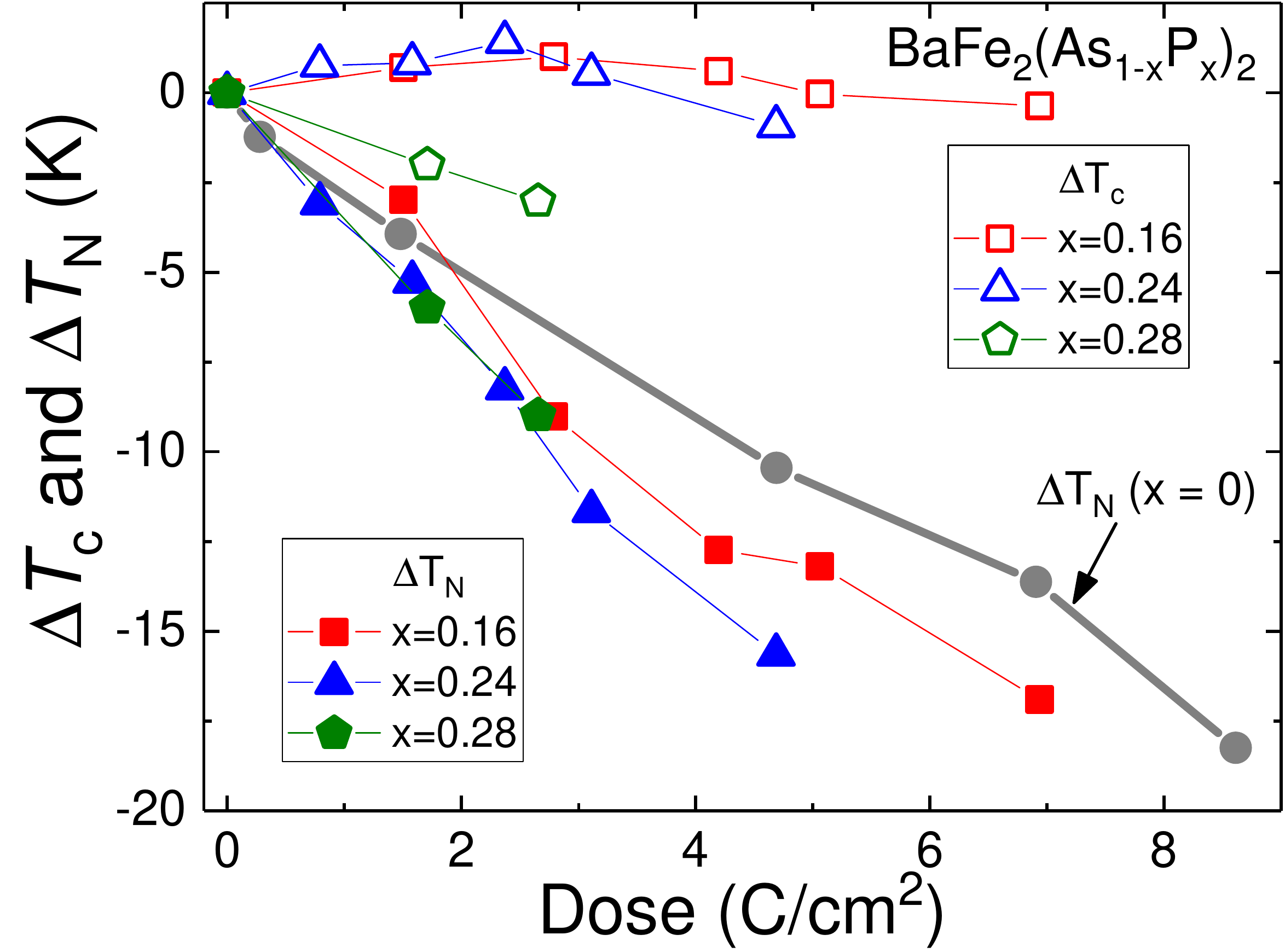}
\caption{(Color online) Comparison between $\Delta T_c$ and $\Delta T_N$ upon electron irradiation. Data from Ref.~\onlinecite{Mizukami2017JPSJ_BaP122_e-irr}. Data of $x$ = 0.25 are directly obtained by authors in Ref.~\onlinecite{Mizukami2017JPSJ_BaP122_e-irr} and presented with permission.}
\label{fig2-2-6_BaP122}
\end{figure}

BaFe$_2$(As$_{1-x}$P$_x$)$_2$ is an isovalently substituted FeSC which has the maximum $T_c$ $\sim$ 30 K~\cite{Kasahara2010PRB_BaP122}. This compound is particularly interesting due to the presence of quantum critical point beneath the superconducting dome as shown in Fig.~\ref{fig2-0-1_122_Family} (b)~\cite{HashimotoCho2012Science_BaP122_dL} and nodal superconducting gaps over all compositions~\cite{Shibauchi2014ARCMP}. Since the nodes can be symmetry imposed (as in d-wave case) or accidental, the origin of nodal gap structure in BaFe$_2$(As$_{1-x}$P$_x$)$_2$ has been a key subject. This question was answered by conducting combined study of electron irradiation and measurement of low-temperature penetration depth by Mizukami \textit{et al.}~\cite{Mizukami2014NatureComm}. As shown in Fig.~\ref{fig2-2-1_BaP122}, the electron irradiation effectively suppresses $T_c$ down to 0.44 $T_{c0}$. Simultaneously, the low-temperature penetration depth shows a non-monotonic evolution of the power-law exponent $n$, from $n \sim 1$ ($T$ - linear) to above $n > 3$ (exponential) and then back to $n \sim 2$ ($T^2$). If the nodes in the gap gap were symmetry-imposed, the monotonic change from linear to $T^2$ with disorder should be expected. Thus, the occurrence of exponential penetration depth during the irradiation clearly supports the presence of accidental nodes~\cite{Mizukami2014NatureComm}. Furthermore, Mizukami \textit{et al.}~\cite{Mizukami2017JPSJ_BaP122_e-irr} carefully investigated how the superconducting dome changes upon irradiation and found the shift of superconducting dome toward lower composition side as shown in Fig.~\ref{fig2-2-2_BaP122}. This implies that the maximum $T_c$ follows the location of quantum-critical point as it also moves toward lower $x$. 

To characterize $T_c$ suppression more quantitatively we calculated the dimensionless scattering parameter ($g^\lambda$). For this purpose, we summarize the zero-temperature London penetration depth of BaFe$_2$(As$_{1-x}$P$_x$)$_2$ from literatures in Fig.~\ref{fig2-2-3_BaP122} (b), and the $T_c$ versus dose of electron irradiation in Fig.~\ref{fig2-2-4_BaP122}. At optimally doped region ($x =$ 0.30 and 0.33), the large doses of electron irradiation were applied up to 11.2 and 15.5 C/cm$^2$, respectively. For both compositions, $T_c$ drops linearly  without any sign of saturation. This is a strong evidence against $s_{++}$ pairing, but consistent with sign-changing $s_{\pm}$ pairing. Among the data in Fig.~\ref{fig2-2-4_BaP122}, only limited data have corresponding resistivity $\Delta \rho$ upon irradiation. For those data, $\Delta t_c = \Delta T_c / T_{c0}$ versus $\Delta \rho$ is plotted in Fig.~\ref{fig2-2-5_BaP122} (a). Following equation~\ref{eq06}, the dimensionless parameter is calculated and plotted in Fig.~\ref{fig2-2-5_BaP122} (b). In general, the suppression of $t_c$ is similar among different compositions. More interestingly, these values are very close to Abrikosov-Gor'kov value. Another interesting fact is shown in Fig.~\ref{fig2-2-6_BaP122} that suppression rates of $\Delta T_N$ of under-doped compositions ($x$ = 0, 0.16, 0.24, 0.28) are very similar. The reason of these similar rates suppression requires further studies. Unlike hole-doped Ba$_{1-x}$K$_x$Fe$_2$As$_2$, $\Delta T_c$ is not comparable to $\Delta T_N$, potentially due to the presence and shift of quantum criticality point.

\subsection{Isovalently substituted Ba(Fe$_{1-x}$Ru$_x$)$_{2}$As$_{2}$}

\begin{figure}[htb]
\includegraphics[width=7cm]{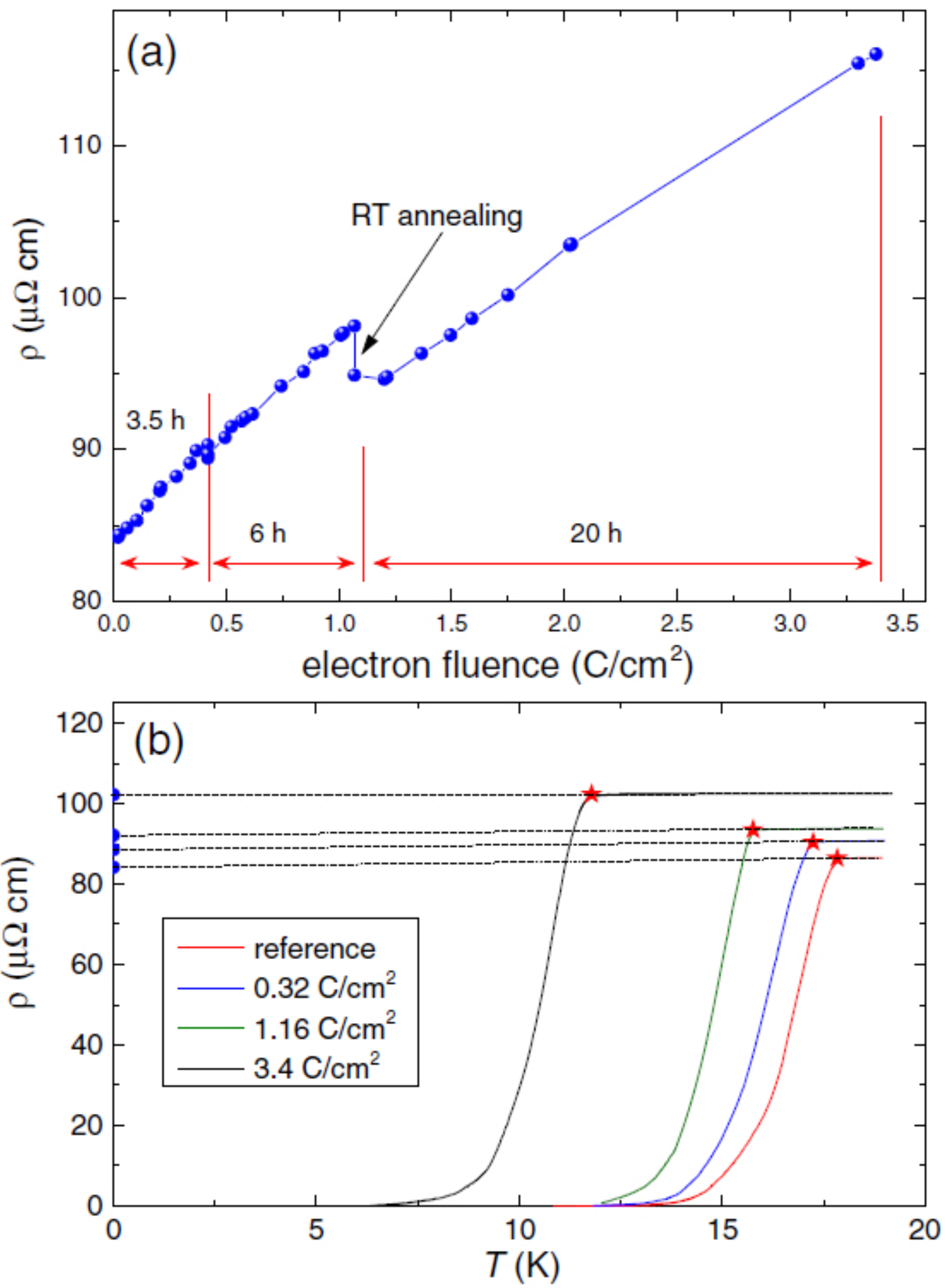}
\caption{(Color online) (a) The $in$ - $situ$ measurements of resistivity in Ba(Fe$_{1-x}$Ru$_x$)$_{2}$As$_{2}$ ($x$ = 0.24) at $T \approx 22 K$ as a function of the irradiation dose. The breaks in the curve correspond to the extraction of the sample and warming it up to room temperature resulting in a partial annealing of the defects. (b) The $ex$ - $situ$ measurements of resistivity versus temperature between the irradiation runs. Dashed lines show linear extrapolation of $\rho (T)$ from above $T_c$ to $T$ = 0 K. Reprinted with permission from Ref.~\onlinecite{Prozorov2014PRX_e-irr}, copyright 2014 APS.}
\label{fig2-3-1_BaRu122} 
\end{figure}

\begin{figure}[htb]
\includegraphics[width=7cm]{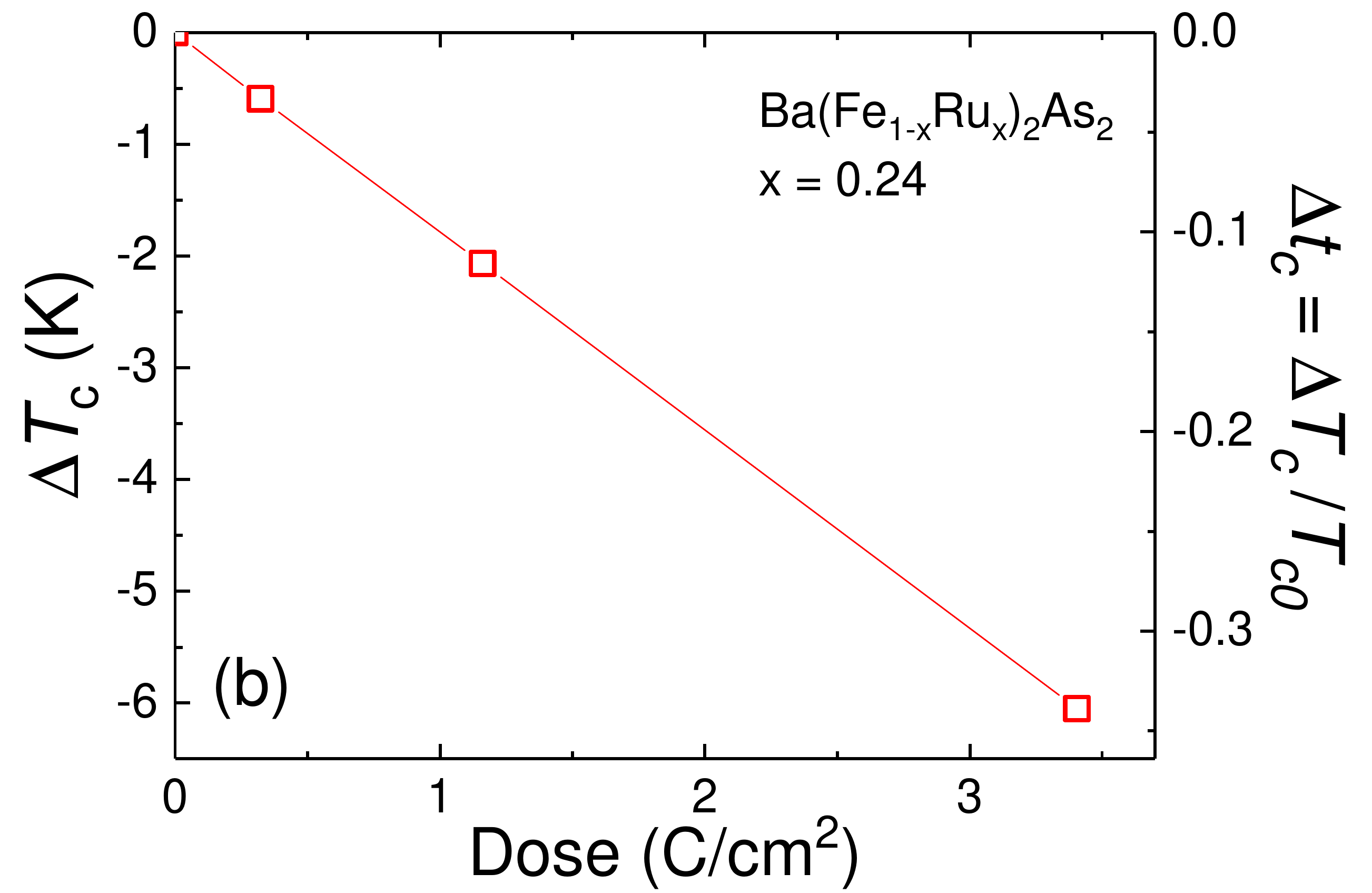}
\caption{(Color online) Suppression of $T_c$: $\Delta T_c$ (left) and $\Delta t_c = \Delta T_c / T_{c0}$ (right) of Ba(Fe$_{1-x}$Ru$_x$)$_{2}$As$_{2}$ ($x$ = 0.24) upon electron irradiation. Data from Ref.~\onlinecite{Prozorov2014PRX_e-irr}.}
\label{fig2-3-2_BaRu122} 
\end{figure}

\begin{figure}[htb]
\includegraphics[width=7cm]{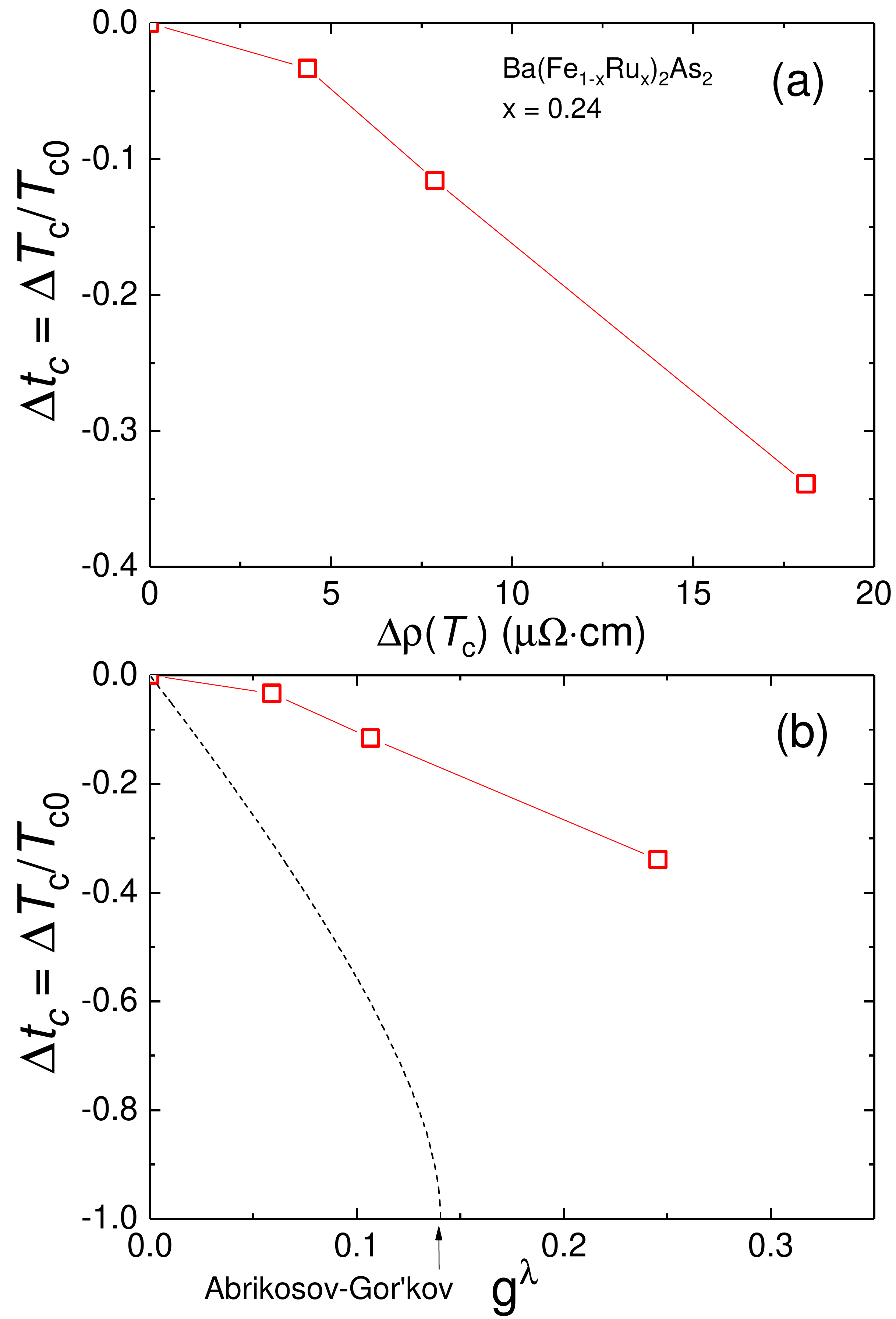}
\caption{(Color online) (a) $\Delta t_c = \Delta T_c / T_{c0}$ versus $\Delta \rho$ of Ba(Fe$_{1-x}$Ru$_x$)$_{2}$As$_{2}$ ($x$ = 0.24) upon electron irradiation. The dimensionless scattering rate $g^\lambda$ is calculated from resistivity and the penetration depth following equation~\ref{eq06}. Here $\lambda_0$ = 200 nm was used. Abrikosov-Gor'kov calculation is also shown for comparison. Data from Ref.~\onlinecite{Prozorov2014PRX_e-irr}.}
\label{fig2-3-3_BaRu122}
\end{figure}

Similar to other substitutions in 122 family FeSCs, isovalent ruthenium substitution on iron-site of  BaFe$_{2}$As$_{2}$ also suppresses long-range magnetic order and induces superconductivity with range of bulk coexistence, see composition phase diagram in  Fig.~\ref{fig2-0-1_122_Family} (c). Unlike the electron-doped FeSCs, the structural and magnetic transitions remain coincident
in temperature. The compensation condition between hole and electron carriers doesn't change in this compound~\cite{Rullier-Albenque2010PRB_BaRu122, DhakaCanfieldKaminski2013PRL_BaRu122, Thaler2010PRB_BaRu122, Xu2012PRB_BaRu122}. Since the quantum critical point was discovered in nodal-gap superconductor BaFe$_2$(As$_{1-x}$P$_x$)$_2$, it is interesting to see the effect of Ru-substitution as another isovalently substituted compound.

Prozorov \textit{et al}. conducted $in$-$situ$ and $ex$-$situ$ measurements of the resistivity in a slightly under-doped single crystal of Ba(Fe$_{1-x}$Ru$_x$)$_{2}$As$_{2}$ ($x$ = 0.24) with increasing dose of 2.5 MeV electron irradiation as shown in Fig.~\ref{fig2-3-1_BaRu122}. The suppression of $T_c$ is summarized in Fig.~\ref{fig2-3-2_BaRu122}. Furthermore, the dimensionless scattering rate $g^\lambda$ is calculated following equation~\ref{eq06} and plotted in Fig.~\ref{fig2-3-3_BaRu122}. In general, the rapid suppression of $T_c$ is observed, which cannot be explained by $s_{++}$ scenario, but supports $s_{\pm}$ pairing mechanism. The rate of suppression is much slower thatn that of BaFe$_2$(As$_{1-x}$P$_x$)$_2$.

\subsection{SrFe$_2$(As$_{1-x}$P$_x$)$_2$}

\begin{figure}[htb]
\includegraphics[width=7cm]{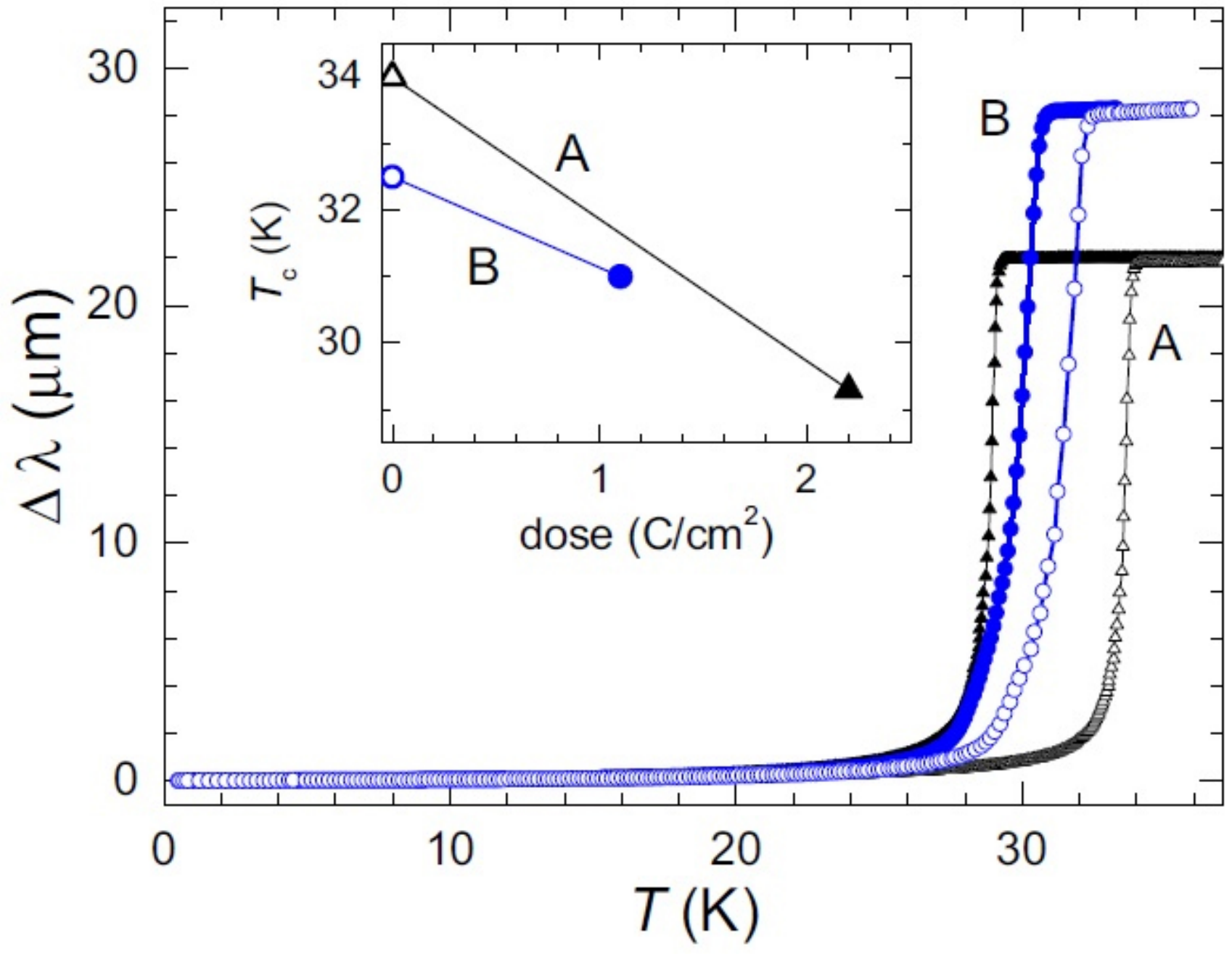}
\caption{(Color online) Full temperature range variation of $\Delta \lambda (T)$, in two single crystals of SrFe$_2$(As$_{1-x}$P$_x$)$_2$, $x$ = 0.35, A (black triangles) and B (blue circles) before (open symbols) and after (solid symbols) electron irradiation with doses of 2.2 and 1.1 C/cm$^2$, respectively. The inset shows the change of $T_c$ as a function of the irradiation dose. Reprinted with permission from Ref.~\onlinecite{StrehlowProzorov2014PRB_SrP122}, copyright 2014 APS.}
\label{fig2-4-1_SrP122} 
\end{figure}

\begin{figure}[htb]
\includegraphics[width=8.5cm]{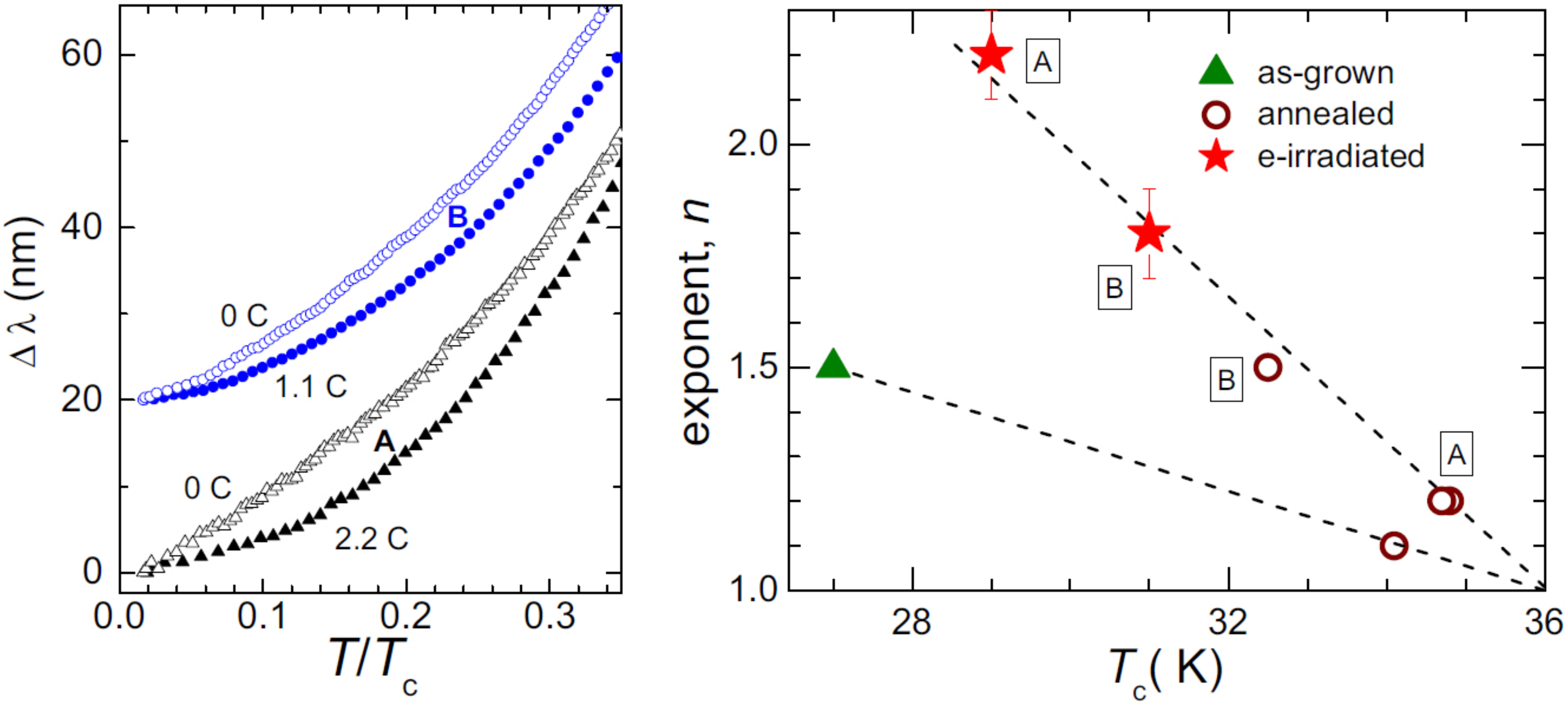}
\caption{(Color online) (a) Low-temperature variation of $\Delta \lambda$ in SrFe$_2$(As$_{1-x}$P$_x$)$_2$ (x = 0.35) versus reduced temperature $T/T_c$. The data before and after irradiation are shown in open and solid symbols, respectively. Offset of 20 nm is applied to avoid overlapping. (b) The exponent $n$ of the power-law fit of $\Delta \lambda$. Note the significantly smaller exponents for as-grown and annealed compared to the samples with irradiation defects. Reprinted with permission from Ref.~\onlinecite{StrehlowProzorov2014PRB_SrP122}, copyright 2014 APS.}
\label{fig2-4-2_SrP122} 
\end{figure}

\begin{figure}[htb]
\includegraphics[width=7cm]{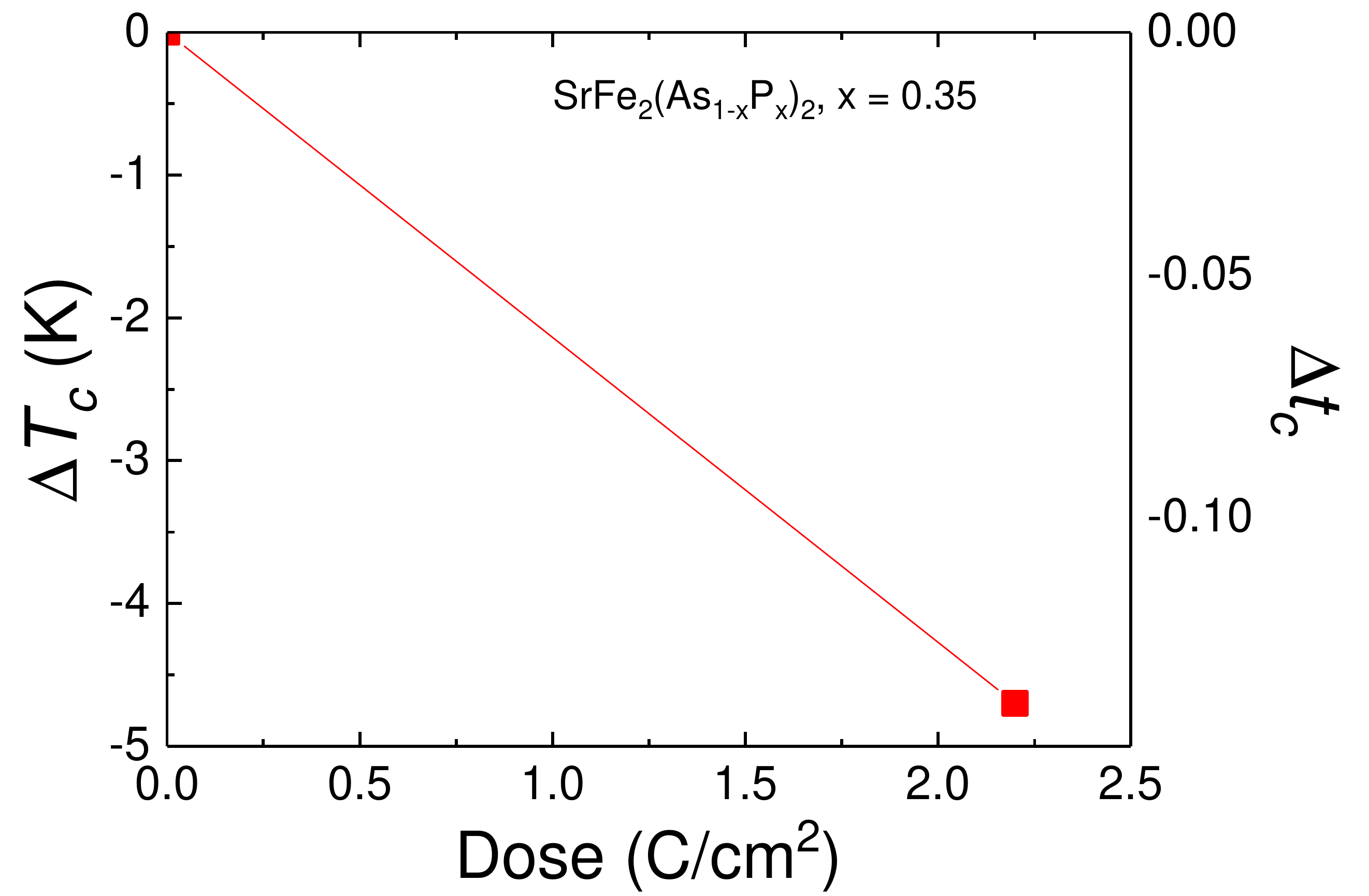}
\caption{(Color online) Suppression of $T_c$: $\Delta T_c$ (left) and $\Delta t_c = \Delta T_c / T_{c0}$ (right) of SrFe$_2$(As$_{1-x}$P$_x$)$_2$ ($x$ = 0.35) upon electron irradiation. The sample A with higher $T_c$ is plotted. Data from Ref.~\onlinecite{StrehlowProzorov2014PRB_SrP122}.}
\label{fig2-4-3_SrP122} 
\end{figure}

The phase diagram of another 122 compound with isovalent substitution, SrFe$_2$(As$_{1-x}$P$_x$)$_2$, with the maximum value of $T_c \sim$30~K, is very similar to BaFe$_2$(As$_{1-x}$P$_x$)$_2$ (Fig.~\ref{fig2-0-1_122_Family} (d)). In particular, it also shows the nodal superconducting gaps~\cite{DulguunTajima2012PRB_SrP122_NMR_SpecificHeat, Murphy2013PRB_SrP122}. Specific heat and NMR studies are consistent with the nodal small gap and nodeless large gaps~\cite{DulguunTajima2012PRB_SrP122_NMR_SpecificHeat}. According to the analysis of the low-temperature behavior of the London penetration depth, the superconducting gap of SrFe$_2$(As$_{1-x}$P$_x$)$_2$ is consistent with the presence of line nodes in the gap~\cite{Murphy2013PRB_SrP122}, very similar to BaFe$_2$(As$_{1-x}$P$_x$)$_2$. 

In order to understand the origin of nodal gap, Strehlow \textit{et al.} studied the effect of electron irradiation by measuring the London penetration depth before and after irradiation~\cite{StrehlowProzorov2014PRB_SrP122}. As shown in Fig.~\ref{fig2-4-1_SrP122}, the electron irradiation effectively suppressed $T_c$ of optimally doped SrFe$_2$(As$_{1-x}$P$_x$)$_2$ (x = 0.35). Upon irradiation, the low-temperature penetration depth shows increase of the power-law exponent ($n$). Interestingly, this exponent exceeds the value of $n =$ 2 (Fig.~\ref{fig2-4-2_SrP122}) suggesting that the nodes in superconducting gap are of accidental type, not symmetry-imposed. In Fig.~\ref{fig2-4-3_SrP122}, the $\Delta T_c$ and $\Delta T_c / T_{c0}$ are plotted against dose of irradiation only for sample A (higher $T_c$, clean sample). This will be compared with other 122 compounds later in the section of discussion. Due to lack of resistivity data, $g^{\lambda}$ is not calculated.

\subsection{Electron-doped Ba(Fe$_{1-x}$Co$_x$)$_{2}$As$_{2}$}

\begin{figure}[htb]
\includegraphics[width=7cm]{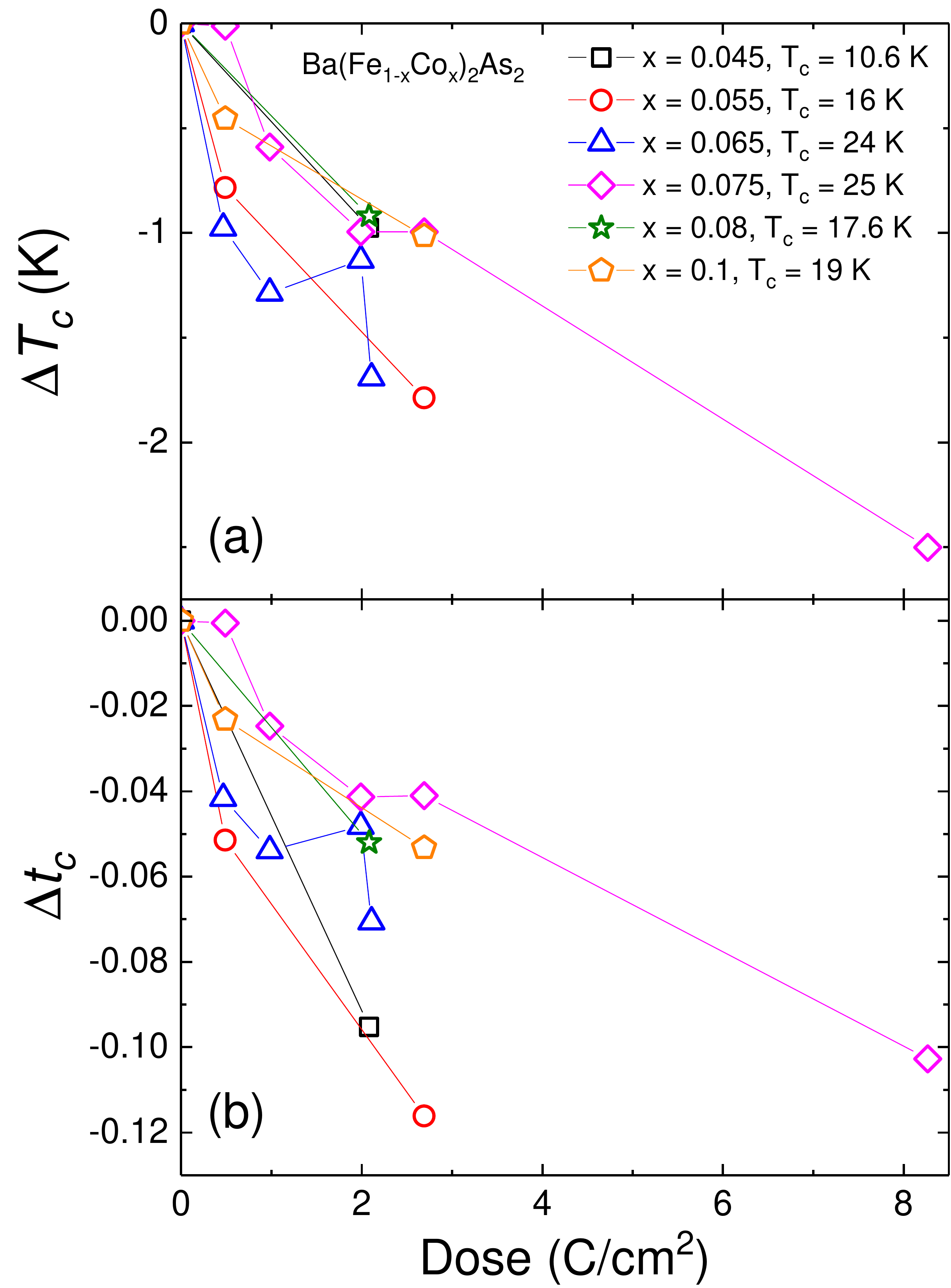}
\caption{(Color online) Suppression of $T_c$: (a) $\Delta T_c$ and (b) $\Delta t_c = \Delta T_c / T_{c0}$ versus irradiation dose in Ba(Fe$_{1-x}$Co$_x$)$_2$As$_2$. Data from Ref.~\onlinecite{VanDerBeek2013JPCS_e-irr_review}.}
\label{fig2-5-1_BaCo122}
\end{figure}

\begin{figure}[htb]
\includegraphics[width=7cm]{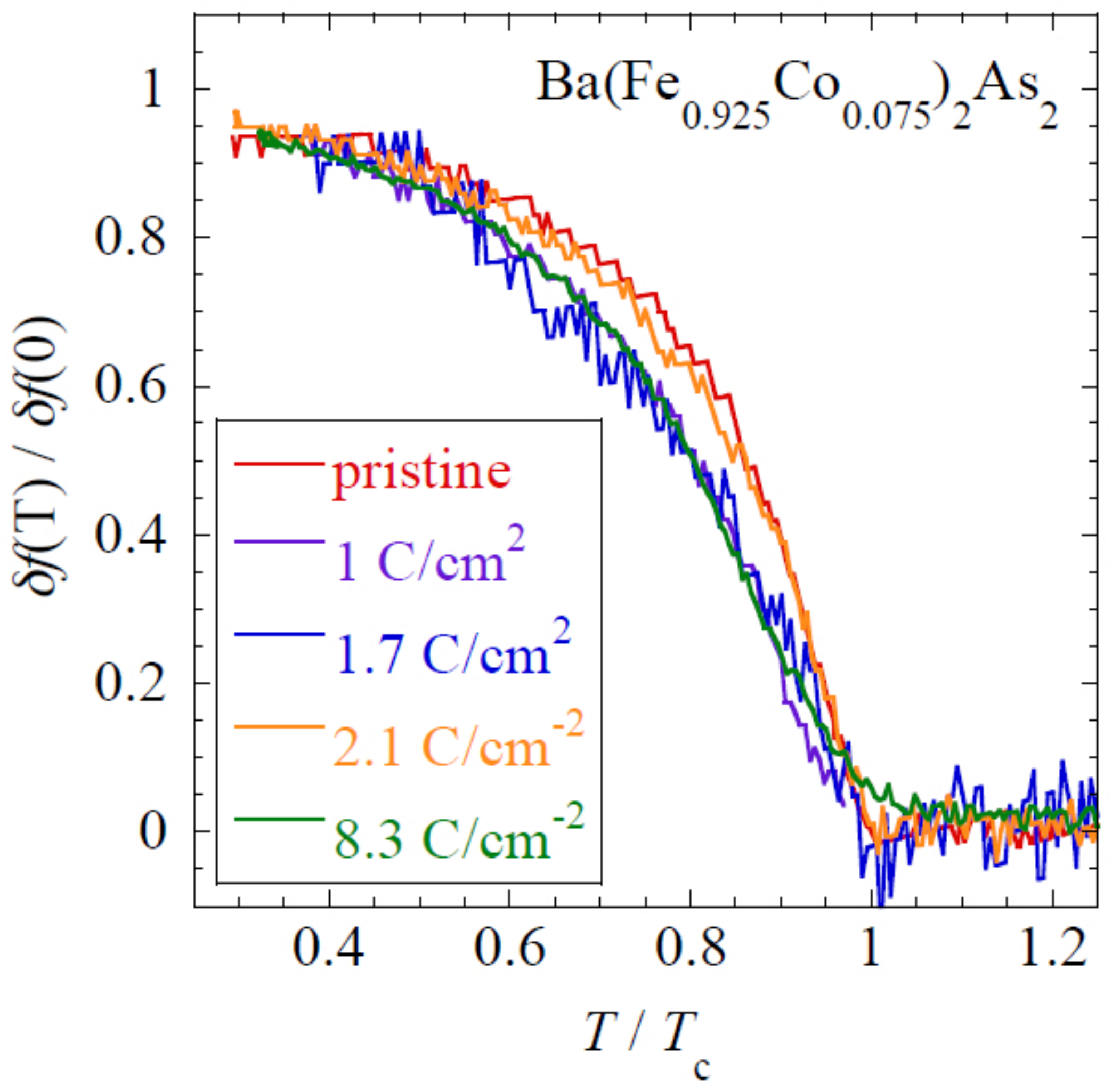}
\caption{(Color online) Superfluid density of Ba(Fe$_{1-x}$Co$_x$)$_{2}$As$_{2}$ ($x$ = 0.075) upon electron irradiation by using microwave cavity perturbation technique. Reprinted with permission from Ref.~\onlinecite{VanDerBeek2013JPCS_e-irr_review}, copyright IOP.}
\label{fig2-5-2_BaCo122}
\end{figure}

The competition between superconductivity and magnetic phase has been intensively investigated  in electron-doped Ba(Fe$_{1-x}$Co$_x$)$_{2}$As$_{2}$  [6-8]. A clear separation between the temperatures of structural transition ($T_s$) and the magnetic phase transition ($T_N$) with doping is found, as shown in Fig.~\ref{fig2-0-1_122_Family} (e)~\cite{NandiGoldman2010PRL}, which is different from the compounds with hole-doping and isovalent substitution.  
 
The effect of electron irradiation on Ba(Fe$_{1-x}$Co$_x$)$_{2}$As$_{2}$ was mainly investigated by van der Beek \textit{et al.}~\cite{VanDerBeek2013JPCS_e-irr_review}. The electron irradiation effectively suppresses $T_c$ of Ba(Fe$_{1-x}$Co$_x$)$_{2}$As$_{2}$ as shown in Fig.~\ref{fig2-5-1_BaCo122}. The largest suppression occurs in heavily under-doped and overdoped regions in which gap is strongly anisotropic and nodal~\cite{TanatarProzorov2010PRL_BaCo122, ReidTanatarProzorovTaillefer2010PRB_BaCo122}. The suppression is the weakest near the optimal doping region similar to other 122 compounds. Since there  are no reports on variation of  resistivity upon electron irradiation for various compositions, van der Beek \textit{et al.} estimated the scattering parameter ($z \Gamma / 2\pi T_c$) based on the density of states, effective mass, atomic point defect density, scattering angle, and so on (See details in Ref.~\onlinecite{VanDerBeek2013JPCS_e-irr_review}). While van der Beek \textit{et al.} mentioned that $\delta R/R \sim 0.05 [\text{C cm}^{-2}]^{-1}$, the actual variations of resistivity ($\Delta \rho$) are not available for all compositions. Nakajima \textit{et al.} estimated the dimensionless parameter based on $\Delta \rho$ by proton irradiation which is likely to result from clusters of defects instead of point defects~\cite{Nakajima2010PRB_BaCo122_proton}. Since we limit our scope to electron irradiation, $g^{\lambda}$ of Ba(Fe$_{1-x}$Co$_x$)$_{2}$As$_{2}$  is not estimated.

Van der Beek \textit{et al.} also used microwave cavity perturbation technique to measure surface impedance, and studied the variation of the superfluid density of the optimally doped Ba(Fe$_{1-x}$Co$_x$)$_{2}$As$_{2}$ ($x$ = 0.075) upon electron irradiation in Fig.~\ref{fig2-5-2_BaCo122}. The normalized frequency shift, which is proportional to the superfluid density $n_s \propto \lambda^{-2}$, shows little to no change upon irradiation while $T_c$ drops by 10 $\%$. This suggests that the isotropic superconducting gaps with s$_\pm$-pairing symmetry are intact upon irradiation.

\subsection{Electron-doped Ba(Fe$_{1-x}$Ni$_x$)$_2$As$_2$}

\begin{figure}[htb]
\includegraphics[width=7cm]{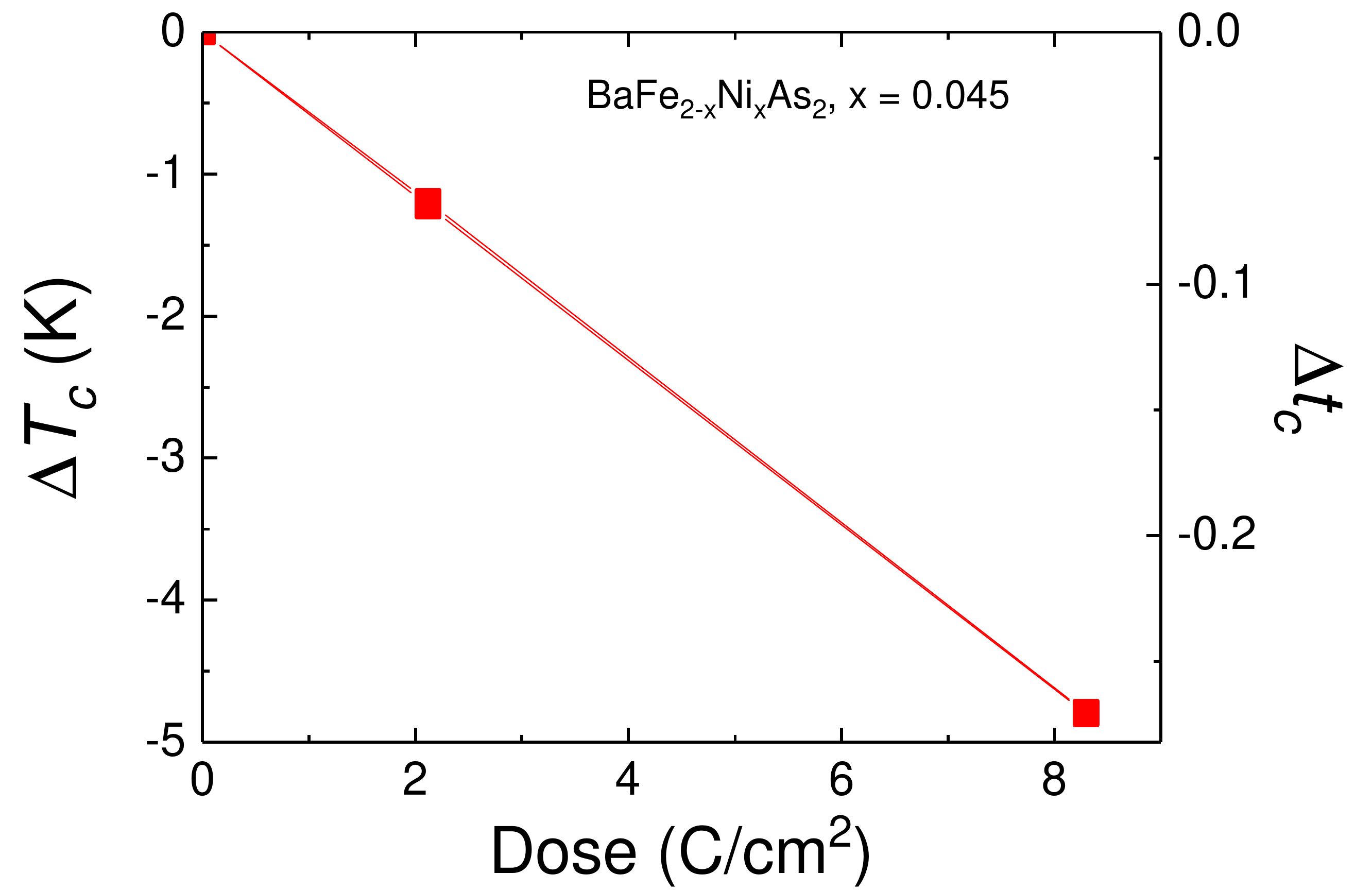}
\caption{(Color online) Suppression of $T_c$: $\Delta T_c$ (left) and $\Delta t_c = \Delta T_c / T_{c0}$ (right) versus dose of BaFe$_{2-x}$Ni$_x$As$_2$. Data from Ref.~\onlinecite{VanDerBeek2013JPCS_e-irr_review}.}
\label{fig2-6-1_BaNi122} 
\end{figure} 

Ba(Fe$_{1-x}$Ni$_x$)$_2$As$_2$ is an electron-doped 122 FeSC~\cite{LiXu2009NJP_BaNi122} in which every Ni donates two electrons in contrast to Co substitution that donates
only one electron~\cite{Sefat2008PRL_BaCo122}. Similar to other iron - based superconductors, Ba(Fe$_{1-x}$Ni$_x$)$_2$As$_2$ system shows a superconducting dome with the total suppression of static AF order near the optimal doping level x $\sim$ 0.5~\cite{Abdel-Hafiez2015PRB_BaNi122}. As commonly found in electron-doped systems, the separation between structural ($T_s$) and magnetic ($T_m$) transitions is also observed in this compound. However, the details vary among various studies. High-resolution synchrotron x-ray and neutron scattering study shows sharp first-order like disappearance of magnetic ordering above the optimally doped region, and the authors interpreted it as an avoidance of quantum criticality~\cite{LuDai2013PRL_BaNI122} as shown in Fig.~\ref{fig2-0-1_122_Family} (f). In NMR study, the separation was interpreted as an evidence of two critical points at $x_{c1}$ = 0.05 and $x_{c2}$ = 0.07, respectively. Since the highest $T_c$ is found around $x_{c1}$, it is claimed that the superconductivity is more closely tied to the magnetic quantum critical point.

The effect of electron irradiation on Ba(Fe$_{1-x}$Ni$_x$)$_{2}$As$_{2}$ ($x$ = 0.045) was studied by van der Beek \textit{et al.}~\cite{VanDerBeek2013JPCS_e-irr_review}. The suppression of $T_c$ is shown in Fig.~\ref{fig2-6-1_BaNi122}. Since the resistivity data are not available, the dimensionless parameter ($g^{\lambda}$) is not estimated.
 
\section{Discussion}
   
\begin{figure}[htb]
\includegraphics[width=8.5cm]{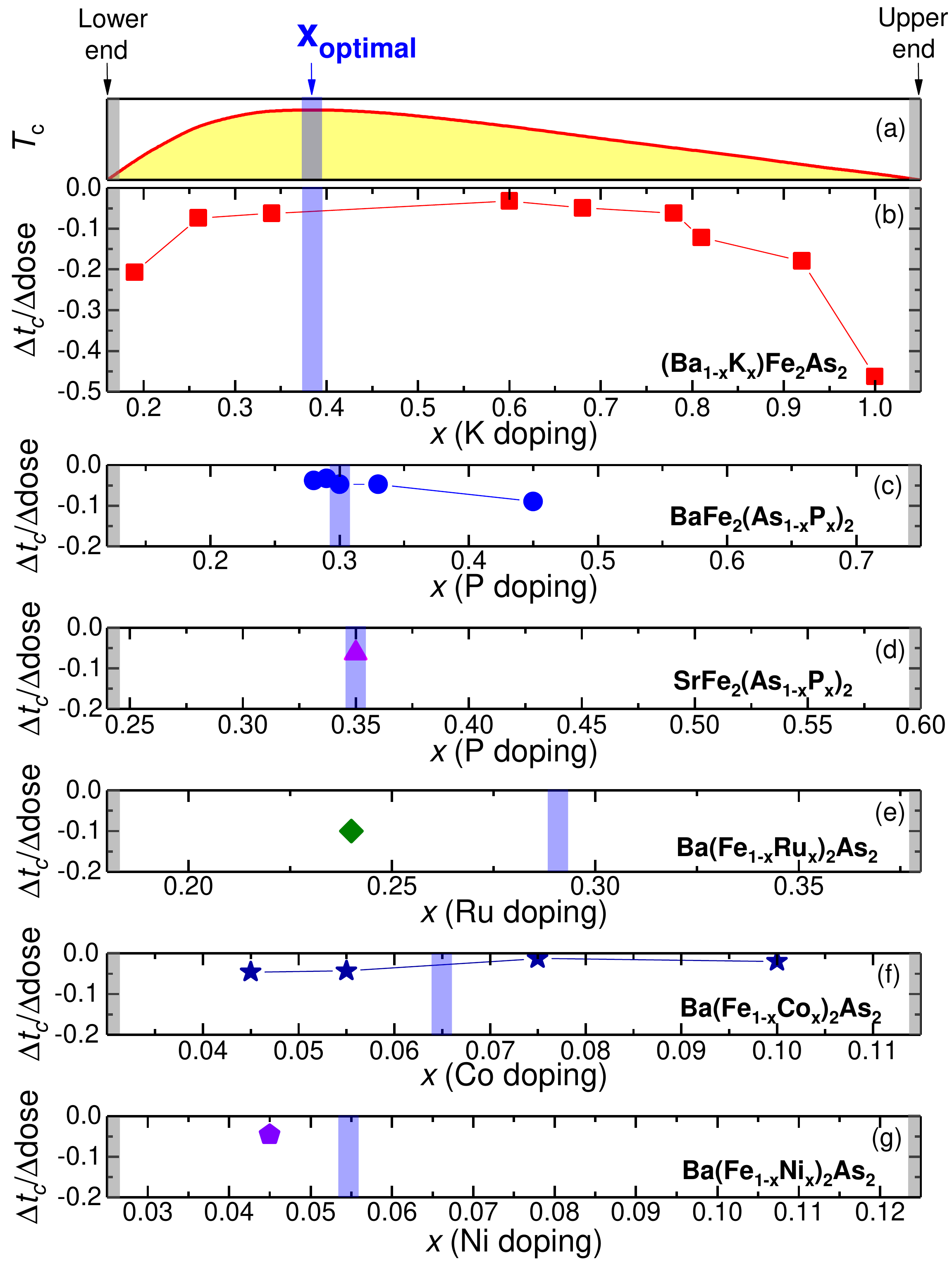}
\caption{(Color online) $\Delta t_c / \Delta \text{dose}$ versus $x$ of 122 FeSCs. (a) Schematic $T_c$ - $x$ phase diagram. (b) - (g) $\Delta t_c / \Delta \text{dose}$ versus $x$ of (Ba$_{1-x}$K$_x$)Fe$_2$As$_2$, BaFe$_2$(As$_{1-x}$P$_x$)$_2$, SrFe$_2$(As$_{1-x}$P$_x$)$_2$, Ba(Fe$_{1-x}$Ru$_x$)$_2$As$_2$, Ba(Fe$_{1-x}$Co$_x$)$_2$As$_2$, and Ba(Fe$_{1-x}$Ni$_x$)$_2$As$_2$, respectively. Note that the blue shaded area indicates the optimally doped composition ($x_{optimal}$) with maximum $T_c$ for all panels. Approximate upper and lower ends of superconducting dome ($T_c$ = 0) are marked as gray shaded area.} 
\label{fig3-1_122_comparison}
\end{figure}

\begin{figure}[htb]
\includegraphics[width=7cm]{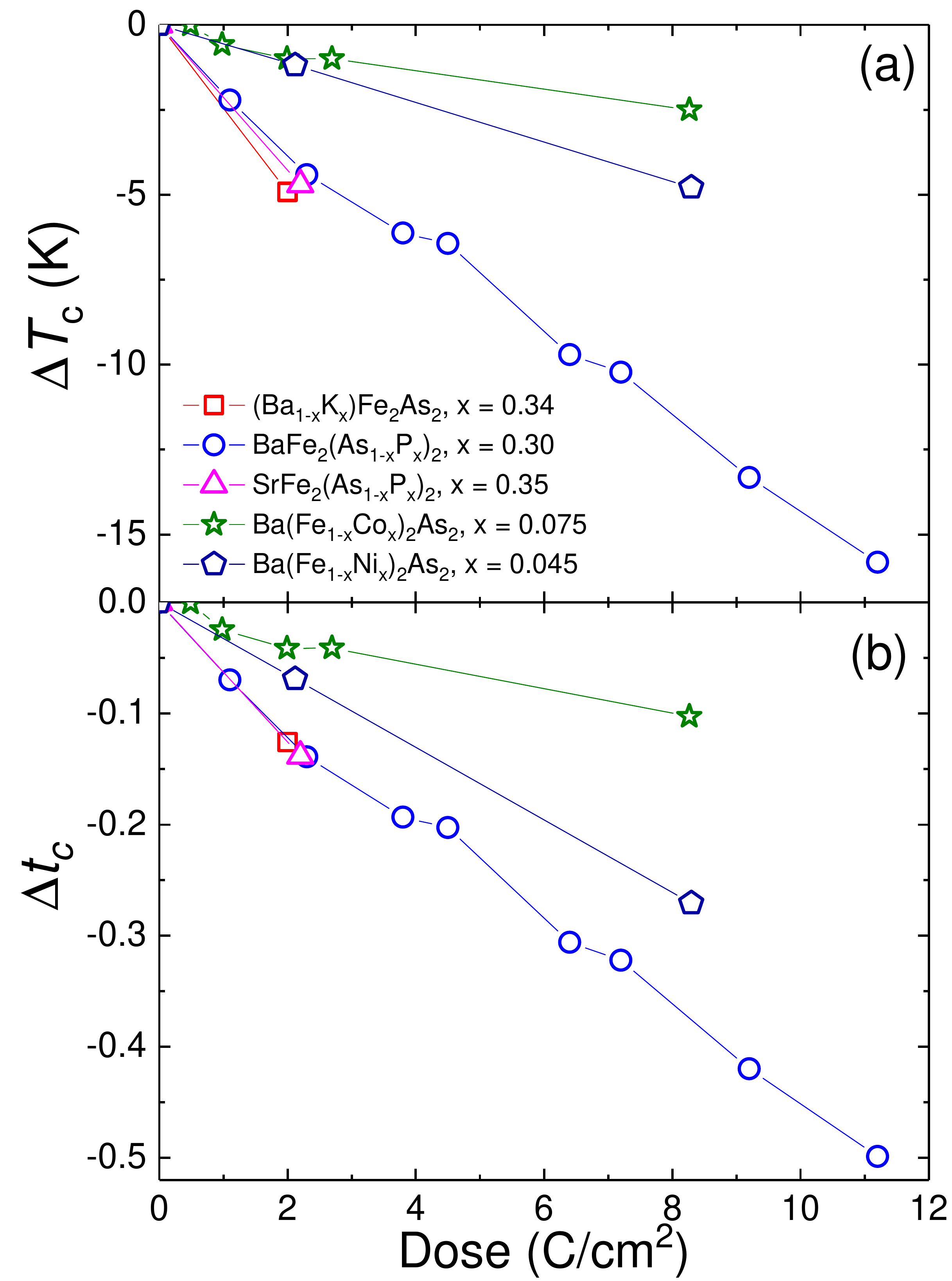}
\caption{(Color online) Suppression of $T_c$: (a) $\Delta T_c$ and (b) $\Delta t_c = \Delta T_c /T_{c0}$ upon electron irradiation in optimally doped 122 FeSCs.}
\label{fig3-2_122_comparison}
\end{figure}

\begin{figure}[htb]
\includegraphics[width=7cm]{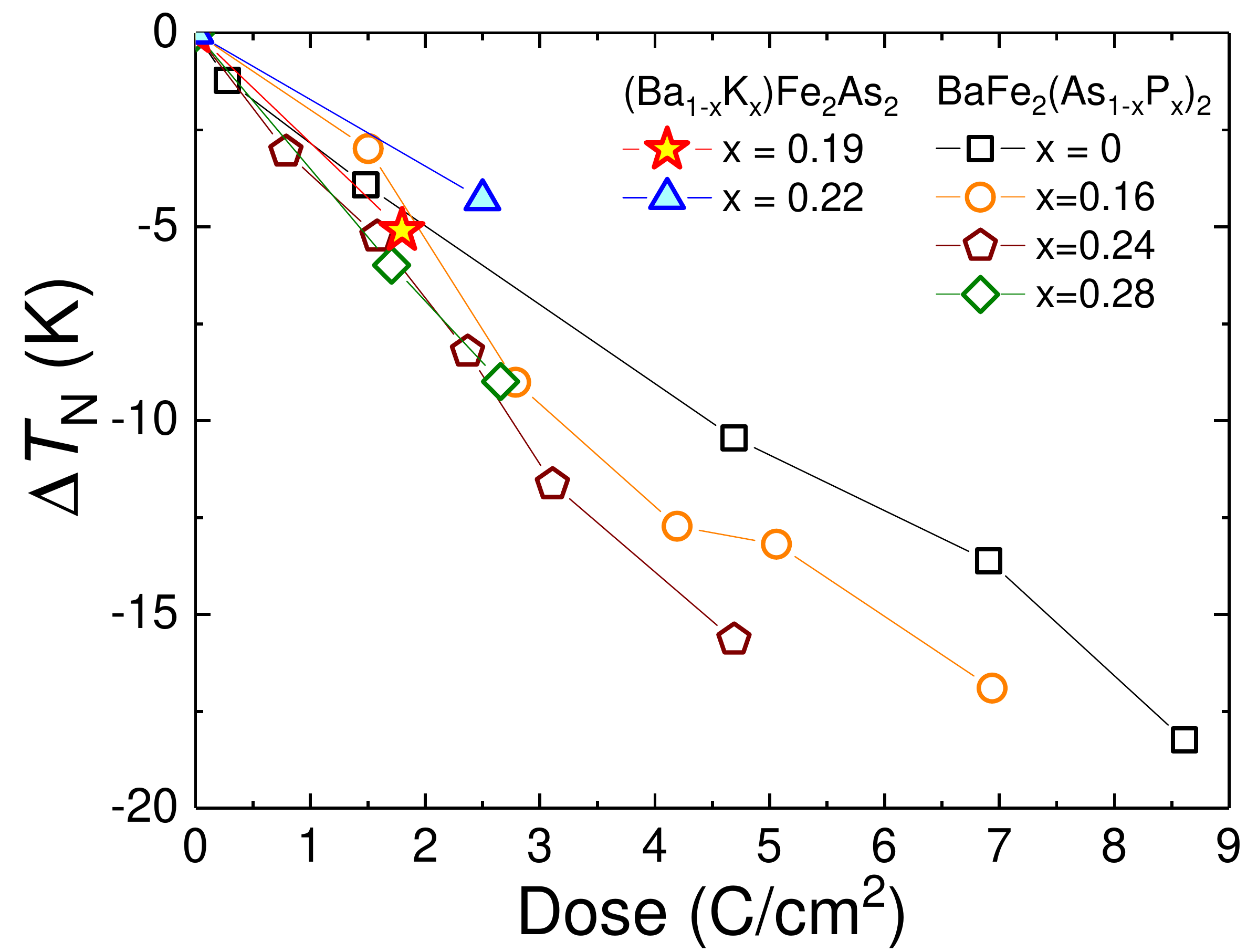}
\caption{(Color online) Comparison of $\Delta T_N$ upon electron irradiation in under-doped (Ba$_{1-x}$K$_x$)Fe$_2$As$_2$ and BaFe$_2$(As$_{1-x}$P$_x$)$_2$.}
\label{fig3-3_122_comparison}
\end{figure}

\begin{figure}[htb]
\includegraphics[width=8.5cm]{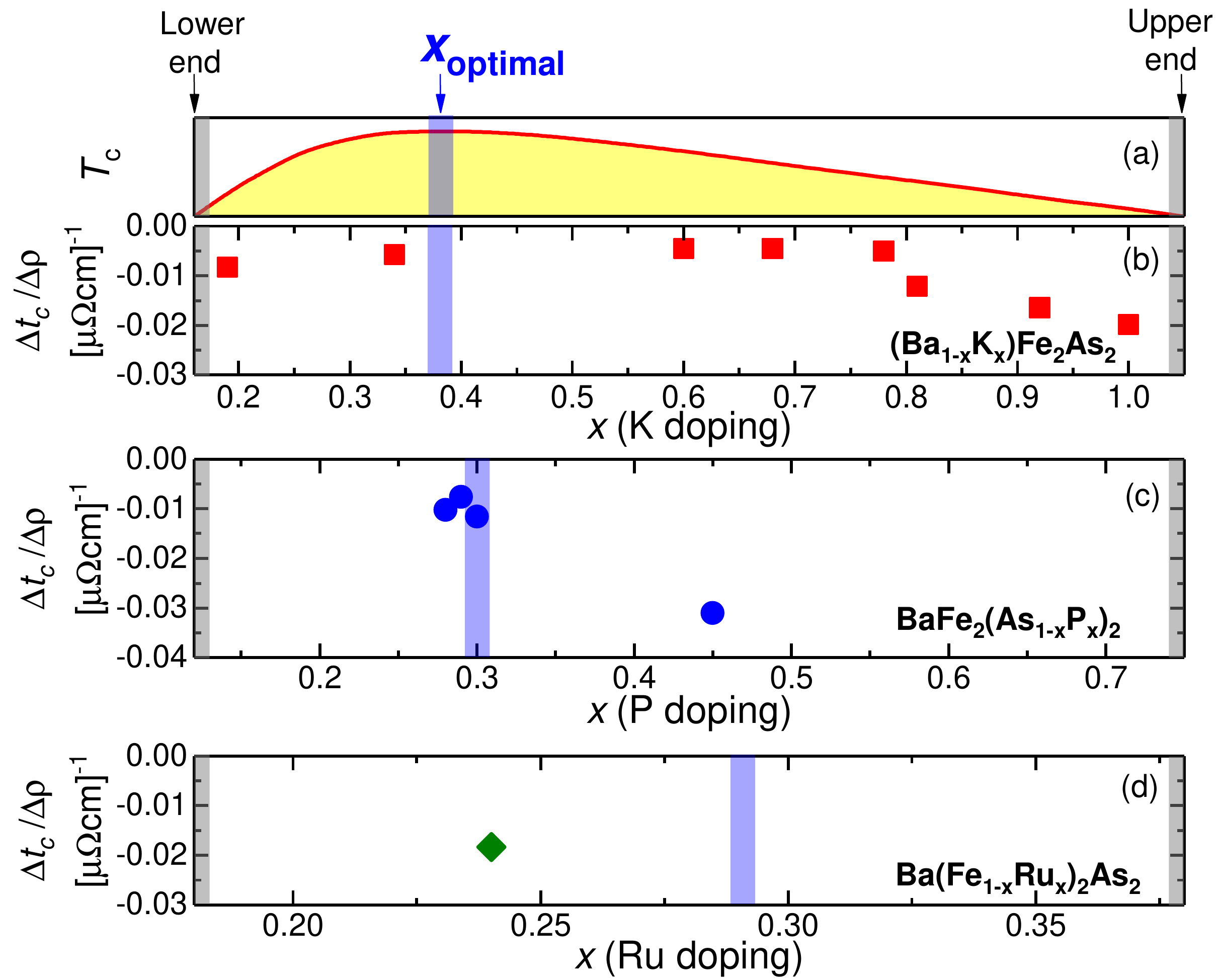}
\caption{(Color online) $\Delta t_c / \Delta \rho$ versus compositions of 122 FeSCs. (a) Schematic $T_c$ - $x$ phase diagram. (b) - (d) $\Delta t_c / \Delta \text{dose}$ versus $x$ in (Ba$_{1-x}$K$_x$)Fe$_2$As$_2$, BaFe$_2$(As$_{1-x}$P$_x$)$_2$, and Ba(Fe$_{1-x}$Ru$_x$)$_2$As$_2$, respectively. Note that the blue shaded area indicates the optimally doped composition ($x_{optimal}$) with maximum $T_c$ for all panels. Approximate upper and lower ends of superconducting dome ($T_c$ = 0) are marked as gray shaded area.}
\label{fig3-4_122_comparison}
\end{figure}

\begin{figure}[htb]
\includegraphics[width=8.5cm]{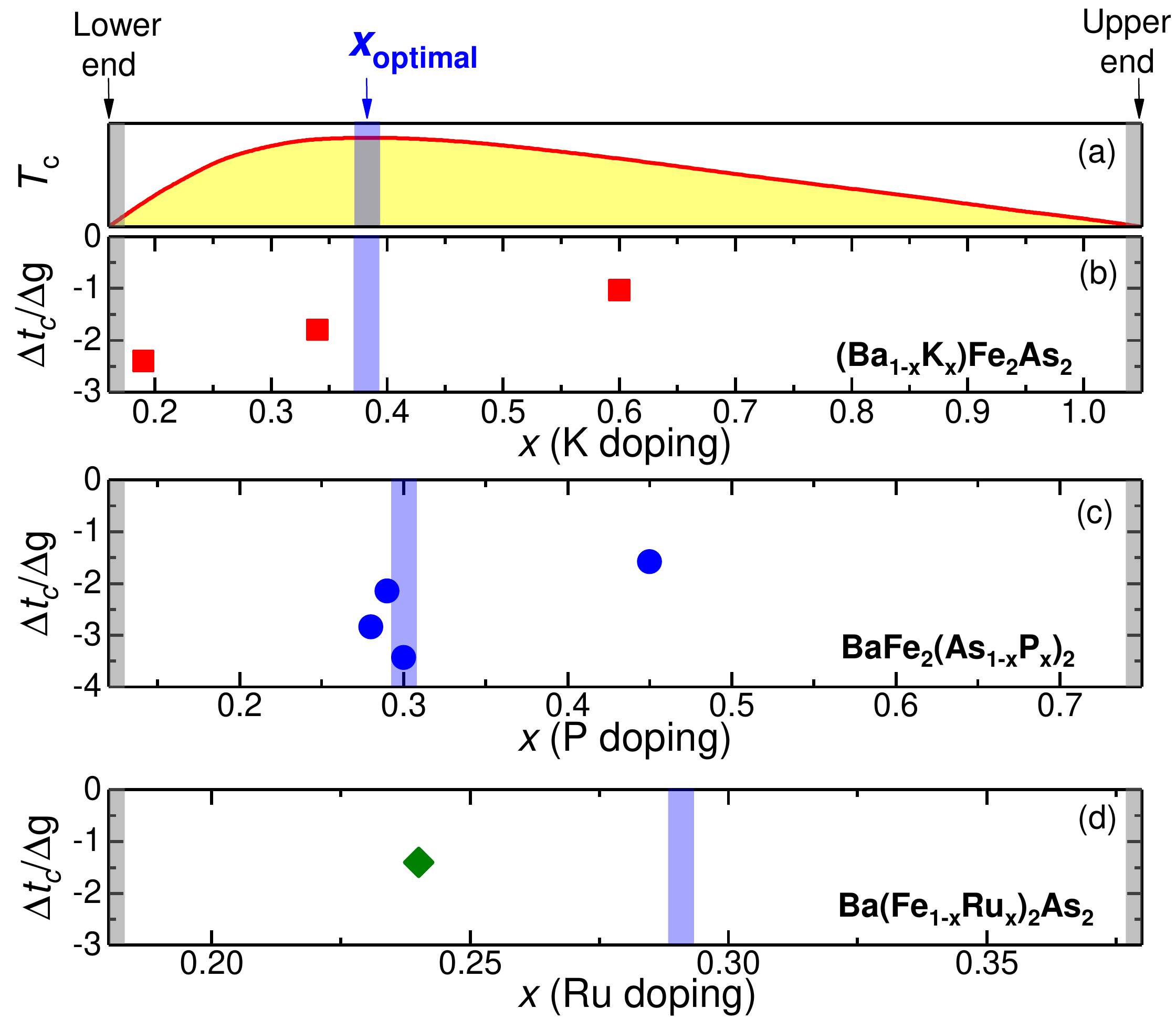}
\caption{(Color online) $\Delta t_c / \Delta \text{g}$ versus $x$ of 122 FeSCs. (a) Schematic $T_c$ versus $x$ phase diagram. (b)-(d) $\Delta t_c / \Delta \text{dose}$ versus $x$ in (Ba$_{1-x}$K$_x$)Fe$_2$As$_2$, BaFe$_2$(As$_{1-x}$P$_x$)$_2$, and Ba(Fe$_{1-x}$Ru$_x$)$_2$As$_2$, respectively. Note that the blue shaded area indicates the optimally doped composition ($x_{optimal}$) with maximum $T_c$ for all panels. Approximate upper and lower ends of superconducting dome ($T_c$ = 0) are marked as gray shaded area.}
\label{fig3-5_122_comparison}
\end{figure}

\begin{figure}[htb]
\includegraphics[width=7cm]{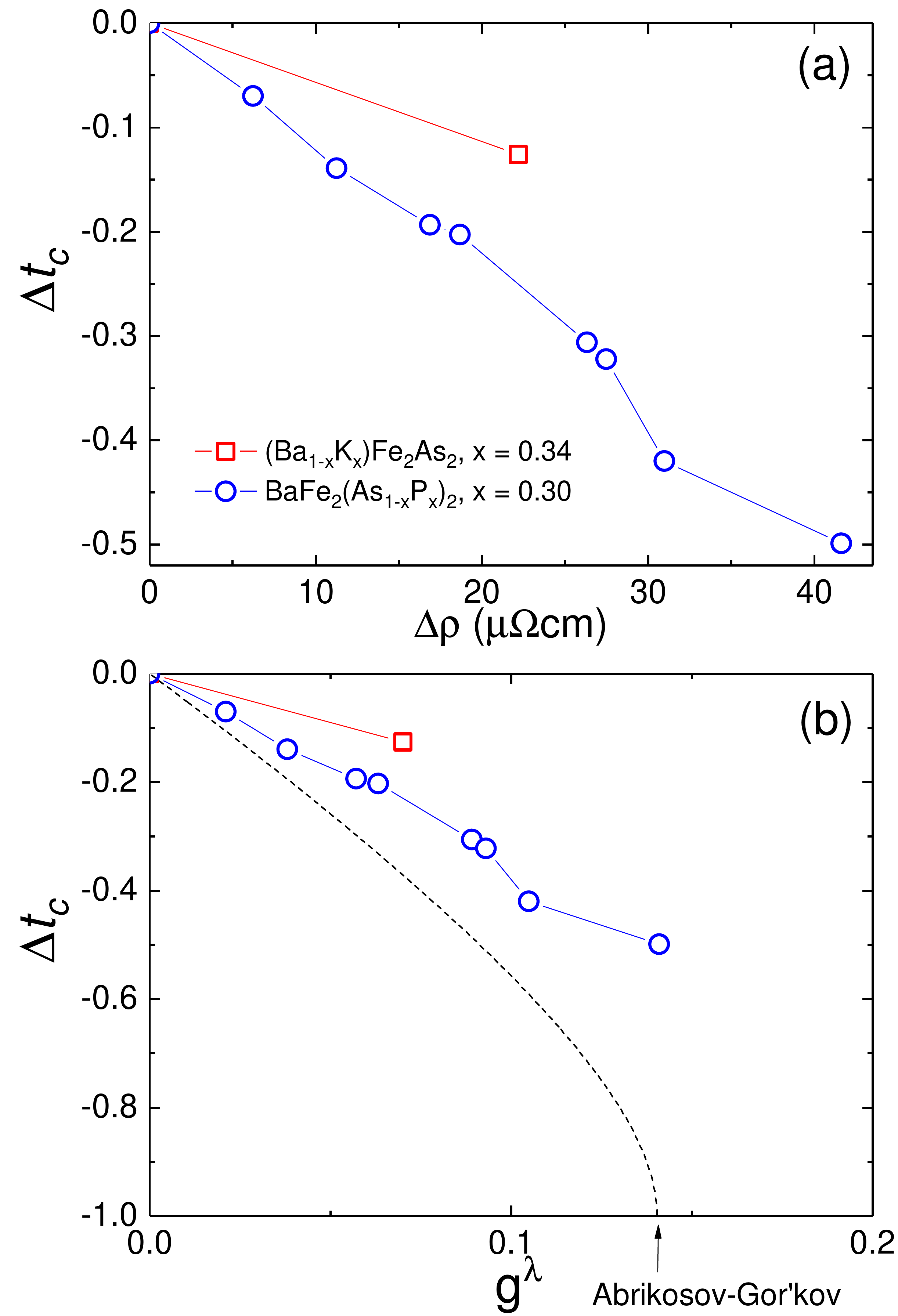}
\caption{(Color online) Comparison of optimally doped Ba$_{1-x}$K$_x$Fe$_2$As$_2$ and BaFe$_2$(As$_{1-x}$P$_x$)$_2$ compounds: (a) $\Delta t_c$ versus $\Delta \rho$ and (b) $\Delta t_c$ versus the dimensionless scattering parameter ($g^\lambda$) calculated following equation~\ref{eq06}.}
\label{fig3-6_122_comparison}
\end{figure}

\begin{figure}[htb]
\includegraphics[width=8.5cm]{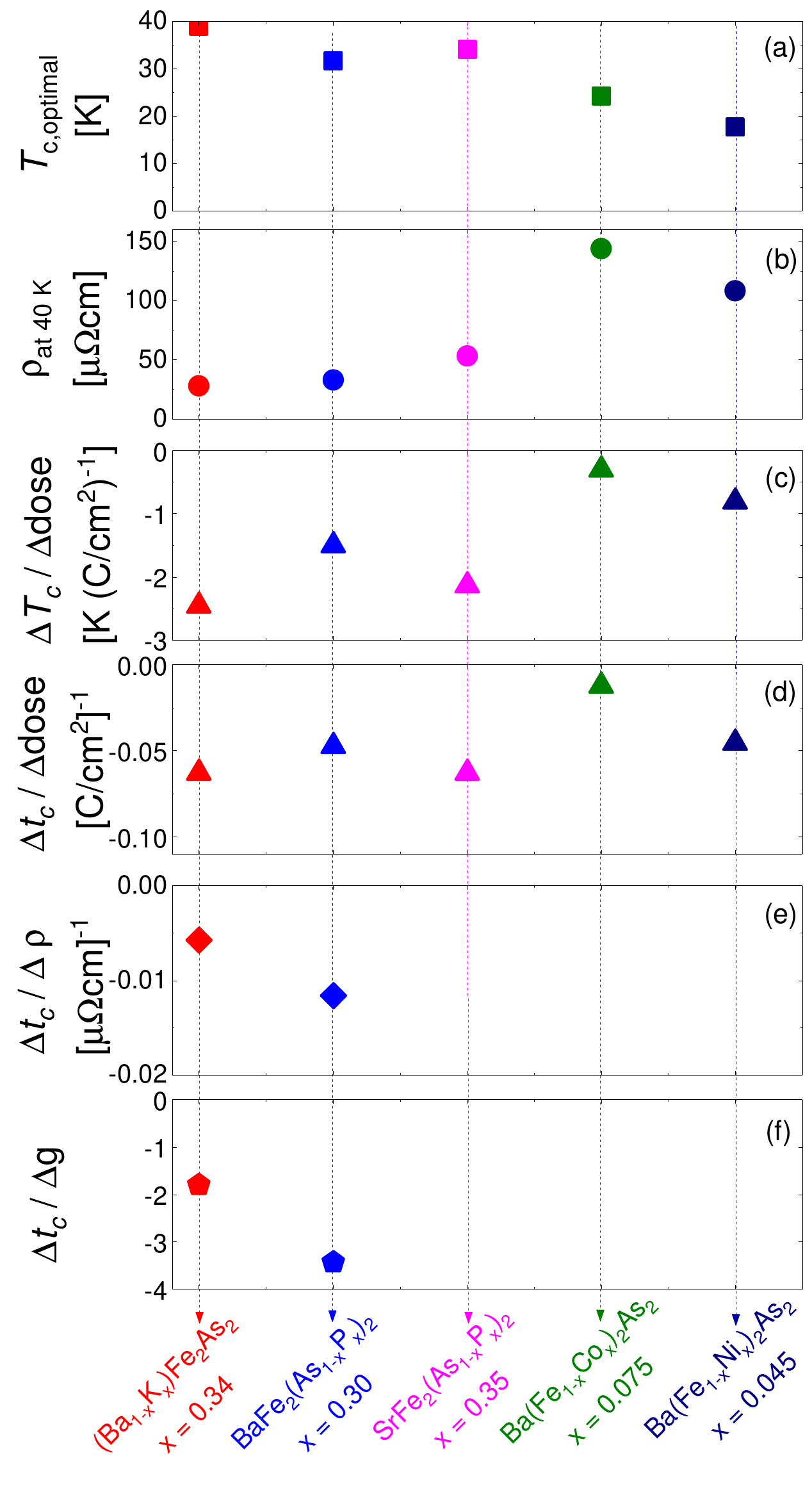}
\caption{(Color online) Summary of key parameters in optimally-doped 122 FeSCs: (a) $T_c$, (b) $\rho$ at 40 K, (c) $\Delta T_c / \Delta \text{dose}$, (d) $\Delta t_c / \Delta \text{dose}$, (e) $\Delta t_c / \Delta \rho$, and (f) $\Delta t_c / \Delta \text{g}$.}
\label{fig3-7_122_comparison}
\end{figure}

In this section, we summarize and compare the $T_c$ suppression rates upon electron irradiation of various 122 FeSCs. We use the normalized transition temperature $t_c=T_c/T_{c0}$, where $T_{c0}$ is the value in pristine samples. In Fig.~\ref{fig3-1_122_comparison} we compare $\Delta t_c / \Delta \text{dose}$ versus $x$ of all six compounds. For convenience of comparison, the range of $x$ is limited to superconducting compositions only. The schematic $T_c$ - $x$ phase diagram is shown in panel (a) with marks of optimally doped region (blue shaded area) and two ends of the superconducting dome (gray shaded area). For panels (b) - (g), both ends with gray area are the end compositions of superconducting dome, and blue area indicates the optimally doped composition. Systematic studies that cover more than 50~$\%$ of superconducting composition are only available for (Ba$_{1-x}$K$_x$)Fe$_2$As$_2$ and Ba(Fe$_{1-x}$Co$_x$)$_2$As$_2$. For the case of (Ba$_{1-x}$K$_x$)Fe$_2$As$_2$, the value of $\Delta t_c / \Delta \text{dose}$ is small near optimal doping, and becomes larger as it goes toward under and over-doped compositions. In particular, the pure KFe$_2$As$_2$ shows the largest value, $\Delta t_c / \Delta \text{dose} \approx 0.5$. However, the $\Delta t_c / \Delta \text{dose}$ of  Ba(Fe$_{1-x}$Co$_x$)$_2$As$_2$ in panel (f) is small ($< 0.05$), and doesn't change much with compositions. These small values can be attributed to the fact that the pristine Ba(Fe$_{1-x}$Co$_x$)$_2$As$_2$ is already in the dirty limit as shown in Fig.~\ref{fig2-0-2_122_Family}, so additional disorder introduced by electron irradiation is less effective in changing its properties. For BaFe$_2$(As$_{1-x}$P$_x$)$_2$, only near optimal and slightly over doped compositions were studied. While the doping dependence of $\Delta t_c / \Delta \text{dose}$ is similar to (Ba$_{1-x}$K$_x$)Fe$_2$As$_2$, the comparison is not conclusive due to the limited range of composition studied. Studies for singular compositions were only performed for Ba(Fe$_{1-x}$Ru$_x$)$_2$As$_2$ and Ba(Fe$_{1-x}$Ni$_x$)$_2$As$_2$, so further studies are needed to find their doping dependence.

Another interesting aspect in Fig.~\ref{fig3-1_122_comparison} is that in general the optimally doped compositions of all 122 FeSCs commonly show lowest suppression with similar low values ($< 0.05$). To get additional insight into this fact,  the data for only optimally doped compositions are replotted in Fig.~\ref{fig3-2_122_comparison} (a) $\Delta T_c$ and (b) $\Delta t_c = \Delta T_c / T_{c0}$ versus increasing dosage. Two different trends of $T_c$ suppression are observed, panel (a). The first group includes hole-doped and isovalent substituted compounds: (Ba$_{1-x}$K$_x$)Fe$_2$As$_2$, BaFe$_2$(As$_{1-x}$P$_x$)$_2$, SrFe$_2$(As$_{1-x}$P$_x$)$_2$, and Ba(Fe$_{1-x}$Ru$_x$)$_2$As$_2$. The second group includes electron doped compounds: Ba(Fe$_{1-x}$Co$_x$)$_2$As$_2$ and Ba(Fe$_{1-x}$Ni$_x$)$_2$As$_2$. The former group commonly shows suppression rate of $\Delta T_c$ = -4 K [C/cm]$^{-1}$, regardless of different $T_c$ and different chemical contents. The reason why these different compounds show similar suppression behavior in absolute Kelvin scale remains unclear. For the latter group of electron doped compounds, the rate is about three times smaller. This slow suppression of $T_c$ in  electron doped compounds can be understood by the fact that the pristine samples are already in the dirty limit as clearly seen in resistivity measurement in the bottom panels of Fig.~\ref{fig2-0-2_122_Family}. 

In addition to $\Delta T_c$,  Fig.~\ref{fig3-3_122_comparison} shows comparison of $\Delta T_N$ for under-doped (Ba$_{1-x}$K$_x$)Fe$_2$As$_2$ and BaFe$_2$(As$_{1-x}$P$_x$)$_2$. Interestingly, these different compounds show similar suppression rates upon electron irradiation even though their $T_{N0}$ are notably different as well as chmical contents. This can be an indication that scattering mechanism in SDW phase are similar producing similar response upon electron irradiation. 

In Fig.~\ref{fig3-4_122_comparison}, the suppression rates per resistivity increase ($t_c / \Delta \rho$) in (Ba$_{1-x}$K$_x$)Fe$_2$As$_2$, BaFe$_2$(As$_{1-x}$P$_x$)$_2$, and Ba(Fe$_{1-x}$Ru$_x$)$_2$As$_2$ are summarized. The panel (a) of (Ba$_{1-x}$K$_x$)Fe$_2$As$_2$ shows a rather complete doping dependence. The largest value of $t_c / \Delta \rho$ is obtained in the pure KFe$_2$As$_2$, and the smallest value in the near-optimally doped compounds. In general, the value of $t_c / \Delta \rho$ increases away from the optimally doped composition. The similar trend is observed in BaFe$_2$(As$_{1-x}$P$_x$)$_2$, but the further experimental data in over and under doped compositions are needed. In Ba(Fe$_{1-x}$Ru$_x$)$_2$As$_2$, only one underdoped data result is available, which is comparable to those of underdoped  (Ba$_{1-x}$K$_x$)Fe$_2$As$_2$ and BaFe$_2$(As$_{1-x}$P$_x$)$_2$.

In Fig.~\ref{fig3-5_122_comparison}, $\Delta t_c / \Delta \text{g}$ of three compounds in Fig.~\ref{fig3-4_122_comparison} is summarized. In (Ba$_{1-x}$K$_x$)Fe$_2$As$_2$ (panel (b)), the most under doped compound shows the largest value, then weakens toward near-optimal and slightly over-doped region. Since $\lambda_0$ values are not available for over doped region, dimensionless parameter $g^{\lambda}$ in over doped region is not estimated. In general, one can find that $\Delta t_c / \Delta \text{g}$ of (Ba$_{1-x}$K$_x$)Fe$_2$As$_2$ has larger values for most of compositions than those of BaFe$_2$(As$_{1-x}$P$_x$)$_2$. For further comparison, the $\Delta t_c$ versus $g^\lambda$ of near optimally doped region is plotted in Fig.~\ref{fig3-6_122_comparison} (b). It is clear that the suppression of $T_c$ is larger in BaFe$_2$(As$_{1-x}$P$_x$)$_2$ . However, in all cases, the suppression rate is slower than the Abrikosov-Gor'kov value.

In Fig.~\ref{fig3-7_122_comparison}, the key parameters of optimally doped 122 FeSCs are summarized. Panel (a) shows $T_c$ of all compounds ranging the highest value in (Ba$_{1-x}$K$_x$)Fe$_2$As$_2$, and the lowest in Ba(Fe$_{1-x}$Ni$_x$)$_2$As$_2$. Since the initial resistivity values (as an initial disorder before irradiation) are important to understand the post-irradiated properties, $\rho_0$ at 40 K for pristine samples are compared in panel (b). (Ba$_{1-x}$K$_x$)Fe$_2$As$_2$, BaFe$_2$(As$_{1-x}$P$_x$)$_2$, and SrFe$_2$(As$_{1-x}$P$_x$)$_2$ have rather low resistivity values ($< 50~\mu \Omega \text{cm}$) while electron-doped Ba(Fe$_{1-x}$Co$_x$)$_2$As$_2$ and Ba(Fe$_{1-x}$Ni$_x$)$_2$As$_2$) have quite large values ($> 100~\mu \Omega \text{cm}$) indicating that they are already in dirty limit at pristine state. In panel (c), the values of $\Delta T_c / \text{dose}$ are compared. In general, the electron-doped compounds (Ba(Fe$_{1-x}$Co$_x$)$_2$As$_2$ and Ba(Fe$_{1-x}$Ni$_x$)$_2$As$_2$)) have the least change upon electron irradiation mainly due to pre-existing disorder (dirty limit), while hole-doped and and isovalent-substituted compounds show larger change. In panel (d), $\Delta t_c / \text{dose}$ of (Ba$_{1-x}$K$_x$)Fe$_2$As$_2$, BaFe$_2$(As$_{1-x}$P$_x$)$_2$, and SrFe$_2$(As$_{1-x}$P$_x$)$_2$ shows similar values, while Ba(Fe$_{1-x}$Co$_x$)$_2$As$_2$ has the smallest value. In panel (e), $\Delta T_c / \text{dose}$ is only shown for (Ba$_{1-x}$K$_x$)Fe$_2$As$_2$ and BaFe$_2$(As$_{1-x}$P$_x$)$_2$. It is clear that the suppression is stronger in BaFe$_2$(As$_{1-x}$P$_x$)$_2$.

\section{Conclusions}
In this article, we reviewed the use of electron irradiation that induces controlled point - like disorders as a phase sensitive probe to study superconductivity in the 122 family of iron - based superconductors. The simultaneous measurements of the changes in the superconducting transition temperature and low - temperature variation of the London penetration depth lead to the experimental conclusion that $s_{\pm}$ pairing is robust and ubiquitous in iron - based superconductors. Substantial sensitivity to non-magnetic disorder also means that all experimental studies of the superconducting gap structure should be analyzed taking the effect of impurity scatting into account. While the material may be ``clean" in terms of comparison of normal mean free path and very short coherence length, the pairbreaking is significant at any concentration of scattering centers. This will affect temperature dependence of all thermodynamic, spectroscopic and transport properties.

\section*{Acknowledgements}
We thank P. C. Canfield, S. L. Bud'ko, P. J. Hirschfeld, V. G. Kogan, T. Shibauchi, Y. Matsuda, V. Mishra, A. V. Chubukov, S. Miyasaka, and C. J. van der Beek for useful discussion. This work was supported by the U.S. Department of Energy (DOE), Office of Science, Basic Energy Sciences, Materials Science and Engineering Division. Ames Laboratory is operated for the U.S. DOE by Iowa State University under contract DE-AC02-07CH11358. We thank the SIRIUS team, O. Cavani, B. Boizot, V. Metayer, and J. Losco, for running electron irradiation at \'Ecole Polytechnique [supported by the EMIR (R\'eseau national \'dacc\'el\'erateurs pour les Etudes des Mat\'eriaux sous Irradiation) network, proposal 11-11-0121].


\begin{thebibliography}{146}%
\makeatletter
\providecommand \@ifxundefined [1]{%
 \@ifx{#1\undefined}
}%
\providecommand \@ifnum [1]{%
 \ifnum #1\expandafter \@firstoftwo
 \else \expandafter \@secondoftwo
 \fi
}%
\providecommand \@ifx [1]{%
 \ifx #1\expandafter \@firstoftwo
 \else \expandafter \@secondoftwo
 \fi
}%
\providecommand \natexlab [1]{#1}%
\providecommand \enquote  [1]{``#1''}%
\providecommand \bibnamefont  [1]{#1}%
\providecommand \bibfnamefont [1]{#1}%
\providecommand \citenamefont [1]{#1}%
\providecommand \href@noop [0]{\@secondoftwo}%
\providecommand \href [0]{\begingroup \@sanitize@url \@href}%
\providecommand \@href[1]{\@@startlink{#1}\@@href}%
\providecommand \@@href[1]{\endgroup#1\@@endlink}%
\providecommand \@sanitize@url [0]{\catcode `\\12\catcode `\$12\catcode
  `\&12\catcode `\#12\catcode `\^12\catcode `\_12\catcode `\%12\relax}%
\providecommand \@@startlink[1]{}%
\providecommand \@@endlink[0]{}%
\providecommand \url  [0]{\begingroup\@sanitize@url \@url }%
\providecommand \@url [1]{\endgroup\@href {#1}{\urlprefix }}%
\providecommand \urlprefix  [0]{URL }%
\providecommand \Eprint [0]{\href }%
\providecommand \doibase [0]{http://dx.doi.org/}%
\providecommand \selectlanguage [0]{\@gobble}%
\providecommand \bibinfo  [0]{\@secondoftwo}%
\providecommand \bibfield  [0]{\@secondoftwo}%
\providecommand \translation [1]{[#1]}%
\providecommand \BibitemOpen [0]{}%
\providecommand \bibitemStop [0]{}%
\providecommand \bibitemNoStop [0]{.\EOS\space}%
\providecommand \EOS [0]{\spacefactor3000\relax}%
\providecommand \BibitemShut  [1]{\csname bibitem#1\endcsname}%
\let\auto@bib@innerbib\@empty
\bibitem [{\citenamefont {Kamihara}\ \emph {et~al.}(2006)\citenamefont
  {Kamihara}, \citenamefont {Hiramatsu}, \citenamefont {Hirano}, \citenamefont
  {Kawamura}, \citenamefont {Yanagi}, \citenamefont {Kamiya},\ and\
  \citenamefont {Hosono}}]{Kamihara2006JACS_LaOFeP}%
  \BibitemOpen
  \bibfield  {author} {\bibinfo {author} {\bibfnamefont {Y.}~\bibnamefont
  {Kamihara}}, \bibinfo {author} {\bibfnamefont {H.}~\bibnamefont {Hiramatsu}},
  \bibinfo {author} {\bibfnamefont {M.}~\bibnamefont {Hirano}}, \bibinfo
  {author} {\bibfnamefont {R.}~\bibnamefont {Kawamura}}, \bibinfo {author}
  {\bibfnamefont {H.}~\bibnamefont {Yanagi}}, \bibinfo {author} {\bibfnamefont
  {T.}~\bibnamefont {Kamiya}}, \ and\ \bibinfo {author} {\bibfnamefont
  {H.}~\bibnamefont {Hosono}},\ }\href {\doibase 10.1021/ja063355c} {\bibfield
  {journal} {\bibinfo  {journal} {Journal of the American Chemical Society}\
  }\textbf {\bibinfo {volume} {128}},\ \bibinfo {pages} {10012} (\bibinfo
  {year} {2006})}\BibitemShut {NoStop}%
\bibitem [{\citenamefont {Kamihara}\ \emph {et~al.}(2008)\citenamefont
  {Kamihara}, \citenamefont {Watanabe}, \citenamefont {Hirano},\ and\
  \citenamefont {Hosono}}]{KamiharaHosono2008JACS}%
  \BibitemOpen
  \bibfield  {author} {\bibinfo {author} {\bibfnamefont {Y.}~\bibnamefont
  {Kamihara}}, \bibinfo {author} {\bibfnamefont {T.}~\bibnamefont {Watanabe}},
  \bibinfo {author} {\bibfnamefont {M.}~\bibnamefont {Hirano}}, \ and\ \bibinfo
  {author} {\bibfnamefont {H.}~\bibnamefont {Hosono}},\ }\href@noop {}
  {\bibfield  {journal} {\bibinfo  {journal} {Journal of the American Chemical
  Society}\ }\textbf {\bibinfo {volume} {130}},\ \bibinfo {pages} {3296}
  (\bibinfo {year} {2008})}\BibitemShut {NoStop}%
\bibitem [{\citenamefont {Johnston}(2010)}]{Johnston2010AP_review}%
  \BibitemOpen
  \bibfield  {author} {\bibinfo {author} {\bibfnamefont {D.~C.}\ \bibnamefont
  {Johnston}},\ }\href {\doibase 10.1080/00018732.2010.513480} {\bibfield
  {journal} {\bibinfo  {journal} {Advances in Physics}\ }\textbf {\bibinfo
  {volume} {59}},\ \bibinfo {pages} {803} (\bibinfo {year} {2010})}\BibitemShut
  {NoStop}%
\bibitem [{\citenamefont {Canfield}\ and\ \citenamefont
  {Bud'ko}(2010)}]{Canfield2010ARCMP}%
  \BibitemOpen
  \bibfield  {author} {\bibinfo {author} {\bibfnamefont {P.~C.}\ \bibnamefont
  {Canfield}}\ and\ \bibinfo {author} {\bibfnamefont {S.~L.}\ \bibnamefont
  {Bud'ko}},\ }\href@noop {} {\bibfield  {journal} {\bibinfo  {journal} {Annual
  Review of Condensed Matter Physics}\ }\textbf {\bibinfo {volume} {1}},\
  \bibinfo {pages} {27} (\bibinfo {year} {2010})}\BibitemShut {NoStop}%
\bibitem [{\citenamefont {Paglione}\ and\ \citenamefont
  {Greene}(2010)}]{Paglione2010NP}%
  \BibitemOpen
  \bibfield  {author} {\bibinfo {author} {\bibfnamefont {J.}~\bibnamefont
  {Paglione}}\ and\ \bibinfo {author} {\bibfnamefont {R.~L.}\ \bibnamefont
  {Greene}},\ }\href@noop {} {\bibfield  {journal} {\bibinfo  {journal} {Nat
  Phys}\ }\textbf {\bibinfo {volume} {6}},\ \bibinfo {pages} {645} (\bibinfo
  {year} {2010})}\BibitemShut {NoStop}%
\bibitem [{\citenamefont {Hirschfeld}\ \emph {et~al.}(2011)\citenamefont
  {Hirschfeld}, \citenamefont {Korshunov},\ and\ \citenamefont
  {Mazin}}]{Hirschfeld2011ROPP}%
  \BibitemOpen
  \bibfield  {author} {\bibinfo {author} {\bibfnamefont {P.~J.}\ \bibnamefont
  {Hirschfeld}}, \bibinfo {author} {\bibfnamefont {M.~M.}\ \bibnamefont
  {Korshunov}}, \ and\ \bibinfo {author} {\bibfnamefont {I.~I.}\ \bibnamefont
  {Mazin}},\ }\href@noop {} {\bibfield  {journal} {\bibinfo  {journal} {Reports
  on Progress in Physics}\ }\textbf {\bibinfo {volume} {74}},\ \bibinfo {pages}
  {124508} (\bibinfo {year} {2011})}\BibitemShut {NoStop}%
\bibitem [{\citenamefont {Stewart}(2011)}]{Stewart2011RevModPhys}%
  \BibitemOpen
  \bibfield  {author} {\bibinfo {author} {\bibfnamefont {G.~R.}\ \bibnamefont
  {Stewart}},\ }\href {\doibase 10.1103/RevModPhys.83.1589} {\bibfield
  {journal} {\bibinfo  {journal} {Rev. Mod. Phys.}\ }\textbf {\bibinfo {volume}
  {83}},\ \bibinfo {pages} {1589} (\bibinfo {year} {2011})}\BibitemShut
  {NoStop}%
\bibitem [{\citenamefont
  {Carrington}(2011{\natexlab{a}})}]{Carrington2011RPP_review}%
  \BibitemOpen
  \bibfield  {author} {\bibinfo {author} {\bibfnamefont {A.}~\bibnamefont
  {Carrington}},\ }\href {http://stacks.iop.org/0034-4885/74/i=12/a=124507}
  {\bibfield  {journal} {\bibinfo  {journal} {Reports on Progress in Physics}\
  }\textbf {\bibinfo {volume} {74}},\ \bibinfo {pages} {124507} (\bibinfo
  {year} {2011}{\natexlab{a}})}\BibitemShut {NoStop}%
\bibitem [{\citenamefont {Abrahams}\ and\ \citenamefont
  {Si}(2011)}]{Abrahams2011JPCM_review}%
  \BibitemOpen
  \bibfield  {author} {\bibinfo {author} {\bibfnamefont {E.}~\bibnamefont
  {Abrahams}}\ and\ \bibinfo {author} {\bibfnamefont {Q.}~\bibnamefont {Si}},\
  }\href@noop {} {\bibfield  {journal} {\bibinfo  {journal} {Journal of
  Physics: Condensed Matter}\ }\textbf {\bibinfo {volume} {23}},\ \bibinfo
  {pages} {223201} (\bibinfo {year} {2011})}\BibitemShut {NoStop}%
\bibitem [{\citenamefont
  {Carrington}(2011{\natexlab{b}})}]{CARRINGTON2011CRPhys_review}%
  \BibitemOpen
  \bibfield  {author} {\bibinfo {author} {\bibfnamefont {A.}~\bibnamefont
  {Carrington}},\ }\href@noop {} {\bibfield  {journal} {\bibinfo  {journal}
  {Comptes Rendus Physique}\ }\textbf {\bibinfo {volume} {12}},\ \bibinfo
  {pages} {502 } (\bibinfo {year} {2011}{\natexlab{b}})}\BibitemShut {NoStop}%
\bibitem [{\citenamefont {Prozorov}\ and\ \citenamefont
  {Kogan}(2011{\natexlab{a}})}]{Prozorov2011RPP_review}%
  \BibitemOpen
  \bibfield  {author} {\bibinfo {author} {\bibfnamefont {R.}~\bibnamefont
  {Prozorov}}\ and\ \bibinfo {author} {\bibfnamefont {V.~G.}\ \bibnamefont
  {Kogan}},\ }\href {http://stacks.iop.org/0034-4885/74/i=12/a=124505}
  {\bibfield  {journal} {\bibinfo  {journal} {Reports on Progress in Physics}\
  }\textbf {\bibinfo {volume} {74}},\ \bibinfo {pages} {124505} (\bibinfo
  {year} {2011}{\natexlab{a}})}\BibitemShut {NoStop}%
\bibitem [{\citenamefont {Wen}\ and\ \citenamefont
  {Li}(2011)}]{Wen2011ARCMP_review}%
  \BibitemOpen
  \bibfield  {author} {\bibinfo {author} {\bibfnamefont {H.-H.}\ \bibnamefont
  {Wen}}\ and\ \bibinfo {author} {\bibfnamefont {S.}~\bibnamefont {Li}},\
  }\href {\doibase 10.1146/annurev-conmatphys-062910-140518} {\bibfield
  {journal} {\bibinfo  {journal} {Annual Review of Condensed Matter Physics}\
  }\textbf {\bibinfo {volume} {2}},\ \bibinfo {pages} {121} (\bibinfo {year}
  {2011})}\BibitemShut {NoStop}%
\bibitem [{\citenamefont {Zhang}\ \emph {et~al.}(2011)\citenamefont {Zhang},
  \citenamefont {Jiao}, \citenamefont {Chen},\ and\ \citenamefont
  {Yuan}}]{Zhang2011FP_review}%
  \BibitemOpen
  \bibfield  {author} {\bibinfo {author} {\bibfnamefont {J.-l.}\ \bibnamefont
  {Zhang}}, \bibinfo {author} {\bibfnamefont {L.}~\bibnamefont {Jiao}},
  \bibinfo {author} {\bibfnamefont {Y.}~\bibnamefont {Chen}}, \ and\ \bibinfo
  {author} {\bibfnamefont {H.-q.}\ \bibnamefont {Yuan}},\ }\href {\doibase
  10.1007/s11467-011-0235-7} {\bibfield  {journal} {\bibinfo  {journal}
  {Frontiers of Physics}\ }\textbf {\bibinfo {volume} {6}},\ \bibinfo {pages}
  {463} (\bibinfo {year} {2011})}\BibitemShut {NoStop}%
\bibitem [{\citenamefont {Chubukov}(2012)}]{Chubukov2012ARCMP}%
  \BibitemOpen
  \bibfield  {author} {\bibinfo {author} {\bibfnamefont {A.}~\bibnamefont
  {Chubukov}},\ }\href@noop {} {\bibfield  {journal} {\bibinfo  {journal}
  {Annual Review of Condensed Matter Physics}\ }\textbf {\bibinfo {volume}
  {3}},\ \bibinfo {pages} {57} (\bibinfo {year} {2012})}\BibitemShut {NoStop}%
\bibitem [{\citenamefont {Kordyuk}(2012)}]{Kordyuk2012LowTempPhys_review}%
  \BibitemOpen
  \bibfield  {author} {\bibinfo {author} {\bibfnamefont {A.~A.}\ \bibnamefont
  {Kordyuk}},\ }\href {\doibase 10.1063/1.4752092} {\bibfield  {journal}
  {\bibinfo  {journal} {Low Temperature Physics}\ }\textbf {\bibinfo {volume}
  {38}},\ \bibinfo {pages} {888} (\bibinfo {year} {2012})}\BibitemShut
  {NoStop}%
\bibitem [{\citenamefont {Song}\ and\ \citenamefont
  {Hoffman}(2013)}]{SONG201339CuOpSSMS_review}%
  \BibitemOpen
  \bibfield  {author} {\bibinfo {author} {\bibfnamefont {C.-L.}\ \bibnamefont
  {Song}}\ and\ \bibinfo {author} {\bibfnamefont {J.~E.}\ \bibnamefont
  {Hoffman}},\ }\href@noop {} {\bibfield  {journal} {\bibinfo  {journal}
  {Current Opinion in Solid State and Materials Science}\ }\textbf {\bibinfo
  {volume} {17}},\ \bibinfo {pages} {39 } (\bibinfo {year} {2013})}\BibitemShut
  {NoStop}%
\bibitem [{\citenamefont {Long}\ and\ \citenamefont
  {Wei-Qiang}(2013)}]{Long2013CPB_review}%
  \BibitemOpen
  \bibfield  {author} {\bibinfo {author} {\bibfnamefont {M.}~\bibnamefont
  {Long}}\ and\ \bibinfo {author} {\bibfnamefont {Y.}~\bibnamefont
  {Wei-Qiang}},\ }\href {http://stacks.iop.org/1674-1056/22/i=8/a=087414}
  {\bibfield  {journal} {\bibinfo  {journal} {Chinese Physics B}\ }\textbf
  {\bibinfo {volume} {22}},\ \bibinfo {pages} {087414} (\bibinfo {year}
  {2013})}\BibitemShut {NoStop}%
\bibitem [{\citenamefont {Hao}\ \emph {et~al.}(2013)\citenamefont {Hao},
  \citenamefont {Yun-Lei}, \citenamefont {Zhu-An},\ and\ \citenamefont
  {Guang-Han}}]{Jiang2013CPB_review}%
  \BibitemOpen
  \bibfield  {author} {\bibinfo {author} {\bibfnamefont {J.}~\bibnamefont
  {Hao}}, \bibinfo {author} {\bibfnamefont {S.}~\bibnamefont {Yun-Lei}},
  \bibinfo {author} {\bibfnamefont {X.}~\bibnamefont {Zhu-An}}, \ and\ \bibinfo
  {author} {\bibfnamefont {C.}~\bibnamefont {Guang-Han}},\ }\href
  {http://stacks.iop.org/1674-1056/22/i=8/a=087410} {\bibfield  {journal}
  {\bibinfo  {journal} {Chinese Physics B}\ }\textbf {\bibinfo {volume} {22}},\
  \bibinfo {pages} {087410} (\bibinfo {year} {2013})}\BibitemShut {NoStop}%
\bibitem [{\citenamefont {Carretta}\ \emph {et~al.}(2013)\citenamefont
  {Carretta}, \citenamefont {Renzi}, \citenamefont {Prando},\ and\
  \citenamefont {Sanna}}]{Carretta2013PhysScr_review}%
  \BibitemOpen
  \bibfield  {author} {\bibinfo {author} {\bibfnamefont {P.}~\bibnamefont
  {Carretta}}, \bibinfo {author} {\bibfnamefont {R.~D.}\ \bibnamefont {Renzi}},
  \bibinfo {author} {\bibfnamefont {G.}~\bibnamefont {Prando}}, \ and\ \bibinfo
  {author} {\bibfnamefont {S.}~\bibnamefont {Sanna}},\ }\href
  {http://stacks.iop.org/1402-4896/88/i=6/a=068504} {\bibfield  {journal}
  {\bibinfo  {journal} {Physica Scripta}\ }\textbf {\bibinfo {volume} {88}},\
  \bibinfo {pages} {068504} (\bibinfo {year} {2013})}\BibitemShut {NoStop}%
\bibitem [{\citenamefont {Zi-Rong}\ \emph {et~al.}(2013)\citenamefont
  {Zi-Rong}, \citenamefont {Yan}, \citenamefont {Bin-Ping},\ and\ \citenamefont
  {Dong-Lai}}]{Ye2013ChPB_review}%
  \BibitemOpen
  \bibfield  {author} {\bibinfo {author} {\bibfnamefont {Y.}~\bibnamefont
  {Zi-Rong}}, \bibinfo {author} {\bibfnamefont {Z.}~\bibnamefont {Yan}},
  \bibinfo {author} {\bibfnamefont {X.}~\bibnamefont {Bin-Ping}}, \ and\
  \bibinfo {author} {\bibfnamefont {F.}~\bibnamefont {Dong-Lai}},\ }\href
  {http://stacks.iop.org/1674-1056/22/i=8/a=087407} {\bibfield  {journal}
  {\bibinfo  {journal} {Chinese Physics B}\ }\textbf {\bibinfo {volume} {22}},\
  \bibinfo {pages} {087407} (\bibinfo {year} {2013})}\BibitemShut {NoStop}%
\bibitem [{\citenamefont {Eremin}\ \emph {et~al.}(2014)\citenamefont {Eremin},
  \citenamefont {Knolle}, \citenamefont {Fernandes}, \citenamefont
  {Schmalian},\ and\ \citenamefont {Chubukov}}]{Eremin2014JPSJ_review}%
  \BibitemOpen
  \bibfield  {author} {\bibinfo {author} {\bibfnamefont {I.}~\bibnamefont
  {Eremin}}, \bibinfo {author} {\bibfnamefont {J.}~\bibnamefont {Knolle}},
  \bibinfo {author} {\bibfnamefont {R.~M.}\ \bibnamefont {Fernandes}}, \bibinfo
  {author} {\bibfnamefont {J.}~\bibnamefont {Schmalian}}, \ and\ \bibinfo
  {author} {\bibfnamefont {A.~V.}\ \bibnamefont {Chubukov}},\ }\href {\doibase
  10.7566/JPSJ.83.061015} {\bibfield  {journal} {\bibinfo  {journal} {Journal
  of the Physical Society of Japan}\ }\textbf {\bibinfo {volume} {83}},\
  \bibinfo {pages} {061015} (\bibinfo {year} {2014})}\BibitemShut {NoStop}%
\bibitem [{\citenamefont {Charnukha}(2014)}]{Charnukha2014JPCM_review}%
  \BibitemOpen
  \bibfield  {author} {\bibinfo {author} {\bibfnamefont {A.}~\bibnamefont
  {Charnukha}},\ }\href {http://stacks.iop.org/0953-8984/26/i=25/a=253203}
  {\bibfield  {journal} {\bibinfo  {journal} {Journal of Physics: Condensed
  Matter}\ }\textbf {\bibinfo {volume} {26}},\ \bibinfo {pages} {253203}
  (\bibinfo {year} {2014})}\BibitemShut {NoStop}%
\bibitem [{\citenamefont {Shibauchi}\ \emph {et~al.}(2014)\citenamefont
  {Shibauchi}, \citenamefont {Carrington},\ and\ \citenamefont
  {Matsuda}}]{Shibauchi2014ARCMP}%
  \BibitemOpen
  \bibfield  {author} {\bibinfo {author} {\bibfnamefont {T.}~\bibnamefont
  {Shibauchi}}, \bibinfo {author} {\bibfnamefont {A.}~\bibnamefont
  {Carrington}}, \ and\ \bibinfo {author} {\bibfnamefont {Y.}~\bibnamefont
  {Matsuda}},\ }\href {\doibase 10.1146/annurev-conmatphys-031113-133921}
  {\bibfield  {journal} {\bibinfo  {journal} {Annual Review of Condensed Matter
  Physics}\ }\textbf {\bibinfo {volume} {5}},\ \bibinfo {pages} {113} (\bibinfo
  {year} {2014})}\BibitemShut {NoStop}%
\bibitem [{\citenamefont {Dai}(2015)}]{Dai2015RMP_review}%
  \BibitemOpen
  \bibfield  {author} {\bibinfo {author} {\bibfnamefont {P.}~\bibnamefont
  {Dai}},\ }\href {\doibase 10.1103/RevModPhys.87.855} {\bibfield  {journal}
  {\bibinfo  {journal} {Rev. Mod. Phys.}\ }\textbf {\bibinfo {volume} {87}},\
  \bibinfo {pages} {855} (\bibinfo {year} {2015})}\BibitemShut {NoStop}%
\bibitem [{\citenamefont {Jasek}\ \emph {et~al.}(2015)\citenamefont {Jasek},
  \citenamefont {Komedera}, \citenamefont {Blachowski}, \citenamefont
  {Ruebenbauer}, \citenamefont {Zukrowski}, \citenamefont {Bukowski},\ and\
  \citenamefont {Karpinski}}]{JasekKarpinski2015PhilMag_review}%
  \BibitemOpen
  \bibfield  {author} {\bibinfo {author} {\bibfnamefont {A.}~\bibnamefont
  {Jasek}}, \bibinfo {author} {\bibfnamefont {K.}~\bibnamefont {Komedera}},
  \bibinfo {author} {\bibfnamefont {A.}~\bibnamefont {Blachowski}}, \bibinfo
  {author} {\bibfnamefont {K.}~\bibnamefont {Ruebenbauer}}, \bibinfo {author}
  {\bibfnamefont {J.}~\bibnamefont {Zukrowski}}, \bibinfo {author}
  {\bibfnamefont {Z.}~\bibnamefont {Bukowski}}, \ and\ \bibinfo {author}
  {\bibfnamefont {J.}~\bibnamefont {Karpinski}},\ }\href {\doibase
  10.1080/14786435.2014.970240} {\bibfield  {journal} {\bibinfo  {journal}
  {Philosophical Magazine}\ }\textbf {\bibinfo {volume} {95}},\ \bibinfo
  {pages} {493} (\bibinfo {year} {2015})}\BibitemShut {NoStop}%
\bibitem [{\citenamefont {Bohmer}\ and\ \citenamefont
  {Meingast}(2016)}]{BOHMER2016CRP_review}%
  \BibitemOpen
  \bibfield  {author} {\bibinfo {author} {\bibfnamefont {A.~E.}\ \bibnamefont
  {Bohmer}}\ and\ \bibinfo {author} {\bibfnamefont {C.}~\bibnamefont
  {Meingast}},\ }\href {\doibase https://doi.org/10.1016/j.crhy.2015.07.001}
  {\bibfield  {journal} {\bibinfo  {journal} {Comptes Rendus Physique}\
  }\textbf {\bibinfo {volume} {17}},\ \bibinfo {pages} {90 } (\bibinfo {year}
  {2016})}\BibitemShut {NoStop}%
\bibitem [{\citenamefont {Pallecchi}\ \emph {et~al.}(2016)\citenamefont
  {Pallecchi}, \citenamefont {Caglieris},\ and\ \citenamefont
  {Putti}}]{PallecchiPutti2016SST_review}%
  \BibitemOpen
  \bibfield  {author} {\bibinfo {author} {\bibfnamefont {I.}~\bibnamefont
  {Pallecchi}}, \bibinfo {author} {\bibfnamefont {F.}~\bibnamefont
  {Caglieris}}, \ and\ \bibinfo {author} {\bibfnamefont {M.}~\bibnamefont
  {Putti}},\ }\href {http://stacks.iop.org/0953-2048/29/i=7/a=073002}
  {\bibfield  {journal} {\bibinfo  {journal} {Superconductor Science and
  Technology}\ }\textbf {\bibinfo {volume} {29}},\ \bibinfo {pages} {073002}
  (\bibinfo {year} {2016})}\BibitemShut {NoStop}%
\bibitem [{\citenamefont {Gallais}\ and\ \citenamefont
  {Paul}(2016)}]{GALLAIS2016ComprenPhys_review}%
  \BibitemOpen
  \bibfield  {author} {\bibinfo {author} {\bibfnamefont {Y.}~\bibnamefont
  {Gallais}}\ and\ \bibinfo {author} {\bibfnamefont {I.}~\bibnamefont {Paul}},\
  }\href {\doibase https://doi.org/10.1016/j.crhy.2015.10.001} {\bibfield
  {journal} {\bibinfo  {journal} {Comptes Rendus Physique}\ }\textbf {\bibinfo
  {volume} {17}},\ \bibinfo {pages} {113 } (\bibinfo {year}
  {2016})}\BibitemShut {NoStop}%
\bibitem [{\citenamefont {Li}\ \emph {et~al.}(2016)\citenamefont {Li},
  \citenamefont {Guo}, \citenamefont {Yang}, \citenamefont {Yamaura},
  \citenamefont {Takayama-Muromachi}, \citenamefont {Wang},\ and\ \citenamefont
  {Wu}}]{LiWu2016SST_review_FeSC}%
  \BibitemOpen
  \bibfield  {author} {\bibinfo {author} {\bibfnamefont {J.}~\bibnamefont
  {Li}}, \bibinfo {author} {\bibfnamefont {Y.-F.}\ \bibnamefont {Guo}},
  \bibinfo {author} {\bibfnamefont {Z.-R.}\ \bibnamefont {Yang}}, \bibinfo
  {author} {\bibfnamefont {K.}~\bibnamefont {Yamaura}}, \bibinfo {author}
  {\bibfnamefont {E.}~\bibnamefont {Takayama-Muromachi}}, \bibinfo {author}
  {\bibfnamefont {H.-B.}\ \bibnamefont {Wang}}, \ and\ \bibinfo {author}
  {\bibfnamefont {P.-H.}\ \bibnamefont {Wu}},\ }\href
  {http://stacks.iop.org/0953-2048/29/i=5/a=053001} {\bibfield  {journal}
  {\bibinfo  {journal} {Superconductor Science and Technology}\ }\textbf
  {\bibinfo {volume} {29}},\ \bibinfo {pages} {053001} (\bibinfo {year}
  {2016})}\BibitemShut {NoStop}%
\bibitem [{\citenamefont {Martinelli}\ \emph {et~al.}(2016)\citenamefont
  {Martinelli}, \citenamefont {Bernardini},\ and\ \citenamefont
  {Massidda}}]{MARTINELLI2016CompRenPhys_review}%
  \BibitemOpen
  \bibfield  {author} {\bibinfo {author} {\bibfnamefont {A.}~\bibnamefont
  {Martinelli}}, \bibinfo {author} {\bibfnamefont {F.}~\bibnamefont
  {Bernardini}}, \ and\ \bibinfo {author} {\bibfnamefont {S.}~\bibnamefont
  {Massidda}},\ }\href {\doibase https://doi.org/10.1016/j.crhy.2015.06.001}
  {\bibfield  {journal} {\bibinfo  {journal} {Comptes Rendus Physique}\
  }\textbf {\bibinfo {volume} {17}},\ \bibinfo {pages} {5 } (\bibinfo {year}
  {2016})}\BibitemShut {NoStop}%
\bibitem [{\citenamefont {Guterding}\ \emph {et~al.}(2017)\citenamefont
  {Guterding}, \citenamefont {Backes}, \citenamefont {Tomic}, \citenamefont
  {Jeschke},\ and\ \citenamefont {Valenti}}]{Guterding2017PhysStatSol_review}%
  \BibitemOpen
  \bibfield  {author} {\bibinfo {author} {\bibfnamefont {D.}~\bibnamefont
  {Guterding}}, \bibinfo {author} {\bibfnamefont {S.}~\bibnamefont {Backes}},
  \bibinfo {author} {\bibfnamefont {M.}~\bibnamefont {Tomic}}, \bibinfo
  {author} {\bibfnamefont {H.~O.}\ \bibnamefont {Jeschke}}, \ and\ \bibinfo
  {author} {\bibfnamefont {R.}~\bibnamefont {Valenti}},\ }\href@noop {}
  {\bibfield  {journal} {\bibinfo  {journal} {physica status solidi (b)}\
  }\textbf {\bibinfo {volume} {254}},\ \bibinfo {pages} {1600164} (\bibinfo
  {year} {2017})}\BibitemShut {NoStop}%
\bibitem [{\citenamefont {Fernandes}\ and\ \citenamefont
  {Chubukov}(2017)}]{Fernandes2017RPP_review}%
  \BibitemOpen
  \bibfield  {author} {\bibinfo {author} {\bibfnamefont {R.~M.}\ \bibnamefont
  {Fernandes}}\ and\ \bibinfo {author} {\bibfnamefont {A.~V.}\ \bibnamefont
  {Chubukov}},\ }\href {http://stacks.iop.org/0034-4885/80/i=1/a=014503}
  {\bibfield  {journal} {\bibinfo  {journal} {Reports on Progress in Physics}\
  }\textbf {\bibinfo {volume} {80}},\ \bibinfo {pages} {014503} (\bibinfo
  {year} {2017})}\BibitemShut {NoStop}%
\bibitem [{\citenamefont {Yi}\ \emph {et~al.}(2017)\citenamefont {Yi},
  \citenamefont {Zhang}, \citenamefont {Shen},\ and\ \citenamefont
  {Lu}}]{Yi2017NPJ_QuanMat_review}%
  \BibitemOpen
  \bibfield  {author} {\bibinfo {author} {\bibfnamefont {M.}~\bibnamefont
  {Yi}}, \bibinfo {author} {\bibfnamefont {Y.}~\bibnamefont {Zhang}}, \bibinfo
  {author} {\bibfnamefont {Z.-X.}\ \bibnamefont {Shen}}, \ and\ \bibinfo
  {author} {\bibfnamefont {D.}~\bibnamefont {Lu}},\ }\href {\doibase
  10.1038/s41535-017-0059-y} {\bibfield  {journal} {\bibinfo  {journal} {npj
  Quantum Materials}\ }\textbf {\bibinfo {volume} {2}},\ \bibinfo {pages} {57}
  (\bibinfo {year} {2017})}\BibitemShut {NoStop}%
\bibitem [{\citenamefont {Putti}\ \emph {et~al.}(2010)\citenamefont {Putti},
  \citenamefont {Pallecchi}, \citenamefont {Bellingeri}, \citenamefont
  {Cimberle}, \citenamefont {Tropeano}, \citenamefont {Ferdeghini},
  \citenamefont {Palenzona}, \citenamefont {Tarantini}, \citenamefont
  {Yamamoto}, \citenamefont {Jiang}, \citenamefont {Jaroszynski}, \citenamefont
  {Kametani}, \citenamefont {Abraimov}, \citenamefont {Polyanskii},
  \citenamefont {Weiss}, \citenamefont {Hellstrom}, \citenamefont {Gurevich},
  \citenamefont {Larbalestier}, \citenamefont {Jin}, \citenamefont {Sales},
  \citenamefont {Sefat}, \citenamefont {McGuire}, \citenamefont {Mandrus},
  \citenamefont {Cheng}, \citenamefont {Jia}, \citenamefont {Wen},
  \citenamefont {Lee},\ and\ \citenamefont
  {Eom}}]{Putti2010SST_FeSC_Application}%
  \BibitemOpen
  \bibfield  {author} {\bibinfo {author} {\bibfnamefont {M.}~\bibnamefont
  {Putti}}, \bibinfo {author} {\bibfnamefont {I.}~\bibnamefont {Pallecchi}},
  \bibinfo {author} {\bibfnamefont {E.}~\bibnamefont {Bellingeri}}, \bibinfo
  {author} {\bibfnamefont {M.~R.}\ \bibnamefont {Cimberle}}, \bibinfo {author}
  {\bibfnamefont {M.}~\bibnamefont {Tropeano}}, \bibinfo {author}
  {\bibfnamefont {C.}~\bibnamefont {Ferdeghini}}, \bibinfo {author}
  {\bibfnamefont {A.}~\bibnamefont {Palenzona}}, \bibinfo {author}
  {\bibfnamefont {C.}~\bibnamefont {Tarantini}}, \bibinfo {author}
  {\bibfnamefont {A.}~\bibnamefont {Yamamoto}}, \bibinfo {author}
  {\bibfnamefont {J.}~\bibnamefont {Jiang}}, \bibinfo {author} {\bibfnamefont
  {J.}~\bibnamefont {Jaroszynski}}, \bibinfo {author} {\bibfnamefont
  {F.}~\bibnamefont {Kametani}}, \bibinfo {author} {\bibfnamefont
  {D.}~\bibnamefont {Abraimov}}, \bibinfo {author} {\bibfnamefont
  {A.}~\bibnamefont {Polyanskii}}, \bibinfo {author} {\bibfnamefont {J.~D.}\
  \bibnamefont {Weiss}}, \bibinfo {author} {\bibfnamefont {E.~E.}\ \bibnamefont
  {Hellstrom}}, \bibinfo {author} {\bibfnamefont {A.}~\bibnamefont {Gurevich}},
  \bibinfo {author} {\bibfnamefont {D.~C.}\ \bibnamefont {Larbalestier}},
  \bibinfo {author} {\bibfnamefont {R.}~\bibnamefont {Jin}}, \bibinfo {author}
  {\bibfnamefont {B.~C.}\ \bibnamefont {Sales}}, \bibinfo {author}
  {\bibfnamefont {A.~S.}\ \bibnamefont {Sefat}}, \bibinfo {author}
  {\bibfnamefont {M.~A.}\ \bibnamefont {McGuire}}, \bibinfo {author}
  {\bibfnamefont {D.}~\bibnamefont {Mandrus}}, \bibinfo {author} {\bibfnamefont
  {P.}~\bibnamefont {Cheng}}, \bibinfo {author} {\bibfnamefont
  {Y.}~\bibnamefont {Jia}}, \bibinfo {author} {\bibfnamefont {H.~H.}\
  \bibnamefont {Wen}}, \bibinfo {author} {\bibfnamefont {S.}~\bibnamefont
  {Lee}}, \ and\ \bibinfo {author} {\bibfnamefont {C.~B.}\ \bibnamefont
  {Eom}},\ }\href {http://stacks.iop.org/0953-2048/23/i=3/a=034003} {\bibfield
  {journal} {\bibinfo  {journal} {Superconductor Science and Technology}\
  }\textbf {\bibinfo {volume} {23}},\ \bibinfo {pages} {034003} (\bibinfo
  {year} {2010})}\BibitemShut {NoStop}%
\bibitem [{\citenamefont {Ma}(2012)}]{Ma2012SST_FeSC_Application}%
  \BibitemOpen
  \bibfield  {author} {\bibinfo {author} {\bibfnamefont {Y.}~\bibnamefont
  {Ma}},\ }\href {http://stacks.iop.org/0953-2048/25/i=11/a=113001} {\bibfield
  {journal} {\bibinfo  {journal} {Superconductor Science and Technology}\
  }\textbf {\bibinfo {volume} {25}},\ \bibinfo {pages} {113001} (\bibinfo
  {year} {2012})}\BibitemShut {NoStop}%
\bibitem [{\citenamefont {Tanabe}\ and\ \citenamefont
  {Hosono}(2012)}]{TanabeHosono2012JJAP_FeSC_Application}%
  \BibitemOpen
  \bibfield  {author} {\bibinfo {author} {\bibfnamefont {K.}~\bibnamefont
  {Tanabe}}\ and\ \bibinfo {author} {\bibfnamefont {H.}~\bibnamefont
  {Hosono}},\ }\href {http://stacks.iop.org/1347-4065/51/i=1R/a=010005}
  {\bibfield  {journal} {\bibinfo  {journal} {Japanese Journal of Applied
  Physics}\ }\textbf {\bibinfo {volume} {51}},\ \bibinfo {pages} {010005}
  (\bibinfo {year} {2012})}\BibitemShut {NoStop}%
\bibitem [{\citenamefont {Ma}(2015)}]{MA2015PhysicaC_FeSC_Application}%
  \BibitemOpen
  \bibfield  {author} {\bibinfo {author} {\bibfnamefont {Y.}~\bibnamefont
  {Ma}},\ }\href {\doibase https://doi.org/10.1016/j.physc.2015.05.006}
  {\bibfield  {journal} {\bibinfo  {journal} {Physica C: Superconductivity and
  its Applications}\ }\textbf {\bibinfo {volume} {516}},\ \bibinfo {pages} {17
  } (\bibinfo {year} {2015})}\BibitemShut {NoStop}%
\bibitem [{\citenamefont {Hosono}\ \emph {et~al.}(2017)\citenamefont {Hosono},
  \citenamefont {Yamamoto}, \citenamefont {Hiramatsu},\ and\ \citenamefont
  {Ma}}]{Hosono2017MatToday_application}%
  \BibitemOpen
  \bibfield  {author} {\bibinfo {author} {\bibfnamefont {H.}~\bibnamefont
  {Hosono}}, \bibinfo {author} {\bibfnamefont {A.}~\bibnamefont {Yamamoto}},
  \bibinfo {author} {\bibfnamefont {H.}~\bibnamefont {Hiramatsu}}, \ and\
  \bibinfo {author} {\bibfnamefont {Y.}~\bibnamefont {Ma}},\ }\href@noop {}
  {\bibfield  {journal} {\bibinfo  {journal} {Materials Today}\ } (\bibinfo
  {year} {2017})}\BibitemShut {NoStop}%
\bibitem [{\citenamefont {Mizukami}\ \emph {et~al.}(2014)\citenamefont
  {Mizukami}, \citenamefont {Konczykowski}, \citenamefont {Kurata},
  \citenamefont {Hashimoto}, \citenamefont {Mishra}, \citenamefont {Kreisel},
  \citenamefont {Hirschfeld}, \citenamefont {Matsuda},\ and\ \citenamefont
  {Shibauchi}}]{Mizukami2014NatureComm}%
  \BibitemOpen
  \bibfield  {author} {\bibinfo {author} {\bibfnamefont {Y.}~\bibnamefont
  {Mizukami}}, \bibinfo {author} {\bibfnamefont {Y.}~\bibnamefont
  {Konczykowski}, \bibfnamefont {M.~andKawamoto}}, \bibinfo {author}
  {\bibfnamefont {S.}~\bibnamefont {Kurata}, \bibfnamefont {S.~andKasahara}},
  \bibinfo {author} {\bibfnamefont {K.}~\bibnamefont {Hashimoto}}, \bibinfo
  {author} {\bibfnamefont {V.}~\bibnamefont {Mishra}}, \bibinfo {author}
  {\bibfnamefont {Y.}~\bibnamefont {Kreisel}, \bibfnamefont {A.~andWang}},
  \bibinfo {author} {\bibfnamefont {P.~J.}\ \bibnamefont {Hirschfeld}},
  \bibinfo {author} {\bibfnamefont {Y.}~\bibnamefont {Matsuda}}, \ and\
  \bibinfo {author} {\bibfnamefont {T.}~\bibnamefont {Shibauchi}},\ }\href
  {\doibase 10.1038/ncomms6657} {\bibfield  {journal} {\bibinfo  {journal}
  {Nat. Comm.}\ }\textbf {\bibinfo {volume} {5}},\ \bibinfo {pages} {5657}
  (\bibinfo {year} {2014})}\BibitemShut {NoStop}%
\bibitem [{\citenamefont {Cho}\ \emph {et~al.}(2016)\citenamefont {Cho},
  \citenamefont {Ko{\'n}czykowski}, \citenamefont {Teknowijoyo}, \citenamefont
  {Tanatar}, \citenamefont {Liu}, \citenamefont {Lograsso}, \citenamefont
  {Straszheim}, \citenamefont {Mishra}, \citenamefont {Maiti}, \citenamefont
  {Hirschfeld},\ and\ \citenamefont
  {Prozorov}}]{Cho2016ScienceAdvances_BaK122_e-irr}%
  \BibitemOpen
  \bibfield  {author} {\bibinfo {author} {\bibfnamefont {K.}~\bibnamefont
  {Cho}}, \bibinfo {author} {\bibfnamefont {M.}~\bibnamefont
  {Ko{\'n}czykowski}}, \bibinfo {author} {\bibfnamefont {S.}~\bibnamefont
  {Teknowijoyo}}, \bibinfo {author} {\bibfnamefont {M.~A.}\ \bibnamefont
  {Tanatar}}, \bibinfo {author} {\bibfnamefont {Y.}~\bibnamefont {Liu}},
  \bibinfo {author} {\bibfnamefont {T.~A.}\ \bibnamefont {Lograsso}}, \bibinfo
  {author} {\bibfnamefont {W.~E.}\ \bibnamefont {Straszheim}}, \bibinfo
  {author} {\bibfnamefont {V.}~\bibnamefont {Mishra}}, \bibinfo {author}
  {\bibfnamefont {S.}~\bibnamefont {Maiti}}, \bibinfo {author} {\bibfnamefont
  {P.~J.}\ \bibnamefont {Hirschfeld}}, \ and\ \bibinfo {author} {\bibfnamefont
  {R.}~\bibnamefont {Prozorov}},\ }\href@noop {} {\bibfield  {journal}
  {\bibinfo  {journal} {Science Advances}\ }\textbf {\bibinfo {volume} {2}}
  (\bibinfo {year} {2016})}\BibitemShut {NoStop}%
\bibitem [{\citenamefont {Anderson}(1959)}]{Anderson1959JPCS}%
  \BibitemOpen
  \bibfield  {author} {\bibinfo {author} {\bibfnamefont {P.~W.}\ \bibnamefont
  {Anderson}},\ }\href@noop {} {\bibfield  {journal} {\bibinfo  {journal} {J.
  Phys. Chem. Solids}\ }\textbf {\bibinfo {volume} {11}},\ \bibinfo {pages}
  {26} (\bibinfo {year} {1959})}\BibitemShut {NoStop}%
\bibitem [{\citenamefont {Abrikosov}\ and\ \citenamefont
  {Gor'kov}(1961)}]{AbrikosovGorkov1961JETP}%
  \BibitemOpen
  \bibfield  {author} {\bibinfo {author} {\bibfnamefont {A.~A.}\ \bibnamefont
  {Abrikosov}}\ and\ \bibinfo {author} {\bibfnamefont {L.~P.}\ \bibnamefont
  {Gor'kov}},\ }\href@noop {} {\bibfield  {journal} {\bibinfo  {journal} {Sov.
  Phys. JETP}\ }\textbf {\bibinfo {volume} {12}},\ \bibinfo {pages} {1243}
  (\bibinfo {year} {1961})}\BibitemShut {NoStop}%
\bibitem [{\citenamefont {Balian}\ and\ \citenamefont
  {Werthamer}(1963)}]{Balian1963PR}%
  \BibitemOpen
  \bibfield  {author} {\bibinfo {author} {\bibfnamefont {R.}~\bibnamefont
  {Balian}}\ and\ \bibinfo {author} {\bibfnamefont {N.~R.}\ \bibnamefont
  {Werthamer}},\ }\href {\doibase 10.1103/PhysRev.131.1553} {\bibfield
  {journal} {\bibinfo  {journal} {Phys. Rev.}\ }\textbf {\bibinfo {volume}
  {131}},\ \bibinfo {pages} {1553} (\bibinfo {year} {1963})}\BibitemShut
  {NoStop}%
\bibitem [{\citenamefont {Openov}(1998)}]{Openov1998PRB_impurity}%
  \BibitemOpen
  \bibfield  {author} {\bibinfo {author} {\bibfnamefont {L.~A.}\ \bibnamefont
  {Openov}},\ }\href {\doibase 10.1103/PhysRevB.58.9468} {\bibfield  {journal}
  {\bibinfo  {journal} {Phys. Rev. B}\ }\textbf {\bibinfo {volume} {58}},\
  \bibinfo {pages} {9468} (\bibinfo {year} {1998})}\BibitemShut {NoStop}%
\bibitem [{\citenamefont {Senga}\ and\ \citenamefont
  {Kontani}(2008)}]{SengaKontani2008JPSJ}%
  \BibitemOpen
  \bibfield  {author} {\bibinfo {author} {\bibfnamefont {Y.}~\bibnamefont
  {Senga}}\ and\ \bibinfo {author} {\bibfnamefont {H.}~\bibnamefont
  {Kontani}},\ }\href {\doibase 10.1143/JPSJ.77.113710} {\bibfield  {journal}
  {\bibinfo  {journal} {Journal of the Physical Society of Japan}\ }\textbf
  {\bibinfo {volume} {77}},\ \bibinfo {pages} {113710} (\bibinfo {year}
  {2008})}\BibitemShut {NoStop}%
\bibitem [{\citenamefont {Onari}\ and\ \citenamefont
  {Kontani}(2009)}]{OnariKontani2009PRL}%
  \BibitemOpen
  \bibfield  {author} {\bibinfo {author} {\bibfnamefont {S.}~\bibnamefont
  {Onari}}\ and\ \bibinfo {author} {\bibfnamefont {H.}~\bibnamefont
  {Kontani}},\ }\href {\doibase 10.1103/PhysRevLett.103.177001} {\bibfield
  {journal} {\bibinfo  {journal} {Phys. Rev. Lett.}\ }\textbf {\bibinfo
  {volume} {103}},\ \bibinfo {pages} {177001} (\bibinfo {year}
  {2009})}\BibitemShut {NoStop}%
\bibitem [{\citenamefont {Efremov}\ \emph {et~al.}(2011)\citenamefont
  {Efremov}, \citenamefont {Korshunov}, \citenamefont {Dolgov}, \citenamefont
  {Golubov},\ and\ \citenamefont {Hirschfeld}}]{EfremovHirschfeld2011PRB}%
  \BibitemOpen
  \bibfield  {author} {\bibinfo {author} {\bibfnamefont {D.~V.}\ \bibnamefont
  {Efremov}}, \bibinfo {author} {\bibfnamefont {M.~M.}\ \bibnamefont
  {Korshunov}}, \bibinfo {author} {\bibfnamefont {O.~V.}\ \bibnamefont
  {Dolgov}}, \bibinfo {author} {\bibfnamefont {A.~A.}\ \bibnamefont {Golubov}},
  \ and\ \bibinfo {author} {\bibfnamefont {P.~J.}\ \bibnamefont {Hirschfeld}},\
  }\href {\doibase 10.1103/PhysRevB.84.180512} {\bibfield  {journal} {\bibinfo
  {journal} {Phys. Rev. B}\ }\textbf {\bibinfo {volume} {84}},\ \bibinfo
  {pages} {180512} (\bibinfo {year} {2011})}\BibitemShut {NoStop}%
\bibitem [{\citenamefont {Fernandes}\ \emph {et~al.}(2012)\citenamefont
  {Fernandes}, \citenamefont {Vavilov},\ and\ \citenamefont
  {Chubukov}}]{FernandesChubukov2012PRB_TcEnhancement}%
  \BibitemOpen
  \bibfield  {author} {\bibinfo {author} {\bibfnamefont {R.~M.}\ \bibnamefont
  {Fernandes}}, \bibinfo {author} {\bibfnamefont {M.~G.}\ \bibnamefont
  {Vavilov}}, \ and\ \bibinfo {author} {\bibfnamefont {A.~V.}\ \bibnamefont
  {Chubukov}},\ }\href {\doibase 10.1103/PhysRevB.85.140512} {\bibfield
  {journal} {\bibinfo  {journal} {Phys. Rev. B}\ }\textbf {\bibinfo {volume}
  {85}},\ \bibinfo {pages} {140512} (\bibinfo {year} {2012})}\BibitemShut
  {NoStop}%
\bibitem [{\citenamefont {Wang}\ \emph {et~al.}(2013)\citenamefont {Wang},
  \citenamefont {Kreisel}, \citenamefont {Hirschfeld},\ and\ \citenamefont
  {Mishra}}]{WangHirschfeldMishra2013PRB}%
  \BibitemOpen
  \bibfield  {author} {\bibinfo {author} {\bibfnamefont {Y.}~\bibnamefont
  {Wang}}, \bibinfo {author} {\bibfnamefont {A.}~\bibnamefont {Kreisel}},
  \bibinfo {author} {\bibfnamefont {P.~J.}\ \bibnamefont {Hirschfeld}}, \ and\
  \bibinfo {author} {\bibfnamefont {V.}~\bibnamefont {Mishra}},\ }\href
  {\doibase 10.1103/PhysRevB.87.094504} {\bibfield  {journal} {\bibinfo
  {journal} {Phys. Rev. B}\ }\textbf {\bibinfo {volume} {87}},\ \bibinfo
  {pages} {094504} (\bibinfo {year} {2013})}\BibitemShut {NoStop}%
\bibitem [{\citenamefont {Chen}\ \emph {et~al.}(2016)\citenamefont {Chen},
  \citenamefont {Mishra}, \citenamefont {Maiti},\ and\ \citenamefont
  {Hirschfeld}}]{ChenMishraHirschfield2016PRB_impurity_theory}%
  \BibitemOpen
  \bibfield  {author} {\bibinfo {author} {\bibfnamefont {X.}~\bibnamefont
  {Chen}}, \bibinfo {author} {\bibfnamefont {V.}~\bibnamefont {Mishra}},
  \bibinfo {author} {\bibfnamefont {S.}~\bibnamefont {Maiti}}, \ and\ \bibinfo
  {author} {\bibfnamefont {P.~J.}\ \bibnamefont {Hirschfeld}},\ }\href
  {\doibase 10.1103/PhysRevB.94.054524} {\bibfield  {journal} {\bibinfo
  {journal} {Phys. Rev. B}\ }\textbf {\bibinfo {volume} {94}},\ \bibinfo
  {pages} {054524} (\bibinfo {year} {2016})}\BibitemShut {NoStop}%
\bibitem [{\citenamefont {Korshunov}\ \emph {et~al.}(2017)\citenamefont
  {Korshunov}, \citenamefont {Togushova},\ and\ \citenamefont
  {Dolgov}}]{Korshunov2017Phy-Usp_impurity_theory_review}%
  \BibitemOpen
  \bibfield  {author} {\bibinfo {author} {\bibfnamefont {M.~M.}\ \bibnamefont
  {Korshunov}}, \bibinfo {author} {\bibfnamefont {Y.~N.}\ \bibnamefont
  {Togushova}}, \ and\ \bibinfo {author} {\bibfnamefont {O.~V.}\ \bibnamefont
  {Dolgov}},\ }\href {http://stacks.iop.org/1063-7869/59/i=12/a=1211}
  {\bibfield  {journal} {\bibinfo  {journal} {Physics-Uspekhi}\ }\textbf
  {\bibinfo {volume} {59}},\ \bibinfo {pages} {1211} (\bibinfo {year}
  {2017})}\BibitemShut {NoStop}%
\bibitem [{\citenamefont {Tsuei}\ and\ \citenamefont
  {Kirtley}(2000)}]{TsueiKirtley2000RMP_review_cuprates}%
  \BibitemOpen
  \bibfield  {author} {\bibinfo {author} {\bibfnamefont {C.~C.}\ \bibnamefont
  {Tsuei}}\ and\ \bibinfo {author} {\bibfnamefont {J.~R.}\ \bibnamefont
  {Kirtley}},\ }\href@noop {} {\bibfield  {journal} {\bibinfo  {journal} {Rev.
  Mod. Phys.}\ }\textbf {\bibinfo {volume} {72}},\ \bibinfo {pages} {969}
  (\bibinfo {year} {2000})}\BibitemShut {NoStop}%
\bibitem [{\citenamefont {Alloul}\ \emph {et~al.}(2009)\citenamefont {Alloul},
  \citenamefont {Bobroff}, \citenamefont {Gabay},\ and\ \citenamefont
  {Hirschfeld}}]{Alloul2009RMP_defect_cuprates}%
  \BibitemOpen
  \bibfield  {author} {\bibinfo {author} {\bibfnamefont {H.}~\bibnamefont
  {Alloul}}, \bibinfo {author} {\bibfnamefont {J.}~\bibnamefont {Bobroff}},
  \bibinfo {author} {\bibfnamefont {M.}~\bibnamefont {Gabay}}, \ and\ \bibinfo
  {author} {\bibfnamefont {P.~J.}\ \bibnamefont {Hirschfeld}},\ }\href
  {\doibase 10.1103/RevModPhys.81.45} {\bibfield  {journal} {\bibinfo
  {journal} {Rev. Mod. Phys.}\ }\textbf {\bibinfo {volume} {81}},\ \bibinfo
  {pages} {45} (\bibinfo {year} {2009})}\BibitemShut {NoStop}%
\bibitem [{\citenamefont {Damask}\ and\ \citenamefont
  {Dienes}(1963)}]{Damask1963PointDefectsInMetals}%
  \BibitemOpen
  \bibfield  {author} {\bibinfo {author} {\bibfnamefont {A.~C.}\ \bibnamefont
  {Damask}}\ and\ \bibinfo {author} {\bibfnamefont {G.~J.}\ \bibnamefont
  {Dienes}},\ }\href@noop {} {\emph {\bibinfo {title} {Point Defects in
  Metals}}}\ (\bibinfo  {publisher} {Gordon \& Breach Science Publishers Ltd,
  London},\ \bibinfo {year} {1963})\BibitemShut {NoStop}%
\bibitem [{\citenamefont {Nakajima}\ \emph {et~al.}(2009)\citenamefont
  {Nakajima}, \citenamefont {Tsuchiya}, \citenamefont {Taen}, \citenamefont
  {Tamegai}, \citenamefont {Okayasu},\ and\ \citenamefont
  {Sasase}}]{Nakajima2009PRB_HeavyIon}%
  \BibitemOpen
  \bibfield  {author} {\bibinfo {author} {\bibfnamefont {Y.}~\bibnamefont
  {Nakajima}}, \bibinfo {author} {\bibfnamefont {Y.}~\bibnamefont {Tsuchiya}},
  \bibinfo {author} {\bibfnamefont {T.}~\bibnamefont {Taen}}, \bibinfo {author}
  {\bibfnamefont {T.}~\bibnamefont {Tamegai}}, \bibinfo {author} {\bibfnamefont
  {S.}~\bibnamefont {Okayasu}}, \ and\ \bibinfo {author} {\bibfnamefont
  {M.}~\bibnamefont {Sasase}},\ }\href {\doibase 10.1103/PhysRevB.80.012510}
  {\bibfield  {journal} {\bibinfo  {journal} {Phys. Rev. B}\ }\textbf {\bibinfo
  {volume} {80}},\ \bibinfo {pages} {012510} (\bibinfo {year}
  {2009})}\BibitemShut {NoStop}%
\bibitem [{\citenamefont {Civale}\ \emph {et~al.}(1991)\citenamefont {Civale},
  \citenamefont {Marwick}, \citenamefont {Worthington}, \citenamefont {Kirk},
  \citenamefont {Thompson}, \citenamefont {Krusin-Elbaum}, \citenamefont {Sun},
  \citenamefont {Clem},\ and\ \citenamefont
  {Holtzberg}}]{CivaleHoltzberg1991PRL_YBCO_columnar_defects}%
  \BibitemOpen
  \bibfield  {author} {\bibinfo {author} {\bibfnamefont {L.}~\bibnamefont
  {Civale}}, \bibinfo {author} {\bibfnamefont {A.~D.}\ \bibnamefont {Marwick}},
  \bibinfo {author} {\bibfnamefont {T.~K.}\ \bibnamefont {Worthington}},
  \bibinfo {author} {\bibfnamefont {M.~A.}\ \bibnamefont {Kirk}}, \bibinfo
  {author} {\bibfnamefont {J.~R.}\ \bibnamefont {Thompson}}, \bibinfo {author}
  {\bibfnamefont {L.}~\bibnamefont {Krusin-Elbaum}}, \bibinfo {author}
  {\bibfnamefont {Y.}~\bibnamefont {Sun}}, \bibinfo {author} {\bibfnamefont
  {J.~R.}\ \bibnamefont {Clem}}, \ and\ \bibinfo {author} {\bibfnamefont
  {F.}~\bibnamefont {Holtzberg}},\ }\href {\doibase 10.1103/PhysRevLett.67.648}
  {\bibfield  {journal} {\bibinfo  {journal} {Phys. Rev. Lett.}\ }\textbf
  {\bibinfo {volume} {67}},\ \bibinfo {pages} {648} (\bibinfo {year}
  {1991})}\BibitemShut {NoStop}%
\bibitem [{\citenamefont {Konczykowski}\ \emph {et~al.}(1991)\citenamefont
  {Konczykowski}, \citenamefont {Rullier-Albenque}, \citenamefont {Yacoby},
  \citenamefont {Shaulov}, \citenamefont {Yeshurun},\ and\ \citenamefont
  {Lejay}}]{KonczykowskiPRB1991_HeavyIon_YBCO}%
  \BibitemOpen
  \bibfield  {author} {\bibinfo {author} {\bibfnamefont {M.}~\bibnamefont
  {Konczykowski}}, \bibinfo {author} {\bibfnamefont {F.}~\bibnamefont
  {Rullier-Albenque}}, \bibinfo {author} {\bibfnamefont {E.~R.}\ \bibnamefont
  {Yacoby}}, \bibinfo {author} {\bibfnamefont {A.}~\bibnamefont {Shaulov}},
  \bibinfo {author} {\bibfnamefont {Y.}~\bibnamefont {Yeshurun}}, \ and\
  \bibinfo {author} {\bibfnamefont {P.}~\bibnamefont {Lejay}},\ }\href
  {\doibase 10.1103/PhysRevB.44.7167} {\bibfield  {journal} {\bibinfo
  {journal} {Phys. Rev. B}\ }\textbf {\bibinfo {volume} {44}},\ \bibinfo
  {pages} {7167} (\bibinfo {year} {1991})}\BibitemShut {NoStop}%
\bibitem [{\citenamefont {Kim}\ \emph {et~al.}(2010)\citenamefont {Kim},
  \citenamefont {Gordon}, \citenamefont {Tanatar}, \citenamefont {Hua},
  \citenamefont {Welp}, \citenamefont {Kwok}, \citenamefont {Ni}, \citenamefont
  {Bud'ko}, \citenamefont {Canfield}, \citenamefont {Vorontsov},\ and\
  \citenamefont {Prozorov}}]{KimProzorov2010PRB_heavyIon}%
  \BibitemOpen
  \bibfield  {author} {\bibinfo {author} {\bibfnamefont {H.}~\bibnamefont
  {Kim}}, \bibinfo {author} {\bibfnamefont {R.~T.}\ \bibnamefont {Gordon}},
  \bibinfo {author} {\bibfnamefont {M.~A.}\ \bibnamefont {Tanatar}}, \bibinfo
  {author} {\bibfnamefont {J.}~\bibnamefont {Hua}}, \bibinfo {author}
  {\bibfnamefont {U.}~\bibnamefont {Welp}}, \bibinfo {author} {\bibfnamefont
  {W.~K.}\ \bibnamefont {Kwok}}, \bibinfo {author} {\bibfnamefont
  {N.}~\bibnamefont {Ni}}, \bibinfo {author} {\bibfnamefont {S.~L.}\
  \bibnamefont {Bud'ko}}, \bibinfo {author} {\bibfnamefont {P.~C.}\
  \bibnamefont {Canfield}}, \bibinfo {author} {\bibfnamefont {A.~B.}\
  \bibnamefont {Vorontsov}}, \ and\ \bibinfo {author} {\bibfnamefont
  {R.}~\bibnamefont {Prozorov}},\ }\href@noop {} {\bibfield  {journal}
  {\bibinfo  {journal} {Phys. Rev. B}\ }\textbf {\bibinfo {volume} {82}},\
  \bibinfo {pages} {060518} (\bibinfo {year} {2010})}\BibitemShut {NoStop}%
\bibitem [{\citenamefont {Prozorov}\ \emph {et~al.}(2010)\citenamefont
  {Prozorov}, \citenamefont {Tanatar}, \citenamefont {Roy}, \citenamefont {Ni},
  \citenamefont {Bud'ko}, \citenamefont {Canfield}, \citenamefont {Hua},
  \citenamefont {Welp},\ and\ \citenamefont {Kwok}}]{Prozorov2010PRB_HeavyIon}%
  \BibitemOpen
  \bibfield  {author} {\bibinfo {author} {\bibfnamefont {R.}~\bibnamefont
  {Prozorov}}, \bibinfo {author} {\bibfnamefont {M.~A.}\ \bibnamefont
  {Tanatar}}, \bibinfo {author} {\bibfnamefont {B.}~\bibnamefont {Roy}},
  \bibinfo {author} {\bibfnamefont {N.}~\bibnamefont {Ni}}, \bibinfo {author}
  {\bibfnamefont {S.~L.}\ \bibnamefont {Bud'ko}}, \bibinfo {author}
  {\bibfnamefont {P.~C.}\ \bibnamefont {Canfield}}, \bibinfo {author}
  {\bibfnamefont {J.}~\bibnamefont {Hua}}, \bibinfo {author} {\bibfnamefont
  {U.}~\bibnamefont {Welp}}, \ and\ \bibinfo {author} {\bibfnamefont {W.~K.}\
  \bibnamefont {Kwok}},\ }\href {\doibase 10.1103/PhysRevB.81.094509}
  {\bibfield  {journal} {\bibinfo  {journal} {Phys. Rev. B}\ }\textbf {\bibinfo
  {volume} {81}},\ \bibinfo {pages} {094509} (\bibinfo {year}
  {2010})}\BibitemShut {NoStop}%
\bibitem [{\citenamefont {Murphy}\ \emph
  {et~al.}(2013{\natexlab{a}})\citenamefont {Murphy}, \citenamefont {Tanatar},
  \citenamefont {Kim}, \citenamefont {Kwok}, \citenamefont {Welp},
  \citenamefont {Graf}, \citenamefont {Brooks}, \citenamefont {Bud'ko},
  \citenamefont {Canfield},\ and\ \citenamefont
  {Prozorov}}]{MurphyProzorov2013PRB_heavy_ion}%
  \BibitemOpen
  \bibfield  {author} {\bibinfo {author} {\bibfnamefont {J.}~\bibnamefont
  {Murphy}}, \bibinfo {author} {\bibfnamefont {M.~A.}\ \bibnamefont {Tanatar}},
  \bibinfo {author} {\bibfnamefont {H.}~\bibnamefont {Kim}}, \bibinfo {author}
  {\bibfnamefont {W.}~\bibnamefont {Kwok}}, \bibinfo {author} {\bibfnamefont
  {U.}~\bibnamefont {Welp}}, \bibinfo {author} {\bibfnamefont {D.}~\bibnamefont
  {Graf}}, \bibinfo {author} {\bibfnamefont {J.~S.}\ \bibnamefont {Brooks}},
  \bibinfo {author} {\bibfnamefont {S.~L.}\ \bibnamefont {Bud'ko}}, \bibinfo
  {author} {\bibfnamefont {P.~C.}\ \bibnamefont {Canfield}}, \ and\ \bibinfo
  {author} {\bibfnamefont {R.}~\bibnamefont {Prozorov}},\ }\href {\doibase
  10.1103/PhysRevB.88.054514} {\bibfield  {journal} {\bibinfo  {journal} {Phys.
  Rev. B}\ }\textbf {\bibinfo {volume} {88}},\ \bibinfo {pages} {054514}
  (\bibinfo {year} {2013}{\natexlab{a}})}\BibitemShut {NoStop}%
\bibitem [{\citenamefont {Nakajima}\ \emph {et~al.}(2010)\citenamefont
  {Nakajima}, \citenamefont {Taen}, \citenamefont {Tsuchiya}, \citenamefont
  {Tamegai}, \citenamefont {Kitamura},\ and\ \citenamefont
  {Murakami}}]{Nakajima2010PRB_BaCo122_proton}%
  \BibitemOpen
  \bibfield  {author} {\bibinfo {author} {\bibfnamefont {Y.}~\bibnamefont
  {Nakajima}}, \bibinfo {author} {\bibfnamefont {T.}~\bibnamefont {Taen}},
  \bibinfo {author} {\bibfnamefont {Y.}~\bibnamefont {Tsuchiya}}, \bibinfo
  {author} {\bibfnamefont {T.}~\bibnamefont {Tamegai}}, \bibinfo {author}
  {\bibfnamefont {H.}~\bibnamefont {Kitamura}}, \ and\ \bibinfo {author}
  {\bibfnamefont {T.}~\bibnamefont {Murakami}},\ }\href {\doibase
  10.1103/PhysRevB.82.220504} {\bibfield  {journal} {\bibinfo  {journal} {Phys.
  Rev. B}\ }\textbf {\bibinfo {volume} {82}},\ \bibinfo {pages} {220504}
  (\bibinfo {year} {2010})}\BibitemShut {NoStop}%
\bibitem [{\citenamefont {Taen}\ \emph {et~al.}(2013)\citenamefont {Taen},
  \citenamefont {Ohtake}, \citenamefont {Akiyama}, \citenamefont {Inoue},
  \citenamefont {Sun}, \citenamefont {Pyon}, \citenamefont {Tamegai},\ and\
  \citenamefont {Kitamura}}]{Taen2013PRB_proton_irr}%
  \BibitemOpen
  \bibfield  {author} {\bibinfo {author} {\bibfnamefont {T.}~\bibnamefont
  {Taen}}, \bibinfo {author} {\bibfnamefont {F.}~\bibnamefont {Ohtake}},
  \bibinfo {author} {\bibfnamefont {H.}~\bibnamefont {Akiyama}}, \bibinfo
  {author} {\bibfnamefont {H.}~\bibnamefont {Inoue}}, \bibinfo {author}
  {\bibfnamefont {Y.}~\bibnamefont {Sun}}, \bibinfo {author} {\bibfnamefont
  {S.}~\bibnamefont {Pyon}}, \bibinfo {author} {\bibfnamefont {T.}~\bibnamefont
  {Tamegai}}, \ and\ \bibinfo {author} {\bibfnamefont {H.}~\bibnamefont
  {Kitamura}},\ }\href {\doibase 10.1103/PhysRevB.88.224514} {\bibfield
  {journal} {\bibinfo  {journal} {Phys. Rev. B}\ }\textbf {\bibinfo {volume}
  {88}},\ \bibinfo {pages} {224514} (\bibinfo {year} {2013})}\BibitemShut
  {NoStop}%
\bibitem [{\citenamefont {Smylie}\ \emph {et~al.}(2016)\citenamefont {Smylie},
  \citenamefont {Leroux}, \citenamefont {Mishra}, \citenamefont {Fang},
  \citenamefont {Taddei}, \citenamefont {Chmaissem}, \citenamefont {Claus},
  \citenamefont {Kayani}, \citenamefont {Snezhko}, \citenamefont {Welp},\ and\
  \citenamefont {Kwok}}]{Smylie2016PRB_BaP122_proton_irr}%
  \BibitemOpen
  \bibfield  {author} {\bibinfo {author} {\bibfnamefont {M.~P.}\ \bibnamefont
  {Smylie}}, \bibinfo {author} {\bibfnamefont {M.}~\bibnamefont {Leroux}},
  \bibinfo {author} {\bibfnamefont {V.}~\bibnamefont {Mishra}}, \bibinfo
  {author} {\bibfnamefont {L.}~\bibnamefont {Fang}}, \bibinfo {author}
  {\bibfnamefont {K.~M.}\ \bibnamefont {Taddei}}, \bibinfo {author}
  {\bibfnamefont {O.}~\bibnamefont {Chmaissem}}, \bibinfo {author}
  {\bibfnamefont {H.}~\bibnamefont {Claus}}, \bibinfo {author} {\bibfnamefont
  {A.}~\bibnamefont {Kayani}}, \bibinfo {author} {\bibfnamefont
  {A.}~\bibnamefont {Snezhko}}, \bibinfo {author} {\bibfnamefont
  {U.}~\bibnamefont {Welp}}, \ and\ \bibinfo {author} {\bibfnamefont {W.-K.}\
  \bibnamefont {Kwok}},\ }\href {\doibase 10.1103/PhysRevB.93.115119}
  {\bibfield  {journal} {\bibinfo  {journal} {Phys. Rev. B}\ }\textbf {\bibinfo
  {volume} {93}},\ \bibinfo {pages} {115119} (\bibinfo {year}
  {2016})}\BibitemShut {NoStop}%
\bibitem [{\citenamefont {Moroni}\ \emph {et~al.}(2017)\citenamefont {Moroni},
  \citenamefont {Gozzelino}, \citenamefont {Ghigo}, \citenamefont {Tanatar},
  \citenamefont {Prozorov}, \citenamefont {Canfield},\ and\ \citenamefont
  {Carretta}}]{Moroni2017PRB_proton-irr}%
  \BibitemOpen
  \bibfield  {author} {\bibinfo {author} {\bibfnamefont {M.}~\bibnamefont
  {Moroni}}, \bibinfo {author} {\bibfnamefont {L.}~\bibnamefont {Gozzelino}},
  \bibinfo {author} {\bibfnamefont {G.}~\bibnamefont {Ghigo}}, \bibinfo
  {author} {\bibfnamefont {M.~A.}\ \bibnamefont {Tanatar}}, \bibinfo {author}
  {\bibfnamefont {R.}~\bibnamefont {Prozorov}}, \bibinfo {author}
  {\bibfnamefont {P.~C.}\ \bibnamefont {Canfield}}, \ and\ \bibinfo {author}
  {\bibfnamefont {P.}~\bibnamefont {Carretta}},\ }\href {\doibase
  10.1103/PhysRevB.96.094523} {\bibfield  {journal} {\bibinfo  {journal} {Phys.
  Rev. B}\ }\textbf {\bibinfo {volume} {96}},\ \bibinfo {pages} {094523}
  (\bibinfo {year} {2017})}\BibitemShut {NoStop}%
\bibitem [{\citenamefont {Tarantini}\ \emph {et~al.}(2010)\citenamefont
  {Tarantini}, \citenamefont {Putti}, \citenamefont {Gurevich}, \citenamefont
  {Shen}, \citenamefont {Singh}, \citenamefont {Rowell}, \citenamefont
  {Newman}, \citenamefont {Larbalestier}, \citenamefont {Cheng}, \citenamefont
  {Jia},\ and\ \citenamefont {Wen}}]{Tarantini2010PRL_alpha_irr}%
  \BibitemOpen
  \bibfield  {author} {\bibinfo {author} {\bibfnamefont {C.}~\bibnamefont
  {Tarantini}}, \bibinfo {author} {\bibfnamefont {M.}~\bibnamefont {Putti}},
  \bibinfo {author} {\bibfnamefont {A.}~\bibnamefont {Gurevich}}, \bibinfo
  {author} {\bibfnamefont {Y.}~\bibnamefont {Shen}}, \bibinfo {author}
  {\bibfnamefont {R.~K.}\ \bibnamefont {Singh}}, \bibinfo {author}
  {\bibfnamefont {J.~M.}\ \bibnamefont {Rowell}}, \bibinfo {author}
  {\bibfnamefont {N.}~\bibnamefont {Newman}}, \bibinfo {author} {\bibfnamefont
  {D.~C.}\ \bibnamefont {Larbalestier}}, \bibinfo {author} {\bibfnamefont
  {P.}~\bibnamefont {Cheng}}, \bibinfo {author} {\bibfnamefont
  {Y.}~\bibnamefont {Jia}}, \ and\ \bibinfo {author} {\bibfnamefont {H.-H.}\
  \bibnamefont {Wen}},\ }\href {\doibase 10.1103/PhysRevLett.104.087002}
  {\bibfield  {journal} {\bibinfo  {journal} {Phys. Rev. Lett.}\ }\textbf
  {\bibinfo {volume} {104}},\ \bibinfo {pages} {087002} (\bibinfo {year}
  {2010})}\BibitemShut {NoStop}%
\bibitem [{\citenamefont {Yamakawa}\ \emph {et~al.}(2013)\citenamefont
  {Yamakawa}, \citenamefont {Onari},\ and\ \citenamefont
  {Kontani}}]{Yamakawa2013PRB_impurity_theory}%
  \BibitemOpen
  \bibfield  {author} {\bibinfo {author} {\bibfnamefont {Y.}~\bibnamefont
  {Yamakawa}}, \bibinfo {author} {\bibfnamefont {S.}~\bibnamefont {Onari}}, \
  and\ \bibinfo {author} {\bibfnamefont {H.}~\bibnamefont {Kontani}},\ }\href
  {\doibase 10.1103/PhysRevB.87.195121} {\bibfield  {journal} {\bibinfo
  {journal} {Phys. Rev. B}\ }\textbf {\bibinfo {volume} {87}},\ \bibinfo
  {pages} {195121} (\bibinfo {year} {2013})}\BibitemShut {NoStop}%
\bibitem [{\citenamefont
  {Mott}(1929)}]{Mott1929ProRoySocLondon_electon_scattering}%
  \BibitemOpen
  \bibfield  {author} {\bibinfo {author} {\bibfnamefont {N.}~\bibnamefont
  {Mott}},\ }\href {\doibase 10.1098/rspa.1929.0127} {\bibfield  {journal}
  {\bibinfo  {journal} {Proceedings of the Royal Society of London A:
  Mathematical, Physical and Engineering Sciences}\ }\textbf {\bibinfo {volume}
  {124}},\ \bibinfo {pages} {425} (\bibinfo {year} {1929})}\BibitemShut
  {NoStop}%
\bibitem [{\citenamefont {Prozorov}\ \emph {et~al.}(2014)\citenamefont
  {Prozorov}, \citenamefont {Ko\ifmmode~\acute{n}\else \'{n}\fi{}czykowski},
  \citenamefont {Tanatar}, \citenamefont {Thaler}, \citenamefont {Bud'ko},
  \citenamefont {Canfield}, \citenamefont {Mishra},\ and\ \citenamefont
  {Hirschfeld}}]{Prozorov2014PRX_e-irr}%
  \BibitemOpen
  \bibfield  {author} {\bibinfo {author} {\bibfnamefont {R.}~\bibnamefont
  {Prozorov}}, \bibinfo {author} {\bibfnamefont {M.}~\bibnamefont
  {Ko\ifmmode~\acute{n}\else \'{n}\fi{}czykowski}}, \bibinfo {author}
  {\bibfnamefont {M.~A.}\ \bibnamefont {Tanatar}}, \bibinfo {author}
  {\bibfnamefont {A.}~\bibnamefont {Thaler}}, \bibinfo {author} {\bibfnamefont
  {S.~L.}\ \bibnamefont {Bud'ko}}, \bibinfo {author} {\bibfnamefont {P.~C.}\
  \bibnamefont {Canfield}}, \bibinfo {author} {\bibfnamefont {V.}~\bibnamefont
  {Mishra}}, \ and\ \bibinfo {author} {\bibfnamefont {P.~J.}\ \bibnamefont
  {Hirschfeld}},\ }\href {\doibase 10.1103/PhysRevX.4.041032} {\bibfield
  {journal} {\bibinfo  {journal} {Phys. Rev. X}\ }\textbf {\bibinfo {volume}
  {4}},\ \bibinfo {pages} {041032} (\bibinfo {year} {2014})}\BibitemShut
  {NoStop}%
\bibitem [{\citenamefont {Strehlow}\ \emph {et~al.}(2014)\citenamefont
  {Strehlow}, \citenamefont {Ko\ifmmode~\acute{n}\else \'{n}\fi{}czykowski},
  \citenamefont {Murphy}, \citenamefont {Teknowijoyo}, \citenamefont {Cho},
  \citenamefont {Tanatar}, \citenamefont {Kobayashi}, \citenamefont {Miyasaka},
  \citenamefont {Tajima},\ and\ \citenamefont
  {Prozorov}}]{StrehlowProzorov2014PRB_SrP122}%
  \BibitemOpen
  \bibfield  {author} {\bibinfo {author} {\bibfnamefont {C.~P.}\ \bibnamefont
  {Strehlow}}, \bibinfo {author} {\bibfnamefont {M.}~\bibnamefont
  {Ko\ifmmode~\acute{n}\else \'{n}\fi{}czykowski}}, \bibinfo {author}
  {\bibfnamefont {J.~A.}\ \bibnamefont {Murphy}}, \bibinfo {author}
  {\bibfnamefont {S.}~\bibnamefont {Teknowijoyo}}, \bibinfo {author}
  {\bibfnamefont {K.}~\bibnamefont {Cho}}, \bibinfo {author} {\bibfnamefont
  {M.~A.}\ \bibnamefont {Tanatar}}, \bibinfo {author} {\bibfnamefont
  {T.}~\bibnamefont {Kobayashi}}, \bibinfo {author} {\bibfnamefont
  {S.}~\bibnamefont {Miyasaka}}, \bibinfo {author} {\bibfnamefont
  {S.}~\bibnamefont {Tajima}}, \ and\ \bibinfo {author} {\bibfnamefont
  {R.}~\bibnamefont {Prozorov}},\ }\href {\doibase 10.1103/PhysRevB.90.020508}
  {\bibfield  {journal} {\bibinfo  {journal} {Phys. Rev. B}\ }\textbf {\bibinfo
  {volume} {90}},\ \bibinfo {pages} {020508} (\bibinfo {year}
  {2014})}\BibitemShut {NoStop}%
\bibitem [{\citenamefont {Mishra}\ \emph {et~al.}(2009)\citenamefont {Mishra},
  \citenamefont {Boyd}, \citenamefont {Graser}, \citenamefont {Maier},
  \citenamefont {Hirschfeld},\ and\ \citenamefont {Scalapino}}]{Mishra2009PRB}%
  \BibitemOpen
  \bibfield  {author} {\bibinfo {author} {\bibfnamefont {V.}~\bibnamefont
  {Mishra}}, \bibinfo {author} {\bibfnamefont {G.}~\bibnamefont {Boyd}},
  \bibinfo {author} {\bibfnamefont {S.}~\bibnamefont {Graser}}, \bibinfo
  {author} {\bibfnamefont {T.}~\bibnamefont {Maier}}, \bibinfo {author}
  {\bibfnamefont {P.~J.}\ \bibnamefont {Hirschfeld}}, \ and\ \bibinfo {author}
  {\bibfnamefont {D.~J.}\ \bibnamefont {Scalapino}},\ }\href {\doibase
  10.1103/PhysRevB.79.094512} {\bibfield  {journal} {\bibinfo  {journal} {Phys.
  Rev. B}\ }\textbf {\bibinfo {volume} {79}},\ \bibinfo {pages} {094512}
  (\bibinfo {year} {2009})}\BibitemShut {NoStop}%
\bibitem [{\citenamefont {Hirschfeld}\ and\ \citenamefont
  {Goldenfeld}(1993)}]{Hirschfeld1993PRB}%
  \BibitemOpen
  \bibfield  {author} {\bibinfo {author} {\bibfnamefont {P.~J.}\ \bibnamefont
  {Hirschfeld}}\ and\ \bibinfo {author} {\bibfnamefont {N.}~\bibnamefont
  {Goldenfeld}},\ }\href@noop {} {\bibfield  {journal} {\bibinfo  {journal}
  {Phys. Rev. B}\ }\textbf {\bibinfo {volume} {48}},\ \bibinfo {pages} {4219}
  (\bibinfo {year} {1993})}\BibitemShut {NoStop}%
\bibitem [{\citenamefont {Bonn}\ \emph {et~al.}(1994)\citenamefont {Bonn},
  \citenamefont {Kamal}, \citenamefont {Zhang}, \citenamefont {Liang},
  \citenamefont {Baar}, \citenamefont {Klein},\ and\ \citenamefont
  {Hardy}}]{BonnHardy1994PRB_YBCO_impurity}%
  \BibitemOpen
  \bibfield  {author} {\bibinfo {author} {\bibfnamefont {D.~A.}\ \bibnamefont
  {Bonn}}, \bibinfo {author} {\bibfnamefont {S.}~\bibnamefont {Kamal}},
  \bibinfo {author} {\bibfnamefont {K.}~\bibnamefont {Zhang}}, \bibinfo
  {author} {\bibfnamefont {R.}~\bibnamefont {Liang}}, \bibinfo {author}
  {\bibfnamefont {D.~J.}\ \bibnamefont {Baar}}, \bibinfo {author}
  {\bibfnamefont {E.}~\bibnamefont {Klein}}, \ and\ \bibinfo {author}
  {\bibfnamefont {W.~N.}\ \bibnamefont {Hardy}},\ }\href {\doibase
  10.1103/PhysRevB.50.4051} {\bibfield  {journal} {\bibinfo  {journal} {Phys.
  Rev. B}\ }\textbf {\bibinfo {volume} {50}},\ \bibinfo {pages} {4051}
  (\bibinfo {year} {1994})}\BibitemShut {NoStop}%
\bibitem [{\citenamefont {Luetkens}\ \emph {et~al.}(2008)\citenamefont
  {Luetkens}, \citenamefont {Klauss}, \citenamefont {Khasanov}, \citenamefont
  {Amato}, \citenamefont {Klingeler}, \citenamefont {Hellmann}, \citenamefont
  {Leps}, \citenamefont {Kondrat}, \citenamefont {Hess}, \citenamefont
  {K\"ohler}, \citenamefont {Behr}, \citenamefont {Werner},\ and\ \citenamefont
  {B\"uchner}}]{Luetkens2008PRL_uSR}%
  \BibitemOpen
  \bibfield  {author} {\bibinfo {author} {\bibfnamefont {H.}~\bibnamefont
  {Luetkens}}, \bibinfo {author} {\bibfnamefont {H.-H.}\ \bibnamefont
  {Klauss}}, \bibinfo {author} {\bibfnamefont {R.}~\bibnamefont {Khasanov}},
  \bibinfo {author} {\bibfnamefont {A.}~\bibnamefont {Amato}}, \bibinfo
  {author} {\bibfnamefont {R.}~\bibnamefont {Klingeler}}, \bibinfo {author}
  {\bibfnamefont {I.}~\bibnamefont {Hellmann}}, \bibinfo {author}
  {\bibfnamefont {N.}~\bibnamefont {Leps}}, \bibinfo {author} {\bibfnamefont
  {A.}~\bibnamefont {Kondrat}}, \bibinfo {author} {\bibfnamefont
  {C.}~\bibnamefont {Hess}}, \bibinfo {author} {\bibfnamefont {A.}~\bibnamefont
  {K\"ohler}}, \bibinfo {author} {\bibfnamefont {G.}~\bibnamefont {Behr}},
  \bibinfo {author} {\bibfnamefont {J.}~\bibnamefont {Werner}}, \ and\ \bibinfo
  {author} {\bibfnamefont {B.}~\bibnamefont {B\"uchner}},\ }\href {\doibase
  10.1103/PhysRevLett.101.097009} {\bibfield  {journal} {\bibinfo  {journal}
  {Phys. Rev. Lett.}\ }\textbf {\bibinfo {volume} {101}},\ \bibinfo {pages}
  {097009} (\bibinfo {year} {2008})}\BibitemShut {NoStop}%
\bibitem [{\citenamefont {Williams}\ \emph {et~al.}(2010)\citenamefont
  {Williams}, \citenamefont {Aczel}, \citenamefont {Baggio-Saitovitch},
  \citenamefont {Bud'ko}, \citenamefont {Canfield}, \citenamefont {Carlo},
  \citenamefont {Goko}, \citenamefont {Kageyama}, \citenamefont {Kitada},
  \citenamefont {Munevar}, \citenamefont {Ni}, \citenamefont {Saha},
  \citenamefont {Kirschenbaum}, \citenamefont {Paglione}, \citenamefont
  {Sanchez-Candela}, \citenamefont {Uemura},\ and\ \citenamefont
  {Luke}}]{Williams2010PRB_uSR}%
  \BibitemOpen
  \bibfield  {author} {\bibinfo {author} {\bibfnamefont {T.~J.}\ \bibnamefont
  {Williams}}, \bibinfo {author} {\bibfnamefont {A.~A.}\ \bibnamefont {Aczel}},
  \bibinfo {author} {\bibfnamefont {E.}~\bibnamefont {Baggio-Saitovitch}},
  \bibinfo {author} {\bibfnamefont {S.~L.}\ \bibnamefont {Bud'ko}}, \bibinfo
  {author} {\bibfnamefont {P.~C.}\ \bibnamefont {Canfield}}, \bibinfo {author}
  {\bibfnamefont {J.~P.}\ \bibnamefont {Carlo}}, \bibinfo {author}
  {\bibfnamefont {T.}~\bibnamefont {Goko}}, \bibinfo {author} {\bibfnamefont
  {H.}~\bibnamefont {Kageyama}}, \bibinfo {author} {\bibfnamefont
  {A.}~\bibnamefont {Kitada}}, \bibinfo {author} {\bibfnamefont
  {J.}~\bibnamefont {Munevar}}, \bibinfo {author} {\bibfnamefont
  {N.}~\bibnamefont {Ni}}, \bibinfo {author} {\bibfnamefont {S.~R.}\
  \bibnamefont {Saha}}, \bibinfo {author} {\bibfnamefont {K.}~\bibnamefont
  {Kirschenbaum}}, \bibinfo {author} {\bibfnamefont {J.}~\bibnamefont
  {Paglione}}, \bibinfo {author} {\bibfnamefont {D.~R.}\ \bibnamefont
  {Sanchez-Candela}}, \bibinfo {author} {\bibfnamefont {Y.~J.}\ \bibnamefont
  {Uemura}}, \ and\ \bibinfo {author} {\bibfnamefont {G.~M.}\ \bibnamefont
  {Luke}},\ }\href {\doibase 10.1103/PhysRevB.82.094512} {\bibfield  {journal}
  {\bibinfo  {journal} {Phys. Rev. B}\ }\textbf {\bibinfo {volume} {82}},\
  \bibinfo {pages} {094512} (\bibinfo {year} {2010})}\BibitemShut {NoStop}%
\bibitem [{\citenamefont {Sonier}\ \emph {et~al.}(2011)\citenamefont {Sonier},
  \citenamefont {Huang}, \citenamefont {Kaiser}, \citenamefont {Cochrane},
  \citenamefont {Pacradouni}, \citenamefont {Sabok-Sayr}, \citenamefont
  {Lumsden}, \citenamefont {Sales}, \citenamefont {McGuire}, \citenamefont
  {Sefat},\ and\ \citenamefont {Mandrus}}]{Sonier2011PRL_uSR}%
  \BibitemOpen
  \bibfield  {author} {\bibinfo {author} {\bibfnamefont {J.~E.}\ \bibnamefont
  {Sonier}}, \bibinfo {author} {\bibfnamefont {W.}~\bibnamefont {Huang}},
  \bibinfo {author} {\bibfnamefont {C.~V.}\ \bibnamefont {Kaiser}}, \bibinfo
  {author} {\bibfnamefont {C.}~\bibnamefont {Cochrane}}, \bibinfo {author}
  {\bibfnamefont {V.}~\bibnamefont {Pacradouni}}, \bibinfo {author}
  {\bibfnamefont {S.~A.}\ \bibnamefont {Sabok-Sayr}}, \bibinfo {author}
  {\bibfnamefont {M.~D.}\ \bibnamefont {Lumsden}}, \bibinfo {author}
  {\bibfnamefont {B.~C.}\ \bibnamefont {Sales}}, \bibinfo {author}
  {\bibfnamefont {M.~A.}\ \bibnamefont {McGuire}}, \bibinfo {author}
  {\bibfnamefont {A.~S.}\ \bibnamefont {Sefat}}, \ and\ \bibinfo {author}
  {\bibfnamefont {D.}~\bibnamefont {Mandrus}},\ }\href {\doibase
  10.1103/PhysRevLett.106.127002} {\bibfield  {journal} {\bibinfo  {journal}
  {Phys. Rev. Lett.}\ }\textbf {\bibinfo {volume} {106}},\ \bibinfo {pages}
  {127002} (\bibinfo {year} {2011})}\BibitemShut {NoStop}%
\bibitem [{\citenamefont {Vald\'es~Aguilar}\ \emph {et~al.}(2010)\citenamefont
  {Vald\'es~Aguilar}, \citenamefont {Bilbro}, \citenamefont {Lee},
  \citenamefont {Bark}, \citenamefont {Jiang}, \citenamefont {Weiss},
  \citenamefont {Hellstrom}, \citenamefont {Larbalestier}, \citenamefont
  {Eom},\ and\ \citenamefont {Armitage}}]{ValdesAguilar2010PRB_terahertz}%
  \BibitemOpen
  \bibfield  {author} {\bibinfo {author} {\bibfnamefont {R.}~\bibnamefont
  {Vald\'es~Aguilar}}, \bibinfo {author} {\bibfnamefont {L.~S.}\ \bibnamefont
  {Bilbro}}, \bibinfo {author} {\bibfnamefont {S.}~\bibnamefont {Lee}},
  \bibinfo {author} {\bibfnamefont {C.~W.}\ \bibnamefont {Bark}}, \bibinfo
  {author} {\bibfnamefont {J.}~\bibnamefont {Jiang}}, \bibinfo {author}
  {\bibfnamefont {J.~D.}\ \bibnamefont {Weiss}}, \bibinfo {author}
  {\bibfnamefont {E.~E.}\ \bibnamefont {Hellstrom}}, \bibinfo {author}
  {\bibfnamefont {D.~C.}\ \bibnamefont {Larbalestier}}, \bibinfo {author}
  {\bibfnamefont {C.~B.}\ \bibnamefont {Eom}}, \ and\ \bibinfo {author}
  {\bibfnamefont {N.~P.}\ \bibnamefont {Armitage}},\ }\href {\doibase
  10.1103/PhysRevB.82.180514} {\bibfield  {journal} {\bibinfo  {journal} {Phys.
  Rev. B}\ }\textbf {\bibinfo {volume} {82}},\ \bibinfo {pages} {180514}
  (\bibinfo {year} {2010})}\BibitemShut {NoStop}%
\bibitem [{\citenamefont {Wu}\ \emph {et~al.}(2010)\citenamefont {Wu},
  \citenamefont {Bari\ifmmode \check{s}\else \v{s}\fi{}i\ifmmode~\acute{c}\else
  \'{c}\fi{}}, \citenamefont {Dressel}, \citenamefont {Cao}, \citenamefont
  {Xu}, \citenamefont {Carbotte},\ and\ \citenamefont
  {Schachinger}}]{Wu2010PRB_infrared}%
  \BibitemOpen
  \bibfield  {author} {\bibinfo {author} {\bibfnamefont {D.}~\bibnamefont
  {Wu}}, \bibinfo {author} {\bibfnamefont {N.}~\bibnamefont {Bari\ifmmode
  \check{s}\else \v{s}\fi{}i\ifmmode~\acute{c}\else \'{c}\fi{}}}, \bibinfo
  {author} {\bibfnamefont {M.}~\bibnamefont {Dressel}}, \bibinfo {author}
  {\bibfnamefont {G.~H.}\ \bibnamefont {Cao}}, \bibinfo {author} {\bibfnamefont
  {Z.~A.}\ \bibnamefont {Xu}}, \bibinfo {author} {\bibfnamefont {J.~P.}\
  \bibnamefont {Carbotte}}, \ and\ \bibinfo {author} {\bibfnamefont
  {E.}~\bibnamefont {Schachinger}},\ }\href {\doibase
  10.1103/PhysRevB.82.184527} {\bibfield  {journal} {\bibinfo  {journal} {Phys.
  Rev. B}\ }\textbf {\bibinfo {volume} {82}},\ \bibinfo {pages} {184527}
  (\bibinfo {year} {2010})}\BibitemShut {NoStop}%
\bibitem [{\citenamefont {Hashimoto}\ \emph
  {et~al.}(2009{\natexlab{a}})\citenamefont {Hashimoto}, \citenamefont
  {Shibauchi}, \citenamefont {Kasahara}, \citenamefont {Ikada}, \citenamefont
  {Tonegawa}, \citenamefont {Kato}, \citenamefont {Okazaki}, \citenamefont
  {van~der Beek}, \citenamefont {Konczykowski}, \citenamefont {Takeya},
  \citenamefont {Hirata}, \citenamefont {Terashima},\ and\ \citenamefont
  {Matsuda}}]{Hashimoto2009PRL_Microwave}%
  \BibitemOpen
  \bibfield  {author} {\bibinfo {author} {\bibfnamefont {K.}~\bibnamefont
  {Hashimoto}}, \bibinfo {author} {\bibfnamefont {T.}~\bibnamefont
  {Shibauchi}}, \bibinfo {author} {\bibfnamefont {S.}~\bibnamefont {Kasahara}},
  \bibinfo {author} {\bibfnamefont {K.}~\bibnamefont {Ikada}}, \bibinfo
  {author} {\bibfnamefont {S.}~\bibnamefont {Tonegawa}}, \bibinfo {author}
  {\bibfnamefont {T.}~\bibnamefont {Kato}}, \bibinfo {author} {\bibfnamefont
  {R.}~\bibnamefont {Okazaki}}, \bibinfo {author} {\bibfnamefont {C.~J.}\
  \bibnamefont {van~der Beek}}, \bibinfo {author} {\bibfnamefont
  {M.}~\bibnamefont {Konczykowski}}, \bibinfo {author} {\bibfnamefont
  {H.}~\bibnamefont {Takeya}}, \bibinfo {author} {\bibfnamefont
  {K.}~\bibnamefont {Hirata}}, \bibinfo {author} {\bibfnamefont
  {T.}~\bibnamefont {Terashima}}, \ and\ \bibinfo {author} {\bibfnamefont
  {Y.}~\bibnamefont {Matsuda}},\ }\href {\doibase
  10.1103/PhysRevLett.102.207001} {\bibfield  {journal} {\bibinfo  {journal}
  {Phys. Rev. Lett.}\ }\textbf {\bibinfo {volume} {102}},\ \bibinfo {pages}
  {207001} (\bibinfo {year} {2009}{\natexlab{a}})}\BibitemShut {NoStop}%
\bibitem [{\citenamefont {Hashimoto}\ \emph
  {et~al.}(2009{\natexlab{b}})\citenamefont {Hashimoto}, \citenamefont
  {Shibauchi}, \citenamefont {Kato}, \citenamefont {Ikada}, \citenamefont
  {Okazaki}, \citenamefont {Shishido}, \citenamefont {Ishikado}, \citenamefont
  {Kito}, \citenamefont {Iyo}, \citenamefont {Eisaki}, \citenamefont
  {Shamoto},\ and\ \citenamefont
  {Matsuda}}]{Hashimoto2009PRL_Microwave_PrFeAsO}%
  \BibitemOpen
  \bibfield  {author} {\bibinfo {author} {\bibfnamefont {K.}~\bibnamefont
  {Hashimoto}}, \bibinfo {author} {\bibfnamefont {T.}~\bibnamefont
  {Shibauchi}}, \bibinfo {author} {\bibfnamefont {T.}~\bibnamefont {Kato}},
  \bibinfo {author} {\bibfnamefont {K.}~\bibnamefont {Ikada}}, \bibinfo
  {author} {\bibfnamefont {R.}~\bibnamefont {Okazaki}}, \bibinfo {author}
  {\bibfnamefont {H.}~\bibnamefont {Shishido}}, \bibinfo {author}
  {\bibfnamefont {M.}~\bibnamefont {Ishikado}}, \bibinfo {author}
  {\bibfnamefont {H.}~\bibnamefont {Kito}}, \bibinfo {author} {\bibfnamefont
  {A.}~\bibnamefont {Iyo}}, \bibinfo {author} {\bibfnamefont {H.}~\bibnamefont
  {Eisaki}}, \bibinfo {author} {\bibfnamefont {S.}~\bibnamefont {Shamoto}}, \
  and\ \bibinfo {author} {\bibfnamefont {Y.}~\bibnamefont {Matsuda}},\ }\href
  {\doibase 10.1103/PhysRevLett.102.017002} {\bibfield  {journal} {\bibinfo
  {journal} {Phys. Rev. Lett.}\ }\textbf {\bibinfo {volume} {102}},\ \bibinfo
  {pages} {017002} (\bibinfo {year} {2009}{\natexlab{b}})}\BibitemShut
  {NoStop}%
\bibitem [{\citenamefont {Bobowski}\ \emph {et~al.}(2010)\citenamefont
  {Bobowski}, \citenamefont {Baglo}, \citenamefont {Day}, \citenamefont
  {Dosanjh}, \citenamefont {Ofer}, \citenamefont {Ramshaw}, \citenamefont
  {Liang}, \citenamefont {Bonn}, \citenamefont {Hardy}, \citenamefont {Luo},
  \citenamefont {Wang}, \citenamefont {Fang},\ and\ \citenamefont
  {Wen}}]{Bobowski2010PRB_Microwave}%
  \BibitemOpen
  \bibfield  {author} {\bibinfo {author} {\bibfnamefont {J.~S.}\ \bibnamefont
  {Bobowski}}, \bibinfo {author} {\bibfnamefont {J.~C.}\ \bibnamefont {Baglo}},
  \bibinfo {author} {\bibfnamefont {J.}~\bibnamefont {Day}}, \bibinfo {author}
  {\bibfnamefont {P.}~\bibnamefont {Dosanjh}}, \bibinfo {author} {\bibfnamefont
  {R.}~\bibnamefont {Ofer}}, \bibinfo {author} {\bibfnamefont {B.~J.}\
  \bibnamefont {Ramshaw}}, \bibinfo {author} {\bibfnamefont {R.}~\bibnamefont
  {Liang}}, \bibinfo {author} {\bibfnamefont {D.~A.}\ \bibnamefont {Bonn}},
  \bibinfo {author} {\bibfnamefont {W.~N.}\ \bibnamefont {Hardy}}, \bibinfo
  {author} {\bibfnamefont {H.}~\bibnamefont {Luo}}, \bibinfo {author}
  {\bibfnamefont {Z.-S.}\ \bibnamefont {Wang}}, \bibinfo {author}
  {\bibfnamefont {L.}~\bibnamefont {Fang}}, \ and\ \bibinfo {author}
  {\bibfnamefont {H.-H.}\ \bibnamefont {Wen}},\ }\href {\doibase
  10.1103/PhysRevB.82.094520} {\bibfield  {journal} {\bibinfo  {journal} {Phys.
  Rev. B}\ }\textbf {\bibinfo {volume} {82}},\ \bibinfo {pages} {094520}
  (\bibinfo {year} {2010})}\BibitemShut {NoStop}%
\bibitem [{\citenamefont {Yong}\ \emph {et~al.}(2011)\citenamefont {Yong},
  \citenamefont {Lee}, \citenamefont {Jiang}, \citenamefont {Bark},
  \citenamefont {Weiss}, \citenamefont {Hellstrom}, \citenamefont
  {Larbalestier}, \citenamefont {Eom},\ and\ \citenamefont
  {Lemberger}}]{YongLemberger2011PRB_Mutual_Induc}%
  \BibitemOpen
  \bibfield  {author} {\bibinfo {author} {\bibfnamefont {J.}~\bibnamefont
  {Yong}}, \bibinfo {author} {\bibfnamefont {S.}~\bibnamefont {Lee}}, \bibinfo
  {author} {\bibfnamefont {J.}~\bibnamefont {Jiang}}, \bibinfo {author}
  {\bibfnamefont {C.~W.}\ \bibnamefont {Bark}}, \bibinfo {author}
  {\bibfnamefont {J.~D.}\ \bibnamefont {Weiss}}, \bibinfo {author}
  {\bibfnamefont {E.~E.}\ \bibnamefont {Hellstrom}}, \bibinfo {author}
  {\bibfnamefont {D.~C.}\ \bibnamefont {Larbalestier}}, \bibinfo {author}
  {\bibfnamefont {C.~B.}\ \bibnamefont {Eom}}, \ and\ \bibinfo {author}
  {\bibfnamefont {T.~R.}\ \bibnamefont {Lemberger}},\ }\href {\doibase
  10.1103/PhysRevB.83.104510} {\bibfield  {journal} {\bibinfo  {journal} {Phys.
  Rev. B}\ }\textbf {\bibinfo {volume} {83}},\ \bibinfo {pages} {104510}
  (\bibinfo {year} {2011})}\BibitemShut {NoStop}%
\bibitem [{\citenamefont {Luan}\ \emph {et~al.}(2010)\citenamefont {Luan},
  \citenamefont {Auslaender}, \citenamefont {Lippman}, \citenamefont {Hicks},
  \citenamefont {Kalisky}, \citenamefont {Chu}, \citenamefont {Analytis},
  \citenamefont {Fisher}, \citenamefont {Kirtley},\ and\ \citenamefont
  {Moler}}]{LuanMoler2010PRB_MFM}%
  \BibitemOpen
  \bibfield  {author} {\bibinfo {author} {\bibfnamefont {L.}~\bibnamefont
  {Luan}}, \bibinfo {author} {\bibfnamefont {O.~M.}\ \bibnamefont
  {Auslaender}}, \bibinfo {author} {\bibfnamefont {T.~M.}\ \bibnamefont
  {Lippman}}, \bibinfo {author} {\bibfnamefont {C.~W.}\ \bibnamefont {Hicks}},
  \bibinfo {author} {\bibfnamefont {B.}~\bibnamefont {Kalisky}}, \bibinfo
  {author} {\bibfnamefont {J.-H.}\ \bibnamefont {Chu}}, \bibinfo {author}
  {\bibfnamefont {J.~G.}\ \bibnamefont {Analytis}}, \bibinfo {author}
  {\bibfnamefont {I.~R.}\ \bibnamefont {Fisher}}, \bibinfo {author}
  {\bibfnamefont {J.~R.}\ \bibnamefont {Kirtley}}, \ and\ \bibinfo {author}
  {\bibfnamefont {K.~A.}\ \bibnamefont {Moler}},\ }\href {\doibase
  10.1103/PhysRevB.81.100501} {\bibfield  {journal} {\bibinfo  {journal} {Phys.
  Rev. B}\ }\textbf {\bibinfo {volume} {81}},\ \bibinfo {pages} {100501}
  (\bibinfo {year} {2010})}\BibitemShut {NoStop}%
\bibitem [{\citenamefont {Luan}\ \emph {et~al.}(2011)\citenamefont {Luan},
  \citenamefont {Lippman}, \citenamefont {Hicks}, \citenamefont {Bert},
  \citenamefont {Auslaender}, \citenamefont {Chu}, \citenamefont {Analytis},
  \citenamefont {Fisher},\ and\ \citenamefont
  {Moler}}]{LuanMoler2011PRL_MFM_SQUID}%
  \BibitemOpen
  \bibfield  {author} {\bibinfo {author} {\bibfnamefont {L.}~\bibnamefont
  {Luan}}, \bibinfo {author} {\bibfnamefont {T.~M.}\ \bibnamefont {Lippman}},
  \bibinfo {author} {\bibfnamefont {C.~W.}\ \bibnamefont {Hicks}}, \bibinfo
  {author} {\bibfnamefont {J.~A.}\ \bibnamefont {Bert}}, \bibinfo {author}
  {\bibfnamefont {O.~M.}\ \bibnamefont {Auslaender}}, \bibinfo {author}
  {\bibfnamefont {J.-H.}\ \bibnamefont {Chu}}, \bibinfo {author} {\bibfnamefont
  {J.~G.}\ \bibnamefont {Analytis}}, \bibinfo {author} {\bibfnamefont {I.~R.}\
  \bibnamefont {Fisher}}, \ and\ \bibinfo {author} {\bibfnamefont {K.~A.}\
  \bibnamefont {Moler}},\ }\href {\doibase 10.1103/PhysRevLett.106.067001}
  {\bibfield  {journal} {\bibinfo  {journal} {Phys. Rev. Lett.}\ }\textbf
  {\bibinfo {volume} {106}},\ \bibinfo {pages} {067001} (\bibinfo {year}
  {2011})}\BibitemShut {NoStop}%
\bibitem [{\citenamefont {Prozorov}\ \emph {et~al.}(2009)\citenamefont
  {Prozorov}, \citenamefont {Tanatar}, \citenamefont {Gordon}, \citenamefont
  {Martin}, \citenamefont {Kim}, \citenamefont {Kogan}, \citenamefont {Ni},
  \citenamefont {Tillman}, \citenamefont {Bud'ko},\ and\ \citenamefont
  {Canfield}}]{Prozorov2009PhysicaC}%
  \BibitemOpen
  \bibfield  {author} {\bibinfo {author} {\bibfnamefont {R.}~\bibnamefont
  {Prozorov}}, \bibinfo {author} {\bibfnamefont {M.}~\bibnamefont {Tanatar}},
  \bibinfo {author} {\bibfnamefont {R.}~\bibnamefont {Gordon}}, \bibinfo
  {author} {\bibfnamefont {C.}~\bibnamefont {Martin}}, \bibinfo {author}
  {\bibfnamefont {H.}~\bibnamefont {Kim}}, \bibinfo {author} {\bibfnamefont
  {V.}~\bibnamefont {Kogan}}, \bibinfo {author} {\bibfnamefont
  {N.}~\bibnamefont {Ni}}, \bibinfo {author} {\bibfnamefont {M.}~\bibnamefont
  {Tillman}}, \bibinfo {author} {\bibfnamefont {S.}~\bibnamefont {Bud'ko}}, \
  and\ \bibinfo {author} {\bibfnamefont {P.}~\bibnamefont {Canfield}},\
  }\href@noop {} {\bibfield  {journal} {\bibinfo  {journal} {Physica C:
  Superconductivity}\ }\textbf {\bibinfo {volume} {469}},\ \bibinfo {pages}
  {582 } (\bibinfo {year} {2009})}\BibitemShut {NoStop}%
\bibitem [{\citenamefont {Song}\ \emph {et~al.}(2011)\citenamefont {Song},
  \citenamefont {Ghim}, \citenamefont {Yoon}, \citenamefont {Lee},
  \citenamefont {Jung}, \citenamefont {Ji}, \citenamefont {Shim}, \citenamefont
  {Bang},\ and\ \citenamefont {Kwon}}]{SongKwon2011EPL_Hc1}%
  \BibitemOpen
  \bibfield  {author} {\bibinfo {author} {\bibfnamefont {Y.~J.}\ \bibnamefont
  {Song}}, \bibinfo {author} {\bibfnamefont {J.~S.}\ \bibnamefont {Ghim}},
  \bibinfo {author} {\bibfnamefont {J.~H.}\ \bibnamefont {Yoon}}, \bibinfo
  {author} {\bibfnamefont {K.~J.}\ \bibnamefont {Lee}}, \bibinfo {author}
  {\bibfnamefont {M.~H.}\ \bibnamefont {Jung}}, \bibinfo {author}
  {\bibfnamefont {H.-S.}\ \bibnamefont {Ji}}, \bibinfo {author} {\bibfnamefont
  {J.~H.}\ \bibnamefont {Shim}}, \bibinfo {author} {\bibfnamefont
  {Y.}~\bibnamefont {Bang}}, \ and\ \bibinfo {author} {\bibfnamefont {Y.~S.}\
  \bibnamefont {Kwon}},\ }\href
  {http://stacks.iop.org/0295-5075/94/i=5/a=57008} {\bibfield  {journal}
  {\bibinfo  {journal} {EPL (Europhysics Letters)}\ }\textbf {\bibinfo {volume}
  {94}},\ \bibinfo {pages} {57008} (\bibinfo {year} {2011})}\BibitemShut
  {NoStop}%
\bibitem [{\citenamefont {Okazaki}\ \emph {et~al.}(2009)\citenamefont
  {Okazaki}, \citenamefont {Konczykowski}, \citenamefont {van~der Beek},
  \citenamefont {Kato}, \citenamefont {Hashimoto}, \citenamefont {Shimozawa},
  \citenamefont {Shishido}, \citenamefont {Yamashita}, \citenamefont
  {Ishikado}, \citenamefont {Kito}, \citenamefont {Iyo}, \citenamefont
  {Eisaki}, \citenamefont {Shamoto}, \citenamefont {Shibauchi},\ and\
  \citenamefont {Matsuda}}]{Okazaki2009PRB_Hc1_hall_sensor}%
  \BibitemOpen
  \bibfield  {author} {\bibinfo {author} {\bibfnamefont {R.}~\bibnamefont
  {Okazaki}}, \bibinfo {author} {\bibfnamefont {M.}~\bibnamefont
  {Konczykowski}}, \bibinfo {author} {\bibfnamefont {C.~J.}\ \bibnamefont
  {van~der Beek}}, \bibinfo {author} {\bibfnamefont {T.}~\bibnamefont {Kato}},
  \bibinfo {author} {\bibfnamefont {K.}~\bibnamefont {Hashimoto}}, \bibinfo
  {author} {\bibfnamefont {M.}~\bibnamefont {Shimozawa}}, \bibinfo {author}
  {\bibfnamefont {H.}~\bibnamefont {Shishido}}, \bibinfo {author}
  {\bibfnamefont {M.}~\bibnamefont {Yamashita}}, \bibinfo {author}
  {\bibfnamefont {M.}~\bibnamefont {Ishikado}}, \bibinfo {author}
  {\bibfnamefont {H.}~\bibnamefont {Kito}}, \bibinfo {author} {\bibfnamefont
  {A.}~\bibnamefont {Iyo}}, \bibinfo {author} {\bibfnamefont {H.}~\bibnamefont
  {Eisaki}}, \bibinfo {author} {\bibfnamefont {S.}~\bibnamefont {Shamoto}},
  \bibinfo {author} {\bibfnamefont {T.}~\bibnamefont {Shibauchi}}, \ and\
  \bibinfo {author} {\bibfnamefont {Y.}~\bibnamefont {Matsuda}},\ }\href
  {\doibase 10.1103/PhysRevB.79.064520} {\bibfield  {journal} {\bibinfo
  {journal} {Phys. Rev. B}\ }\textbf {\bibinfo {volume} {79}},\ \bibinfo
  {pages} {064520} (\bibinfo {year} {2009})}\BibitemShut {NoStop}%
\bibitem [{\citenamefont {Klein}\ \emph {et~al.}(2010)\citenamefont {Klein},
  \citenamefont {Braithwaite}, \citenamefont {Demuer}, \citenamefont {Knafo},
  \citenamefont {Lapertot}, \citenamefont {Marcenat}, \citenamefont
  {Rodi\`ere}, \citenamefont {Sheikin}, \citenamefont {Strobel}, \citenamefont
  {Sulpice},\ and\ \citenamefont {Toulemonde}}]{Klein2010PRB_hall_probe}%
  \BibitemOpen
  \bibfield  {author} {\bibinfo {author} {\bibfnamefont {T.}~\bibnamefont
  {Klein}}, \bibinfo {author} {\bibfnamefont {D.}~\bibnamefont {Braithwaite}},
  \bibinfo {author} {\bibfnamefont {A.}~\bibnamefont {Demuer}}, \bibinfo
  {author} {\bibfnamefont {W.}~\bibnamefont {Knafo}}, \bibinfo {author}
  {\bibfnamefont {G.}~\bibnamefont {Lapertot}}, \bibinfo {author}
  {\bibfnamefont {C.}~\bibnamefont {Marcenat}}, \bibinfo {author}
  {\bibfnamefont {P.}~\bibnamefont {Rodi\`ere}}, \bibinfo {author}
  {\bibfnamefont {I.}~\bibnamefont {Sheikin}}, \bibinfo {author} {\bibfnamefont
  {P.}~\bibnamefont {Strobel}}, \bibinfo {author} {\bibfnamefont
  {A.}~\bibnamefont {Sulpice}}, \ and\ \bibinfo {author} {\bibfnamefont
  {P.}~\bibnamefont {Toulemonde}},\ }\href {\doibase
  10.1103/PhysRevB.82.184506} {\bibfield  {journal} {\bibinfo  {journal} {Phys.
  Rev. B}\ }\textbf {\bibinfo {volume} {82}},\ \bibinfo {pages} {184506}
  (\bibinfo {year} {2010})}\BibitemShut {NoStop}%
\bibitem [{\citenamefont {Joshi}\ \emph {et~al.}(2017)\citenamefont {Joshi}, ,
  \citenamefont {Nusran}, \citenamefont {Cho}, \citenamefont {Tanatar},
  \citenamefont {Bud'ko}, \citenamefont {Canfield},\ and\ \citenamefont
  {Prozorov}}]{JoshiProzorov2017_NV_Hc1}%
  \BibitemOpen
  \bibfield  {author} {\bibinfo {author} {\bibfnamefont {K.~R.}\ \bibnamefont
  {Joshi}}, , \bibinfo {author} {\bibfnamefont {N.~M.}\ \bibnamefont {Nusran}},
  \bibinfo {author} {\bibfnamefont {K.}~\bibnamefont {Cho}}, \bibinfo {author}
  {\bibfnamefont {M.~A.}\ \bibnamefont {Tanatar}}, \bibinfo {author}
  {\bibfnamefont {S.~L.}\ \bibnamefont {Bud'ko}}, \bibinfo {author}
  {\bibfnamefont {P.~C.}\ \bibnamefont {Canfield}}, \ and\ \bibinfo {author}
  {\bibfnamefont {R.}~\bibnamefont {Prozorov}},\ }\href@noop {} {\bibfield
  {journal} {\bibinfo  {journal} {To be submitted}\ } (\bibinfo {year}
  {2017})}\BibitemShut {NoStop}%
\bibitem [{\citenamefont {Shibauchi}\ \emph {et~al.}(2009)\citenamefont
  {Shibauchi}, \citenamefont {Hashimoto}, \citenamefont {Okazaki},\ and\
  \citenamefont {Matsuda}}]{Shibauchi2009PhysicaC}%
  \BibitemOpen
  \bibfield  {author} {\bibinfo {author} {\bibfnamefont {T.}~\bibnamefont
  {Shibauchi}}, \bibinfo {author} {\bibfnamefont {K.}~\bibnamefont
  {Hashimoto}}, \bibinfo {author} {\bibfnamefont {R.}~\bibnamefont {Okazaki}},
  \ and\ \bibinfo {author} {\bibfnamefont {Y.}~\bibnamefont {Matsuda}},\
  }\href@noop {} {\bibfield  {journal} {\bibinfo  {journal} {Physica C:
  Superconductivity}\ }\textbf {\bibinfo {volume} {469}},\ \bibinfo {pages}
  {590 } (\bibinfo {year} {2009})}\BibitemShut {NoStop}%
\bibitem [{\citenamefont {Malone}\ \emph {et~al.}(2009)\citenamefont {Malone},
  \citenamefont {Fletcher}, \citenamefont {Serafin}, \citenamefont
  {Carrington}, \citenamefont {Zhigadlo}, \citenamefont {Bukowski},
  \citenamefont {Katrych},\ and\ \citenamefont
  {Karpinski}}]{MaloneKarpinski2009PRB_dL}%
  \BibitemOpen
  \bibfield  {author} {\bibinfo {author} {\bibfnamefont {L.}~\bibnamefont
  {Malone}}, \bibinfo {author} {\bibfnamefont {J.~D.}\ \bibnamefont
  {Fletcher}}, \bibinfo {author} {\bibfnamefont {A.}~\bibnamefont {Serafin}},
  \bibinfo {author} {\bibfnamefont {A.}~\bibnamefont {Carrington}}, \bibinfo
  {author} {\bibfnamefont {N.~D.}\ \bibnamefont {Zhigadlo}}, \bibinfo {author}
  {\bibfnamefont {Z.}~\bibnamefont {Bukowski}}, \bibinfo {author}
  {\bibfnamefont {S.}~\bibnamefont {Katrych}}, \ and\ \bibinfo {author}
  {\bibfnamefont {J.}~\bibnamefont {Karpinski}},\ }\href {\doibase
  10.1103/PhysRevB.79.140501} {\bibfield  {journal} {\bibinfo  {journal} {Phys.
  Rev. B}\ }\textbf {\bibinfo {volume} {79}},\ \bibinfo {pages} {140501}
  (\bibinfo {year} {2009})}\BibitemShut {NoStop}%
\bibitem [{\citenamefont {Gordon}\ \emph {et~al.}(2009)\citenamefont {Gordon},
  \citenamefont {Martin}, \citenamefont {Kim}, \citenamefont {Ni},
  \citenamefont {Tanatar}, \citenamefont {Schmalian}, \citenamefont {Mazin},
  \citenamefont {Bud'ko}, \citenamefont {Canfield},\ and\ \citenamefont
  {Prozorov}}]{Gordon2009PRB}%
  \BibitemOpen
  \bibfield  {author} {\bibinfo {author} {\bibfnamefont {R.~T.}\ \bibnamefont
  {Gordon}}, \bibinfo {author} {\bibfnamefont {C.}~\bibnamefont {Martin}},
  \bibinfo {author} {\bibfnamefont {H.}~\bibnamefont {Kim}}, \bibinfo {author}
  {\bibfnamefont {N.}~\bibnamefont {Ni}}, \bibinfo {author} {\bibfnamefont
  {M.~A.}\ \bibnamefont {Tanatar}}, \bibinfo {author} {\bibfnamefont
  {J.}~\bibnamefont {Schmalian}}, \bibinfo {author} {\bibfnamefont {I.~I.}\
  \bibnamefont {Mazin}}, \bibinfo {author} {\bibfnamefont {S.~L.}\ \bibnamefont
  {Bud'ko}}, \bibinfo {author} {\bibfnamefont {P.~C.}\ \bibnamefont
  {Canfield}}, \ and\ \bibinfo {author} {\bibfnamefont {R.}~\bibnamefont
  {Prozorov}},\ }\href {\doibase 10.1103/PhysRevB.79.100506} {\bibfield
  {journal} {\bibinfo  {journal} {Phys. Rev. B}\ }\textbf {\bibinfo {volume}
  {79}},\ \bibinfo {pages} {100506} (\bibinfo {year} {2009})}\BibitemShut
  {NoStop}%
\bibitem [{\citenamefont {Prozorov}\ and\ \citenamefont
  {Kogan}(2011{\natexlab{b}})}]{ProzorovKogan2011RPP}%
  \BibitemOpen
  \bibfield  {author} {\bibinfo {author} {\bibfnamefont {R.}~\bibnamefont
  {Prozorov}}\ and\ \bibinfo {author} {\bibfnamefont {V.~G.}\ \bibnamefont
  {Kogan}},\ }\href@noop {} {\bibfield  {journal} {\bibinfo  {journal} {Reports
  on Progress in Physics}\ }\textbf {\bibinfo {volume} {74}},\ \bibinfo {pages}
  {124505} (\bibinfo {year} {2011}{\natexlab{b}})}\BibitemShut {NoStop}%
\bibitem [{\citenamefont {Prozorov}\ and\ \citenamefont
  {Giannetta}(2006)}]{Prozorov2006SST}%
  \BibitemOpen
  \bibfield  {author} {\bibinfo {author} {\bibfnamefont {R.}~\bibnamefont
  {Prozorov}}\ and\ \bibinfo {author} {\bibfnamefont {R.~W.}\ \bibnamefont
  {Giannetta}},\ }\href@noop {} {\bibfield  {journal} {\bibinfo  {journal}
  {Superconductor Science and Technology}\ }\textbf {\bibinfo {volume} {19}},\
  \bibinfo {pages} {R41} (\bibinfo {year} {2006})}\BibitemShut {NoStop}%
\bibitem [{\citenamefont {Prozorov}\ \emph {et~al.}(2000)\citenamefont
  {Prozorov}, \citenamefont {Giannetta}, \citenamefont {Carrington},\ and\
  \citenamefont {Araujo-Moreira}}]{Prozorov2000PRB}%
  \BibitemOpen
  \bibfield  {author} {\bibinfo {author} {\bibfnamefont {R.}~\bibnamefont
  {Prozorov}}, \bibinfo {author} {\bibfnamefont {R.~W.}\ \bibnamefont
  {Giannetta}}, \bibinfo {author} {\bibfnamefont {A.}~\bibnamefont
  {Carrington}}, \ and\ \bibinfo {author} {\bibfnamefont {F.~M.}\ \bibnamefont
  {Araujo-Moreira}},\ }\href {\doibase 10.1103/PhysRevB.62.115} {\bibfield
  {journal} {\bibinfo  {journal} {Phys. Rev. B}\ }\textbf {\bibinfo {volume}
  {62}},\ \bibinfo {pages} {115} (\bibinfo {year} {2000})}\BibitemShut
  {NoStop}%
\bibitem [{\citenamefont {Avci}\ \emph {et~al.}(2012)\citenamefont {Avci},
  \citenamefont {Chmaissem}, \citenamefont {Chung}, \citenamefont {Rosenkranz},
  \citenamefont {Goremychkin}, \citenamefont {Castellan}, \citenamefont
  {Todorov}, \citenamefont {Schlueter}, \citenamefont {Claus}, \citenamefont
  {Daoud-Aladine}, \citenamefont {Khalyavin}, \citenamefont {Kanatzidis},\ and\
  \citenamefont {Osborn}}]{Avci2012PRB}%
  \BibitemOpen
  \bibfield  {author} {\bibinfo {author} {\bibfnamefont {S.}~\bibnamefont
  {Avci}}, \bibinfo {author} {\bibfnamefont {O.}~\bibnamefont {Chmaissem}},
  \bibinfo {author} {\bibfnamefont {D.~Y.}\ \bibnamefont {Chung}}, \bibinfo
  {author} {\bibfnamefont {S.}~\bibnamefont {Rosenkranz}}, \bibinfo {author}
  {\bibfnamefont {E.~A.}\ \bibnamefont {Goremychkin}}, \bibinfo {author}
  {\bibfnamefont {J.~P.}\ \bibnamefont {Castellan}}, \bibinfo {author}
  {\bibfnamefont {I.~S.}\ \bibnamefont {Todorov}}, \bibinfo {author}
  {\bibfnamefont {J.~A.}\ \bibnamefont {Schlueter}}, \bibinfo {author}
  {\bibfnamefont {H.}~\bibnamefont {Claus}}, \bibinfo {author} {\bibfnamefont
  {A.}~\bibnamefont {Daoud-Aladine}}, \bibinfo {author} {\bibfnamefont {D.~D.}\
  \bibnamefont {Khalyavin}}, \bibinfo {author} {\bibfnamefont {M.~G.}\
  \bibnamefont {Kanatzidis}}, \ and\ \bibinfo {author} {\bibfnamefont
  {R.}~\bibnamefont {Osborn}},\ }\href {\doibase 10.1103/PhysRevB.85.184507}
  {\bibfield  {journal} {\bibinfo  {journal} {Phys. Rev. B}\ }\textbf {\bibinfo
  {volume} {85}},\ \bibinfo {pages} {184507} (\bibinfo {year}
  {2012})}\BibitemShut {NoStop}%
\bibitem [{\citenamefont {Hashimoto}\ \emph {et~al.}(2012)\citenamefont
  {Hashimoto}, \citenamefont {Cho}, \citenamefont {Shibauchi}, \citenamefont
  {Kasahara}, \citenamefont {Mizukami}, \citenamefont {Katsumata},
  \citenamefont {Tsuruhara}, \citenamefont {Terashima}, \citenamefont {Ikeda},
  \citenamefont {Tanatar}, \citenamefont {Kitano}, \citenamefont {Salovich},
  \citenamefont {Giannetta}, \citenamefont {Walmsley}, \citenamefont
  {Carrington}, \citenamefont {Prozorov},\ and\ \citenamefont
  {Matsuda}}]{HashimotoCho2012Science_BaP122_dL}%
  \BibitemOpen
  \bibfield  {author} {\bibinfo {author} {\bibfnamefont {K.}~\bibnamefont
  {Hashimoto}}, \bibinfo {author} {\bibfnamefont {K.}~\bibnamefont {Cho}},
  \bibinfo {author} {\bibfnamefont {T.}~\bibnamefont {Shibauchi}}, \bibinfo
  {author} {\bibfnamefont {S.}~\bibnamefont {Kasahara}}, \bibinfo {author}
  {\bibfnamefont {Y.}~\bibnamefont {Mizukami}}, \bibinfo {author}
  {\bibfnamefont {R.}~\bibnamefont {Katsumata}}, \bibinfo {author}
  {\bibfnamefont {Y.}~\bibnamefont {Tsuruhara}}, \bibinfo {author}
  {\bibfnamefont {T.}~\bibnamefont {Terashima}}, \bibinfo {author}
  {\bibfnamefont {H.}~\bibnamefont {Ikeda}}, \bibinfo {author} {\bibfnamefont
  {M.~A.}\ \bibnamefont {Tanatar}}, \bibinfo {author} {\bibfnamefont
  {H.}~\bibnamefont {Kitano}}, \bibinfo {author} {\bibfnamefont
  {N.}~\bibnamefont {Salovich}}, \bibinfo {author} {\bibfnamefont {R.~W.}\
  \bibnamefont {Giannetta}}, \bibinfo {author} {\bibfnamefont {P.}~\bibnamefont
  {Walmsley}}, \bibinfo {author} {\bibfnamefont {A.}~\bibnamefont
  {Carrington}}, \bibinfo {author} {\bibfnamefont {R.}~\bibnamefont
  {Prozorov}}, \ and\ \bibinfo {author} {\bibfnamefont {Y.}~\bibnamefont
  {Matsuda}},\ }\href {\doibase 10.1126/science.1219821} {\bibfield  {journal}
  {\bibinfo  {journal} {Science}\ }\textbf {\bibinfo {volume} {336}},\ \bibinfo
  {pages} {1554} (\bibinfo {year} {2012})}\BibitemShut {NoStop}%
\bibitem [{\citenamefont {Kim}\ \emph {et~al.}(2011)\citenamefont {Kim},
  \citenamefont {Pratt}, \citenamefont {Rustan}, \citenamefont {Tian},
  \citenamefont {Zarestky}, \citenamefont {Thaler}, \citenamefont {Bud'ko},
  \citenamefont {Canfield}, \citenamefont {McQueeney}, \citenamefont
  {Kreyssig},\ and\ \citenamefont {Goldman}}]{KimGoldman2011PRB_BaRu122}%
  \BibitemOpen
  \bibfield  {author} {\bibinfo {author} {\bibfnamefont {M.~G.}\ \bibnamefont
  {Kim}}, \bibinfo {author} {\bibfnamefont {D.~K.}\ \bibnamefont {Pratt}},
  \bibinfo {author} {\bibfnamefont {G.~E.}\ \bibnamefont {Rustan}}, \bibinfo
  {author} {\bibfnamefont {W.}~\bibnamefont {Tian}}, \bibinfo {author}
  {\bibfnamefont {J.~L.}\ \bibnamefont {Zarestky}}, \bibinfo {author}
  {\bibfnamefont {A.}~\bibnamefont {Thaler}}, \bibinfo {author} {\bibfnamefont
  {S.~L.}\ \bibnamefont {Bud'ko}}, \bibinfo {author} {\bibfnamefont {P.~C.}\
  \bibnamefont {Canfield}}, \bibinfo {author} {\bibfnamefont {R.~J.}\
  \bibnamefont {McQueeney}}, \bibinfo {author} {\bibfnamefont {A.}~\bibnamefont
  {Kreyssig}}, \ and\ \bibinfo {author} {\bibfnamefont {A.~I.}\ \bibnamefont
  {Goldman}},\ }\href {\doibase 10.1103/PhysRevB.83.054514} {\bibfield
  {journal} {\bibinfo  {journal} {Phys. Rev. B}\ }\textbf {\bibinfo {volume}
  {83}},\ \bibinfo {pages} {054514} (\bibinfo {year} {2011})}\BibitemShut
  {NoStop}%
\bibitem [{\citenamefont {Kobayashi}\ \emph {et~al.}(2014)\citenamefont
  {Kobayashi}, \citenamefont {Miyasaka}, \citenamefont {Tajima},\ and\
  \citenamefont {Chikumoto}}]{Kobayashi2014JPSJ_SrP122}%
  \BibitemOpen
  \bibfield  {author} {\bibinfo {author} {\bibfnamefont {T.}~\bibnamefont
  {Kobayashi}}, \bibinfo {author} {\bibfnamefont {S.}~\bibnamefont {Miyasaka}},
  \bibinfo {author} {\bibfnamefont {S.}~\bibnamefont {Tajima}}, \ and\ \bibinfo
  {author} {\bibfnamefont {N.}~\bibnamefont {Chikumoto}},\ }\href {\doibase
  10.7566/JPSJ.83.104702} {\bibfield  {journal} {\bibinfo  {journal} {Journal
  of the Physical Society of Japan}\ }\textbf {\bibinfo {volume} {83}},\
  \bibinfo {pages} {104702} (\bibinfo {year} {2014})}\BibitemShut {NoStop}%
\bibitem [{\citenamefont {Nandi}\ \emph {et~al.}(2010)\citenamefont {Nandi},
  \citenamefont {Kim}, \citenamefont {Kreyssig}, \citenamefont {Fernandes},
  \citenamefont {Pratt}, \citenamefont {Thaler}, \citenamefont {Ni},
  \citenamefont {Bud'ko}, \citenamefont {Canfield}, \citenamefont {Schmalian},
  \citenamefont {McQueeney},\ and\ \citenamefont
  {Goldman}}]{NandiGoldman2010PRL}%
  \BibitemOpen
  \bibfield  {author} {\bibinfo {author} {\bibfnamefont {S.}~\bibnamefont
  {Nandi}}, \bibinfo {author} {\bibfnamefont {M.~G.}\ \bibnamefont {Kim}},
  \bibinfo {author} {\bibfnamefont {A.}~\bibnamefont {Kreyssig}}, \bibinfo
  {author} {\bibfnamefont {R.~M.}\ \bibnamefont {Fernandes}}, \bibinfo {author}
  {\bibfnamefont {D.~K.}\ \bibnamefont {Pratt}}, \bibinfo {author}
  {\bibfnamefont {A.}~\bibnamefont {Thaler}}, \bibinfo {author} {\bibfnamefont
  {N.}~\bibnamefont {Ni}}, \bibinfo {author} {\bibfnamefont {S.~L.}\
  \bibnamefont {Bud'ko}}, \bibinfo {author} {\bibfnamefont {P.~C.}\
  \bibnamefont {Canfield}}, \bibinfo {author} {\bibfnamefont {J.}~\bibnamefont
  {Schmalian}}, \bibinfo {author} {\bibfnamefont {R.~J.}\ \bibnamefont
  {McQueeney}}, \ and\ \bibinfo {author} {\bibfnamefont {A.~I.}\ \bibnamefont
  {Goldman}},\ }\href {\doibase 10.1103/PhysRevLett.104.057006} {\bibfield
  {journal} {\bibinfo  {journal} {Phys. Rev. Lett.}\ }\textbf {\bibinfo
  {volume} {104}},\ \bibinfo {pages} {057006} (\bibinfo {year}
  {2010})}\BibitemShut {NoStop}%
\bibitem [{\citenamefont {Abdel-Hafiez}\ \emph {et~al.}(2015)\citenamefont
  {Abdel-Hafiez}, \citenamefont {Zhang}, \citenamefont {He}, \citenamefont
  {Zhao}, \citenamefont {Bergmann}, \citenamefont {Krellner}, \citenamefont
  {Duan}, \citenamefont {Lu}, \citenamefont {Luo}, \citenamefont {Dai},\ and\
  \citenamefont {Chen}}]{Abdel-Hafiez2015PRB_BaNi122}%
  \BibitemOpen
  \bibfield  {author} {\bibinfo {author} {\bibfnamefont {M.}~\bibnamefont
  {Abdel-Hafiez}}, \bibinfo {author} {\bibfnamefont {Y.}~\bibnamefont {Zhang}},
  \bibinfo {author} {\bibfnamefont {Z.}~\bibnamefont {He}}, \bibinfo {author}
  {\bibfnamefont {J.}~\bibnamefont {Zhao}}, \bibinfo {author} {\bibfnamefont
  {C.}~\bibnamefont {Bergmann}}, \bibinfo {author} {\bibfnamefont
  {C.}~\bibnamefont {Krellner}}, \bibinfo {author} {\bibfnamefont {C.-G.}\
  \bibnamefont {Duan}}, \bibinfo {author} {\bibfnamefont {X.}~\bibnamefont
  {Lu}}, \bibinfo {author} {\bibfnamefont {H.}~\bibnamefont {Luo}}, \bibinfo
  {author} {\bibfnamefont {P.}~\bibnamefont {Dai}}, \ and\ \bibinfo {author}
  {\bibfnamefont {X.-J.}\ \bibnamefont {Chen}},\ }\href@noop {} {\bibfield
  {journal} {\bibinfo  {journal} {Phys. Rev. B}\ }\textbf {\bibinfo {volume}
  {91}},\ \bibinfo {pages} {024510} (\bibinfo {year} {2015})}\BibitemShut
  {NoStop}%
\bibitem [{\citenamefont {Tanatar}\ \emph
  {et~al.}(2014{\natexlab{a}})\citenamefont {Tanatar}, \citenamefont
  {Straszheim}, \citenamefont {Kim}, \citenamefont {Murphy}, \citenamefont
  {Spyrison}, \citenamefont {Blomberg}, \citenamefont {Cho}, \citenamefont
  {Reid}, \citenamefont {Shen}, \citenamefont {Taillefer}, \citenamefont
  {Wen},\ and\ \citenamefont {Prozorov}}]{TanatarProzorov2014PRB_BaK122}%
  \BibitemOpen
  \bibfield  {author} {\bibinfo {author} {\bibfnamefont {M.~A.}\ \bibnamefont
  {Tanatar}}, \bibinfo {author} {\bibfnamefont {W.~E.}\ \bibnamefont
  {Straszheim}}, \bibinfo {author} {\bibfnamefont {H.}~\bibnamefont {Kim}},
  \bibinfo {author} {\bibfnamefont {J.}~\bibnamefont {Murphy}}, \bibinfo
  {author} {\bibfnamefont {N.}~\bibnamefont {Spyrison}}, \bibinfo {author}
  {\bibfnamefont {E.~C.}\ \bibnamefont {Blomberg}}, \bibinfo {author}
  {\bibfnamefont {K.}~\bibnamefont {Cho}}, \bibinfo {author} {\bibfnamefont
  {J.-P.}\ \bibnamefont {Reid}}, \bibinfo {author} {\bibfnamefont
  {B.}~\bibnamefont {Shen}}, \bibinfo {author} {\bibfnamefont {L.}~\bibnamefont
  {Taillefer}}, \bibinfo {author} {\bibfnamefont {H.-H.}\ \bibnamefont {Wen}},
  \ and\ \bibinfo {author} {\bibfnamefont {R.}~\bibnamefont {Prozorov}},\
  }\href {\doibase http://dx.doi.org/10.1103/PhysRevB.89.144514} {\bibfield
  {journal} {\bibinfo  {journal} {Phys. Rev. B}\ }\textbf {\bibinfo {volume}
  {89}},\ \bibinfo {pages} {144514} (\bibinfo {year}
  {2014}{\natexlab{a}})}\BibitemShut {NoStop}%
\bibitem [{\citenamefont {Tanatar}\ \emph
  {et~al.}(2014{\natexlab{b}})\citenamefont {Tanatar}, \citenamefont
  {Torikachvili}, \citenamefont {Thaler}, \citenamefont {Bud'ko}, \citenamefont
  {Canfield},\ and\ \citenamefont {Prozorov}}]{TanatarProzorov2014PRB_BaRu122}%
  \BibitemOpen
  \bibfield  {author} {\bibinfo {author} {\bibfnamefont {M.~A.}\ \bibnamefont
  {Tanatar}}, \bibinfo {author} {\bibfnamefont {M.~S.}\ \bibnamefont
  {Torikachvili}}, \bibinfo {author} {\bibfnamefont {A.}~\bibnamefont
  {Thaler}}, \bibinfo {author} {\bibfnamefont {S.~L.}\ \bibnamefont {Bud'ko}},
  \bibinfo {author} {\bibfnamefont {P.~C.}\ \bibnamefont {Canfield}}, \ and\
  \bibinfo {author} {\bibfnamefont {R.}~\bibnamefont {Prozorov}},\ }\href
  {\doibase 10.1103/PhysRevB.90.104518} {\bibfield  {journal} {\bibinfo
  {journal} {Phys. Rev. B}\ }\textbf {\bibinfo {volume} {90}},\ \bibinfo
  {pages} {104518} (\bibinfo {year} {2014}{\natexlab{b}})}\BibitemShut
  {NoStop}%
\bibitem [{\citenamefont {Murphy}\ \emph
  {et~al.}(2013{\natexlab{b}})\citenamefont {Murphy}, \citenamefont {Tanatar},
  \citenamefont {Graf}, \citenamefont {Brooks}, \citenamefont {Bud'ko},
  \citenamefont {Canfield}, \citenamefont {Kogan},\ and\ \citenamefont
  {Prozorov}}]{MurphyProzorov2013PRB_over-doped_BaNi122}%
  \BibitemOpen
  \bibfield  {author} {\bibinfo {author} {\bibfnamefont {J.}~\bibnamefont
  {Murphy}}, \bibinfo {author} {\bibfnamefont {M.~A.}\ \bibnamefont {Tanatar}},
  \bibinfo {author} {\bibfnamefont {D.}~\bibnamefont {Graf}}, \bibinfo {author}
  {\bibfnamefont {J.~S.}\ \bibnamefont {Brooks}}, \bibinfo {author}
  {\bibfnamefont {S.~L.}\ \bibnamefont {Bud'ko}}, \bibinfo {author}
  {\bibfnamefont {P.~C.}\ \bibnamefont {Canfield}}, \bibinfo {author}
  {\bibfnamefont {V.~G.}\ \bibnamefont {Kogan}}, \ and\ \bibinfo {author}
  {\bibfnamefont {R.}~\bibnamefont {Prozorov}},\ }\href {\doibase
  10.1103/PhysRevB.87.094505} {\bibfield  {journal} {\bibinfo  {journal} {Phys.
  Rev. B}\ }\textbf {\bibinfo {volume} {87}},\ \bibinfo {pages} {094505}
  (\bibinfo {year} {2013}{\natexlab{b}})}\BibitemShut {NoStop}%
\bibitem [{\citenamefont {Mathur}\ \emph {et~al.}(1998)\citenamefont {Mathur},
  \citenamefont {Grosche}, \citenamefont {Julian}, \citenamefont {Walker},
  \citenamefont {Freye}, \citenamefont {Haselwimmer},\ and\ \citenamefont
  {Lonzarich}}]{Mathur1998Nature_HeavyFermion}%
  \BibitemOpen
  \bibfield  {author} {\bibinfo {author} {\bibfnamefont {N.~D.}\ \bibnamefont
  {Mathur}}, \bibinfo {author} {\bibfnamefont {F.~M.}\ \bibnamefont {Grosche}},
  \bibinfo {author} {\bibfnamefont {S.~R.}\ \bibnamefont {Julian}}, \bibinfo
  {author} {\bibfnamefont {I.~R.}\ \bibnamefont {Walker}}, \bibinfo {author}
  {\bibfnamefont {D.~M.}\ \bibnamefont {Freye}}, \bibinfo {author}
  {\bibfnamefont {R.~K.~W.}\ \bibnamefont {Haselwimmer}}, \ and\ \bibinfo
  {author} {\bibfnamefont {G.~G.}\ \bibnamefont {Lonzarich}},\ }\href@noop {}
  {\bibfield  {journal} {\bibinfo  {journal} {Nature}\ }\textbf {\bibinfo
  {volume} {394}},\ \bibinfo {pages} {39 EP } (\bibinfo {year}
  {1998})}\BibitemShut {NoStop}%
\bibitem [{\citenamefont {Tanatar}\ \emph
  {et~al.}(2010{\natexlab{a}})\citenamefont {Tanatar}, \citenamefont {Reid},
  \citenamefont {Shakeripour}, \citenamefont {Luo}, \citenamefont
  {Doiron-Leyraud}, \citenamefont {Ni}, \citenamefont {Bud'ko}, \citenamefont
  {Canfield}, \citenamefont {Prozorov},\ and\ \citenamefont
  {Taillefer}}]{Tanatar2010PRL}%
  \BibitemOpen
  \bibfield  {author} {\bibinfo {author} {\bibfnamefont {M.~A.}\ \bibnamefont
  {Tanatar}}, \bibinfo {author} {\bibfnamefont {J.-P.}\ \bibnamefont {Reid}},
  \bibinfo {author} {\bibfnamefont {H.}~\bibnamefont {Shakeripour}}, \bibinfo
  {author} {\bibfnamefont {X.~G.}\ \bibnamefont {Luo}}, \bibinfo {author}
  {\bibfnamefont {N.}~\bibnamefont {Doiron-Leyraud}}, \bibinfo {author}
  {\bibfnamefont {N.}~\bibnamefont {Ni}}, \bibinfo {author} {\bibfnamefont
  {S.~L.}\ \bibnamefont {Bud'ko}}, \bibinfo {author} {\bibfnamefont {P.~C.}\
  \bibnamefont {Canfield}}, \bibinfo {author} {\bibfnamefont {R.}~\bibnamefont
  {Prozorov}}, \ and\ \bibinfo {author} {\bibfnamefont {L.}~\bibnamefont
  {Taillefer}},\ }\href {\doibase 10.1103/PhysRevLett.104.067002} {\bibfield
  {journal} {\bibinfo  {journal} {Phys. Rev. Lett.}\ }\textbf {\bibinfo
  {volume} {104}},\ \bibinfo {pages} {067002} (\bibinfo {year}
  {2010}{\natexlab{a}})}\BibitemShut {NoStop}%
\bibitem [{\citenamefont {Xu}\ \emph {et~al.}(2013)\citenamefont {Xu},
  \citenamefont {Richard}, \citenamefont {Shi}, \citenamefont {van Roekeghem},
  \citenamefont {Qian}, \citenamefont {Razzoli}, \citenamefont {Rienks},
  \citenamefont {Chen}, \citenamefont {Ieki}, \citenamefont {Nakayama},
  \citenamefont {Sato}, \citenamefont {Takahashi}, \citenamefont {Shi},\ and\
  \citenamefont {Ding}}]{Xu2013PRB_BaK122}%
  \BibitemOpen
  \bibfield  {author} {\bibinfo {author} {\bibfnamefont {N.}~\bibnamefont
  {Xu}}, \bibinfo {author} {\bibfnamefont {P.}~\bibnamefont {Richard}},
  \bibinfo {author} {\bibfnamefont {X.}~\bibnamefont {Shi}}, \bibinfo {author}
  {\bibfnamefont {A.}~\bibnamefont {van Roekeghem}}, \bibinfo {author}
  {\bibfnamefont {T.}~\bibnamefont {Qian}}, \bibinfo {author} {\bibfnamefont
  {E.}~\bibnamefont {Razzoli}}, \bibinfo {author} {\bibfnamefont
  {E.}~\bibnamefont {Rienks}}, \bibinfo {author} {\bibfnamefont {G.-F.}\
  \bibnamefont {Chen}}, \bibinfo {author} {\bibfnamefont {E.}~\bibnamefont
  {Ieki}}, \bibinfo {author} {\bibfnamefont {K.}~\bibnamefont {Nakayama}},
  \bibinfo {author} {\bibfnamefont {T.}~\bibnamefont {Sato}}, \bibinfo {author}
  {\bibfnamefont {T.}~\bibnamefont {Takahashi}}, \bibinfo {author}
  {\bibfnamefont {M.}~\bibnamefont {Shi}}, \ and\ \bibinfo {author}
  {\bibfnamefont {H.}~\bibnamefont {Ding}},\ }\href {\doibase
  10.1103/PhysRevB.88.220508} {\bibfield  {journal} {\bibinfo  {journal} {Phys.
  Rev. B}\ }\textbf {\bibinfo {volume} {88}},\ \bibinfo {pages} {220508}
  (\bibinfo {year} {2013})}\BibitemShut {NoStop}%
\bibitem [{\citenamefont {Richard}\ \emph {et~al.}(2015)\citenamefont
  {Richard}, \citenamefont {Qian},\ and\ \citenamefont
  {Ding}}]{RichardDing2015JPCM_APRES_IBS}%
  \BibitemOpen
  \bibfield  {author} {\bibinfo {author} {\bibfnamefont {P.}~\bibnamefont
  {Richard}}, \bibinfo {author} {\bibfnamefont {T.}~\bibnamefont {Qian}}, \
  and\ \bibinfo {author} {\bibfnamefont {H.}~\bibnamefont {Ding}},\ }\href
  {http://stacks.iop.org/0953-8984/27/i=29/a=293203} {\bibfield  {journal}
  {\bibinfo  {journal} {Journal of Physics: Condensed Matter}\ }\textbf
  {\bibinfo {volume} {27}},\ \bibinfo {pages} {293203} (\bibinfo {year}
  {2015})}\BibitemShut {NoStop}%
\bibitem [{\citenamefont {Cho}\ \emph {et~al.}(2014)\citenamefont {Cho},
  \citenamefont {Ko\ifmmode~\acute{n}\else \'{n}\fi{}czykowski}, \citenamefont
  {Murphy}, \citenamefont {Kim}, \citenamefont {Tanatar}, \citenamefont
  {Straszheim}, \citenamefont {Shen}, \citenamefont {Wen},\ and\ \citenamefont
  {Prozorov}}]{Cho2014PRB_e-irr}%
  \BibitemOpen
  \bibfield  {author} {\bibinfo {author} {\bibfnamefont {K.}~\bibnamefont
  {Cho}}, \bibinfo {author} {\bibfnamefont {M.}~\bibnamefont
  {Ko\ifmmode~\acute{n}\else \'{n}\fi{}czykowski}}, \bibinfo {author}
  {\bibfnamefont {J.}~\bibnamefont {Murphy}}, \bibinfo {author} {\bibfnamefont
  {H.}~\bibnamefont {Kim}}, \bibinfo {author} {\bibfnamefont {M.~A.}\
  \bibnamefont {Tanatar}}, \bibinfo {author} {\bibfnamefont {W.~E.}\
  \bibnamefont {Straszheim}}, \bibinfo {author} {\bibfnamefont
  {B.}~\bibnamefont {Shen}}, \bibinfo {author} {\bibfnamefont {H.~H.}\
  \bibnamefont {Wen}}, \ and\ \bibinfo {author} {\bibfnamefont
  {R.}~\bibnamefont {Prozorov}},\ }\href {\doibase 10.1103/PhysRevB.90.104514}
  {\bibfield  {journal} {\bibinfo  {journal} {Phys. Rev. B}\ }\textbf {\bibinfo
  {volume} {90}},\ \bibinfo {pages} {104514} (\bibinfo {year}
  {2014})}\BibitemShut {NoStop}%
\bibitem [{\citenamefont {Konczykowski}()}]{Konczykowski_unpublished}%
  \BibitemOpen
  \bibfield  {author} {\bibinfo {author} {\bibfnamefont {M.}~\bibnamefont
  {Konczykowski}},\ }\href@noop {} {\bibinfo  {journal} {Unpublished data}\
  }\BibitemShut {NoStop}%
\bibitem [{\citenamefont {Almoalem}\ \emph {et~al.}(2017)\citenamefont
  {Almoalem}, \citenamefont {Yagil}, \citenamefont {Cho}, \citenamefont
  {Teknowijoyo}, \citenamefont {Tanatar}, \citenamefont {Prozorov},
  \citenamefont {Liu}, \citenamefont {Lograsso},\ and\ \citenamefont
  {Auslaender}}]{AlmoalemChoProzorovAuslaender2017arXiv_BaK122}%
  \BibitemOpen
\bibfield  {journal} {  }\bibfield  {author} {\bibinfo {author} {\bibfnamefont
  {A.}~\bibnamefont {Almoalem}}, \bibinfo {author} {\bibfnamefont
  {A.}~\bibnamefont {Yagil}}, \bibinfo {author} {\bibfnamefont
  {K.}~\bibnamefont {Cho}}, \bibinfo {author} {\bibfnamefont {S.}~\bibnamefont
  {Teknowijoyo}}, \bibinfo {author} {\bibfnamefont {M.~A.}\ \bibnamefont
  {Tanatar}}, \bibinfo {author} {\bibfnamefont {R.}~\bibnamefont {Prozorov}},
  \bibinfo {author} {\bibfnamefont {Y.}~\bibnamefont {Liu}}, \bibinfo {author}
  {\bibfnamefont {T.~A.}\ \bibnamefont {Lograsso}}, \ and\ \bibinfo {author}
  {\bibfnamefont {O.~M.}\ \bibnamefont {Auslaender}},\ }\href@noop {}
  {\bibfield  {journal} {\bibinfo  {journal} {e-print 1708.00683}\ } (\bibinfo
  {year} {2017})}\BibitemShut {NoStop}%
\bibitem [{\citenamefont {Evtushinsky}\ \emph {et~al.}(2009)\citenamefont
  {Evtushinsky}, \citenamefont {Inosov}, \citenamefont {Zabolotnyy},
  \citenamefont {Viazovska}, \citenamefont {Khasanov}, \citenamefont {Amato},
  \citenamefont {Klauss}, \citenamefont {Luetkens}, \citenamefont
  {Niedermayer}, \citenamefont {Sun}, \citenamefont {Hinkov}, \citenamefont
  {Lin}, \citenamefont {Varykhalov}, \citenamefont {Koitzsch}, \citenamefont
  {Knupfer}, \citenamefont {Buchner}, \citenamefont {Kordyuk},\ and\
  \citenamefont {Borisenko}}]{EvtushinskyBoriesenko2009NJP_BaK122}%
  \BibitemOpen
  \bibfield  {author} {\bibinfo {author} {\bibfnamefont {D.~V.}\ \bibnamefont
  {Evtushinsky}}, \bibinfo {author} {\bibfnamefont {D.~S.}\ \bibnamefont
  {Inosov}}, \bibinfo {author} {\bibfnamefont {V.~B.}\ \bibnamefont
  {Zabolotnyy}}, \bibinfo {author} {\bibfnamefont {M.~S.}\ \bibnamefont
  {Viazovska}}, \bibinfo {author} {\bibfnamefont {R.}~\bibnamefont {Khasanov}},
  \bibinfo {author} {\bibfnamefont {A.}~\bibnamefont {Amato}}, \bibinfo
  {author} {\bibfnamefont {H.-H.}\ \bibnamefont {Klauss}}, \bibinfo {author}
  {\bibfnamefont {H.}~\bibnamefont {Luetkens}}, \bibinfo {author}
  {\bibfnamefont {C.}~\bibnamefont {Niedermayer}}, \bibinfo {author}
  {\bibfnamefont {G.~L.}\ \bibnamefont {Sun}}, \bibinfo {author} {\bibfnamefont
  {V.}~\bibnamefont {Hinkov}}, \bibinfo {author} {\bibfnamefont {C.~T.}\
  \bibnamefont {Lin}}, \bibinfo {author} {\bibfnamefont {A.}~\bibnamefont
  {Varykhalov}}, \bibinfo {author} {\bibfnamefont {A.}~\bibnamefont
  {Koitzsch}}, \bibinfo {author} {\bibfnamefont {M.}~\bibnamefont {Knupfer}},
  \bibinfo {author} {\bibfnamefont {B.}~\bibnamefont {Buchner}}, \bibinfo
  {author} {\bibfnamefont {A.~A.}\ \bibnamefont {Kordyuk}}, \ and\ \bibinfo
  {author} {\bibfnamefont {S.~V.}\ \bibnamefont {Borisenko}},\ }\href
  {http://stacks.iop.org/1367-2630/11/i=5/a=055069} {\bibfield  {journal}
  {\bibinfo  {journal} {New Journal of Physics}\ }\textbf {\bibinfo {volume}
  {11}},\ \bibinfo {pages} {055069} (\bibinfo {year} {2009})}\BibitemShut
  {NoStop}%
\bibitem [{\citenamefont {Welp}\ \emph {et~al.}(2009)\citenamefont {Welp},
  \citenamefont {Xie}, \citenamefont {Koshelev}, \citenamefont {Kwok},
  \citenamefont {Luo}, \citenamefont {Wang}, \citenamefont {Mu},\ and\
  \citenamefont {Wen}}]{Welp2009PRB_BaK122}%
  \BibitemOpen
  \bibfield  {author} {\bibinfo {author} {\bibfnamefont {U.}~\bibnamefont
  {Welp}}, \bibinfo {author} {\bibfnamefont {R.}~\bibnamefont {Xie}}, \bibinfo
  {author} {\bibfnamefont {A.~E.}\ \bibnamefont {Koshelev}}, \bibinfo {author}
  {\bibfnamefont {W.~K.}\ \bibnamefont {Kwok}}, \bibinfo {author}
  {\bibfnamefont {H.~Q.}\ \bibnamefont {Luo}}, \bibinfo {author} {\bibfnamefont
  {Z.~S.}\ \bibnamefont {Wang}}, \bibinfo {author} {\bibfnamefont
  {G.}~\bibnamefont {Mu}}, \ and\ \bibinfo {author} {\bibfnamefont {H.~H.}\
  \bibnamefont {Wen}},\ }\href {\doibase 10.1103/PhysRevB.79.094505} {\bibfield
   {journal} {\bibinfo  {journal} {Phys. Rev. B}\ }\textbf {\bibinfo {volume}
  {79}},\ \bibinfo {pages} {094505} (\bibinfo {year} {2009})}\BibitemShut
  {NoStop}%
\bibitem [{\citenamefont {Li}\ \emph {et~al.}(2008)\citenamefont {Li},
  \citenamefont {Hu}, \citenamefont {Dong}, \citenamefont {Li}, \citenamefont
  {Zheng}, \citenamefont {Chen}, \citenamefont {Luo},\ and\ \citenamefont
  {Wang}}]{LiWang2008PRL_BaK122}%
  \BibitemOpen
  \bibfield  {author} {\bibinfo {author} {\bibfnamefont {G.}~\bibnamefont
  {Li}}, \bibinfo {author} {\bibfnamefont {W.~Z.}\ \bibnamefont {Hu}}, \bibinfo
  {author} {\bibfnamefont {J.}~\bibnamefont {Dong}}, \bibinfo {author}
  {\bibfnamefont {Z.}~\bibnamefont {Li}}, \bibinfo {author} {\bibfnamefont
  {P.}~\bibnamefont {Zheng}}, \bibinfo {author} {\bibfnamefont {G.~F.}\
  \bibnamefont {Chen}}, \bibinfo {author} {\bibfnamefont {J.~L.}\ \bibnamefont
  {Luo}}, \ and\ \bibinfo {author} {\bibfnamefont {N.~L.}\ \bibnamefont
  {Wang}},\ }\href {\doibase 10.1103/PhysRevLett.101.107004} {\bibfield
  {journal} {\bibinfo  {journal} {Phys. Rev. Lett.}\ }\textbf {\bibinfo
  {volume} {101}},\ \bibinfo {pages} {107004} (\bibinfo {year}
  {2008})}\BibitemShut {NoStop}%
\bibitem [{\citenamefont {Luo}\ \emph {et~al.}(2009)\citenamefont {Luo},
  \citenamefont {Tanatar}, \citenamefont {Reid}, \citenamefont {Shakeripour},
  \citenamefont {Doiron-Leyraud}, \citenamefont {Ni}, \citenamefont {Bud'ko},
  \citenamefont {Canfield}, \citenamefont {Luo}, \citenamefont {Wang},
  \citenamefont {Wen}, \citenamefont {Prozorov},\ and\ \citenamefont
  {Taillefer}}]{LuoTaillefer2009PRB}%
  \BibitemOpen
  \bibfield  {author} {\bibinfo {author} {\bibfnamefont {X.~G.}\ \bibnamefont
  {Luo}}, \bibinfo {author} {\bibfnamefont {M.~A.}\ \bibnamefont {Tanatar}},
  \bibinfo {author} {\bibfnamefont {J.-P.}\ \bibnamefont {Reid}}, \bibinfo
  {author} {\bibfnamefont {H.}~\bibnamefont {Shakeripour}}, \bibinfo {author}
  {\bibfnamefont {N.}~\bibnamefont {Doiron-Leyraud}}, \bibinfo {author}
  {\bibfnamefont {N.}~\bibnamefont {Ni}}, \bibinfo {author} {\bibfnamefont
  {S.~L.}\ \bibnamefont {Bud'ko}}, \bibinfo {author} {\bibfnamefont {P.~C.}\
  \bibnamefont {Canfield}}, \bibinfo {author} {\bibfnamefont {H.}~\bibnamefont
  {Luo}}, \bibinfo {author} {\bibfnamefont {Z.}~\bibnamefont {Wang}}, \bibinfo
  {author} {\bibfnamefont {H.-H.}\ \bibnamefont {Wen}}, \bibinfo {author}
  {\bibfnamefont {R.}~\bibnamefont {Prozorov}}, \ and\ \bibinfo {author}
  {\bibfnamefont {L.}~\bibnamefont {Taillefer}},\ }\href {\doibase
  10.1103/PhysRevB.80.140503} {\bibfield  {journal} {\bibinfo  {journal} {Phys.
  Rev. B}\ }\textbf {\bibinfo {volume} {80}},\ \bibinfo {pages} {140503}
  (\bibinfo {year} {2009})}\BibitemShut {NoStop}%
\bibitem [{\citenamefont {Ding}\ \emph {et~al.}(2008)\citenamefont {Ding},
  \citenamefont {Richard}, \citenamefont {Nakayama}, \citenamefont {Sugawara},
  \citenamefont {Arakane}, \citenamefont {Sekiba}, \citenamefont {Takayama},
  \citenamefont {Souma}, \citenamefont {Sato}, \citenamefont {Takahashi},
  \citenamefont {Wang}, \citenamefont {Dai}, \citenamefont {Fang},
  \citenamefont {Chen}, \citenamefont {Luo},\ and\ \citenamefont
  {Wang}}]{Ding2008EPL}%
  \BibitemOpen
  \bibfield  {author} {\bibinfo {author} {\bibfnamefont {H.}~\bibnamefont
  {Ding}}, \bibinfo {author} {\bibfnamefont {P.}~\bibnamefont {Richard}},
  \bibinfo {author} {\bibfnamefont {K.}~\bibnamefont {Nakayama}}, \bibinfo
  {author} {\bibfnamefont {K.}~\bibnamefont {Sugawara}}, \bibinfo {author}
  {\bibfnamefont {T.}~\bibnamefont {Arakane}}, \bibinfo {author} {\bibfnamefont
  {Y.}~\bibnamefont {Sekiba}}, \bibinfo {author} {\bibfnamefont
  {A.}~\bibnamefont {Takayama}}, \bibinfo {author} {\bibfnamefont
  {S.}~\bibnamefont {Souma}}, \bibinfo {author} {\bibfnamefont
  {T.}~\bibnamefont {Sato}}, \bibinfo {author} {\bibfnamefont {T.}~\bibnamefont
  {Takahashi}}, \bibinfo {author} {\bibfnamefont {Z.}~\bibnamefont {Wang}},
  \bibinfo {author} {\bibfnamefont {X.}~\bibnamefont {Dai}}, \bibinfo {author}
  {\bibfnamefont {Z.}~\bibnamefont {Fang}}, \bibinfo {author} {\bibfnamefont
  {G.~F.}\ \bibnamefont {Chen}}, \bibinfo {author} {\bibfnamefont {J.~L.}\
  \bibnamefont {Luo}}, \ and\ \bibinfo {author} {\bibfnamefont {N.~L.}\
  \bibnamefont {Wang}},\ }\href
  {http://stacks.iop.org/0295-5075/83/i=4/a=47001} {\bibfield  {journal}
  {\bibinfo  {journal} {EPL (Europhysics Letters)}\ }\textbf {\bibinfo {volume}
  {83}},\ \bibinfo {pages} {47001} (\bibinfo {year} {2008})}\BibitemShut
  {NoStop}%
\bibitem [{\citenamefont {Nakayama}\ \emph {et~al.}(2011)\citenamefont
  {Nakayama}, \citenamefont {Sato}, \citenamefont {Richard}, \citenamefont
  {Xu}, \citenamefont {Kawahara}, \citenamefont {Umezawa}, \citenamefont
  {Qian}, \citenamefont {Neupane}, \citenamefont {Chen}, \citenamefont {Ding},\
  and\ \citenamefont {Takahashi}}]{Nakayama2011PRB}%
  \BibitemOpen
  \bibfield  {author} {\bibinfo {author} {\bibfnamefont {K.}~\bibnamefont
  {Nakayama}}, \bibinfo {author} {\bibfnamefont {T.}~\bibnamefont {Sato}},
  \bibinfo {author} {\bibfnamefont {P.}~\bibnamefont {Richard}}, \bibinfo
  {author} {\bibfnamefont {Y.-M.}\ \bibnamefont {Xu}}, \bibinfo {author}
  {\bibfnamefont {T.}~\bibnamefont {Kawahara}}, \bibinfo {author}
  {\bibfnamefont {K.}~\bibnamefont {Umezawa}}, \bibinfo {author} {\bibfnamefont
  {T.}~\bibnamefont {Qian}}, \bibinfo {author} {\bibfnamefont {M.}~\bibnamefont
  {Neupane}}, \bibinfo {author} {\bibfnamefont {G.~F.}\ \bibnamefont {Chen}},
  \bibinfo {author} {\bibfnamefont {H.}~\bibnamefont {Ding}}, \ and\ \bibinfo
  {author} {\bibfnamefont {T.}~\bibnamefont {Takahashi}},\ }\href {\doibase
  10.1103/PhysRevB.83.020501} {\bibfield  {journal} {\bibinfo  {journal} {Phys.
  Rev. B}\ }\textbf {\bibinfo {volume} {83}},\ \bibinfo {pages} {020501}
  (\bibinfo {year} {2011})}\BibitemShut {NoStop}%
\bibitem [{\citenamefont {Ota}\ \emph {et~al.}(2014)\citenamefont {Ota},
  \citenamefont {Okazaki}, \citenamefont {Kotani}, \citenamefont {Shimojima},
  \citenamefont {Malaeb}, \citenamefont {Watanabe}, \citenamefont {Chen},
  \citenamefont {Kihou}, \citenamefont {Lee}, \citenamefont {Iyo},
  \citenamefont {Eisaki}, \citenamefont {Saito}, \citenamefont {Fukazawa},
  \citenamefont {Kohori},\ and\ \citenamefont
  {Shin}}]{OtaOkazaki2014PRB_ARPES_BaK122}%
  \BibitemOpen
  \bibfield  {author} {\bibinfo {author} {\bibfnamefont {Y.}~\bibnamefont
  {Ota}}, \bibinfo {author} {\bibfnamefont {K.}~\bibnamefont {Okazaki}},
  \bibinfo {author} {\bibfnamefont {Y.}~\bibnamefont {Kotani}}, \bibinfo
  {author} {\bibfnamefont {T.}~\bibnamefont {Shimojima}}, \bibinfo {author}
  {\bibfnamefont {W.}~\bibnamefont {Malaeb}}, \bibinfo {author} {\bibfnamefont
  {S.}~\bibnamefont {Watanabe}}, \bibinfo {author} {\bibfnamefont {C.-T.}\
  \bibnamefont {Chen}}, \bibinfo {author} {\bibfnamefont {K.}~\bibnamefont
  {Kihou}}, \bibinfo {author} {\bibfnamefont {C.~H.}\ \bibnamefont {Lee}},
  \bibinfo {author} {\bibfnamefont {A.}~\bibnamefont {Iyo}}, \bibinfo {author}
  {\bibfnamefont {H.}~\bibnamefont {Eisaki}}, \bibinfo {author} {\bibfnamefont
  {T.}~\bibnamefont {Saito}}, \bibinfo {author} {\bibfnamefont
  {H.}~\bibnamefont {Fukazawa}}, \bibinfo {author} {\bibfnamefont
  {Y.}~\bibnamefont {Kohori}}, \ and\ \bibinfo {author} {\bibfnamefont
  {S.}~\bibnamefont {Shin}},\ }\href {\doibase 10.1103/PhysRevB.89.081103}
  {\bibfield  {journal} {\bibinfo  {journal} {Phys. Rev. B}\ }\textbf {\bibinfo
  {volume} {89}},\ \bibinfo {pages} {081103} (\bibinfo {year}
  {2014})}\BibitemShut {NoStop}%
\bibitem [{\citenamefont {Reid}\ \emph
  {et~al.}(2012{\natexlab{a}})\citenamefont {Reid}, \citenamefont {Tanatar},
  \citenamefont {Juneau-Fecteau}, \citenamefont {Gordon}, \citenamefont
  {de~Cotret}, \citenamefont {Doiron-Leyraud}, \citenamefont {Saito},
  \citenamefont {Fukazawa}, \citenamefont {Kohori}, \citenamefont {Kihou},
  \citenamefont {Lee}, \citenamefont {Iyo}, \citenamefont {Eisaki},
  \citenamefont {Prozorov},\ and\ \citenamefont
  {Taillefer}}]{ReidTaillefer2012PRL_KFe2As2_d-wave}%
  \BibitemOpen
  \bibfield  {author} {\bibinfo {author} {\bibfnamefont {J.-P.}\ \bibnamefont
  {Reid}}, \bibinfo {author} {\bibfnamefont {M.~A.}\ \bibnamefont {Tanatar}},
  \bibinfo {author} {\bibfnamefont {A.}~\bibnamefont {Juneau-Fecteau}},
  \bibinfo {author} {\bibfnamefont {R.~T.}\ \bibnamefont {Gordon}}, \bibinfo
  {author} {\bibfnamefont {S.~R.}\ \bibnamefont {de~Cotret}}, \bibinfo {author}
  {\bibfnamefont {N.}~\bibnamefont {Doiron-Leyraud}}, \bibinfo {author}
  {\bibfnamefont {T.}~\bibnamefont {Saito}}, \bibinfo {author} {\bibfnamefont
  {H.}~\bibnamefont {Fukazawa}}, \bibinfo {author} {\bibfnamefont
  {Y.}~\bibnamefont {Kohori}}, \bibinfo {author} {\bibfnamefont
  {K.}~\bibnamefont {Kihou}}, \bibinfo {author} {\bibfnamefont {C.~H.}\
  \bibnamefont {Lee}}, \bibinfo {author} {\bibfnamefont {A.}~\bibnamefont
  {Iyo}}, \bibinfo {author} {\bibfnamefont {H.}~\bibnamefont {Eisaki}},
  \bibinfo {author} {\bibfnamefont {R.}~\bibnamefont {Prozorov}}, \ and\
  \bibinfo {author} {\bibfnamefont {L.}~\bibnamefont {Taillefer}},\ }\href
  {\doibase 10.1103/PhysRevLett.109.087001} {\bibfield  {journal} {\bibinfo
  {journal} {Phys. Rev. Lett.}\ }\textbf {\bibinfo {volume} {109}},\ \bibinfo
  {pages} {087001} (\bibinfo {year} {2012}{\natexlab{a}})}\BibitemShut
  {NoStop}%
\bibitem [{\citenamefont {Reid}\ \emph
  {et~al.}(2012{\natexlab{b}})\citenamefont {Reid}, \citenamefont
  {Juneau-Fecteau}, \citenamefont {Gordon}, \citenamefont {de~Cotret},
  \citenamefont {Doiron-Leyraud}, \citenamefont {Luo}, \citenamefont
  {Shakeripour}, \citenamefont {Chang}, \citenamefont {Tanatar}, \citenamefont
  {Kim}, \citenamefont {Prozorov}, \citenamefont {Saito}, \citenamefont
  {Fukazawa}, \citenamefont {Kohori}, \citenamefont {Kihou}, \citenamefont
  {Lee}, \citenamefont {Iyo}, \citenamefont {Eisaki}, \citenamefont {Shen},
  \citenamefont {Wen},\ and\ \citenamefont
  {Taillefer}}]{ReidTanatarProzorovTaillefer2012SST_KFe2As2}%
  \BibitemOpen
  \bibfield  {author} {\bibinfo {author} {\bibfnamefont {J.-P.}\ \bibnamefont
  {Reid}}, \bibinfo {author} {\bibfnamefont {A.}~\bibnamefont
  {Juneau-Fecteau}}, \bibinfo {author} {\bibfnamefont {R.~T.}\ \bibnamefont
  {Gordon}}, \bibinfo {author} {\bibfnamefont {S.~R.}\ \bibnamefont
  {de~Cotret}}, \bibinfo {author} {\bibfnamefont {N.}~\bibnamefont
  {Doiron-Leyraud}}, \bibinfo {author} {\bibfnamefont {X.~G.}\ \bibnamefont
  {Luo}}, \bibinfo {author} {\bibfnamefont {H.}~\bibnamefont {Shakeripour}},
  \bibinfo {author} {\bibfnamefont {J.}~\bibnamefont {Chang}}, \bibinfo
  {author} {\bibfnamefont {M.~A.}\ \bibnamefont {Tanatar}}, \bibinfo {author}
  {\bibfnamefont {H.}~\bibnamefont {Kim}}, \bibinfo {author} {\bibfnamefont
  {R.}~\bibnamefont {Prozorov}}, \bibinfo {author} {\bibfnamefont
  {T.}~\bibnamefont {Saito}}, \bibinfo {author} {\bibfnamefont
  {H.}~\bibnamefont {Fukazawa}}, \bibinfo {author} {\bibfnamefont
  {Y.}~\bibnamefont {Kohori}}, \bibinfo {author} {\bibfnamefont
  {K.}~\bibnamefont {Kihou}}, \bibinfo {author} {\bibfnamefont {C.~H.}\
  \bibnamefont {Lee}}, \bibinfo {author} {\bibfnamefont {A.}~\bibnamefont
  {Iyo}}, \bibinfo {author} {\bibfnamefont {H.}~\bibnamefont {Eisaki}},
  \bibinfo {author} {\bibfnamefont {B.}~\bibnamefont {Shen}}, \bibinfo {author}
  {\bibfnamefont {H.-H.}\ \bibnamefont {Wen}}, \ and\ \bibinfo {author}
  {\bibfnamefont {L.}~\bibnamefont {Taillefer}},\ }\href
  {http://stacks.iop.org/0953-2048/25/i=8/a=084013} {\bibfield  {journal}
  {\bibinfo  {journal} {Superconductor Science and Technology}\ }\textbf
  {\bibinfo {volume} {25}},\ \bibinfo {pages} {084013} (\bibinfo {year}
  {2012}{\natexlab{b}})}\BibitemShut {NoStop}%
\bibitem [{\citenamefont {Xiao-Chen}\ \emph {et~al.}(2015)\citenamefont
  {Xiao-Chen}, \citenamefont {Ai-Feng}, \citenamefont {Zhen}, \citenamefont
  {Jian}, \citenamefont {Lan-Po}, \citenamefont {Xi-Gang}, \citenamefont
  {Xian-Hui},\ and\ \citenamefont
  {Shi-Yan}}]{Hong2015CPL_BaK122_HeatTransport}%
  \BibitemOpen
  \bibfield  {author} {\bibinfo {author} {\bibfnamefont {H.}~\bibnamefont
  {Xiao-Chen}}, \bibinfo {author} {\bibfnamefont {W.}~\bibnamefont {Ai-Feng}},
  \bibinfo {author} {\bibfnamefont {Z.}~\bibnamefont {Zhen}}, \bibinfo {author}
  {\bibfnamefont {P.}~\bibnamefont {Jian}}, \bibinfo {author} {\bibfnamefont
  {H.}~\bibnamefont {Lan-Po}}, \bibinfo {author} {\bibfnamefont
  {L.}~\bibnamefont {Xi-Gang}}, \bibinfo {author} {\bibfnamefont
  {C.}~\bibnamefont {Xian-Hui}}, \ and\ \bibinfo {author} {\bibfnamefont
  {L.}~\bibnamefont {Shi-Yan}},\ }\href
  {http://stacks.iop.org/0256-307X/32/i=12/a=127403} {\bibfield  {journal}
  {\bibinfo  {journal} {Chinese Physics Letters}\ }\textbf {\bibinfo {volume}
  {32}},\ \bibinfo {pages} {127403} (\bibinfo {year} {2015})}\BibitemShut
  {NoStop}%
\bibitem [{\citenamefont {Watanabe}\ \emph {et~al.}(2014)\citenamefont
  {Watanabe}, \citenamefont {Yamashita}, \citenamefont {Kawamoto},
  \citenamefont {Kurata}, \citenamefont {Mizukami}, \citenamefont {Ohta},
  \citenamefont {Kasahara}, \citenamefont {Yamashita}, \citenamefont {Saito},
  \citenamefont {Fukazawa}, \citenamefont {Kohori}, \citenamefont {Ishida},
  \citenamefont {Kihou}, \citenamefont {Lee}, \citenamefont {Iyo},
  \citenamefont {Eisaki}, \citenamefont {Vorontsov}, \citenamefont
  {Shibauchi},\ and\ \citenamefont {Matsuda}}]{WatanabeMatsuda2014PRB_BaK122}%
  \BibitemOpen
  \bibfield  {author} {\bibinfo {author} {\bibfnamefont {D.}~\bibnamefont
  {Watanabe}}, \bibinfo {author} {\bibfnamefont {T.}~\bibnamefont {Yamashita}},
  \bibinfo {author} {\bibfnamefont {Y.}~\bibnamefont {Kawamoto}}, \bibinfo
  {author} {\bibfnamefont {S.}~\bibnamefont {Kurata}}, \bibinfo {author}
  {\bibfnamefont {Y.}~\bibnamefont {Mizukami}}, \bibinfo {author}
  {\bibfnamefont {T.}~\bibnamefont {Ohta}}, \bibinfo {author} {\bibfnamefont
  {S.}~\bibnamefont {Kasahara}}, \bibinfo {author} {\bibfnamefont
  {M.}~\bibnamefont {Yamashita}}, \bibinfo {author} {\bibfnamefont
  {T.}~\bibnamefont {Saito}}, \bibinfo {author} {\bibfnamefont
  {H.}~\bibnamefont {Fukazawa}}, \bibinfo {author} {\bibfnamefont
  {Y.}~\bibnamefont {Kohori}}, \bibinfo {author} {\bibfnamefont
  {S.}~\bibnamefont {Ishida}}, \bibinfo {author} {\bibfnamefont
  {K.}~\bibnamefont {Kihou}}, \bibinfo {author} {\bibfnamefont {C.~H.}\
  \bibnamefont {Lee}}, \bibinfo {author} {\bibfnamefont {A.}~\bibnamefont
  {Iyo}}, \bibinfo {author} {\bibfnamefont {H.}~\bibnamefont {Eisaki}},
  \bibinfo {author} {\bibfnamefont {A.~B.}\ \bibnamefont {Vorontsov}}, \bibinfo
  {author} {\bibfnamefont {T.}~\bibnamefont {Shibauchi}}, \ and\ \bibinfo
  {author} {\bibfnamefont {Y.}~\bibnamefont {Matsuda}},\ }\href {\doibase
  10.1103/PhysRevB.89.115112} {\bibfield  {journal} {\bibinfo  {journal} {Phys.
  Rev. B}\ }\textbf {\bibinfo {volume} {89}},\ \bibinfo {pages} {115112}
  (\bibinfo {year} {2014})}\BibitemShut {NoStop}%
\bibitem [{\citenamefont {Hashimoto}\ \emph {et~al.}(2010)\citenamefont
  {Hashimoto}, \citenamefont {Serafin}, \citenamefont {Tonegawa}, \citenamefont
  {Katsumata}, \citenamefont {Okazaki}, \citenamefont {Saito}, \citenamefont
  {Fukazawa}, \citenamefont {Kohori}, \citenamefont {Kihou}, \citenamefont
  {Lee}, \citenamefont {Iyo}, \citenamefont {Eisaki}, \citenamefont {Ikeda},
  \citenamefont {Matsuda}, \citenamefont {Carrington},\ and\ \citenamefont
  {Shibauchi}}]{Hashimoto2010PRB_KFe2As2}%
  \BibitemOpen
  \bibfield  {author} {\bibinfo {author} {\bibfnamefont {K.}~\bibnamefont
  {Hashimoto}}, \bibinfo {author} {\bibfnamefont {A.}~\bibnamefont {Serafin}},
  \bibinfo {author} {\bibfnamefont {S.}~\bibnamefont {Tonegawa}}, \bibinfo
  {author} {\bibfnamefont {R.}~\bibnamefont {Katsumata}}, \bibinfo {author}
  {\bibfnamefont {R.}~\bibnamefont {Okazaki}}, \bibinfo {author} {\bibfnamefont
  {T.}~\bibnamefont {Saito}}, \bibinfo {author} {\bibfnamefont
  {H.}~\bibnamefont {Fukazawa}}, \bibinfo {author} {\bibfnamefont
  {Y.}~\bibnamefont {Kohori}}, \bibinfo {author} {\bibfnamefont
  {K.}~\bibnamefont {Kihou}}, \bibinfo {author} {\bibfnamefont {C.~H.}\
  \bibnamefont {Lee}}, \bibinfo {author} {\bibfnamefont {A.}~\bibnamefont
  {Iyo}}, \bibinfo {author} {\bibfnamefont {H.}~\bibnamefont {Eisaki}},
  \bibinfo {author} {\bibfnamefont {H.}~\bibnamefont {Ikeda}}, \bibinfo
  {author} {\bibfnamefont {Y.}~\bibnamefont {Matsuda}}, \bibinfo {author}
  {\bibfnamefont {A.}~\bibnamefont {Carrington}}, \ and\ \bibinfo {author}
  {\bibfnamefont {T.}~\bibnamefont {Shibauchi}},\ }\href {\doibase
  10.1103/PhysRevB.82.014526} {\bibfield  {journal} {\bibinfo  {journal} {Phys.
  Rev. B}\ }\textbf {\bibinfo {volume} {82}},\ \bibinfo {pages} {014526}
  (\bibinfo {year} {2010})}\BibitemShut {NoStop}%
\bibitem [{\citenamefont {Maiti}\ \emph {et~al.}(2015)\citenamefont {Maiti},
  \citenamefont {Sigrist},\ and\ \citenamefont {Chubukov}}]{Maiti2015PRB_s+is}%
  \BibitemOpen
  \bibfield  {author} {\bibinfo {author} {\bibfnamefont {S.}~\bibnamefont
  {Maiti}}, \bibinfo {author} {\bibfnamefont {M.}~\bibnamefont {Sigrist}}, \
  and\ \bibinfo {author} {\bibfnamefont {A.}~\bibnamefont {Chubukov}},\ }\href
  {\doibase 10.1103/PhysRevB.91.161102} {\bibfield  {journal} {\bibinfo
  {journal} {Phys. Rev. B}\ }\textbf {\bibinfo {volume} {91}},\ \bibinfo
  {pages} {161102} (\bibinfo {year} {2015})}\BibitemShut {NoStop}%
\bibitem [{\citenamefont {Thomale}\ \emph {et~al.}(2011)\citenamefont
  {Thomale}, \citenamefont {Platt}, \citenamefont {Hanke}, \citenamefont {Hu},\
  and\ \citenamefont {Bernevig}}]{Thomale2011PRL}%
  \BibitemOpen
  \bibfield  {author} {\bibinfo {author} {\bibfnamefont {R.}~\bibnamefont
  {Thomale}}, \bibinfo {author} {\bibfnamefont {C.}~\bibnamefont {Platt}},
  \bibinfo {author} {\bibfnamefont {W.}~\bibnamefont {Hanke}}, \bibinfo
  {author} {\bibfnamefont {J.}~\bibnamefont {Hu}}, \ and\ \bibinfo {author}
  {\bibfnamefont {B.~A.}\ \bibnamefont {Bernevig}},\ }\href {\doibase
  10.1103/PhysRevLett.107.117001} {\bibfield  {journal} {\bibinfo  {journal}
  {Phys. Rev. Lett.}\ }\textbf {\bibinfo {volume} {107}},\ \bibinfo {pages}
  {117001} (\bibinfo {year} {2011})}\BibitemShut {NoStop}%
\bibitem [{\citenamefont {Platt}\ \emph {et~al.}(2012)\citenamefont {Platt},
  \citenamefont {Thomale}, \citenamefont {Honerkamp}, \citenamefont {Zhang},\
  and\ \citenamefont {Hanke}}]{Platt2012PRB_IBS}%
  \BibitemOpen
  \bibfield  {author} {\bibinfo {author} {\bibfnamefont {C.}~\bibnamefont
  {Platt}}, \bibinfo {author} {\bibfnamefont {R.}~\bibnamefont {Thomale}},
  \bibinfo {author} {\bibfnamefont {C.}~\bibnamefont {Honerkamp}}, \bibinfo
  {author} {\bibfnamefont {S.-C.}\ \bibnamefont {Zhang}}, \ and\ \bibinfo
  {author} {\bibfnamefont {W.}~\bibnamefont {Hanke}},\ }\href {\doibase
  10.1103/PhysRevB.85.180502} {\bibfield  {journal} {\bibinfo  {journal} {Phys.
  Rev. B}\ }\textbf {\bibinfo {volume} {85}},\ \bibinfo {pages} {180502}
  (\bibinfo {year} {2012})}\BibitemShut {NoStop}%
\bibitem [{\citenamefont {Fernandes}\ and\ \citenamefont
  {Millis}(2013)}]{FernandesMillis2013PRL}%
  \BibitemOpen
  \bibfield  {author} {\bibinfo {author} {\bibfnamefont {R.~M.}\ \bibnamefont
  {Fernandes}}\ and\ \bibinfo {author} {\bibfnamefont {A.~J.}\ \bibnamefont
  {Millis}},\ }\href {\doibase 10.1103/PhysRevLett.111.127001} {\bibfield
  {journal} {\bibinfo  {journal} {Phys. Rev. Lett.}\ }\textbf {\bibinfo
  {volume} {111}},\ \bibinfo {pages} {127001} (\bibinfo {year}
  {2013})}\BibitemShut {NoStop}%
\bibitem [{\citenamefont {Hardy}\ \emph {et~al.}(2014)\citenamefont {Hardy},
  \citenamefont {Eder}, \citenamefont {Jackson}, \citenamefont {Aoki},
  \citenamefont {Paulsen}, \citenamefont {Wolf}, \citenamefont {Burger},
  \citenamefont {Bohmer}, \citenamefont {Schweiss}, \citenamefont {Adelmann},
  \citenamefont {Fisher},\ and\ \citenamefont {Meingast}}]{Hardy2014JPSJ}%
  \BibitemOpen
  \bibfield  {author} {\bibinfo {author} {\bibfnamefont {F.}~\bibnamefont
  {Hardy}}, \bibinfo {author} {\bibfnamefont {R.}~\bibnamefont {Eder}},
  \bibinfo {author} {\bibfnamefont {M.}~\bibnamefont {Jackson}}, \bibinfo
  {author} {\bibfnamefont {D.}~\bibnamefont {Aoki}}, \bibinfo {author}
  {\bibfnamefont {C.}~\bibnamefont {Paulsen}}, \bibinfo {author} {\bibfnamefont
  {T.}~\bibnamefont {Wolf}}, \bibinfo {author} {\bibfnamefont {P.}~\bibnamefont
  {Burger}}, \bibinfo {author} {\bibfnamefont {A.}~\bibnamefont {Bohmer}},
  \bibinfo {author} {\bibfnamefont {P.}~\bibnamefont {Schweiss}}, \bibinfo
  {author} {\bibfnamefont {P.}~\bibnamefont {Adelmann}}, \bibinfo {author}
  {\bibfnamefont {R.~A.}\ \bibnamefont {Fisher}}, \ and\ \bibinfo {author}
  {\bibfnamefont {C.}~\bibnamefont {Meingast}},\ }\href {\doibase
  10.7566/JPSJ.83.014711} {\bibfield  {journal} {\bibinfo  {journal} {Journal
  of the Physical Society of Japan}\ }\textbf {\bibinfo {volume} {83}},\
  \bibinfo {pages} {014711} (\bibinfo {year} {2014})}\BibitemShut {NoStop}%
\bibitem [{\citenamefont {Hardy}\ \emph {et~al.}(2016)\citenamefont {Hardy},
  \citenamefont {B\"ohmer}, \citenamefont {de' Medici}, \citenamefont {Capone},
  \citenamefont {Giovannetti}, \citenamefont {Eder}, \citenamefont {Wang},
  \citenamefont {He}, \citenamefont {Wolf}, \citenamefont {Schweiss},
  \citenamefont {Heid}, \citenamefont {Herbig}, \citenamefont {Adelmann},
  \citenamefont {Fisher},\ and\ \citenamefont
  {Meingast}}]{Hardy2016PRB_BaK122}%
  \BibitemOpen
  \bibfield  {author} {\bibinfo {author} {\bibfnamefont {F.}~\bibnamefont
  {Hardy}}, \bibinfo {author} {\bibfnamefont {A.~E.}\ \bibnamefont {B\"ohmer}},
  \bibinfo {author} {\bibfnamefont {L.}~\bibnamefont {de' Medici}}, \bibinfo
  {author} {\bibfnamefont {M.}~\bibnamefont {Capone}}, \bibinfo {author}
  {\bibfnamefont {G.}~\bibnamefont {Giovannetti}}, \bibinfo {author}
  {\bibfnamefont {R.}~\bibnamefont {Eder}}, \bibinfo {author} {\bibfnamefont
  {L.}~\bibnamefont {Wang}}, \bibinfo {author} {\bibfnamefont {M.}~\bibnamefont
  {He}}, \bibinfo {author} {\bibfnamefont {T.}~\bibnamefont {Wolf}}, \bibinfo
  {author} {\bibfnamefont {P.}~\bibnamefont {Schweiss}}, \bibinfo {author}
  {\bibfnamefont {R.}~\bibnamefont {Heid}}, \bibinfo {author} {\bibfnamefont
  {A.}~\bibnamefont {Herbig}}, \bibinfo {author} {\bibfnamefont
  {P.}~\bibnamefont {Adelmann}}, \bibinfo {author} {\bibfnamefont {R.~A.}\
  \bibnamefont {Fisher}}, \ and\ \bibinfo {author} {\bibfnamefont
  {C.}~\bibnamefont {Meingast}},\ }\href {\doibase 10.1103/PhysRevB.94.205113}
  {\bibfield  {journal} {\bibinfo  {journal} {Phys. Rev. B}\ }\textbf {\bibinfo
  {volume} {94}},\ \bibinfo {pages} {205113} (\bibinfo {year}
  {2016})}\BibitemShut {NoStop}%
\bibitem [{\citenamefont {Kogan}(2009)}]{Kogan2009PRB-2}%
  \BibitemOpen
  \bibfield  {author} {\bibinfo {author} {\bibfnamefont {V.~G.}\ \bibnamefont
  {Kogan}},\ }\href {\doibase 10.1103/PhysRevB.80.214532} {\bibfield  {journal}
  {\bibinfo  {journal} {Phys. Rev. B}\ }\textbf {\bibinfo {volume} {80}},\
  \bibinfo {pages} {214532} (\bibinfo {year} {2009})}\BibitemShut {NoStop}%
\bibitem [{\citenamefont {Kim}\ \emph {et~al.}(2014)\citenamefont {Kim},
  \citenamefont {Tanatar}, \citenamefont {Straszheim}, \citenamefont {Cho},
  \citenamefont {Murphy}, \citenamefont {Spyrison}, \citenamefont {Reid},
  \citenamefont {Shen}, \citenamefont {Wen}, \citenamefont {Fernandes},\ and\
  \citenamefont {Prozorov}}]{KimProzorov2014PRB_underdoped_BaK122}%
  \BibitemOpen
  \bibfield  {author} {\bibinfo {author} {\bibfnamefont {H.}~\bibnamefont
  {Kim}}, \bibinfo {author} {\bibfnamefont {M.~A.}\ \bibnamefont {Tanatar}},
  \bibinfo {author} {\bibfnamefont {W.~E.}\ \bibnamefont {Straszheim}},
  \bibinfo {author} {\bibfnamefont {K.}~\bibnamefont {Cho}}, \bibinfo {author}
  {\bibfnamefont {J.}~\bibnamefont {Murphy}}, \bibinfo {author} {\bibfnamefont
  {N.}~\bibnamefont {Spyrison}}, \bibinfo {author} {\bibfnamefont {J.-P.}\
  \bibnamefont {Reid}}, \bibinfo {author} {\bibfnamefont {B.}~\bibnamefont
  {Shen}}, \bibinfo {author} {\bibfnamefont {H.-H.}\ \bibnamefont {Wen}},
  \bibinfo {author} {\bibfnamefont {R.~M.}\ \bibnamefont {Fernandes}}, \ and\
  \bibinfo {author} {\bibfnamefont {R.}~\bibnamefont {Prozorov}},\ }\href
  {\doibase 10.1103/PhysRevB.90.014517} {\bibfield  {journal} {\bibinfo
  {journal} {Phys. Rev. B}\ }\textbf {\bibinfo {volume} {90}},\ \bibinfo
  {pages} {014517} (\bibinfo {year} {2014})}\BibitemShut {NoStop}%
\bibitem [{\citenamefont {Reid}\ \emph {et~al.}(2016)\citenamefont {Reid},
  \citenamefont {Tanatar}, \citenamefont {Luo}, \citenamefont {Shakeripour},
  \citenamefont {de~Cotret}, \citenamefont {Juneau-Fecteau}, \citenamefont
  {Chang}, \citenamefont {Shen}, \citenamefont {Wen}, \citenamefont {Kim},
  \citenamefont {Prozorov}, \citenamefont {Doiron-Leyraud},\ and\ \citenamefont
  {Taillefer}}]{ReidTanatarProzorov2016PRB_Underdoped_BaK122}%
  \BibitemOpen
  \bibfield  {author} {\bibinfo {author} {\bibfnamefont {J.-P.}\ \bibnamefont
  {Reid}}, \bibinfo {author} {\bibfnamefont {M.~A.}\ \bibnamefont {Tanatar}},
  \bibinfo {author} {\bibfnamefont {X.~G.}\ \bibnamefont {Luo}}, \bibinfo
  {author} {\bibfnamefont {H.}~\bibnamefont {Shakeripour}}, \bibinfo {author}
  {\bibfnamefont {S.~R.}\ \bibnamefont {de~Cotret}}, \bibinfo {author}
  {\bibfnamefont {A.}~\bibnamefont {Juneau-Fecteau}}, \bibinfo {author}
  {\bibfnamefont {J.}~\bibnamefont {Chang}}, \bibinfo {author} {\bibfnamefont
  {B.}~\bibnamefont {Shen}}, \bibinfo {author} {\bibfnamefont {H.-H.}\
  \bibnamefont {Wen}}, \bibinfo {author} {\bibfnamefont {H.}~\bibnamefont
  {Kim}}, \bibinfo {author} {\bibfnamefont {R.}~\bibnamefont {Prozorov}},
  \bibinfo {author} {\bibfnamefont {N.}~\bibnamefont {Doiron-Leyraud}}, \ and\
  \bibinfo {author} {\bibfnamefont {L.}~\bibnamefont {Taillefer}},\ }\href@noop
  {} {\bibfield  {journal} {\bibinfo  {journal} {Phys. Rev. B}\ }\textbf
  {\bibinfo {volume} {93}},\ \bibinfo {pages} {214519} (\bibinfo {year}
  {2016})}\BibitemShut {NoStop}%
\bibitem [{\citenamefont {Mizukami}\ \emph {et~al.}(2017)\citenamefont
  {Mizukami}, \citenamefont {Konczykowski}, \citenamefont {Matsuura},
  \citenamefont {Watashige}, \citenamefont {Kasahara}, \citenamefont
  {Matsuda},\ and\ \citenamefont {Shibauchi}}]{Mizukami2017JPSJ_BaP122_e-irr}%
  \BibitemOpen
  \bibfield  {author} {\bibinfo {author} {\bibfnamefont {Y.}~\bibnamefont
  {Mizukami}}, \bibinfo {author} {\bibfnamefont {M.}~\bibnamefont
  {Konczykowski}}, \bibinfo {author} {\bibfnamefont {K.}~\bibnamefont
  {Matsuura}}, \bibinfo {author} {\bibfnamefont {T.}~\bibnamefont {Watashige}},
  \bibinfo {author} {\bibfnamefont {S.}~\bibnamefont {Kasahara}}, \bibinfo
  {author} {\bibfnamefont {Y.}~\bibnamefont {Matsuda}}, \ and\ \bibinfo
  {author} {\bibfnamefont {T.}~\bibnamefont {Shibauchi}},\ }\href {\doibase
  10.7566/JPSJ.86.083706} {\bibfield  {journal} {\bibinfo  {journal} {Journal
  of the Physical Society of Japan}\ }\textbf {\bibinfo {volume} {86}},\
  \bibinfo {pages} {083706} (\bibinfo {year} {2017})}\BibitemShut {NoStop}%
\bibitem [{\citenamefont {Lamhot}\ \emph {et~al.}(2015)\citenamefont {Lamhot},
  \citenamefont {Yagil}, \citenamefont {Shapira}, \citenamefont {Kasahara},
  \citenamefont {Watashige}, \citenamefont {Shibauchi}, \citenamefont
  {Matsuda},\ and\ \citenamefont {Auslaender}}]{Lamhot2015PRB_BaP122_L(0)}%
  \BibitemOpen
  \bibfield  {author} {\bibinfo {author} {\bibfnamefont {Y.}~\bibnamefont
  {Lamhot}}, \bibinfo {author} {\bibfnamefont {A.}~\bibnamefont {Yagil}},
  \bibinfo {author} {\bibfnamefont {N.}~\bibnamefont {Shapira}}, \bibinfo
  {author} {\bibfnamefont {S.}~\bibnamefont {Kasahara}}, \bibinfo {author}
  {\bibfnamefont {T.}~\bibnamefont {Watashige}}, \bibinfo {author}
  {\bibfnamefont {T.}~\bibnamefont {Shibauchi}}, \bibinfo {author}
  {\bibfnamefont {Y.}~\bibnamefont {Matsuda}}, \ and\ \bibinfo {author}
  {\bibfnamefont {O.~M.}\ \bibnamefont {Auslaender}},\ }\href {\doibase
  10.1103/PhysRevB.91.060504} {\bibfield  {journal} {\bibinfo  {journal} {Phys.
  Rev. B}\ }\textbf {\bibinfo {volume} {91}},\ \bibinfo {pages} {060504}
  (\bibinfo {year} {2015})}\BibitemShut {NoStop}%
\bibitem [{\citenamefont {Kasahara}\ \emph {et~al.}(2010)\citenamefont
  {Kasahara}, \citenamefont {Shibauchi}, \citenamefont {Hashimoto},
  \citenamefont {Ikada}, \citenamefont {Tonegawa}, \citenamefont {Okazaki},
  \citenamefont {Shishido}, \citenamefont {Ikeda}, \citenamefont {Takeya},
  \citenamefont {Hirata}, \citenamefont {Terashima},\ and\ \citenamefont
  {Matsuda}}]{Kasahara2010PRB_BaP122}%
  \BibitemOpen
  \bibfield  {author} {\bibinfo {author} {\bibfnamefont {S.}~\bibnamefont
  {Kasahara}}, \bibinfo {author} {\bibfnamefont {T.}~\bibnamefont {Shibauchi}},
  \bibinfo {author} {\bibfnamefont {K.}~\bibnamefont {Hashimoto}}, \bibinfo
  {author} {\bibfnamefont {K.}~\bibnamefont {Ikada}}, \bibinfo {author}
  {\bibfnamefont {S.}~\bibnamefont {Tonegawa}}, \bibinfo {author}
  {\bibfnamefont {R.}~\bibnamefont {Okazaki}}, \bibinfo {author} {\bibfnamefont
  {H.}~\bibnamefont {Shishido}}, \bibinfo {author} {\bibfnamefont
  {H.}~\bibnamefont {Ikeda}}, \bibinfo {author} {\bibfnamefont
  {H.}~\bibnamefont {Takeya}}, \bibinfo {author} {\bibfnamefont
  {K.}~\bibnamefont {Hirata}}, \bibinfo {author} {\bibfnamefont
  {T.}~\bibnamefont {Terashima}}, \ and\ \bibinfo {author} {\bibfnamefont
  {Y.}~\bibnamefont {Matsuda}},\ }\href {\doibase 10.1103/PhysRevB.81.184519}
  {\bibfield  {journal} {\bibinfo  {journal} {Phys. Rev. B}\ }\textbf {\bibinfo
  {volume} {81}},\ \bibinfo {pages} {184519} (\bibinfo {year}
  {2010})}\BibitemShut {NoStop}%
\bibitem [{\citenamefont {Rullier-Albenque}\ \emph {et~al.}(2010)\citenamefont
  {Rullier-Albenque}, \citenamefont {Colson}, \citenamefont {Forget},
  \citenamefont {Thu\'ery},\ and\ \citenamefont
  {Poissonnet}}]{Rullier-Albenque2010PRB_BaRu122}%
  \BibitemOpen
  \bibfield  {author} {\bibinfo {author} {\bibfnamefont {F.}~\bibnamefont
  {Rullier-Albenque}}, \bibinfo {author} {\bibfnamefont {D.}~\bibnamefont
  {Colson}}, \bibinfo {author} {\bibfnamefont {A.}~\bibnamefont {Forget}},
  \bibinfo {author} {\bibfnamefont {P.}~\bibnamefont {Thu\'ery}}, \ and\
  \bibinfo {author} {\bibfnamefont {S.}~\bibnamefont {Poissonnet}},\ }\href
  {\doibase 10.1103/PhysRevB.81.224503} {\bibfield  {journal} {\bibinfo
  {journal} {Phys. Rev. B}\ }\textbf {\bibinfo {volume} {81}},\ \bibinfo
  {pages} {224503} (\bibinfo {year} {2010})}\BibitemShut {NoStop}%
\bibitem [{\citenamefont {Dhaka}\ \emph {et~al.}(2013)\citenamefont {Dhaka},
  \citenamefont {Hahn}, \citenamefont {Razzoli}, \citenamefont {Jiang},
  \citenamefont {Shi}, \citenamefont {Harmon}, \citenamefont {Thaler},
  \citenamefont {Bud'ko}, \citenamefont {Canfield},\ and\ \citenamefont
  {Kaminski}}]{DhakaCanfieldKaminski2013PRL_BaRu122}%
  \BibitemOpen
  \bibfield  {author} {\bibinfo {author} {\bibfnamefont {R.~S.}\ \bibnamefont
  {Dhaka}}, \bibinfo {author} {\bibfnamefont {S.~E.}\ \bibnamefont {Hahn}},
  \bibinfo {author} {\bibfnamefont {E.}~\bibnamefont {Razzoli}}, \bibinfo
  {author} {\bibfnamefont {R.}~\bibnamefont {Jiang}}, \bibinfo {author}
  {\bibfnamefont {M.}~\bibnamefont {Shi}}, \bibinfo {author} {\bibfnamefont
  {B.~N.}\ \bibnamefont {Harmon}}, \bibinfo {author} {\bibfnamefont
  {A.}~\bibnamefont {Thaler}}, \bibinfo {author} {\bibfnamefont {S.~L.}\
  \bibnamefont {Bud'ko}}, \bibinfo {author} {\bibfnamefont {P.~C.}\
  \bibnamefont {Canfield}}, \ and\ \bibinfo {author} {\bibfnamefont
  {A.}~\bibnamefont {Kaminski}},\ }\href@noop {} {\bibfield  {journal}
  {\bibinfo  {journal} {Phys. Rev. Lett.}\ }\textbf {\bibinfo {volume} {110}},\
  \bibinfo {pages} {067002} (\bibinfo {year} {2013})}\BibitemShut {NoStop}%
\bibitem [{\citenamefont {Thaler}\ \emph {et~al.}(2010)\citenamefont {Thaler},
  \citenamefont {Ni}, \citenamefont {Kracher}, \citenamefont {Yan},
  \citenamefont {Bud'ko},\ and\ \citenamefont
  {Canfield}}]{Thaler2010PRB_BaRu122}%
  \BibitemOpen
  \bibfield  {author} {\bibinfo {author} {\bibfnamefont {A.}~\bibnamefont
  {Thaler}}, \bibinfo {author} {\bibfnamefont {N.}~\bibnamefont {Ni}}, \bibinfo
  {author} {\bibfnamefont {A.}~\bibnamefont {Kracher}}, \bibinfo {author}
  {\bibfnamefont {J.~Q.}\ \bibnamefont {Yan}}, \bibinfo {author} {\bibfnamefont
  {S.~L.}\ \bibnamefont {Bud'ko}}, \ and\ \bibinfo {author} {\bibfnamefont
  {P.~C.}\ \bibnamefont {Canfield}},\ }\href {\doibase
  10.1103/PhysRevB.82.014534} {\bibfield  {journal} {\bibinfo  {journal} {Phys.
  Rev. B}\ }\textbf {\bibinfo {volume} {82}},\ \bibinfo {pages} {014534}
  (\bibinfo {year} {2010})}\BibitemShut {NoStop}%
\bibitem [{\citenamefont {Xu}\ \emph {et~al.}(2012)\citenamefont {Xu},
  \citenamefont {Qian}, \citenamefont {Richard}, \citenamefont {Shi},
  \citenamefont {Wang}, \citenamefont {Zhang}, \citenamefont {Huang},
  \citenamefont {Xu}, \citenamefont {Miao}, \citenamefont {Xu}, \citenamefont
  {Xuan}, \citenamefont {Jiao}, \citenamefont {Xu}, \citenamefont {Cao},\ and\
  \citenamefont {Ding}}]{Xu2012PRB_BaRu122}%
  \BibitemOpen
  \bibfield  {author} {\bibinfo {author} {\bibfnamefont {N.}~\bibnamefont
  {Xu}}, \bibinfo {author} {\bibfnamefont {T.}~\bibnamefont {Qian}}, \bibinfo
  {author} {\bibfnamefont {P.}~\bibnamefont {Richard}}, \bibinfo {author}
  {\bibfnamefont {Y.-B.}\ \bibnamefont {Shi}}, \bibinfo {author} {\bibfnamefont
  {X.-P.}\ \bibnamefont {Wang}}, \bibinfo {author} {\bibfnamefont
  {P.}~\bibnamefont {Zhang}}, \bibinfo {author} {\bibfnamefont {Y.-B.}\
  \bibnamefont {Huang}}, \bibinfo {author} {\bibfnamefont {Y.-M.}\ \bibnamefont
  {Xu}}, \bibinfo {author} {\bibfnamefont {H.}~\bibnamefont {Miao}}, \bibinfo
  {author} {\bibfnamefont {G.}~\bibnamefont {Xu}}, \bibinfo {author}
  {\bibfnamefont {G.-F.}\ \bibnamefont {Xuan}}, \bibinfo {author}
  {\bibfnamefont {W.-H.}\ \bibnamefont {Jiao}}, \bibinfo {author}
  {\bibfnamefont {Z.-A.}\ \bibnamefont {Xu}}, \bibinfo {author} {\bibfnamefont
  {G.-H.}\ \bibnamefont {Cao}}, \ and\ \bibinfo {author} {\bibfnamefont
  {H.}~\bibnamefont {Ding}},\ }\href {\doibase 10.1103/PhysRevB.86.064505}
  {\bibfield  {journal} {\bibinfo  {journal} {Phys. Rev. B}\ }\textbf {\bibinfo
  {volume} {86}},\ \bibinfo {pages} {064505} (\bibinfo {year}
  {2012})}\BibitemShut {NoStop}%
\bibitem [{\citenamefont {Dulguun}\ \emph {et~al.}(2012)\citenamefont
  {Dulguun}, \citenamefont {Mukuda}, \citenamefont {Kobayashi}, \citenamefont
  {Engetsu}, \citenamefont {Kinouchi}, \citenamefont {Yashima}, \citenamefont
  {Kitaoka}, \citenamefont {Miyasaka},\ and\ \citenamefont
  {Tajima}}]{DulguunTajima2012PRB_SrP122_NMR_SpecificHeat}%
  \BibitemOpen
  \bibfield  {author} {\bibinfo {author} {\bibfnamefont {T.}~\bibnamefont
  {Dulguun}}, \bibinfo {author} {\bibfnamefont {H.}~\bibnamefont {Mukuda}},
  \bibinfo {author} {\bibfnamefont {T.}~\bibnamefont {Kobayashi}}, \bibinfo
  {author} {\bibfnamefont {F.}~\bibnamefont {Engetsu}}, \bibinfo {author}
  {\bibfnamefont {H.}~\bibnamefont {Kinouchi}}, \bibinfo {author}
  {\bibfnamefont {M.}~\bibnamefont {Yashima}}, \bibinfo {author} {\bibfnamefont
  {Y.}~\bibnamefont {Kitaoka}}, \bibinfo {author} {\bibfnamefont
  {S.}~\bibnamefont {Miyasaka}}, \ and\ \bibinfo {author} {\bibfnamefont
  {S.}~\bibnamefont {Tajima}},\ }\href {\doibase 10.1103/PhysRevB.85.144515}
  {\bibfield  {journal} {\bibinfo  {journal} {Phys. Rev. B}\ }\textbf {\bibinfo
  {volume} {85}},\ \bibinfo {pages} {144515} (\bibinfo {year}
  {2012})}\BibitemShut {NoStop}%
\bibitem [{\citenamefont {Murphy}\ \emph
  {et~al.}(2013{\natexlab{c}})\citenamefont {Murphy}, \citenamefont {Strehlow},
  \citenamefont {Cho}, \citenamefont {Tanatar}, \citenamefont {Salovich},
  \citenamefont {Giannetta}, \citenamefont {Kobayashi}, \citenamefont
  {Miyasaka}, \citenamefont {Tajima},\ and\ \citenamefont
  {Prozorov}}]{Murphy2013PRB_SrP122}%
  \BibitemOpen
  \bibfield  {author} {\bibinfo {author} {\bibfnamefont {J.}~\bibnamefont
  {Murphy}}, \bibinfo {author} {\bibfnamefont {C.~P.}\ \bibnamefont
  {Strehlow}}, \bibinfo {author} {\bibfnamefont {K.}~\bibnamefont {Cho}},
  \bibinfo {author} {\bibfnamefont {M.~A.}\ \bibnamefont {Tanatar}}, \bibinfo
  {author} {\bibfnamefont {N.}~\bibnamefont {Salovich}}, \bibinfo {author}
  {\bibfnamefont {R.~W.}\ \bibnamefont {Giannetta}}, \bibinfo {author}
  {\bibfnamefont {T.}~\bibnamefont {Kobayashi}}, \bibinfo {author}
  {\bibfnamefont {S.}~\bibnamefont {Miyasaka}}, \bibinfo {author}
  {\bibfnamefont {S.}~\bibnamefont {Tajima}}, \ and\ \bibinfo {author}
  {\bibfnamefont {R.}~\bibnamefont {Prozorov}},\ }\href@noop {} {\bibfield
  {journal} {\bibinfo  {journal} {Phys. Rev. B}\ }\textbf {\bibinfo {volume}
  {87}},\ \bibinfo {pages} {140505} (\bibinfo {year}
  {2013}{\natexlab{c}})}\BibitemShut {NoStop}%
\bibitem [{\citenamefont {van~der Beek}\ \emph {et~al.}(2013)\citenamefont
  {van~der Beek}, \citenamefont {Demirdis}, \citenamefont {Colson},
  \citenamefont {Rullier-Albenque}, \citenamefont {Fasano}, \citenamefont
  {Shibauchi}, \citenamefont {Matsuda}, \citenamefont {Kasahara}, \citenamefont
  {Gierlowski},\ and\ \citenamefont
  {Konczykowski}}]{VanDerBeek2013JPCS_e-irr_review}%
  \BibitemOpen
  \bibfield  {author} {\bibinfo {author} {\bibfnamefont {C.~J.}\ \bibnamefont
  {van~der Beek}}, \bibinfo {author} {\bibfnamefont {S.}~\bibnamefont
  {Demirdis}}, \bibinfo {author} {\bibfnamefont {D.}~\bibnamefont {Colson}},
  \bibinfo {author} {\bibfnamefont {F.}~\bibnamefont {Rullier-Albenque}},
  \bibinfo {author} {\bibfnamefont {Y.}~\bibnamefont {Fasano}}, \bibinfo
  {author} {\bibfnamefont {T.}~\bibnamefont {Shibauchi}}, \bibinfo {author}
  {\bibfnamefont {Y.}~\bibnamefont {Matsuda}}, \bibinfo {author} {\bibfnamefont
  {S.}~\bibnamefont {Kasahara}}, \bibinfo {author} {\bibfnamefont
  {P.}~\bibnamefont {Gierlowski}}, \ and\ \bibinfo {author} {\bibfnamefont
  {M.}~\bibnamefont {Konczykowski}},\ }\href
  {http://stacks.iop.org/1742-6596/449/i=1/a=012023} {\bibfield  {journal}
  {\bibinfo  {journal} {Journal of Physics: Conference Series}\ }\textbf
  {\bibinfo {volume} {449}},\ \bibinfo {pages} {012023} (\bibinfo {year}
  {2013})}\BibitemShut {NoStop}%
\bibitem [{\citenamefont {Tanatar}\ \emph
  {et~al.}(2010{\natexlab{b}})\citenamefont {Tanatar}, \citenamefont {Reid},
  \citenamefont {Shakeripour}, \citenamefont {Luo}, \citenamefont
  {Doiron-Leyraud}, \citenamefont {Ni}, \citenamefont {Bud'ko}, \citenamefont
  {Canfield}, \citenamefont {Prozorov},\ and\ \citenamefont
  {Taillefer}}]{TanatarProzorov2010PRL_BaCo122}%
  \BibitemOpen
  \bibfield  {author} {\bibinfo {author} {\bibfnamefont {M.~A.}\ \bibnamefont
  {Tanatar}}, \bibinfo {author} {\bibfnamefont {J.-P.}\ \bibnamefont {Reid}},
  \bibinfo {author} {\bibfnamefont {H.}~\bibnamefont {Shakeripour}}, \bibinfo
  {author} {\bibfnamefont {X.~G.}\ \bibnamefont {Luo}}, \bibinfo {author}
  {\bibfnamefont {N.}~\bibnamefont {Doiron-Leyraud}}, \bibinfo {author}
  {\bibfnamefont {N.}~\bibnamefont {Ni}}, \bibinfo {author} {\bibfnamefont
  {S.~L.}\ \bibnamefont {Bud'ko}}, \bibinfo {author} {\bibfnamefont {P.~C.}\
  \bibnamefont {Canfield}}, \bibinfo {author} {\bibfnamefont {R.}~\bibnamefont
  {Prozorov}}, \ and\ \bibinfo {author} {\bibfnamefont {L.}~\bibnamefont
  {Taillefer}},\ }\href@noop {} {\bibfield  {journal} {\bibinfo  {journal}
  {Phys. Rev. Lett.}\ }\textbf {\bibinfo {volume} {104}},\ \bibinfo {pages}
  {067002} (\bibinfo {year} {2010}{\natexlab{b}})}\BibitemShut {NoStop}%
\bibitem [{\citenamefont {Reid}\ \emph {et~al.}(2010)\citenamefont {Reid},
  \citenamefont {Tanatar}, \citenamefont {Luo}, \citenamefont {Shakeripour},
  \citenamefont {Doiron-Leyraud}, \citenamefont {Ni}, \citenamefont {Bud'ko},
  \citenamefont {Canfield}, \citenamefont {Prozorov},\ and\ \citenamefont
  {Taillefer}}]{ReidTanatarProzorovTaillefer2010PRB_BaCo122}%
  \BibitemOpen
  \bibfield  {author} {\bibinfo {author} {\bibfnamefont {J.-P.}\ \bibnamefont
  {Reid}}, \bibinfo {author} {\bibfnamefont {M.~A.}\ \bibnamefont {Tanatar}},
  \bibinfo {author} {\bibfnamefont {X.~G.}\ \bibnamefont {Luo}}, \bibinfo
  {author} {\bibfnamefont {H.}~\bibnamefont {Shakeripour}}, \bibinfo {author}
  {\bibfnamefont {N.}~\bibnamefont {Doiron-Leyraud}}, \bibinfo {author}
  {\bibfnamefont {N.}~\bibnamefont {Ni}}, \bibinfo {author} {\bibfnamefont
  {S.~L.}\ \bibnamefont {Bud'ko}}, \bibinfo {author} {\bibfnamefont {P.~C.}\
  \bibnamefont {Canfield}}, \bibinfo {author} {\bibfnamefont {R.}~\bibnamefont
  {Prozorov}}, \ and\ \bibinfo {author} {\bibfnamefont {L.}~\bibnamefont
  {Taillefer}},\ }\href@noop {} {\bibfield  {journal} {\bibinfo  {journal}
  {Phys. Rev. B}\ }\textbf {\bibinfo {volume} {82}},\ \bibinfo {pages} {064501}
  (\bibinfo {year} {2010})}\BibitemShut {NoStop}%
\bibitem [{\citenamefont {Li}\ \emph {et~al.}(2009)\citenamefont {Li},
  \citenamefont {Luo}, \citenamefont {Wang}, \citenamefont {Chen},
  \citenamefont {Ren}, \citenamefont {Tao}, \citenamefont {Li}, \citenamefont
  {Lin}, \citenamefont {He}, \citenamefont {Zhu}, \citenamefont {Cao},\ and\
  \citenamefont {Xu}}]{LiXu2009NJP_BaNi122}%
  \BibitemOpen
  \bibfield  {author} {\bibinfo {author} {\bibfnamefont {L.~J.}\ \bibnamefont
  {Li}}, \bibinfo {author} {\bibfnamefont {Y.~K.}\ \bibnamefont {Luo}},
  \bibinfo {author} {\bibfnamefont {Q.~B.}\ \bibnamefont {Wang}}, \bibinfo
  {author} {\bibfnamefont {H.}~\bibnamefont {Chen}}, \bibinfo {author}
  {\bibfnamefont {Z.}~\bibnamefont {Ren}}, \bibinfo {author} {\bibfnamefont
  {Q.}~\bibnamefont {Tao}}, \bibinfo {author} {\bibfnamefont {Y.~K.}\
  \bibnamefont {Li}}, \bibinfo {author} {\bibfnamefont {X.}~\bibnamefont
  {Lin}}, \bibinfo {author} {\bibfnamefont {M.}~\bibnamefont {He}}, \bibinfo
  {author} {\bibfnamefont {Z.~W.}\ \bibnamefont {Zhu}}, \bibinfo {author}
  {\bibfnamefont {G.~H.}\ \bibnamefont {Cao}}, \ and\ \bibinfo {author}
  {\bibfnamefont {Z.~A.}\ \bibnamefont {Xu}},\ }\href
  {http://stacks.iop.org/1367-2630/11/i=2/a=025008} {\bibfield  {journal}
  {\bibinfo  {journal} {New Journal of Physics}\ }\textbf {\bibinfo {volume}
  {11}},\ \bibinfo {pages} {025008} (\bibinfo {year} {2009})}\BibitemShut
  {NoStop}%
\bibitem [{\citenamefont {Sefat}\ \emph {et~al.}(2008)\citenamefont {Sefat},
  \citenamefont {Jin}, \citenamefont {McGuire}, \citenamefont {Sales},
  \citenamefont {Singh},\ and\ \citenamefont {Mandrus}}]{Sefat2008PRL_BaCo122}%
  \BibitemOpen
  \bibfield  {author} {\bibinfo {author} {\bibfnamefont {A.~S.}\ \bibnamefont
  {Sefat}}, \bibinfo {author} {\bibfnamefont {R.}~\bibnamefont {Jin}}, \bibinfo
  {author} {\bibfnamefont {M.~A.}\ \bibnamefont {McGuire}}, \bibinfo {author}
  {\bibfnamefont {B.~C.}\ \bibnamefont {Sales}}, \bibinfo {author}
  {\bibfnamefont {D.~J.}\ \bibnamefont {Singh}}, \ and\ \bibinfo {author}
  {\bibfnamefont {D.}~\bibnamefont {Mandrus}},\ }\href {\doibase
  10.1103/PhysRevLett.101.117004} {\bibfield  {journal} {\bibinfo  {journal}
  {Phys. Rev. Lett.}\ }\textbf {\bibinfo {volume} {101}},\ \bibinfo {pages}
  {117004} (\bibinfo {year} {2008})}\BibitemShut {NoStop}%
\bibitem [{\citenamefont {Lu}\ \emph {et~al.}(2013)\citenamefont {Lu},
  \citenamefont {Gretarsson}, \citenamefont {Zhang}, \citenamefont {Liu},
  \citenamefont {Luo}, \citenamefont {Tian}, \citenamefont {Laver},
  \citenamefont {Yamani}, \citenamefont {Kim}, \citenamefont {Nevidomskyy},
  \citenamefont {Si},\ and\ \citenamefont {Dai}}]{LuDai2013PRL_BaNI122}%
  \BibitemOpen
  \bibfield  {author} {\bibinfo {author} {\bibfnamefont {X.}~\bibnamefont
  {Lu}}, \bibinfo {author} {\bibfnamefont {H.}~\bibnamefont {Gretarsson}},
  \bibinfo {author} {\bibfnamefont {R.}~\bibnamefont {Zhang}}, \bibinfo
  {author} {\bibfnamefont {X.}~\bibnamefont {Liu}}, \bibinfo {author}
  {\bibfnamefont {H.}~\bibnamefont {Luo}}, \bibinfo {author} {\bibfnamefont
  {W.}~\bibnamefont {Tian}}, \bibinfo {author} {\bibfnamefont {M.}~\bibnamefont
  {Laver}}, \bibinfo {author} {\bibfnamefont {Z.}~\bibnamefont {Yamani}},
  \bibinfo {author} {\bibfnamefont {Y.-J.}\ \bibnamefont {Kim}}, \bibinfo
  {author} {\bibfnamefont {A.~H.}\ \bibnamefont {Nevidomskyy}}, \bibinfo
  {author} {\bibfnamefont {Q.}~\bibnamefont {Si}}, \ and\ \bibinfo {author}
  {\bibfnamefont {P.}~\bibnamefont {Dai}},\ }\href {\doibase
  10.1103/PhysRevLett.110.257001} {\bibfield  {journal} {\bibinfo  {journal}
  {Phys. Rev. Lett.}\ }\textbf {\bibinfo {volume} {110}},\ \bibinfo {pages}
  {257001} (\bibinfo {year} {2013})}\BibitemShut {NoStop}%
\end{thebibliography}
\end{document}